\documentclass[12pt,a4paper,usenatbib]{book}

\usepackage{a4wide}
\usepackage{fancyhdr}
\usepackage{graphicx}
\usepackage[english]{babel}
\usepackage[utf8]{inputenc}
\usepackage[T1]{fontenc}
\usepackage{lmodern}
\usepackage{natbib}
\usepackage{amsmath}
\usepackage{xcolor} 
\definecolor{customblue}{HTML}{002db3}
\usepackage[colorlinks=true, linkcolor=customblue, citecolor=customblue, urlcolor=customblue]{hyperref}
\usepackage{url}
\usepackage{newtxtext,newtxmath}
\usepackage[T1]{fontenc}
\usepackage[normalem]{ulem} 
\usepackage{graphicx}	
\usepackage{amsmath}	
\usepackage{siunitx}
\usepackage{orcidlink}

\renewcommand{\chaptermark}[1]%
         {\markboth{\thechapter.\ #1}{}}
\renewcommand{\sectionmark}[1]%
         {\markright{\thesection\ #1}}
\newcommand{\getTitle}{Tracing The Start and End of Cosmic Reionization}
\newcommand{\getSubtitle}{Exploring The Role of Ionizing Sources as drivers}
\newcommand*{\getAuthor}{Arghyadeep Basu}
\newcommand*{\getPrintLocation}{München}       
\newcommand*{\getPrintYear}{\the\year}         
\newcommand*{\getPlaceOfBirth}{Kolkata, India}
\newcommand*{\getSubmissionDate}{28 April 2025}
\newcommand*{\getExpertOne}{Dr. Volker Springel}
\newcommand*{\getExpertTwo}{Dr. Ariel Sanchez}
\newcommand*{\getExamDate}{3 July 2025}

\newcommand*{\lang}{en-US}

\hypersetup {
pdfpagemode = {UseNone},
pdftitle = {\getTitle},
pdfauthor = {\getAuthor},
pdflang = {\lang},
}

\begin{document}

  \frontmatter

  \begin{titlepage}

   {\sffamily
    \vspace*{\stretch{1}}
    {\parindent0cm
    \rule{\linewidth}{.7ex}}  
  \begin{flushright}
    \vspace*{\stretch{1}}
    {\LARGE \bfseries \getTitle{}}
    
    \vspace*{2ex}
    {\Large \bfseries \getSubtitle{}}
    
    \vspace*{\stretch{1}}
    {\large\bfseries \getAuthor{}}
    
    \vspace*{\stretch{1}}
  \end{flushright}
    \rule{\linewidth}{.7ex}
    
    \vspace*{\stretch{5}}  
  \begin{center}
    \includegraphics[width=2in]{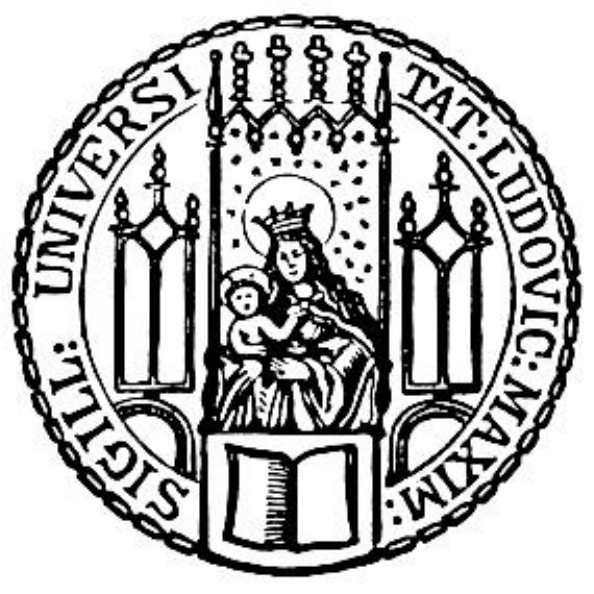}
    
    \vspace*{\stretch{1}} 
    {\Large \getPrintLocation{} \getPrintYear{}}
  \end{center}
    }

\end{titlepage}

\thispagestyle{empty}
\cleardoublepage

  \begin{titlepage}   
   {\sffamily
    \vspace*{\stretch{1}}
    {\parindent0cm
    \rule{\linewidth}{.7ex}}  
  \begin{flushright}
    \vspace*{\stretch{1}}
    {\LARGE \bfseries \getTitle{}}
    
    \vspace*{2ex}
    {\Large \bfseries \getSubtitle{}}
    
    \vspace*{\stretch{1}}
    {\large\bfseries \getAuthor{}}
    
    \vspace*{\stretch{1}}
  \end{flushright}
    \rule{\linewidth}{.7ex}
    \vspace*{\stretch{3}}
    
  \begin{center}
    {\large
     Dissertation\\
     der Fakultät für Physik\\
     der Ludwig-Maximilians-Universität\\
     München
     
    \vspace*{\stretch{1}}    
     vorgelegt von\\
     \getAuthor{}\\
     aus \getPlaceOfBirth{}
     
    \vspace*{\stretch{2}}   
     \getPrintLocation{}, den \getSubmissionDate{}
     }
  \end{center}
   }
\end{titlepage}

  \thispagestyle{empty}

   \vspace*{\stretch{1}}
  \begin{flushleft}
    {\large\sffamily
     Erstgutachter:  \getExpertOne{} \\[1mm]
     Zweitgutachter: \getExpertTwo{} \\[1mm]
     Tag der mündlichen Prüfung: \getExamDate{}\\
     }
  \end{flushleft}

\cleardoublepage{}

  \tableofcontents
  \markboth{Inhaltsverzeichnis}{Inhaltsverzeichnis}

  \listoffigures
  \markboth{Abbildungsverzeichnis}{Abbildungsverzeichnis}

  \listoftables
  \markboth{Tabellenverzeichnis}{Tabellenverzeichnis}
  \cleardoublepage

  \markboth{Zusammenfassung}{Zusammenfassung}
  \addcontentsline{toc}{chapter}{\protect Zusammenfassung}

\chapter*{Zusammenfassung}

Diese Dissertation untersucht die Epoche der kosmischen Reionisation (EoR), eine Schlüsselperiode im frühen Universum, in der die ersten leuchtkräftigen Quellen entstanden und ihre Strahlung das intergalaktische Medium (IGM) von einem neutralen in einen ionisierten Zustand überführte. Das Verständnis der Reionisation hilft, grundlegende Fragen zu beantworten, etwa wie die ersten Sterne und Galaxien entstanden, wie sie ihre Umgebung beeinflussten und wie sich großskalige kosmische Strukturen entwickelten. Der zentrale Fokus dieser Arbeit liegt auf der Untersuchung des Beginns und Endes der Reionisation von Wasserstoff und Helium durch die Identifikation der Quellen ionisierender Strahlung sowie der Analyse ihres Einflusses auf die thermische und ionisationsgeschichtliche Entwicklung des IGM. Ein weiterer Schwerpunkt dieser Arbeit liegt auf der Verbesserung beobachtungsstrategischer Methoden zur Detektion von EoR-Merkmalen bei verschiedenen Rotverschiebungen.

Der erste Teil der Dissertation legt das theoretische Fundament durch eine Erklärung des Standardmodells der Kosmologie ($\Lambda$CDM), wie kleine anfängliche Fluktuationen in der Materie zu Galaxien und Haufen heranwuchsen, und welche physikalischen Prozesse die Sternentstehung und Rückkopplung in Galaxien antreiben. Diese Prozesse bestimmen, wie viel ultraviolette (UV) Strahlung in das IGM entweicht und zur Reionisation beiträgt. Die Studie konzentriert sich auf die UV-Leuchtkraft hochrotverschobener Galaxien unter Verwendung der \texttt{SPICE} kosmologischen Strahlungshydrodynamiksimulationen. Untersucht wird, wie sich die UV-Leuchtkraftfunktion (UVLF), ein statistisches Maß für die Helligkeit von Galaxien, über die Zeit hinweg verändert, beeinflusst durch Sternentstehungsgeschichten. Unterschiedliche Modelle der Supernova(SN)-Rückkopplung führen zu unterschiedlicher Variabilität in verschiedenen Massenbereichen und bei verschiedenen Rotverschiebungen. Die Ergebnisse unterstreichen den signifikanten Einfluss der SN-Rückkopplung auf die Form der UVLF und zeigen die Massen- und Rotverschiebungsabhängigkeit ihrer Variabilität, was nahelegt, dass die UVLF-Variabilität helfen kann, die von \textit{James Webb Space Telescope} (\texttt{JWST}) beobachtete Diskrepanz bei hellen Galaxien zu erklären.

Der zweite Teil der Dissertation konzentriert sich darauf, wie ionisierende Quellen das IGM beeinflussen, insbesondere gegen Ende der Reionisation von Wasserstoff. Hierzu werden Strahlungstransportsimulationen mit verschiedenen Modellen von Spektralenergieverteilungen (SEDs) durchgeführt, die unterschiedliche ionisierende Quellen repräsentieren, darunter Einzel- und Doppelsterne, Röntgendoppelsterne, Emission aus dem interstellaren Medium sowie aktive galaktische Kerne (AGN). Analysiert wird, wie diese Quellen die Ionisation und Temperatur des Wasserstoffs im IGM sowie beobachtbare Signaturen wie den Lyman-$\alpha$-Wald beeinflussen. Die Ergebnisse zeigen, dass das ionisierende Spektrum einer Quelle den Zeitpunkt und die Topologie der Reionisation signifikant beeinflusst. Einige Modelle führen zu einer früheren oder ausgedehnteren Reionisation und erzeugen unterschiedliche thermische Entwicklungen im IGM, die mit aktuellen Beobachtungen verglichen werden können. Die Auswirkungen auf beobachtbare Größen des Lyman-$\alpha$-Waldes sind zwar marginal, eröffnen jedoch neue Ansätze für zukünftige Teleskopbeobachtungen.

Im nächsten Teil der Arbeit richtet sich der Fokus auf die Reionisation von Helium bei viel niedrigeren Rotverschiebungen, die später als die Wasserstoffreionisation stattfand und hauptsächlich von Quasaren angetrieben wurde. Mithilfe von Strahlungstransportsimulationen, die auf Schnappschüssen einer kosmologischen hydrodynamischen Simulation basieren, wird untersucht, wie die He\,\textsc{ii}-Reionisation voranschritt, wie sie das IGM beeinflusste und welche Signaturen des ausgedehnten Prozesses im He\,\textsc{ii}-Ly$\alpha$-Wald beobachtbar sind. Ein parametrisiertes Modell der Quasar-Leuchtkraftfunktion (QLF) wird verwendet, um Quasarpopulationen im Simulationsvolumen zu definieren und deren Einfluss zu untersuchen. Die simulierten Ergebnisse werden mit Beobachtungen des He\,\textsc{ii}-Ly$\alpha$-Waldes verglichen. Die Simulationen zeigen außerdem, wie die Temperatur des IGM davon abhängt, wann und wie He\,\textsc{ii} reionisiert wurde. Darüber hinaus wird untersucht, wie \texttt{JWST}-Beobachtungen von Quasaren bei $z > 5$ unser Verständnis der Quellen und des Zeitpunkts der He\,\textsc{ii}-Reionisation verfeinern könnten.

Im letzten Teil der Dissertation wird eine neue Beobachtungsmethode für die Heliumreionisation untersucht: die Hyperfeinstruktur-Übergangslinie von einfach ionisiertem Helium-3 ($^3$He$^+$) bei einer Wellenlänge von 3{,}5\,cm. Dieses Signal wurde bisher noch nie detektiert, könnte jedoch einzigartige Informationen über den Zustand des IGM nach Abschluss der Wasserstoffreionisation liefern. Es werden Vorhersagen zur Stärke dieses Signals und seiner Entwicklung mit der Rotverschiebung präsentiert. Zudem wird analysiert, wie das $^3$He$^+$-Signal mit den Positionen von Quasaren korreliert, was zukünftigen Radiosurveys helfen könnte, es zu entdecken. Dies stellt eine neuartige Methode zur Untersuchung der Reionisation dar, die über konventionelle Techniken hinausgeht.

Abschließend vereint diese Dissertation Galaxienentstehung, Strahlungsprozesse und die Entwicklung des IGM in einem einheitlichen Rahmen, um die kosmische Wasserstoff- und Heliumreionisation und deren Antriebskräfte besser zu verstehen. Die Arbeit liefert neue Erkenntnisse darüber, wie verschiedene Quellen zur Ionisierung des Universums beitrugen, wie ihre Eigenschaften das IGM prägten und wie wir mit kommenden Beobachtungen diese Modelle überprüfen können. Durch die Verbindung von Theorie, Simulation und Beobachtung trägt diese Dissertation dazu bei, die physikalischen Prozesse zu klären, die einen der bedeutendsten Phasen\"ubergänge im Universum bestimmten.

  \markboth{Abstract}{Abstract}
  \addcontentsline{toc}{chapter}{\protect Abstract}

\chapter*{Abstract}

This thesis investigates the Epoch of Cosmic Reionization (EoR), a key period in the early Universe when the first luminous sources formed and their radiation transformed the intergalactic medium (IGM) from a neutral to an ionized state. Understanding reionization helps us answer fundamental questions about how the first stars and galaxies formed, how they influenced their surroundings, and how large-scale cosmic structures evolved. The central focus of this work is to trace both the beginning and end of hydrogen and helium reionization by identifying the sources that produced ionizing radiation, and to understand how this radiation impacted the thermal and ionization history of the IGM. This thesis also focuses on improving observational strategies to detect features of EoR at different redshifts.

The first part of the thesis lays the theoretical foundation by explaining the standard cosmological model ($\Lambda$CDM), how small initial fluctuations in matter grew into galaxies and clusters, and what physical processes drive star formation and feedback inside galaxies. These processes determine how much ultraviolet (UV) radiation escapes into the IGM and contribute to reionization. This study focuses on the UV luminosity of high-redshift galaxies, using the \texttt{SPICE} cosmological radiation hydrodynamic simulation suite. I study how the UV luminosity function (UVLF), a statistical measure of galaxy brightness, varies over time due to impact from star formation histories. Different supernova (SN) feedback models lead to different levels of variability in different mass ranges and at different redshifts. These results underscore the critical role of SN feedback in shaping the UVLF, and highlights the mass and redshift dependence of its variability, suggesting that UVLF variability may alleviate the bright galaxy tension observed by \textit{James Webb Space Telescope} (\texttt{JWST}) at high redshifts.

The second part of the thesis focuses on how ionizing sources affect the IGM focusing more on to the ending phase of cosmic hydrogen reionization. Towards this goal, I use radiative transfer simulations with various models of spectral energy distributions (SEDs) representing different types of ionizing sources consisting single stars, binary stars, X-ray binaries, emission from interstellar medium and active galactic nuclei (AGN). I analyze how these sources affect the ionization and temperature of hydrogen in the IGM, as well as how they influence observable signatures such as the Lyman-$\alpha$ forest. The results show that the ionizing spectrum of a source significantly impacts the timing and topology of reionization. Some models lead to earlier or more extended reionization, and produce different thermal histories in the IGM, which can be compared to current observational constraints. In terms of observable quantities of Lyman-$\alpha$ forest, the impact is slightly marginal, however it opens up ideas to explore with future upcoming telescopes.

In the next part of the thesis, I turn to much lower redshift Universe focusing on helium reionization, which occurred later than hydrogen reionization and was mainly driven by quasars. Using radiative transfer simulations post processing the snapshots of cosmological hydrodynamical simulation, I study how He\,\textsc{ii} reionization progressed and how it affected the IGM and the signatures of extended helium reionization process detected with the  He\,\textsc{ii} Ly$\alpha$ forest. I explore the recent parametrized model of quasar luminosity function (QLF) to define quasar populations in the simulation volume and how these quasars drive the process and compare the simulated results with observations of the He\,\textsc{ii} Ly$\alpha$ forest. The simulations also reveal how the IGM temperature depends on when and how He\,\textsc{ii} was reionized. I also assess how the \texttt{JWST} observations of quasars at $z > 5$ could help refine our understanding of the sources and timing of He\,\textsc{ii} reionization.

Finally, in the last part of the thesis, I explore a new way to observe helium reionization through the hyperfine transition line of singly ionized helium-3 ($^3$He$^+$) at a wavelength of 3.5\,cm. This signal has never been detected, but could provide unique information about the state of the IGM after hydrogen reionization is complete. I present the predictions for the strength of this signal and how it evolves with redshift. I also explore how the $^3$He$^+$ signal correlates with the positions of quasars, which may help future radio surveys detect it. This offers a novel probe of reionization beyond the standard techniques generally used.

In conclusion, this thesis brings together galaxy formation, radiative processes, and IGM evolution in a single framework to better understand cosmic hydrogen and helium reionization and the drivers of these epochs. The work provides new insights into how different sources contributed to ionizing the Universe, how their properties shaped the IGM, and how we can use upcoming observations to test these models. By combining theory, simulation, and observation, this thesis helps clarify the physical processes that governed one of the Universe’s most important transitions.

  \mainmatter\setcounter{page}{1}
  \chapter{The Beginning}

\begin{flushright}
\begin{minipage}{0.9\textwidth}
\raggedleft
\textbf{\textit{``What doesn't kill you, makes you stronger."}}\\[1ex]
\noindent\rule{0.5\textwidth}{0.4pt}\\[-0.2ex]
Hitesh Kishore Das (\textit{before doing something stupid})
\end{minipage}
\end{flushright}

\begin{flushright}
\begin{minipage}{0.7\textwidth}
\raggedleft
\textbf{\textit{``Keep moving, keep laughing, keep dancing—life is a rhythm of fleeting moments, and the best ones leave you breathless."}}\\[1ex]
\noindent\rule{0.5\textwidth}{0.4pt}\\[-0.2ex]
Eishica Chand
\end{minipage}
\end{flushright}

I begin this thesis by introducing the key ideas that form the foundation of the work presented in the subsequent chapters. My aim is to establish a comprehensive framework that allows us to explore the fundamental themes shaping our understanding of the early Universe, its evolution over time and the observational campaigns focussing on them. By doing so, I provide the necessary context for the research that follows, ensuring a coherent progression of ideas. At its core, this thesis revolves around three fundamental questions:

\begin{itemize}
    \item How did the Universe evolve over time?
    \item What were the properties of early sources that influenced the early Universe?
    \item What are the strategies to observe the early Universe? And how to refine them?
\end{itemize}

While these questions may appear simple at first glance, their answers are not particularly straightforward. Addressing them requires an in-depth exploration of a pivotal cosmic era that began a few hundred thousand years after the \textit{Big Bang} and extended for nearly one billion years. The timing, duration, and key events of this period are intricately linked to our fundamental inquiries, making it a crucial focus of this thesis. This epoch unfolds in distinct phases, each marking significant transitions in the Universe’s history. It begins with the so-called \textit{Dark Ages (DA)}, a time when the first structures had yet to form, and the Universe remained largely featureless. As the first stars and galaxies started to appear, their radiation triggered the \textit{Epoch of Heating (EoH)} during the \textit{Cosmic Dawn (CD)}. This transformative period ultimately led to the \textit{Epoch of Reionization (EoR)} and \textit{Epoch of Helium Reionization (HeEoR)}, during which the radiation from early objects ionized the surrounding neutral gas, reshaping the cosmic features and marking the emergence of large-scale structures as we observe today. When referring to the ‘Universe’ in this context, I primarily focus on its diffuse component, particularly the gas that exists between galaxies, known as the intergalactic medium (IGM). This medium, composed predominantly of hydrogen and helium \citep{Meiksin2009}, plays a fundamental role in cosmic evolution. Understanding how it was heated, ionized, and evolved over time is key to unraveling the broader processes that governed the formation and evolution of the first galaxies, stars, and black holes.

In Section \ref{lambdacdmcosmo}, I provide a concise overview of the $\Lambda$CDM cosmological framework, setting the stage for understanding the early Universe. Section \ref{hierarchicalstructureformation} delves into the physics of structure formation, describing how matter gradually coalesced into the large-scale cosmic structures we observe today. In Section \ref{eorheeor}, I introduce the key epochs relevant to this thesis, highlighting their significance in the broader context of cosmic evolution. Section \ref{sourceseor} examines the primary sources that shaped the early Universe and their role in driving reionization. Section \ref{galaxyphysics} focuses on the internal physical processes within galaxies, while Section \ref{igmphysics} discusses the physics governing the IGM. Observational probes that reveal the features and evolution of the EoR and HeEoR are explored in Section \ref{eorobserve}. Finally, Section \ref{outline} presents an outline of the thesis, summarizing its key components and objectives.

\section{The $\Lambda$CDM Cosmology}
\label{lambdacdmcosmo}

I begin this thesis at a time of great progress in astronomy, though some big questions remain unsolved. We now live in the era of precision cosmology, where many of the mysteries that puzzled scientists in the 20th century have been solved. The long debate between an expanding Universe and the Steady-State theory has been settled down, with strong evidence showing that the Universe is indeed expanding. The redshift of light, once a puzzling phenomenon, is now understood as a direct result of cosmic expansion, where light stretches as space itself grows. Despite these advances, major mysteries remain, such as the nature of matter components of the Universe, the details of cosmic inflation, and how the first structures in the Universe formed. This thesis focuses on a key period in cosmic history to help shed light on many of these fundamental queries.

Since the time of Galileo \citep{Galilei1962}, our effort to understand the Universe has started with looking at the night sky. From these observations, we try to make sense of all the different astronomical events and fit them into a clear and organized picture. 
Explaining the complex physics behind how stars, galaxies, and cosmic structures form deepens our scientific understanding of the Universe’s evolution. Before the 1920s, most people thought the Universe was static and unchanging. However, the work of astronomers like Vesto Slipher and Edwin Hubble helped change this view by showing that the Universe is actually expanding (see \citealt{peacock2013} for historical overview). Between 1914 and 1917, Slipher observed that the light from many galaxies was redshifted \citep{slipher1917}, meaning they were moving away from us. This was one of the first clues supporting the idea of an expanding Universe. Building on this, astronomers like Knut Lundmark and Edwin Hubble looked for a direct relationship between a galaxy’s distance and its redshift \citep{lundmark1924,hubble1929,hubble1931}. Their findings, along with theoretical models proposed by Willem de Sitter and others, provided strong evidence for cosmic expansion. Meanwhile, Einstein's general theory of relativity \citep{einstein1917} introduced a revolutionary shift in our understanding of space and time, laying the foundation for modern cosmology. During the mid-20th century, while many cosmologists continued to interpret galaxies as point sources, \citealt{hoyle1948} proposed the continuous creation of matter throughout the Universe, which implied the presence of an IGM as a continuous medium between galaxies. However, observational confirmation remained elusive for decades. One of the earliest indications of IGM's existence came from the accidental discovery of the cosmic microwave background (CMB) spectrum by \citealt{penzias1965,penzias1969}. Follow-up observations and theoretical studies led to the belief that the Universe contains an energy component, now referred to as \textit{dark energy (DE)}, which fits naturally into Einstein's field equations as a cosmological constant, $\Lambda$ \citep{einstein1917}. Additionally, other evidence suggested that the Universe also contains large amounts of undetected matter, which helped explain galaxy rotation curves \citep{zwicky1933,vandehulst1957,rubin1980} and the clustering of large-scale structures \citep{einasto1980}. The formation of these structures appears to be a biased tracer of the underlying invisible \textit{dark matter (DM)} field \citep{davis1985,frenk1988,wang2020}, which is likely cold, as warm or hot dark matter would suppress the formation of small-scale structures \citep{viel2013,strucker2018}. These discoveries have led to the development of the current cosmological model of choice, the $\Lambda$CDM.

The evolution of the $\Lambda$CDM cosmology is described through the expansion of the Universe, which is represented by a dimensionless scale factor, denoted as \( a \). This expansion is driven primarily by two key components: the cosmological constant \( \Lambda \) and cold dark matter (CDM). The scale factor, which measures how the size of the Universe changes over time, is set to 1 at the present time, and it is related to the redshift \( z \) by the equation \( a = (1+z)^{-1} \). The redshift \( z \) essentially describes the amount of stretching (or time dilation) of light due to the expansion of the Universe, which increases with both the distance of objects and the time it takes for light to reach us. 
The cosmological principle states that, when viewed on large scales, the Universe is both homogeneous and isotropic. This means that, on vast distances, the Universe looks the same in every direction and has a uniform composition. Observational evidence, such as the uniformity of the CMB \citep{planck2013,planck2016,planck18,planck2018a} and the consistent distribution of galaxies across great distances, supports this idea \citep{trodden2004,sarkar2009,migkas2025}. Alternatively, the cosmological principle can be derived by combining the observed symmetry of the sky with the Copernican principle \footnote{The Copernican principle asserts that we do not occupy a unique or special position in the Universe.}. By applying basic geometry, it can be shown that these observations lead to the conclusion that matter is evenly spread throughout the Universe \citep{peacock1999}.

To chart galaxies in the Universe, we first lay out the metric, $ds^2 = g_{\mu\nu} dx^{\mu} dx^{\nu}$,
where $ds$ represents the spacetime interval, $g_{\mu\nu}$ is the metric tensor, and $x^{\mu}$ denotes the spacetime coordinates. By imposing symmetry constraints based on the cosmological principle \citep{trodden2004} the metric takes the form of the Robertson-Walker (RW) metric. Expressed in spherical coordinates $(r,\theta,\phi)$ around an observer, it reads:
\begin{equation}
    ds^2 = c^2 dt^2 - a^2(t) \left( \frac{dr^2}{1 - kr^2} + r^2 (d\theta^2 + \sin^2\theta d\phi^2) \right),
\end{equation}
where $t$ is the cosmological time and $a(t)$ is the scale factor. The physical radial interval between two events is given by $dl_p = \frac{a(t) dr}{\sqrt{1 - kr^2}}$, where the comoving coordinate $r$ factors out the effect of cosmological expansion. The parameter $k$ defines the spatial geometry, where the Universe can be flat ($k=0$), closed ($k=1$), or open ($k=-1$).

The RW metric links the observed redshift of galaxies to their spacetime positions. Observationally, distant galaxies exhibit redshifted light, defining the redshift as, $1 + z \equiv \lambda_{\text{obs}} / \lambda_{\text{em}}$, where $\lambda_{\text{obs}}$ is the observed wavelength and $\lambda_{\text{em}}$ is the emitted wavelength. Light from a galaxy follows null geodesics, $c dt = a(t) dr / \sqrt{1 - kr^2}$. For a galaxy emitting at time $t_{\text{em}}$ at radial distance $r$, two successive wavefronts emitted at intervals $\Delta t_{\text{em}}$, corresponding to wavelength $\lambda_{\text{em}} = c \Delta t_{\text{em}}$, are received at times $t_{\text{obs}}$ and $t_{\text{obs}} + \Delta t_{\text{obs}}$ as $\lambda_{\text{obs}} = c \Delta t_{\text{obs}}$. The relation between emission and observation times follows:
\begin{equation}
    \int_{t_{\text{em}}}^{t_{\text{obs}}} \frac{c dt}{a(t)} = \int_0^r \frac{dr}{\sqrt{1 - kr^2}}.
\end{equation}
Since the right-hand side remains unchanged between two wavefronts, $\int_{t_{\text{em}}}^{t_{\text{obs}}} \frac{c dt}{a(t)} = \int_{t_{\text{em}} + \Delta t_{\text{em}}}^{t_{\text{obs}} + \Delta t_{\text{obs}}} \frac{c dt}{a(t)}$, 
which leads to the fundamental redshift relation, $1 + z = \frac{a(t_{\text{obs}})}{a(t_{\text{em}})}$.
Normalizing the scale factor to unity today, redshift directly corresponds to cosmological time as $1 + z = 1/a(t)$, irrespective of geometry. This provides a direct mapping of galaxy redshifts to their 3D distribution in the Universe.

\begin{figure*}
    \centering
    \includegraphics[width=120mm]{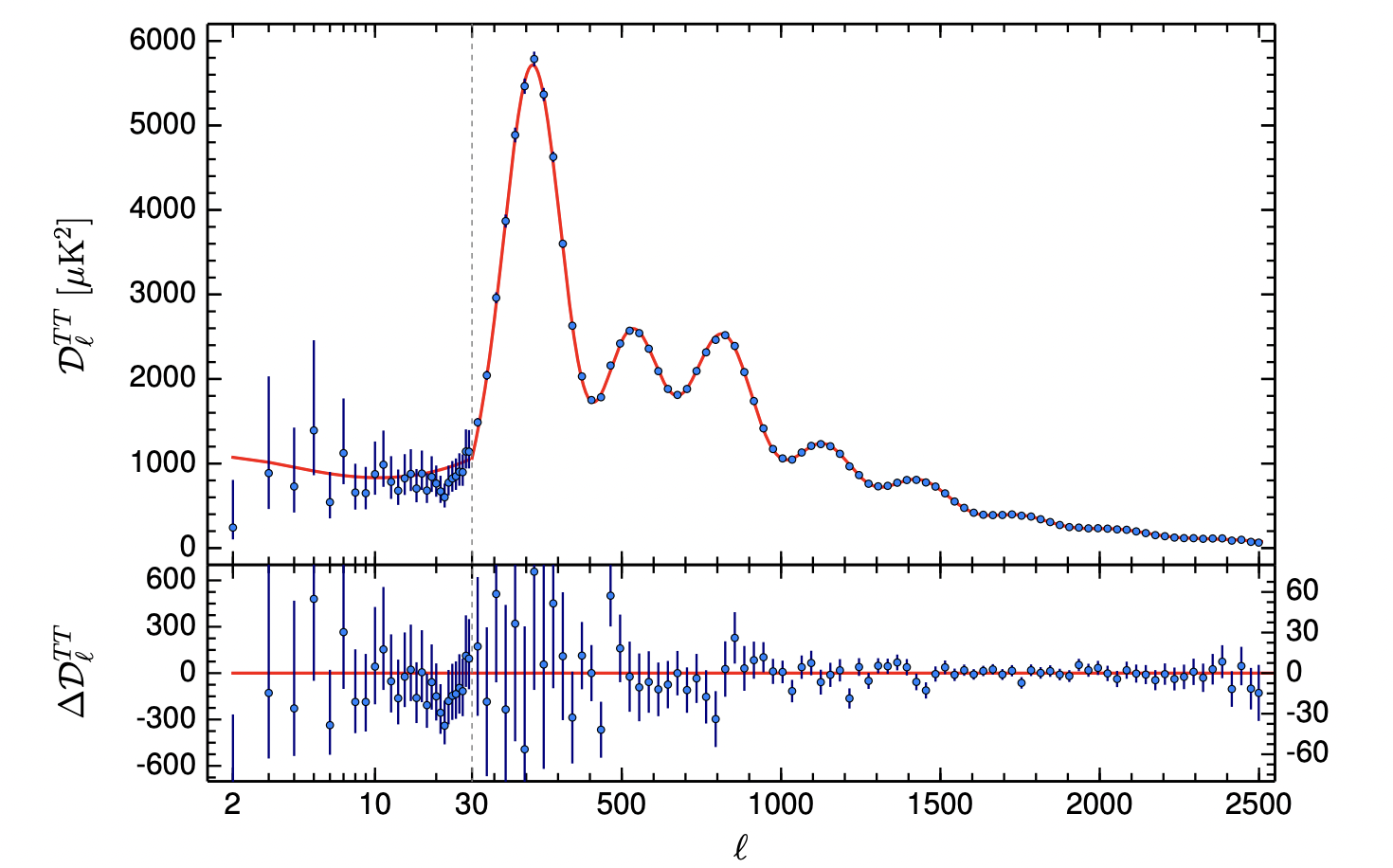}
    \caption{Power spectrum of the temperature fluctuations of the CMB as measured by Planck satellite (solid data points). The solid line represents a fit based on a six-parameter \(\Lambda\)CDM cosmology. This figure is adapted from \citealt{planck2016a}.}
    \label{CMB_planck}
\end{figure*}

These fundamental concepts, the RW metric and cosmological redshift, derive solely from the cosmological principle and spacetime geometry, without requiring a specific gravity theory like general relativity. The physical assumptions made include homogeneity (Copernican principle), isotropy on large scales, and the invariance of the speed of light. To describe cosmic expansion history, we introduce the law of gravity governing spacetime dynamics. In general relativity, the Einstein field equation describes the dynamics of a metric tensor $g_{\mu\nu}$ as,
\begin{equation}
    G_{\mu\nu} + \Lambda g_{\mu\nu} = \frac{8 \pi G}{c^4} T_{\mu\nu},
\end{equation}
where the Einstein tensor, $G_{\mu\nu} = R_{\mu\nu} - \frac{1}{2} g_{\mu\nu} R$ is related to the Ricci tensor $R_{\mu\nu}$ and Ricci scalar $R$ and as defined earlier, $\Lambda$ is the cosmological constant. The total density $\rho$ and pressure $p$ in the Universe contribute to the dynamics via the energy-momentum tensor, $T_{\mu\nu} = (\rho + p c^{-2}) U_{\mu} U_{\nu} + p g_{\mu\nu},$ where $U^{\mu}$ is the four-velocity. Finally, substituting the RW metric into the Einstein equation gives the Friedmann equation governing the Universe’s expansion rate (see \citealt{dodelson2003} for detailed calculations):
\begin{equation}
    H(z) = H_0 \left[ \Omega_{\Lambda} + \Omega_k (1+z)^2 + \Omega_m (1+z)^3 + \Omega_r (1+z)^4 \right]^{1/2},
    \label{Ha}
\end{equation}
where the density parameters $\Omega_i = \rho_i(0)/\rho_{\text{crit}}(0)$ relate to the present-day critical density $\rho_{\text{crit}}(0) = \frac{3 H_0^2}{8 \pi G}$. We define $\Omega_{\Lambda}$ for dark energy, $\Omega_m$ for total matter (dark matter and baryons), $\Omega_r$ for radiation, and curvature term as $\Omega_k = 1 - \Omega_{\Lambda} - \Omega_m - \Omega_r$. The Hubble constant $H_0$ includes the dimensionless parameter $h$. The total matter content, dark energy, and spatial curvature fundamentally determine the expansion history, which is crucial for extragalactic astronomy in mapping galaxies and intergalactic gas in the Universe. Figure \ref{CMB_planck} shows that the predicted power spectrum from the $\Lambda$CDM model closely matches Planck’s observations \citep{planck2016a}. This supports $\Lambda$CDM as the standard cosmological model which has been used in this thesis.

The matter in Equation \ref{Ha} is predominantly dark, with luminous baryonic matter comprising less than 10\% of the total inferred mass \citep{zwicky1933, vandehulst1957,rubin1980, davis1985, frenk1988, wang2020}. Despite its small contribution, this visible matter is crucial to understanding cosmic structures, as it directly interacts with radiation. In this thesis, my focus is on the evolution and properties of this observable matter, its role in shaping the early Universe, and its imprint on cosmic history.

\section{Hierarchical Structure Formation}
\label{hierarchicalstructureformation}

The large-scale structure of the Universe originates from small density fluctuations that formed during the cosmic inflationary period. According to the standard inflationary model, these fluctuations were initially nearly Gaussian, with only slight deviations from perfect uniformity. The initial power spectrum of these fluctuations is often modeled as a scale-invariant power-law, given by $\mathrm{\mathit{P(k)} \propto \mathit{k}^{\mathit{n}_{s}}}$, where the spectral index $n_{\rm s}$ is close to 1. Over time, gravitational forces caused density variations to grow, shaping the structure of the Universe as these early fluctuations evolved during expansion, influenced by factors like radiation pressure and the interaction between dark matter and ordinary matter (baryons). The linear growth of these fluctuations is described by the transfer function, $\mathrm{\mathit{T(k)}}$, which determines how different scales of density variations evolve over time. The shape of the transfer function depends on which component of the Universe dominated at different times—the radiation-dominated era at $z \gtrsim 3500$, the matter-dominated era at $0.5 \lesssim z \lesssim 3500$, and the dark energy-dominated era at $z \lesssim 0.5$. At very large scales, when density variations were smaller than the cosmic horizon during the radiation-dominated era, their growth was suppressed due to the pressure exerted by radiation. However, once the Universe transitioned into the matter-dominated era, these fluctuations grew rapidly. The transfer function plays a crucial role in predicting the power spectrum of today's large-scale structure and helps cosmologists link the Universe's initial conditions to its present structure. 

\begin{figure*}
    \centering
    \includegraphics[width=85mm]{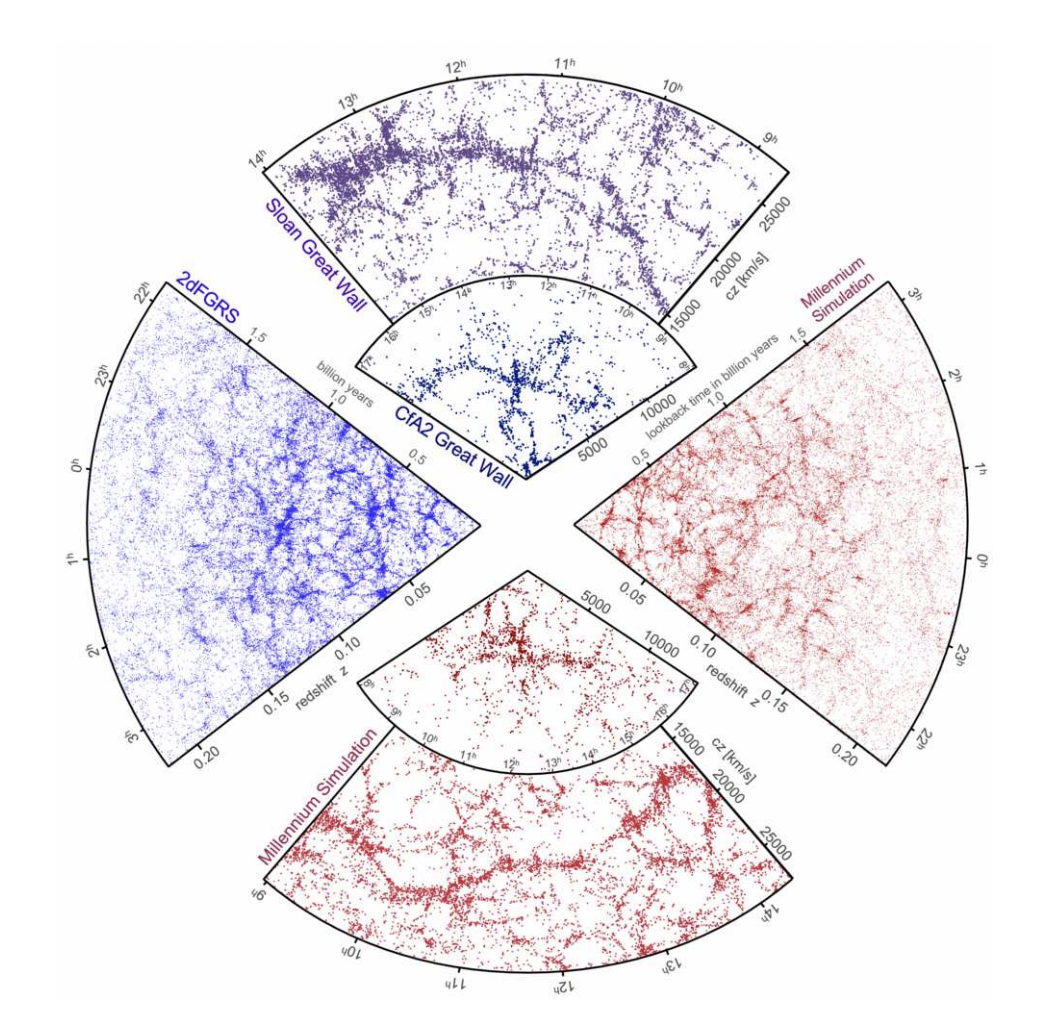}
    \caption{The panels display the large-scale structure of the galaxy distribution, with blue points representing observed data and red points representing simulation. The observed and simulated galaxy distributions show a strikingly similar pattern.  This figure is adapted from \citealt{springel2006}.}
    \label{structure_formation}
\end{figure*}

The formation and evolution of structures can be well described by the collisionless Boltzmann equation and general theory of relativity (see \citealt{bertschinger1995,mo2010} for more detailed discussion), however, solving these equations is complex. As an alternative, in the early Universe, when density contrasts were small ($\delta << 1$), the theory of structure formation is well described by the linear perturbation theory. Essentially we take the velocity moments of the distribution function and that reduces the collisionless Boltzmann equation to a fluid description which follows linear perturbation theory. The evolution of the linear perturbations can be expressed as the mathematical formula: $\ddot{\delta} + 2H \dot{\delta} = 4\pi G \bar{\rho} \delta$. In the linear regime, the perturbation simply grows as $\delta \propto D(z)$, where the growth factor $D(z)$ follows (e.g. \citealt{lahav2004}):
\begin{equation}
    D(z) \propto H(z) \int_z^{\infty} \frac{(1+z')}{H^3(z')} dz'.
\end{equation}
During the matter-dominated era, $D(z) \propto a(z)$. The matter power spectrum then evolves as, $P(k, z) \propto k^{n_s} D^2(z) T^2(k)$. However, as the Universe transitions to the dark energy-dominated era, the growth rate slows significantly. This slowing of growth is a critical feature, as it influences the formation of large-scale structures and the distribution of galaxies in the modern Universe. The growth of cosmic structures can thus be tracked by how the amplitude of density fluctuations evolves with time, showing how the small initial density variations become the vast cosmic web observed today. Figure \ref{structure_formation} displays the similarity between the observations of large scale structures with the predicted results from theory as quoted in \citealt{springel2006}.

When the fluctuation grows to $\delta > 1$, the gravitational collapse leads to the formation of virialized objects, called as DM halos. According to the spherical top hat model, the collapse corresponds to the time when a linear perturbation reaches a critical values of $\delta_c \approx 1.69$. One key characteristic of these haloes is the \textit{virial radius}, which defines the boundary within which the halo's mass is gravitationally bound. The virial radius is determined by the balance between the halo's kinetic energy and its gravitational potential. The virial radius of a halo of mass $M_{\rm h}$ then follows:
\begin{equation}
    R_{\text{vir}} = 52.3 \left(\frac{M_{\rm h}}{10^{10} {\rm M_{\odot}}}\right)^{1/3} \left(\frac{\Omega_m}{0.3} \frac{200}{\Delta_c}\right)^{-1/3} h^{-1} \text{kpc}.
\end{equation}
The formation and clustering of haloes are of my central interest because in the standard paradigm the formation of galaxies is related to halo clustering. The Press-Schechter formalism provides us a useful analytical framework to understand the properties of haloes i.e. mass function, merger tree and clustering. It relates the fraction of overdensity greater than $\delta_c \approx 1.69$ smoothed on a filtering scale  $R_{f} = (\frac{3 M_{h}}{4 \pi \bar{\rho_{m}}(0)})^{\frac{1}{3}}$ to the number of haloes with mass $M_{h}$. So, the Press-Schechter formalism estimates the number of halos per mass interval as:
\begin{equation}
    \frac{dn}{d\ln M_{\rm h}} = \sqrt{\frac{2}{\pi}} \frac{\bar{\rho}_m(0)}{M_{\rm h}} \nu e^{-\nu^2/2} \left| \frac{d\ln \sigma(M_{\rm h}, z)}{d\ln M_{\rm h}} \right|,
\end{equation}
where $\nu = \delta_c / \sigma(M_{\rm h}, z)$ and $\sigma(M_{\rm h}, z)$ is the density fluctuation smoothed on a scale $M_{\rm h}$.

N-body simulations provide a detailed approach to study structure formation by solving the Boltzmann equation numerically. 
These simulations use discrete particles to approximate the phase-space distribution, solving gravity with methods like the Particle Mesh, Tree, or hybrid TreePM solvers (for review, see \citealt{couchman1999,springel2021}). At low redshifts, structure formation follows a hierarchical pattern, where smaller structures merge to form larger ones. This is due to the effective power spectrum slope, $n_{\text{eff}} = d\ln P(k)/d\ln k$, being in the range $-2 < n_{\text{eff}} < 0$ on relevant scales. However, at higher redshifts, the slope approaches $n_{\text{eff}} \to -3$, causing a more synchronized collapse of structures, although the smaller objects that formed then were involve in merging events. 
This non-hierarchical formation suggests that early galaxies formed rapidly within dense regions that later became galaxy clusters. The large-scale structure at high redshifts was therefore more coherent compared to today, where hierarchical merging dominates \citep{kauffmann1999,kitzbichler2007}.

\section{Physical Processes in Galaxies} 
\label{galaxyphysics}

Extragalactic astronomy began with debates about the nature of fuzzy nebulae, with the Shapley-Curtis Debate in 1920 questioning whether they were galaxies outside our Milky Way. Before 1980, galaxies were viewed as isolated islands, forming independently through stellar evolution. From 1980 to 1995, the paradigm shifted to hierarchical galaxy formation within the $\Lambda$CDM cosmological model, where galaxies formed through gas cooling in dark matter halos and merged, altering their morphology. Since 1995, emphasis has grown on the role of feedback and environmental factors in galaxy evolution, with theories now incorporating stellar and AGN feedback, gas accretion, and star formation regulation. The study of high-redshift galaxies and reionization has spurred new observational surveys, and current research focuses on the physical processes that drive galaxy formation and evolution, particularly during the epoch of reionization (for a review, see \citealt{ciardi2005,ellis2008}).

\subsubsection{Cooling of Gas and gas accretion}  

\begin{figure*}
    \centering
    \includegraphics[width=120mm]{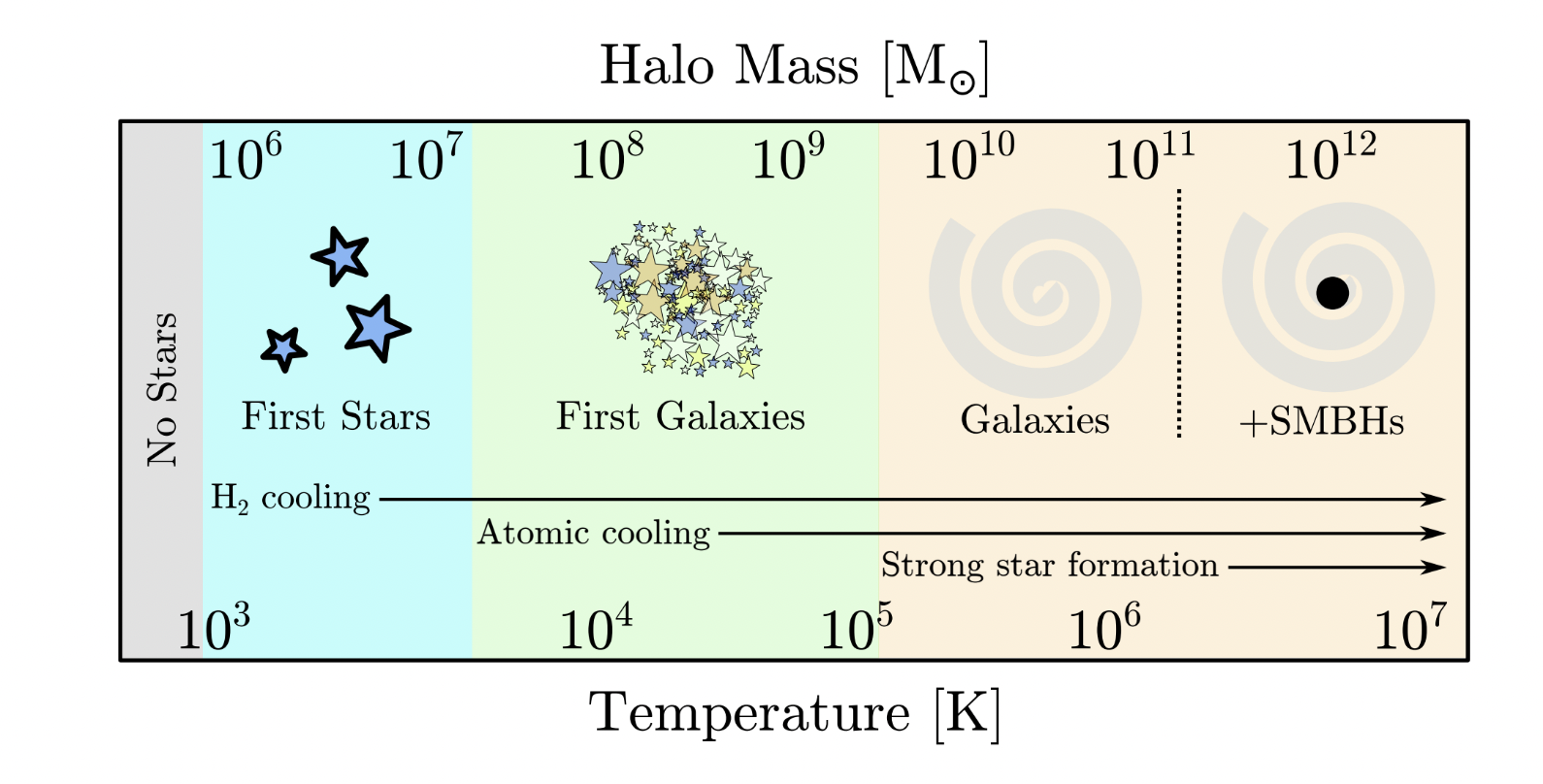}
    \caption{Virial temperatures and mass scales of dark matter halos that facilitate the formation of the first stars, primordial galaxies, and the most massive galaxies in the Universe. The cooling mechanism shifts from molecular hydrogen in smaller halos to atomic hydrogen in more massive halos, driving cosmic structure formation. The figure is taken from \citealt{wise2019}.}
    \label{object_halomass}
\end{figure*}

The cooling of gas in dark matter halos is essential for star formation, as the gas must become cold enough for gravity to overcome thermal pressure. Cooling efficiency depends on temperature: at very high temperatures (above $10^7$ K), bremsstrahlung dominates, while at intermediate temperatures ($10^4$ K–$10^7$ K), collisional ionization and atomic excitation are most effective \citep{schmutzler1993,gnat2007,gnat2009,schure2009,richings2014}. Below $10^4$ K, molecular and metal line cooling become important \citep{jappsen2007,vasiliev2013,ripamonti2007}. In the early Universe ($z \sim 10{-}15$), the first star-forming halos had masses as low as $10^5 {\rm M_{\odot}}$, where cooling was primarily driven by molecular hydrogen (H$_2$) \citep{bovino2014,nebrin2023,hou2024,nadler2025}. This allowed gas to collapse even at temperatures below $10^4$ K. However, H$_2$ is fragile and can be easily dissociated by Lyman-Werner (LW) radiation from the first stars \citep{glover2001,incata2023}. As LW radiation built up, it suppressed H$2$ cooling, raising the minimum halo mass required for efficient gas cooling to around $10^7 {\rm M{\odot}}$. At this stage, star formation proceeded in more massive halos with virial temperatures above $10^4$ K, where atomic hydrogen cooling becomes effective. These halos are less sensitive to LW feedback and played a key role in forming the first galaxies. Figure \ref{object_halomass} illustrates the types of early galaxies hosted by dark matter halos of different masses.

The fate of gas in a dark matter halo is governed by three key timescales: the cooling time ($t_{\text{cool}}$), the free-fall time ($t_{\text{ff}}$), and the Hubble time ($t_{\text{H}}$). The Hubble time represents the age of the Universe at a given redshift and sets the maximum duration over which gravitational collapse can occur before cosmic expansion becomes dominant. It is given by, $t_{\text{H}} = \frac{1}{H(z)}$, where $H(z)$ is defined in Equation \ref{Ha}.
For a gas cloud with number density $n$ and temperature $T$, these timescales are given by: $t_{\text{cool}} = \frac{3k_{\rm B}T}{2n\Lambda(T)}$ and $t_{\text{ff}} = \left(\frac{3\pi}{32G\rho}\right)^{1/2}$, where $k_{\rm B}$ is the Boltzmann constant, $\Lambda(T)$ is the cooling rate, $G$ is the gravitational constant, and $\rho$ is the gas density. Since $t_{\text{cool}} \propto n^{-1}$ and $t_{\text{ff}} \propto n^{-1/2}$, if $t_{\text{cool}} < t_{\text{ff}}$, the gas can collapse to the central regions of the halo, triggering star formation.
When primordial gas falls into dark matter halos, it is shock-heated to the virial temperature: $T_{\text{vir}} = \frac{\mu m_{p}}{2k_{\rm B}} \frac{GM}{R_{\text{vir}}}$, where $\mu$ is the mean molecular weight of the gas, $m_{p}$ is the proton mass, $M$ is the halo mass, and $R_{\text{vir}}$ is the virial radius. At this stage, the cooling of metal-free gas is mainly controlled by H and $\rm{H_{2}}$. Molecular hydrogen can only form efficiently after redshift $z \approx 100$ when the CMB radiation intensity weakens enough to allow its survival.

Gas accretion onto galaxies occurs via two modes. In the first mode, when $t_{\text{cool}} > t_{\text{ff}}$, a quasi-hydrostatic atmosphere forms, and gas cools at the center. Since cooling efficiency scales as $\propto n_{\text{gas}}^{2}$, it is faster in denser regions, leading to gradual gas accretion onto the central star-forming region. This is known as hot-mode accretion. In the second mode, when $t_{\text{cool}} < t_{\text{ff}}$, intergalactic gas can flow into galaxies without being shock-heated, following the filamentary structure of the cosmic web. This cold-stream accretion efficiently feeds galaxies. Additionally, major mergers can drive violent gas inflows, resulting in starburst activity and rapid galaxy growth.

\subsubsection{Star formation}

As gas accumulates in the central regions of dark matter halos, its self-gravity starts to dominate, making the gas cloud unstable to collapse. If the cloud gathers enough mass such that the sound-crossing time exceeds the free-fall time (satisfying the Jeans criterion), it will collapse and trigger star formation. The Jeans mass, which determines the smallest collapsing mass, is given by \citep{jeans1928}:

\begin{equation}
    M_{\text{Jeans}} =
    \left( \frac{5kT}{G \mu m_p} \right)^{\frac{3}{2}}
    \left( \frac{3}{4\pi\rho} \right)^{\frac{1}{2}}
    \approx 2{\rm M_{\odot}}
    \left( \frac{c_s}{0.2 \text{ km/s}} \right)^3
    \left( \frac{n}{10^3 \text{ cm}^{-3}} \right)^{-1/2}
\end{equation}

where $c_s$ is the sound speed, $n$ is the gas number density, $T$ is the temperature, and $G$ is the gravitational constant. As the cloud collapses, it can fragment into smaller pieces, each potentially forming a star. The rate of collapse depends on the cloud’s chemical composition and its ability to cool.

Early metal-free (primordial) gas heats up more than metal-enriched gas, leading to higher Jeans masses. Thus, the first stars (Population III or Pop III stars) were likely much more massive than modern stars, with masses between $10^2 - 10^3 {\rm M_{\odot}}$. However, how fragmentation occurs remains unclear, making the initial mass function (IMF) of Pop III stars uncertain. A commonly used IMF parameterization is: $\frac{dN}{d\log M_*} \propto (1 + M_*/M_c)^{-1.35}$, where $M_c$ is a characteristic mass related to the Jeans mass. This results in a top-heavy IMF, favoring massive stars. As Pop III stars evolve, they enrich the surrounding gas with metals, introducing new cooling channels that allow smaller fragments to form. This leads to the transition from Pop III to Population II (Pop II) star formation. The metallicity threshold for this transition is estimated to be $Z_{\text{crit}} = 10^{-4} - 10^{-3} Z_{\odot}$.

In the broader context of galaxy formation and cosmic reionization (more discussion in Section \ref{eorheeor}), understanding the timing and efficiency of star formation is crucial. The expected star formation rate, assuming direct gravitational collapse, is: $\dot{M}_* = \frac{M_{\text{gas}}}{t_{\text{ff}}}$. However, observations show that galaxies form stars much more slowly, requiring a correction factor—the "star-formation efficiency" ($\epsilon_{\text{ff}}$). This suggests that turbulence and stellar feedback regulate star formation, keeping clouds in a dynamic, quasi-equilibrium state where collapse, turbulence, and feedback interact continuously.

\subsubsection{Feedback}

Feedback plays a crucial role in galaxy formation by regulating star formation through the injection of energy, mass, and radiation into the ISM and IGM \citep{hopkins2012,grisdale2017,emerick2018,hopkins2020,katz2024}. It introduces a self-regulating mechanism that can either enhance or suppress star formation, shaping the observed galaxy population. The stellar mass function (SMF), which describes the number density of galaxies as a function of stellar mass (for a detailed description of how observed galaxy luminosities are converted to SMFs, see \citealt{song2016,weaver2023}), is well represented by the Schechter function \citep{Schechter1976}:

\begin{equation}
\phi(M)dM = \phi_{\rm ch} \left( \frac{M}{M_{\rm ch}} \right)^{\alpha} e^{- M/M_{\rm ch}} dM,
\end{equation}

where $M_{\rm ch}$ is the characteristic mass marking the transition from a power-law to an exponential decline. The observed SMF deviates significantly from the halo mass function (HMF), suggesting that additional processes regulate star formation. Feedback mechanisms explain the suppression of star formation in massive galaxies and the relative efficiency of star formation in low-mass galaxies, preventing unrealistic overproduction of stars.

Mechanical feedback from core-collapse supernovae (SN) plays a dominant role in shaping the ISM by injecting mass, momentum, and thermal energy into the surrounding gas \citep{dekel1986,navarro1996}. The explosion of massive stars releases ejecta at velocities of approximately $6000 \ \text{km s}^{-1}$ \citep{janka2012}, driving shockwaves that heat the ISM to temperatures above $10^6$ K \citep{mckee1977,walch2015}. This process influences the multiphase structure of the ISM, driving turbulence that can support against gravitational collapse or trigger new star formation \citep{Chevalier1985}. The evolution of SN remnants follows distinct phases \citep{chevalier1982,ostriker2010,blondin1998,draine2011}: an early free-expansion phase, the adiabatic Sedov-Taylor phase where momentum and thermal energy are efficiently transferred \citep{taylor1950,sedov1959}, and a later momentum-conserving phase once radiative cooling dominates. These outflows regulate the gas supply, influencing star formation rates by either expelling material from galaxies or redistributing it through fountain flows and galactic winds \citep{Chevalier1985,norman1989,scalo2004,ostriker2011,kim_ostriker2015}.

Radiative feedback from young stars is another essential regulator of star formation \citep{leitherer1999}. Before supernovae explode, massive stars emit ionizing radiation, depositing energy into the ISM and altering cooling rates. Photoionization increases the temperature of surrounding gas, suppressing molecular cooling and evaporating low-mass clouds \citep{efstathiou1992,gnedin2000,okamoto2008}. The impact of radiation pressure, both direct and through dust scattering, further drives turbulence and outflows, affecting the efficiency of gas retention in star-forming regions \citep{murray2005,hopkins2012,agertz2013,roskar2014,agertz2016,costa2019}. The escape of ionizing photons is also a crucial factor in cosmic reionization, influencing the transition from molecular to atomic cooling halos and setting the conditions for the formation of subsequent stellar populations.

\begin{figure*}
    \centering
    \includegraphics[width=80mm]{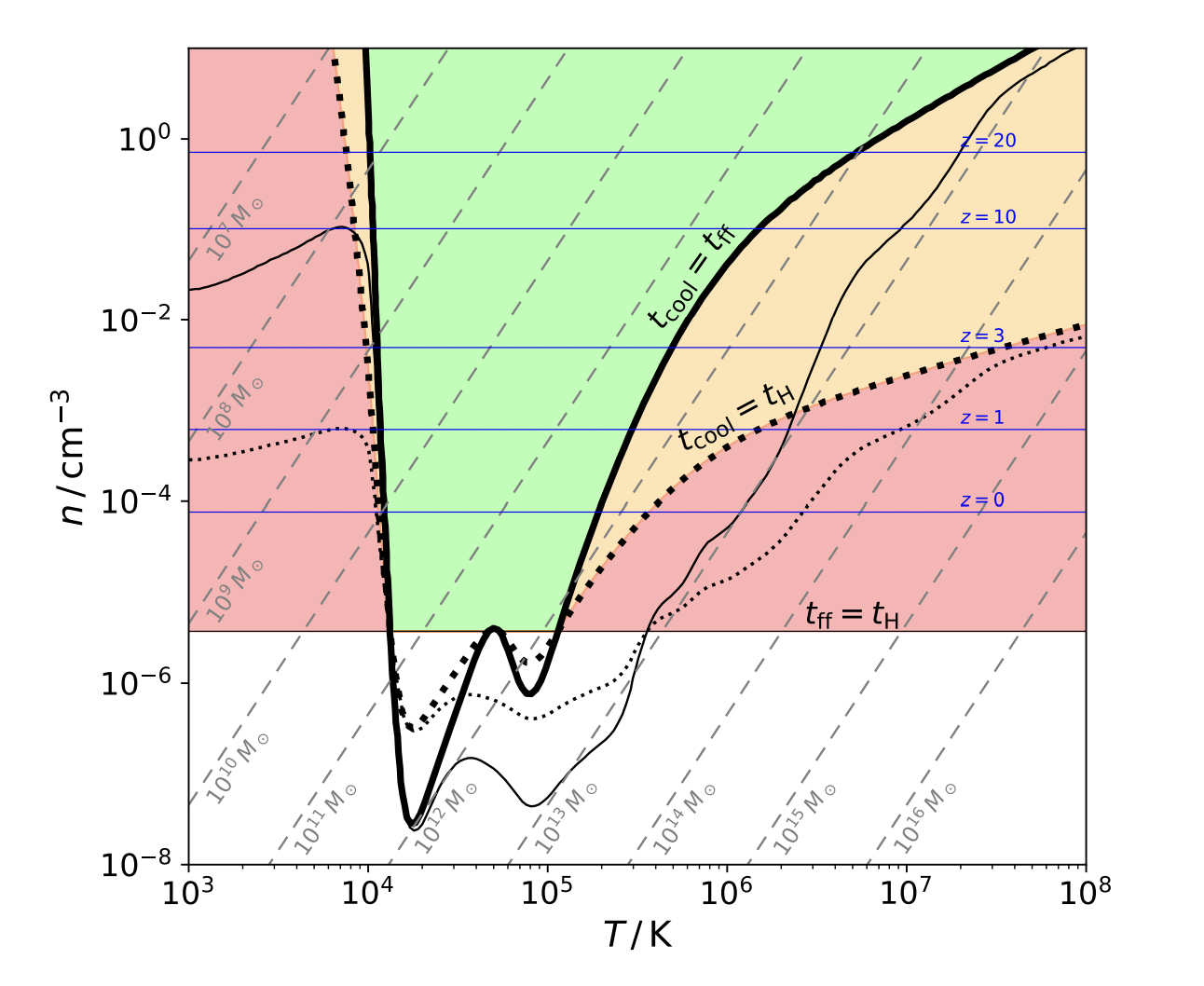}
    \includegraphics[width=80mm]{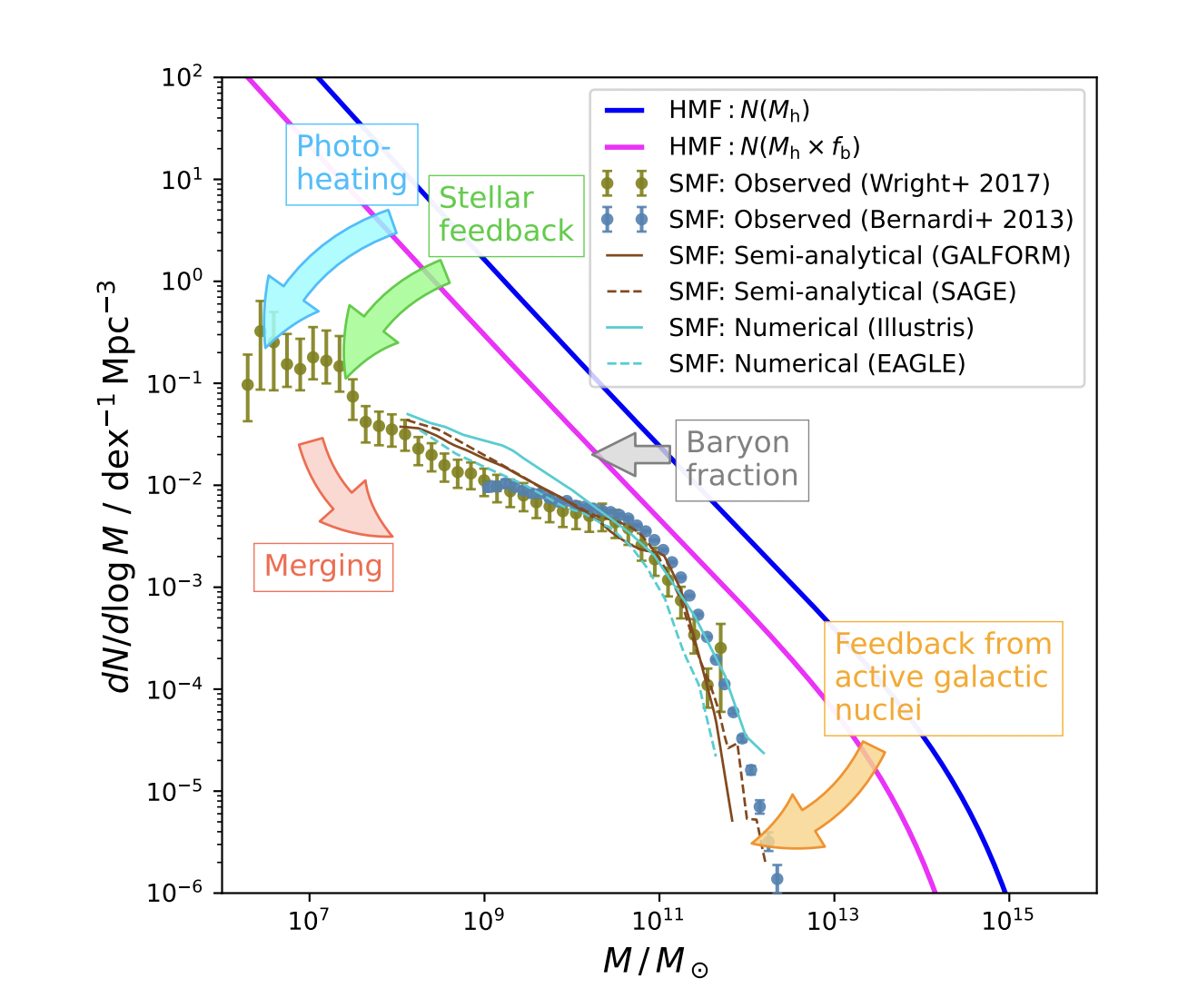}
    \caption{\textit{Left panel:} Relation between gas number density \textit{n} and gas temperature \textit{T} for a cloud of gas. \textit{Right panel:} Number density of galaxies and halos as a function of respective mass showcasing the importance of feedbacks in shaping that in different mass regime. The figures are taken from \citealt{laursen2023}.}
    \label{cooling_feedback}
\end{figure*}

Chemical feedback results from the enrichment of the ISM with heavier elements produced inside stars. Supernovae and stellar winds distribute metals into the surrounding gas, modifying cooling efficiencies and affecting the formation of new stars \citep{karlsson2013}. Metal-line cooling enhances fragmentation, leading to more efficient star formation in metal-enriched environments. The transition from metal-free (Pop III) to metal-enriched (Pop II) star formation is dictated by this feedback, influencing the early chemical evolution of galaxies and the distribution of heavy elements in the IGM \citep{schneider2002,mackey2003,maio2010,maio2011}.

These feedback mechanisms collectively shape galaxy evolution across cosmic time. Without feedback, galaxies would convert their gas into stars too efficiently, leading to unrealistic mass and luminosity distributions. The observed SMF, which differs significantly from the HMF, is well reproduced in simulations that incorporate feedback, demonstrating its necessity in models of galaxy formation. By regulating star formation, redistributing mass and energy, and influencing cosmic reionization, feedback remains an essential ingredient in explaining the observed diversity and evolution of galaxies.
Figure \ref{cooling_feedback} illustrates the interaction between these timescales relevant for cooling and star formation and the models that invoke these feedback mechanisms are able to capture the correct shape of the SMF.

\subsubsection{Black hole formation and growth and AGN feedback}

Supermassive black holes (SMBHs) with masses around \( \sim 10^{6-9} {\rm M_{\odot}} \) are commonly found at the centers of galaxies. In the unified model of AGN, a SMBH powers an accretion disk that emits vast amounts of radiation through non-thermal processes, often outshining the stellar light from the host galaxy during its active phase. The formation of SMBHs is thought to be seeded by several processes, including the death of massive stars as remnants of Population III stars, the collapse of compact star clusters, and the direct collapse of a gas cloud \citep{rees1984,volonteri2010}. Once formed, these seed black holes grow by accreting gas and through mergers with other black holes, eventually reaching supermassive sizes.

The feedback from AGN significantly influences star formation in galaxies. UV and X-ray radiation emitted from the accretion disk and hot corona, along with Compton scattering of high-energy photons by ionized gas, heat up the ISM, preventing further star formation by raising the temperature of the surrounding gas \citep{ciotti2001}. Additionally, the radiation from AGNs can drive powerful winds that further suppress star formation \citep{page2012,brennan2018,bollati2024,scharre2024}. In the low redshift regime, AGN feedback is considered a leading explanation for the exponential break in the galaxy luminosity function and the quenching of star formation in massive galaxies \citep{beckmann2017,piotrowska2022,mulcahey2022}. These insights from lower redshift studies help us understand the formation of the first quasars and their relationship with host galaxies at redshifts \( z > 6 \). During the epoch of reionization, AGNs and mini-quasars play a crucial role by emitting intense UV and X-ray radiation, helping to ionize the hydrogen gas in the early Universe (discussed more in Section \ref{sourceseor}).

\section{Epoch of Hydrogen and Helium Reionization}
\label{eorheeor}

Around 380,000 years after the Big Bang, the Universe underwent a major transformation as free electrons and protons combined to form the first neutral hydrogen and helium atoms. This event, known as recombination, marked the transition of the Universe from being ionized and opaque to neutral and transparent, allowing light to travel freely for the first time. It occurred around redshift \( z \approx 1100 \), when the temperature of the Universe had cooled sufficiently for hydrogen to recombine (H~{\sc ii} \( \to \) H~{\sc i}), allowing electrons to bind with protons without being immediately re-ionized by high-energy photons. The formation of neutral hydrogen atoms led to the decoupling of photons from matter, giving rise to the CMB, which we observe today as a faint relic radiation permeating the Universe. Prior to this, helium recombined in two stages. The first recombination of helium occurred at a much earlier epoch, around redshift \( z \approx 6000 \), when triply ionized helium (He~{\sc iii}) captured an electron to become singly ionized helium (He~{\sc ii}). This transition required an energy of 54.4 eV \citep{Switzer2008b}. Later, as the Universe continued to expand and cool, the second recombination of helium took place at \( z \approx 1800 \), when singly ionized helium (He~{\sc ii}) captured another electron to form neutral helium (He~{\sc i}) \citep{Switzer2008a}. Following the completion of helium and hydrogen recombination, the Universe entered a prolonged phase known as the "dark ages". During this period, there were no significant sources of light, as no stars or galaxies had yet formed. The only radiation present was the residual CMB and faint emission from atomic transitions. 
Over time, these neutral atoms were gradually reionized by photons emitted by the first stars and galaxies. 
The formation of the first galaxies and stars triggered the emission of photons with enough energy to ionize the surrounding IGM and initiate the epoch of (hydrogen) reionization (traditionally called as EoR) \citep{Loeb2001}. Observations of high-redshift QSO absorption spectra \citep{Fan2003,Fan2006, Gunn1965, McGreer2015, bosman2018, Davies2018, Bosman2022}, and Lyman-\(\alpha\) emitters (LAEs; \citealt{Pentericci2014, Tilvi2014, Barros2017, Mason2018, Jung2022, Tang2023, Tang2024, nakane2024, napolitano2024, saxena2024, chen2024, ning2024}), together with the determination of the IGM thermal properties \citep{Bolton2012, Raskutti2012, Gaikwad2020}, and the Thompson scattering optical depth \citep{Komatsu2011, planck2016, Pagano2020} show that reionization was a gradual process, mostly completed by \( z \approx 5.5 \), although small patches of neutral hydrogen remain at lower redshifts \citep{kulkarni2019,nasir2020}.

This reionization process is believed to have occurred in an "inside-out" fashion, starting in the densest regions near galaxies and then spreading outward to the less dense areas. This process can be divided into three stages: pre-overlap, overlap, and post-overlap. During the pre-overlap phase, young galaxies formed localized ionized bubbles around themselves. These bubbles remained isolated due to the distance between galaxies, leaving the surrounding IGM mostly neutral. In the overlap phase, ionized bubbles expanded and merged, allowing high-energy photons to travel longer distances and speeding up reionization. By the post-overlap phase, most of the IGM had been fully ionized, except for small pockets of neutral gas in dense regions.

The first stars and galaxies emitted high-energy photons capable of ionizing hydrogen (\( E_\gamma > 13.6 \, \rm{eV} \)) and singly ionizing helium (\( E_\gamma > 24.6 \, \rm{eV} \)). However, the second ionization of helium (He~{\sc ii}~$\to$~He~{\sc iii}), which marks the epoch of helium reionization (HeEoR), required even more energetic photons (\( E_\gamma > 54.4 \, \rm{eV} \)) that were not abundantly produced by these early sources. Because the first ionization potential of helium is close to that of hydrogen, He~{\sc i} is expected to have been ionized into He~{\sc ii} by redshift \( z \sim 5.5 \) \citep{Fan2006, Becker2015, bosman2018, Bosman2022}. In contrast, the full reionization of helium, transitioning from He~{\sc ii} to He~{\sc iii}, occurred later and was driven by the harder ultraviolet (UV) radiation from quasars \citep{Compostella2013, Garaldi2019, Basu2024}.

Sophisticated modeling of the reionization process employ a variety of approaches, ranging from semi-analytic and semi-numeric models \citep{zhou2013,mesinger2007,santos2010,choudhury2018,hutter2021,choudhury2022}, to full radiation hydrodynamic simulations \citep{Baek2010,ocvirk2016,ockvirk2020, gnedin2014,finlator2018,Rosdahl2018,Trebitsch2021,kannan2022,garaldi2022,bhagwat2024}. The choice of method depends on the scale and complexity of the problem. These simulations track the movement of matter under the influence of gravity, and hydrodynamic simulations also incorporate the effects of gas dynamics, pressure, and shocks.
A simple way to model reionization is by tracking the balance between ionizing photon production and recombinations, which restore neutral hydrogen. This is captured in the equation:  

\begin{equation}
N_{\text{ion}}(t) = N_{\text{HII}}(t) + \int_0^t \frac{N_{\text{HII}}(t')}{t_{\text{rec}}(t')} dt'
\end{equation}

where \(N_{\text{ion}}\) is the number of ionizing photons, \(N_{\text{HII}}\) is the number of ionized hydrogen atoms, and \(t_{\text{rec}}\) is the recombination time. Averaging over a representative cosmic volume leads to the `reionization equation' \citep{madau1999}:

\begin{equation}
\frac{dQ}{dt} = \frac{\dot{n}_{\text{ion}}}{\langle n_{\rm H} \rangle} - \frac{Q}{t_{\text{rec}}}
\label{dqdt}
\end{equation}

where \(Q\) is the ionized fraction, \(\dot{n}_{\text{ion}}\) is the ionizing photon production rate, and \(\langle n_{\rm H} \rangle\) is the mean hydrogen density. This equation is widely used as it efficiently estimates the photon budget needed for reionization, making it a powerful tool for studying different reionization models. Another form of Equation \ref{dqdt} can be expressed as:
\begin{equation}
\frac{dI(\nu)}{d\tau(\nu)} = S(\nu) - I(\nu)
\label{RTeq}
\end{equation}
where \( I(\nu) \) is the monochromatic specific intensity and \( \tau(\nu) \) is the optical depth. The source term \( S(\nu) \) is given in units of erg s\(^{-1}\) Hz\(^{-1}\) cm\(^{-2}\) sr\(^{-2}\), and represents the radiation produced by sources. In the following section (see Section \ref{sourceseor}), we discuss about potential sources in detail.

Both hydrogen and helium reionization were influenced by factors like the growth of cosmic structures, the formation of stars in galaxies, the properties of those stars, and the conditions in the circumgalactic medium surrounding galaxies. The amount of ionizing radiation escaping from galaxies and how it spread across the Universe played a critical role in determining the timeline of reionization. While much progress has been made in understanding this phase, many questions remain. Scientists are still investigating the types of sources responsible for reionization, how numerous they were, how effectively they emitted ionizing radiation, and how this radiation escaped into the surrounding space. Moreover, understanding the distribution and shape of ionized regions provides valuable insights into the properties of the first galaxies and their role in the evolution of the Universe, which is one of the main goals of this thesis.

\section{Sources of Cosmic Reionization}
\label{sourceseor}

The so-called ``Dark Ages'' were not entirely inactive, while there was no visible light yet, the first structures were beginning to form. Some of these eventually produced light, ending this era. Although earlier sections discussed possible formation pathways of these first bound objects, the exact timing  of their formation remain uncertain. With the launch of the James Webb Space Telescope (\texttt{JWST}) in July 2022, our understanding of early galaxy formation has been significantly challenged. Unlike the Hubble Space Telescope (\texttt{HST}), which enabled observations out to $z \approx 10$ \citep{zheng2012,coe2013,oesch2016,morishita@ARTICLE,bagley2022}, \texttt{JWST} has pushed this limit further, detecting an unexpected overabundance of massive galaxies at $z > 10$ \citep{castellano2022,finkelstein2022,naidu2022,adams2023,morishita_stiavelli2023,bouwens2023a,bouwens2023b,donnan2023,atek2023,perez2023,willott2024,Donnan2024}. One notable example is the recent detection of a spectroscopically confirmed galaxy at $z=14.32$ by \citealt{helton2024}. These discoveries suggest that massive systems may have formed earlier than expected, which is difficult to explain under standard models where structure formation begins with small objects. In the early Universe, the lack of metals and dust made gas cooling inefficient, but molecular hydrogen and hydrogen deuteride (HD) allowed some cooling, leading to the formation of the first stars, known as Population III (Pop~III) stars \citep[see e.g.][]{ciardi2005}. In regions exposed to external radiation that destroyed H$_2$ and HD, the gas could collapse more rapidly, possibly forming massive black holes directly \citep{dijkstra2014}. These black holes, along with Pop~III stars, were sources of ionizing radiation, but likely not numerous enough to ionize the full IGM. Therefore, this section now moves beyond the first 100 million years of cosmological evolution to explore the role of later sources of heat and light, such as second- and third-generation stars and galaxies, in shaping the IGM. In the following, I outline the main contributors to cosmic reionization that are the focus of this thesis.

\subsubsection{Stars}

Stars residing within galaxies are widely considered the primary drivers of cosmic reionization, as first suggested by \citealt{madau1999}. The recent low Thomson scattering optical depth reported by the \citealt{planck2016a}, along with constraints from observations with \texttt{HST} and \texttt{JWST}, has established interest in understanding their role in this process \citep{robertson2015,robertson2022}.
They not only contribute ionizing photons but also influence structure formation through thermal and ionizing feedback, as discussed by \citealt{couchman1986,cen1992,fukugita1994}. The efficiency of stars in reionizing the IGM depends on various factors, including their mass and metallicity. These properties determine the number of ionizing photons they produce, as extensively reviewed by \citealt{ciardi2005}. In particular, Population III stars are expected to have emitted large quantities of ionizing radiation, though their exact contribution remains uncertain.

For stars to drive cosmic reionization effectively, a sufficiently high fraction of their ionizing radiation must escape from their host galaxies into the IGM. This fraction, known as the escape fraction ($f_{\text{esc}}$), is a key parameter in reionization models. However, observations indicate that escape fractions vary significantly. While some studies suggest that high-redshift galaxies may have had large $f_{\text{esc}}$ values, other research raises concerns that it may have been too low for stellar sources alone to complete reionization \citep{sokasian2003,mitra2013,ma2015,chisholm2018}. 
In nearby galaxies, low escape fractions are often observed, leading to questions about whether the situation was dramatically different at early times. Several studies have explored these uncertainties, including \citealt{vanzella2016,vanzella2018,matthee2018,naidu2018,steidel2018,fletcher2019}, each providing inferences on $f_{\text{esc}}$ based on different observational techniques.

Although galaxies are widely considered the primary drivers of reionization in most of the literature (e.g., \citealt{BeckerBolton2013,bouwens2015,robertson2015,Madau2017,dayal2020,hassan2018,Marius2018,Marius2020,kannan2022}), the specific properties of the galaxies that contributed most remain uncertain. Key open questions include whether reionization was predominantly driven by numerous low-mass galaxies or by fewer, more massive ones \citep{Naidu2020,Naidu2021,Matthee2022}. An important factor in this discussion is the role of binary stars. Observations show that at least 50\% of stars in the local Universe are in binary systems \citep{han2020}. When stars in binaries interact, they can lose their outer layers, which makes them emit more UV light and harder spectra over longer periods \citep{elridge2012, gotberg2019, berzin2021}. These photons can even be energetic enough to ionize helium twice \citep{gotberg2020}. Models of stellar populations suggest that including binaries increases the total ionizing radiation by up to 60\% in low-metallicity environments and by 10--20\% at near-solar metallicity \citep{stanway2016, gotberg2020}. Binary stars can also boost the escape of ionizing photons from galaxies. Their strong feedback can clear away surrounding gas, making it easier for radiation to reach the intergalactic medium \citep{ma-x2016, ma-x2020, secunda2020}. If many stars during the Epoch of Reionization formed in binaries, these effects could significantly increase the number of photons available for reionization, speeding up the process \citep{Rosdahl2018}. Simulations that include radiative transfer support this idea, showing that a high fraction of binary stars can lead to faster hydrogen reionization, although it may also reduce star formation in small galaxies due to heating \citep{doughty2021}.

In addition to the various types of stars and their properties discussed so far, as introduced before Pop III stars are believed to be the first generation of stars formed in the Universe (see also Section \ref{galaxyphysics}). These stars likely formed at very high redshifts, around \( z \sim 20 - 30 \), when the Universe was only a few hundred million years old \citep{bromm2013, klessen2023}. Pop III stars were unique in that they formed from pristine gas composed almost entirely of hydrogen and helium, with no metals (elements heavier than helium). As the Universe evolved and stars began enriching the gas with metals through supernova explosions, the conditions for Pop III star formation became increasingly rare. Because of this, the formation rate of Pop III stars is expected to have decreased rapidly over time. However, recent simulations suggest that Pop III stars may have continued forming much later than previously thought, potentially even down to the EoR, around \( z \sim 6 \) \citep{xu2016, sarmento2018, jacks2019, liu2020, skinner2020, visbal2020, sarmento2022, venditti2023}. These findings imply that some regions of the Universe may have remained metal-poor long enough to allow Pop III star formation well into the period when galaxies were reionizing the intergalactic medium. If true, this late Pop III formation could have contributed additional ionizing photons during reionization, influencing the timing and structure of the process.

\subsubsection{Quasars / AGNs }

Another significant source of ionization is active galactic nuclei (AGN) or quasars. Since the early search for the IGM (e.g., \citealt{field1959}), QSOs and AGNs have been considered potential contributors to cosmic reionization \citep{rees1970,arons1970}. The recent detection of a larger than expected number density of faint quasars at intermediate (i.e. $z\approx4$; \citealt{Giallongo2015,Giallongo2019,Boutsia2018}) and high (i.e. $z>6$;  \citealt{Weigel2015,McGreer2018,Parsha2018,Akiyama2018,harikane2023,Maiolino2023,Goulding2023,Larson2023,Juodzbalis2023,greene2023}) redshift, has renewed the interest in quasars as possible strong contributors to the ionization budget \citep{grazian2024,madau2024}. Theoretically, QSOs and AGNs could complement stellar ionization, particularly in scenarios with low escape fractions \citealt{koki2017}.

A more extreme scenario, where AGNs alone drive reionization, was explored by \citealt{madau2015}. While their model successfully reproduces the evolution of the ionized volume filling factor ($Q_{\text{HII}}$), it fails to match key observations. It predicts excessively high IGM temperatures \citep{Garaldi2019}, premature HeII reionization at $z \sim 4.2$ \citep{worseck2016,worseck2019,makan2021,makan2022}, and inconsistencies in the Lyman-$\alpha$ (Ly$\alpha$) forest opacity \citep{chardin2015}. Observational constraints \citep{onoue2017,Parsha2018,matsouka2018,kulkarni2019} and theoretical studies \citep{finkelstein2019} suggest that AGNs contribute to the ionizing budget but remain subdominant to stellar sources.

The formation and growth of black holes remain an open question. Simulations suggest that massive black holes could have existed as early as redshift 7.5 \citep{banados2018,fan2019}, and their growth can be well modeled \citep{degraf2012,sijacki2015,weinberger2018}. However, how these black holes initially formed is still uncertain \citep{regan2009,volonteri2012}. One common approach in simulations is to seed galaxies with black holes of a certain mass (essentially seed in massive halos, see, \citealt{dimatteo2012,khandai2015,crain2015},) which helps match observations at lower redshifts \citep{kormendy2013} but does not provide insight into the early abundance and properties of faint or small-mass black holes. These smaller black holes might have played a crucial role in the early stages of reionization \citep{grazian2024,madau2024}.

\subsubsection{X-ray Binaries}

Another potential contributor to ionization and heating was galactic X-ray binary systems (XRBs). These systems consist of a neutron star or black hole consuming material from a companion star. At high redshifts, the primary contributors to ionizing radiation in these systems were high-mass X-ray binaries (HMXBs) rather than low-mass X-ray binaries (LMXBs) \citep{mirabel2011,fragos2013,fragosa2013,madaufragos2017,sazonov2017}. Their spectra are generally hard, meaning they emit high-energy radiation that is more effective at heating the IGM rather than ionizing it \citep{fialkov2014}. While X-ray binaries dominate the X-ray output in gas-poor galaxies, their role in reionization is still debated \citep{fabbiano2006,mineo2012,Marius2018,Marius2020}.

\begin{figure}
    \centering
    \includegraphics[width=70mm]{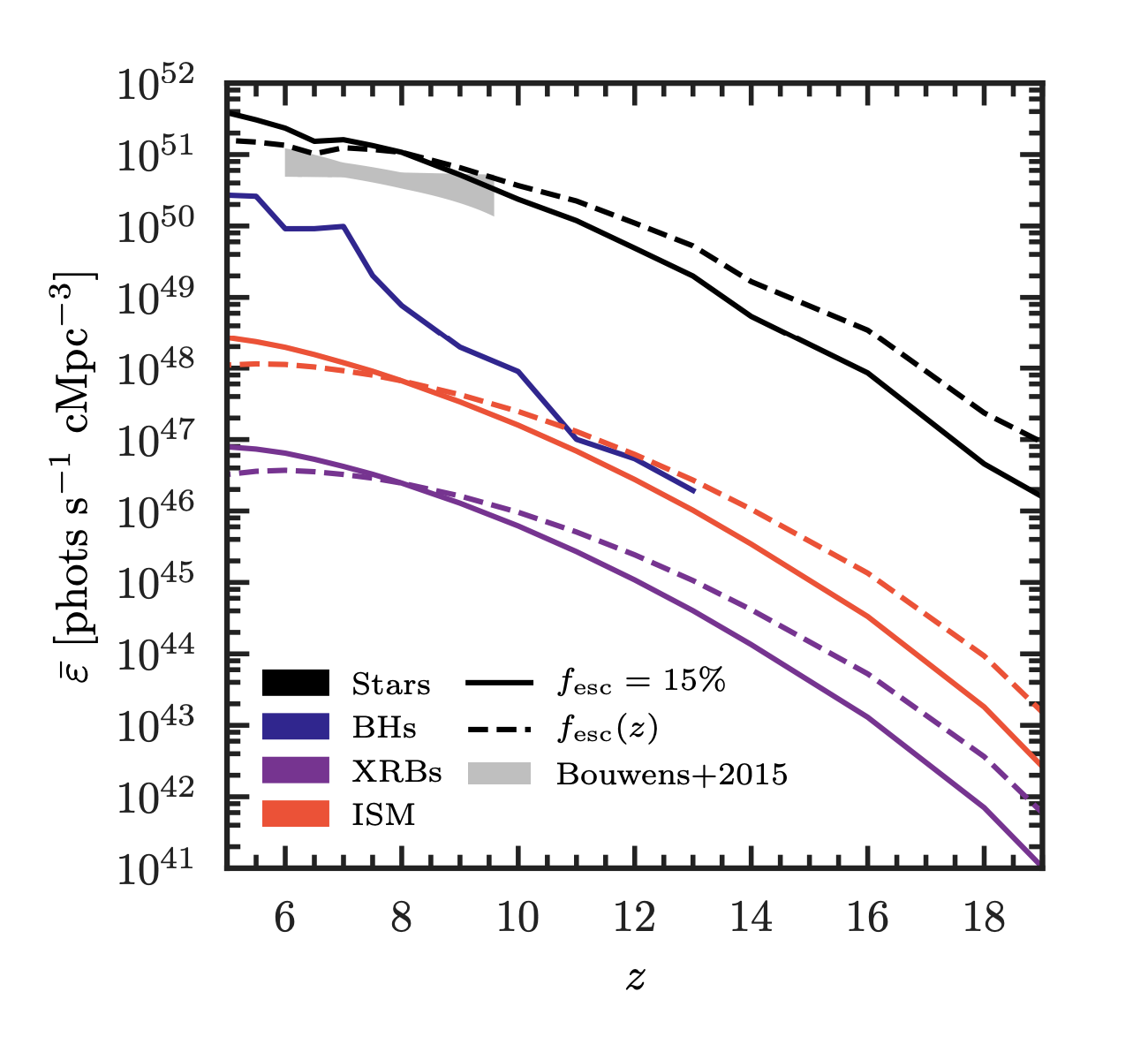}
    \includegraphics[width=70mm]{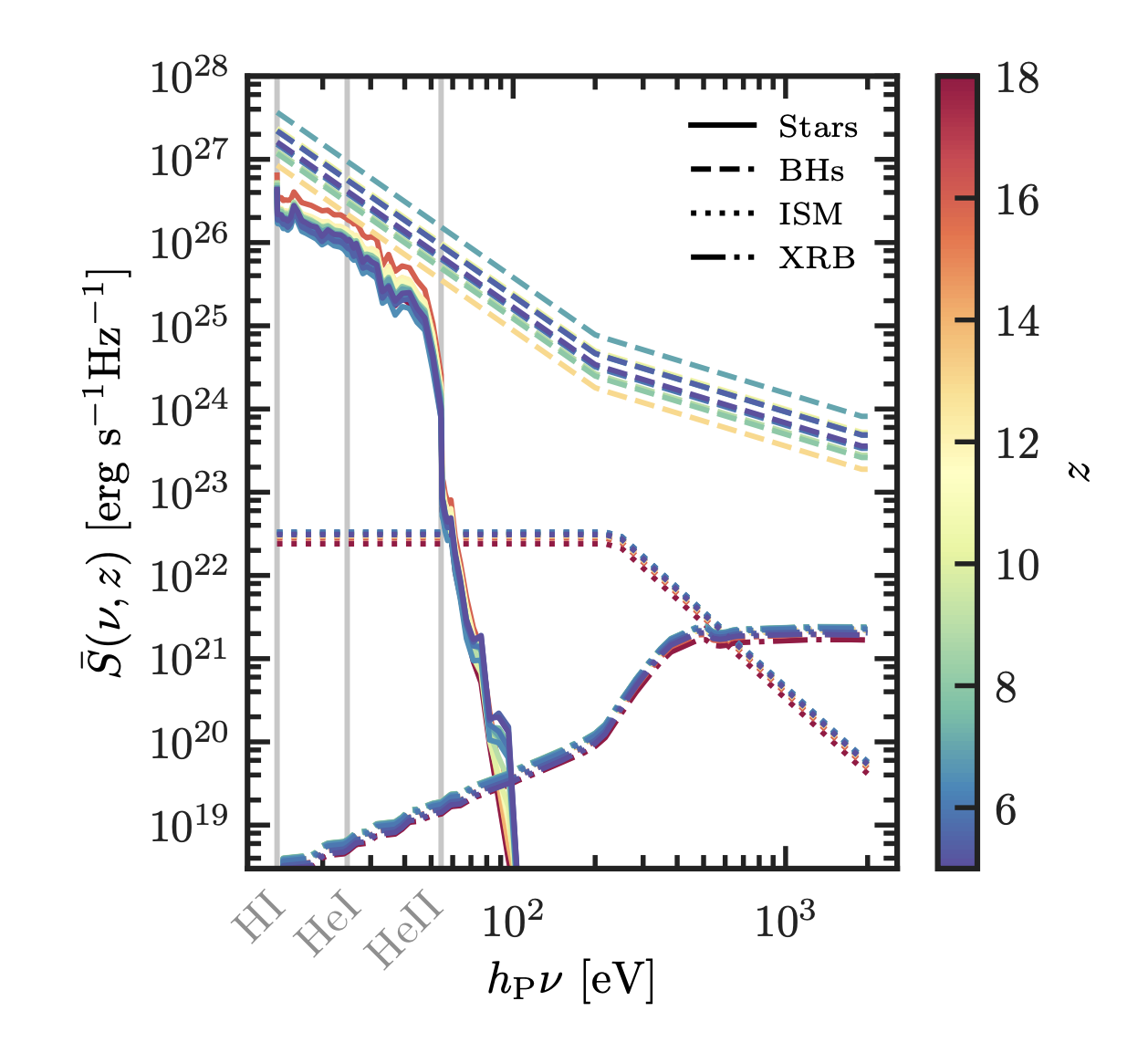}
    \caption{\textit{Left panel:} Redshift evolution of the comoving volume averaged ionizing emissivity per source type assuming either a constant UV escape fraction $f_{esc}$ = 15$\%$ (solid lines) or redshift-dependent $f_{esc}(z)$ (dashed lines). From top to bottom the sets of curves correspond to: stars (black), BHs (blue), XRBs (purple) and ISM (red).
    \textit{Right panel:} Average SEDs for each source type, stars (solid lines), BHs (dashed), ISM (dotted) and XRBs (dash dot-dotted) as function of photon energy plotted at various redshifts (indicated by the line colour). The ionization thresholds for HI, HeI and HeII are plotted as vertical grey lines The figure is taken from \citealt{Marius2020}.
    }
    \label{SED_sources}
\end{figure}

\subsubsection{Thermal Emission from Interstellar Medium}
Thermal bremsstrahlung from interstellar medium (ISM) is another ionizing and heating mechanism that has been studied for many years (see the review by \citealt{fabbiano1989}). Supernova-driven galactic winds interacting with interstellar clouds \citep{Chevalier1985} can produce soft X-rays, which could contribute to the heating of the IGM \citep{pacucci2014}. Other potential mechanisms include shocked gas in the galactic halo and disk \citep{suchkov1994}, as well as hot galactic winds \citep{strickland2000}.

\subsubsection{Additional Candidates}
Finally, there are other possible contributors to heating and ionization, but their role is considered secondary. These include low energy cosmic rays \citep{nath1993,sazonov2015,leite2017,owen2019,bera2023,yokoyama2023,gessey2023,carvalho2024}, self-annihilation or decay of dark matter \citep{liu2016,cang2023} and plasma beam instabilities in TeV blazars \citep{chang2012,puchwein2012,lamberts2022}. While these processes may have played a minor role, they are not the focus of this discussion.

Figure \ref{SED_sources} illustrates the emissivity and spectral energy distributions (SEDs) of various primary sources discussed in this section highlighting a comparison between the sources contributing to the reionization budget. The modest redshift evolution observed is a result of two factors: the averaging effect, which preferentially emphasizes the contribution from brighter sources, and the slow evolution of the underlying physical properties that govern the spectra. For more details about modelling of these SEDs, please refer to \citealt{Marius2018,Marius2020}.

\section{Physical processes in the IGM}
\label{igmphysics}

In the present-day Universe, galaxies account for only about 7\% of the total baryonic matter \citep{fukugita2004,shull2012}, with the vast majority of baryons (approximately 90\%) residing in the IGM, the diffuse gas that fills the space between galaxies. The IGM's evolution is effectively linked to the process of galaxy formation, primarily through mechanisms such as gas inflow and the outflows driven by galactic feedback. The interface between galaxies and the IGM, known as the circum-galactic medium (CGM), serves as a dynamic zone where material is exchanged, playing a vital role in the baryon cycle that regulates galaxy growth and evolution. On cosmic scales, both galaxies and AGNs (along with other potential contributors discussed in Section~\ref{sourceseor}), influence the IGM via the process of cosmic reionization. This epoch began with the initial reionization of hydrogen and helium at redshifts as high as $z \sim 15$ - 20, and extended through the second helium reionization phase ending around $z \sim 2$ - 4. Reionization significantly affects the thermal and structural properties of the Universe; the heating of the gas increases the cosmological Jeans mass, thereby supressing the formation of low-mass dwarf galaxies. Spatial and temporal fluctuations in the UV background and IGM temperature, driven by the progression of reionization process, introduce astrophysical variations that can be utilized to extract cosmological information, particularly through studies of the Ly\(\alpha\) forest at \( z \sim 2\text{--}4 \) \citep{lai2006,mcquinn2011,pontzen2014,greig2015}. As such, the physics of the IGM is central to our understanding of galaxy evolution, the reionization era, and broader cosmological processes.

The ionization state of hydrogen in the IGM is determined by the photoionization equilibrium with the metagalactic UV background, produced by ionizing sources. The photoionization rate of hydrogen, \(\Gamma = 4\pi \int \sigma_{\text{HI}} \frac{J_{\nu}} {h \nu} d \nu\), and the case-A recombination rate of hydrogen, $\alpha_A(T) = 4.063 \times 10^{-13} (T / 10^4 \, \text{K})^{-0.72} \, \text{cm}^3 \, \text{s}^{-1}$ together establish the photoionization equilibrium, which results in the relation, $\Gamma n_{\text{HI}} = \alpha(T) n_e n_{\text{HII}}$, where $n_{\rm H} = \rho_{\text{crit}}(0) \Delta_b \left( \frac{\Omega_b}{m_p} \right) (1 - Y) (1+z)^3$ and \(Y \approx 0.25\) is the primordial helium fraction by mass. The neutral fraction of hydrogen, \(x_{\text{HI}}\), is then given by:

\[
x_{\text{HI}} \approx \frac{\alpha_A(T) n_{\rm H} f_e} {\Gamma} \approx 9.5 \times 10^{-6} \Delta_b f_e (\frac{T}{10^4 \, \text{K}})^{-0.72} (\frac{\Gamma}{10^{-12} \, \text{s}^{-1}})^{-1} (\frac{\Omega_b h^2}{0.022}) (\frac{1 + z}{5})^3,
\]

where \(f_e\) is the fraction of electrons per hydrogen atom, which is \(f_e \approx 1.174\) after He~{\sc II} reionization and \(f_e \approx 1.087\) before He~{\sc II} reionization. The redshift dependence comes from cosmic expansion. At redshifts \(2 < z < 5\), hydrogen in the IGM is highly ionized by the metagalactic UV background, but a small fraction of neutral hydrogen, \(x_{\text{HI}} \sim 10^{-5}\), remains.

The temperature of the IGM is primarily regulated by photoionization heating, adiabatic cooling due to cosmic expansion, and Compton cooling with the CMB. This thermodynamic behavior resembles cooling processes in galaxy formation, though in that context, neither Compton nor adiabatic cooling tends to dominate. In gas within halos, elevated temperatures from shock heating and higher densities allow for efficient cooling via collisional processes. In contrast, the IGM, with its low density and primordial composition, experiences heating and cooling mainly through one-body processes like Compton cooling and photoionization. The temperature evolution for a Lagrangian gas element can be described as (see \citep{mcquinn2016} for a review):

\begin{equation}
\frac{dT}{dt} = -2 H T + \frac{2T}{3\Delta_b} \frac{d \Delta_b}{dt} + \frac{2(\mathcal{H}-\mathcal{C})}{3 k_{\rm B} n_b}.
\end{equation}

where $\mathcal{H}$ and $\mathcal{C}$ are the heating and cooling rates per volume (in units of erg \(\text{s}^{-1} \text{cm}^{-3}\)). The first term represents adiabatic cooling due to cosmological expansion, while the second term represents adiabatic heating or cooling from local compression or expansion. The temperature quickly establishes a power-law relation between temperature and density, $T = T_0 \Delta_b^{\gamma - 1}$, due to the combined effects of photoionization, Compton cooling, and adiabatic cooling. The energy per photoionization, \(\langle E_{\text{HI}} \rangle\), is then, $\langle E_{\text{HI}} \rangle \propto h \nu_{\rm L}$, where \(\nu_{\rm L}\) is the frequency at the ionization threshold. The heating rate due to photoionization is, $H = \langle E_{\text{HI}} \rangle \Gamma n_{\text{HI}}$. Since the neutral hydrogen fraction is set by photoionization equilibrium, $n_{\text{HI}} \approx \frac{\alpha_A(T) n_{\rm H}^2}{\Gamma}$, the heating of the IGM by the photoionization of residual neutral hydrogen leads to the evolution of temperature:
where  and \(C\) are the heating and cooling rates per volume (in units of erg \(\text{s}^{-1} \text{cm}^{-3}\)). The first term represents adiabatic cooling due to cosmological expansion, while the second term represents adiabatic heating or cooling from local compression or expansion. The temperature quickly establishes a power-law relation between temperature and density, $T = T_0 \Delta_b^{\gamma - 1}$, due to the combined effects of photoionization, Compton cooling, and adiabatic cooling. The energy per photoionization, \(\langle E_{\text{HI}} \rangle\), is then, $\langle E_{\text{HI}} \rangle \propto h \nu_{\rm L}$, where \(\nu_{\rm L}\) is the frequency at the ionization threshold. The heating rate due to photoionization is, $H = \langle E_{\text{HI}} \rangle \Gamma n_{\text{HI}}$. Since the neutral hydrogen fraction is set by photoionization equilibrium, $n_{\text{HI}} \approx \frac{\alpha_A(T) n_{\rm H}^2}{\Gamma}$, the heating of the IGM by the photoionization of residual neutral hydrogen leads to the evolution of temperature: $\frac{dT}{dt} = \frac{2\mathcal{H}}{3 k_{\rm B} n_b} \approx \frac{2 \langle E_{\text{HI}} \rangle \alpha_A(T) n_{\rm H}}{3 k_{\rm B}}$.

The asymptotic temperature--density relation arises from the competition between cosmological adiabatic cooling and photoionization heating of residual neutral hydrogen. This interplay yields the scaling \( T \propto T^{-0.72} \Delta_b \), which translates into a temperature--density relation of the form \( T = T_0 \Delta_b^{\gamma - 1} \propto \Delta_b^{1/1.72} \). Consequently, the asymptotic value of the slope is \( \gamma = \frac{1}{1.72} + 1 \approx 1.58 \), a value that aligns well with both observational data and hydrodynamical simulations \citep{bolton2014,boera2016}. However, this derivation is not fully comprehensive, as it omits the local adiabatic heating and cooling experienced by gas elements. During the process of structure formation, low-density regions can undergo preferential adiabatic cooling, which follows a scaling of \( T \propto \Delta_b^{2/3} \). This additional effect contributes to shaping the slope of the temperature--density relation. Fortunately, more rigorous analyses produce results that are very close to the simplified estimate provided above (see \citealt{mcquinn2016} for further elaboration). Although adiabatic expansion and photoheating set the asymptotic slope, the rate at which the intergalactic medium reaches this equilibrium is influenced significantly by Compton cooling. The timescale for this process is given by \( t_{\text{comp}} = \frac{3 m_e c}{4 \sigma_T a_R T_{\text{CMB}} (1+z)^4} \), and can be compared to the characteristic timescale for adiabatic cooling due to the Hubble expansion, \( t_{\text{ad}} = \frac{1}{2 H(z)} \), with their ratio being approximately \( \frac{t_{\text{comp}}}{t_{\text{ad}}} \approx 1.3 \left( \frac{\Omega_m h^2}{0.142} \right)^{1/2} \left( \frac{1+z}{7} \right)^{-5/2} \).

Compton cooling is comparable to adiabatic cooling at \(z = 6\). The combined effects of Compton scattering and cosmological expansion quickly cool the IGM and erase the memory of photoionization heating by HI reionization (i.e., the first passage of HI ionization fronts), when the slope was nearly isothermal, \(\gamma \approx 1\), with dispersion. The IGM temperature eventually approaches the tight asymptotic \(T - \Delta_b\) relation. Photoionization heating by the UV background causes Jeans pressure smoothing in the IGM. Before photoionization heating by reionization, the temperature of the IGM is low, \(T \sim 10\) K in the absence of pre-heating \citep{seager1999}. Therefore, the Jeans length is as low as \(\sim 3.4\) pkpc at \(z = 10\). This means that the IGM is more clumpy before the passage of ionization fronts. The photoionization heating helps to keep the Universe reionized by reducing the clumping factor \citep{pawlik2010}.

\section{Observational Probes of Cosmic Reionization}
\label{eorobserve}

Some of the first major clues about reionization come from two key observations: the CMB and high-redshift quasar spectra. The CMB provides the first evidence through the measurement of the Thomson optical depth (\(\tau\)), which tells us how much ionized gas lies between us (at \(z = 0\)) and the surface of last scattering (at \(z \approx 1100\)). This is determined by the scattering of CMB photons off free electrons, with the latest Planck data giving \(\tau = 0.066 \pm 0.016\), translating to a reionization redshift of $z_{\text{re}} = 8.8^{+1.7}_{-1.4}$ assuming reionization happened suddenly \citep{planck2016}. This result agrees with earlier measurements from WMAP, which found slightly higher values \citep{hinshaw2013}.  The second key evidence comes from studying the light from distant quasars. As UV photons from QSOs travel toward us, they are redshifted and absorbed by neutral hydrogen through Ly\(\alpha\) scattering at a wavelength of 1216 Å. The Gunn-Peterson optical depth (\(\tau_{\alpha}\)), introduced by \citealt{Gunn1965}, quantifies this absorption and helps estimate the amount of neutral hydrogen in the IGM. Observations of QSO spectra between \(5 < z < 6.5\) show that the Gunn-Peterson optical depth increases sharply around \(z \sim 5 - 6\), suggesting that neutral hydrogen was becoming more abundant at this time. Since Ly\(\alpha\) absorption is extremely sensitive, even a tiny neutral fraction of $x_{\text{HI}} \sim 10^{-4}$ is enough to cause complete absorption, meaning that these observations only reveal the final stages of reionization rather than its full history. Additionally, absorption features that appear redward of the Gunn-Peterson trough in QSO spectra, known as the Ly$\alpha$ forest, caused by neutral hydrogen clouds along the line of sight, serve as valuable tools for studying the evolution of the neutral hydrogen fraction and the distribution of residual neutral regions \citep{mesinger2010,greig2017}. Since these clouds respond to the UV background (UVB) radiation produced by the ionizing sources, they can act as indicators of the thermal state of the IGM during the late stages of reionization. By measuring the thermal properties of the Ly$\alpha$ forest, astronomers can gain insight into the timing and spectral characteristics of ionizing sources \citep{hui&gnedin1997,Schaye2000,Bolton2010,BeckerBolton2013}.

Astronomers study cosmic reionization by observing Ly$\alpha$-emitting galaxies, whose light interacts with the surrounding IGM \citep{dijkstra2014,dijkstra2016}. As Ly$\alpha$ photons travel through space, they are absorbed and scattered by neutral gas, making them valuable tracers of the Universe’s evolving ionization state (\citealt{chang2023,chang2024}, but see also \citealt{almada2024}). By comparing the expected intrinsic Ly$\alpha$ flux from galaxies to the observed flux after traveling through the IGM, scientists estimate the fraction of neutral hydrogen present at different cosmic epochs. Such studies suggest that at redshift $z \sim 7$, approximately 40\% of the IGM remained neutral, implying that reionization was a rapid and relatively late process \citep{dijkstra2011,jensen2012,hoag2019}. The spatial distribution and clustering of Ly$\alpha$ emitters provide additional insights into the patchiness of reionization. UV-bright galaxies are thought to reside in large ionized bubbles, while UV-faint galaxies are more likely to be found in regions with residual neutral gas \citep{mcquinn2007}. If reionization were uniform, the clustering pattern of Ly$\alpha$ emitters would follow that of the underlying galaxy distribution. However, deviations from this pattern could indicate large-scale variations in the ionization state of the IGM. The evolution of galaxy populations also sheds light on reionization. The UV luminosity function of Lyman-break galaxies (LBGs) steadily increases from $z \sim 8$ to $z \sim 4$, reflecting the gradual hierarchical buildup of star-forming galaxies over cosmic time \citep{bouwens2015}. In contrast, the Ly$\alpha$ luminosity function remains nearly constant between $z \sim 3$ and $z \sim 5.5$ but declines at $z \sim 7$, likely due to increased absorption by neutral hydrogen \citep{ouchi2008,konno2014,santos2016,dela2019,taylor2020,khusanova2020,taylor2021,morales2021,runnholm2025}. 
Interestingly, despite a decrease in the overall number of galaxies at higher redshifts, the fraction of Ly$\alpha$ photons escaping from galaxies appears to increase with redshift \citep{hayes2011}. This suggests a complex interplay between galaxy formation, dust content, and the surrounding IGM, where Ly$\alpha$ photons become more visible as dust levels decrease, potentially balancing the effects of a more neutral IGM.  The diversity of Ly$\alpha$ line profiles in LBGs provides another crucial piece of the puzzle. Some galaxies exhibit strong Ly$\alpha$ emission, while others show Ly$\alpha$ absorption, revealing variations in the physical conditions of galaxies and their environments. Observations indicate that the fraction of LBGs showing Ly$\alpha$ emission increases from $z = 2$ to $z = 6$ but then declines beyond $z > 6$, coinciding with a rise in IGM opacity \cite{dijkstra2016}. Moreover, UV-fainter galaxies tend to exhibit stronger Ly$\alpha$ emission than their brighter counterparts, likely because they contain less dust, allowing Ly$\alpha$ photons to escape more easily \citep{ando2006,vanzella2009}. These findings reinforce the idea that the transmission of Ly$\alpha$ photons is influenced by both the internal properties of galaxies and the large-scale ionization state of the Universe.

A more direct method for studying reionization is the 21 cm line of neutral hydrogen, which originates from the spin-flip transition of hydrogen atoms \citep{ewen1951,field1959,scott1990,furnaletto2006}. This spectral line serves as a unique probe of the neutral component of the IGM, as it remains observable even at the highest redshifts. 
By examining the differential brightness temperature of the 21 cm line, which depends on the spin temperature relative to the CMB temperature, as well as the local density and neutral fraction, astronomers can place stringent constraints on the evolution of the neutral hydrogen fraction and the nature of ionizing sources \citep{zaroubi2013}. In particular, variations in the 21 cm brightness temperature can reveal information about the X-ray and Ly$\alpha$ emissivity of the earliest sources, including stars, galaxies, and AGNs \citep{mirocha2019,qingbo2022} . Several ongoing and upcoming radio experiments, such as PAPER \citep{parsons2010}, LOFAR \citep{vanharleem2013}, MWA \citep{bowman2013}, HERA \citep{neben2016}, and SKA \citep{dewdney2009}, are dedicated to detecting the 21 cm signal and advancing our understanding of the early Universe.

The \texttt{JWST} has significantly advanced our knowledge of the early Universe, particularly in the study of galaxies and AGN up to redshift \( z \approx 14 \). Its broad wavelength coverage (\( 1 - 28 \, \mu m \)) and high resolution enables precise observations of rest-frame UV and optical emission lines, making galaxy redshift confirmation easier. \texttt{JWST} is providing detailed insights into galaxy properties, such as stellar mass \citep{barrufet2023,gottumukkala2024,wang2024}, star formation histories \citep{whitler2024,endsley2024,atek2023,simmonds2024,looser2023}, ionization states of ISM \citep{rinaldi2023,reddy2023,sanders2023}, metallicities \citep{vanzella2023,nakajima2023,morishita2023,curti2024,deugenio2025}, outflow characteristics \citep{fujimoto2023,carniani2024}, ionizing efficiencies \citep{prieto2023,endsley2024,simmonds2024,borsani2024} and structural and kinematic properties \citep{kartaltepe2023,treu2023,huertas2024,ito2024,degraaff2024}, all at unprecedented resolutions for galaxies at \( z > 6 \). These observations are critical for understanding the role of galaxies in cosmic reionization, particularly through measurements of ionizing photon production down to \( M_{UV} = -15.5 \) \citep{atek2023}. Additionally, \texttt{JWST}'s ability to probe ionized bubbles in the IGM through galaxy-IGM correlations \citep{garaldi2022,kashino2023} and Ly\( \alpha \) emitters \citep{saxena2024,napolitano2024,chen2024,Tang2024} is shedding light on the size and distribution of these bubbles. However, there are still uncertainties due to instrument systematics and the interpretation of some observables \citep{bhagwat2024a,narayanan2024,narayanan2025}, underscoring the need for comparisons with simulations to better constrain our understanding of the high-redshift Universe and its evolution.

\citealt{zahn2012} suggested an alternative method to constrain the reionization history by examining the kinetic Sunyaev-Zel'dovich (kSZ) effect in the CMB (see \citealt{reichardt2016} for a review). This approach is particularly useful because it is independent of the Ly$\alpha$ transfer physics that influence the interpretation of both Ly$\alpha$ emitting galaxies and QSO absorption spectra discussed earlier.
The Sunyaev-Zel'dovich effect introduces secondary anisotropies and spectral distortions in the CMB due to the scattering of CMB photons by free electrons after recombination. The kSZ effect, specifically, is driven by the motion of ionized gas relative to the CMB rest-frame, creating hotspots or cold spots depending on whether the gas is moving toward or away from us.

\section{Outline of the Thesis}
\label{outline}

This thesis explores key astrophysical processes governing the ionization and thermal evolution of the Universe through four interconnected projects. These investigations collectively examine how the properties of ionizing sources shape both early galaxies and the IGM, ultimately influencing observable signatures throughout cosmic reionization. By combining theoretical modeling with observational data, this work addresses the variability of the UV luminosity function (UVLF) of early galaxies, the influence of spectral energy distribution (SED) assumptions on the Ly$\alpha$ forest, the role of quasars in helium reionization, and the observational potential of the $^{3}\mathrm{He}^{+}$ hyperfine transition line. Together, these topics offer critical insights into the formation mechanisms of the first galaxies, the structure of the IGM, and observational strategies for probing the epoch of reionization.

The first part of the thesis (Chapter \ref{chap:chapter2}) focuses on the impact of ionizing source characteristics, particularly SN driven stellar feedback mechanisms, on high-redshift galaxy observables. Using a suite of radiation-hydrodynamic simulations with varying SN feedback models, it investigates how feedback affects star formation histories and UV luminosity over time. This approach allows for a better understanding of the variability in the UVLF, a possible explanation to alleviate the bright galaxy tension observed by JWST at high redshifts, and it contributes to refining models of galaxy formation and evolution during the early Universe.

Building on this, the second part (Chapter \ref{chap:chapter3}) examines how the properties of ionizing sources imprint themselves on the IGM during the final phases of hydrogen reionization. By analyzing how different SED assumptions influence ionization structures and absorption features in the Ly$\alpha$ forest, the study reveals the importance of SED in interpreting the observational signatures of late phase of hydrogen reionization.

The next section of this thesis (Chapter \ref{chap:chapter4}) shifts focus to helium reionization, specifically the role of quasars in driving the transition from He~{\sc II} to He~{\sc III}. Incorporating recent constraints on the quasar luminosity function (QLF), the thesis employs post-processed cosmological radiative transfer simulations to trace the spatial and temporal progression of helium reionization, offering a clearer picture of the second major reionization event in the Universe. This chapter also focuses on impact of higher abundance of quasars (detected by JWST at $z>5$) onto helium reionization and establish their importance in driving this epoch.

Finally, the final part of the thesis (Chapter \ref{chap:chapter5}) explores the prospects of directly observing the $^{3}\mathrm{He}^{+}$ hyperfine transition line as a probe to constrain the properties of IGM during and at the end of cosmic helium reionization. This section assesses the detectability of this faint signal with future instruments and discusses its potential to complement other observational probes of cosmic reionization.

By tracing the impact of ionizing source properties from galaxy-scale observables to large-scale IGM signatures, this thesis provides an integrated view of cosmic reionization and the early Universe. The findings not only enhance our theoretical understanding but also support the development of strategies for interpreting high-redshift observations in the coming years.

\chapter{Variability of the UV luminosity function of high $z$ galaxies}
\label{chap:chapter2}

\begin{flushright}
\begin{minipage}{0.5\textwidth}
\raggedleft
\textbf{\textit{``All simulations are wrong, until they're not."}}\\[1ex]
\noindent\rule{0.5\textwidth}{0.4pt}\\[-0.2ex]
Anshuman Acharya
\end{minipage}
\end{flushright}

\begin{flushright}
\begin{minipage}{0.7\textwidth}
\raggedleft
\textbf{\textit{``May the flowers remind us why rain is necessary."}}\\[1ex]
\noindent\rule{0.5\textwidth}{0.4pt}\\[-0.2ex]
Katyayani Trivedi
\end{minipage}
\end{flushright}


\textit{This work has been accepted for publication in the Monthly Notices of the Royal Astronomical Society. \citep{Basu2025a}}

\hspace{1cm}

The variability of the UVLF of early galaxies, as observed by \texttt{JWST}, has become a focal point in understanding galaxy formation. \texttt{JWST}’s high-redshift observations \citep{castellano2022,finkelstein2022,naidu2022,adams2023,morishita_stiavelli2023,bouwens2023a,bouwens2023b,donnan2023,atek2023,perez2023,willott2024,Donnan2024,helton2024} reveal a highly irregular and clumpy nature of galaxies \citep{bournard2007,elmegreen2009,forster2011,true2023}, consistent with stochastic star formation histories, where star formation rates vary significantly over time \citep{sparre2017,smit2016,emami2019,iyer2020,tachhella2020,flores2021,fukushima2022,inayoshi2022,hopkins2023,harikane2023}. This variability in the UVLF is driven by a combination of factors such as gas inflows and outflows, gravitational instabilities, galaxy mergers, and intense feedback processes \citep{dekel2009,ceverino2010,angles2017, elbadry2016,tacchella2016}. The feedback, whether due to active galactic nuclei or supernovae, can disrupt star formation or trigger bursts of star formation, further contributing to fluctuations in the UVLF \citep{sparre2017,furlanetto2022,sun2023}. Additionally, some galaxies, particularly at high redshifts, exhibit feedback-free starbursts, leading to even more pronounced variability in their UV output \citep{faucher2018,Dekel2023}. The interplay of these physical processes, highlighted by \texttt{JWST}'s observations, is crucial in understanding the dynamic nature of early galaxy evolution and the resulting variability in the UVLF. In this chapter, I investigate the impact of SFR variability on the UVLF of high-$z$ galaxies using the suite of radiation hydrodynamic simulations \texttt{SPICE} \citep{bhagwat2024} (hereafter B24), which comprises three different implementations of SN feedback (while the rest of the simulation setup is the same), resulting in a different level of burstiness and  in a variety of star formation histories. This allows me to systematically quantify the SFR and UVLF variability.

I introduce the simulations  in Section \ref{section:2}, the results are presented in Section \ref{section:3}, while in Section \ref{section:4} I summarise my conclusions and future prospects. Throughout the chapter, I adopt a flat $\Lambda \rm{CDM}$ cosmology consistent with \cite{planck2016} with $\Omega_{\rm m}=0.3099$, $\Omega_{\Lambda}=0.6901$, $\Omega_{\rm b}=0.0489$, $h=0.6774$, $\sigma_{8}=0.8159$ and $n_{\rm s}=0.9682$, where the symbols have their usual meaning.

\section{The \texttt{SPICE} simulation suite}
\label{section:2}
In order to investigate the impact of SN feedback onto galaxy properties, I post-process the suite of radiation-hydrodynamical simulations \texttt{SPICE}, which I briefly introduce below.

\texttt{SPICE} has been performed with \texttt{RAMSES-RT} \citep{rosdahl2013,rosdahl2015}, the radiation-hydrodynamics extension of the Eulerian Adaptive Mesh Refinement (AMR) code \texttt{RAMSES} \citep{teyssier2002}. The simulations target a cubic box of size 10 $\rm{\mathit{h}^{-1} cMpc}$ with $512^{3}$ dark matter particles of mean mass $m\rm{_{dm}} = 6.38 \times 10^{5} \mathrm{M_{\odot}}$. The AMR allows each parent gas cell to split into 8 smaller cells when certain conditions are met (see \citealt{bhagwat2024}). 
As a reference, this provides a physical resolution at $z=5$ ranging from 4.8 kpc for the coarsest level ($\ell_{\text{min}} = 9$) to approximately 28 pc for the finest level ($\ell_{\text{max}} = 16$). 
The hydrodynamical equations are solved using a second order Godonov scheme and the dynamics of collisionless dark matter and stellar particles are computed by solving the Poisson equation with a particle-mesh solver, projecting onto a grid using the Cloud-in-Cell  scheme \citep{guilet2011}. Gas cooling and heating are accounted for as in \citealt{rosdahl2013}. Star-formation has been modelled following \citealt{kretschmer2020}, which adopt a subgrid turbulence model depending on the physical conditions of each computational cell including parameters like the local gas turbulent Mach number and the virial parameter, which acts as a measure of the gravitational balance in the region, and which results in a spatially-variable star formation efficiency. Additionally, \texttt{SPICE} includes a radiative transfer scheme with five frequency groups, i.e. infrared (0.1-1 $\rm{eV}$), optical (1-13.6 $\rm{eV}$), and three UV (13.6-24.59 $\rm{eV}$; 24.59-54.42 $\rm{eV}$; 54.42-$\infty$ $\rm{eV}$).

To correctly capture the momentum injection into the cells due to supernova feedback, the scheme presented in \citealt{kimm2015} is adopted. Keeping all the parameters for momentum injection constant, different simulations have been run based on the timing of supernova events and the amount of energy injected. I briefly present the different feedback models as follows:
\begin{itemize}
    \item "\textbf{\texttt{bursty-sn}}": each stellar particle experiences a single event of SN explosions at a fixed timescale of 10 Myr, equivalent to the mean time at which SN occurs for the adopted \citealt{chabrier2003} initial mass function. An energy of $2 \times 10^{51}$ ergs per SN explosion is injected into the neighbouring cells.
    \item "\textbf{\texttt{smooth-sn}}": for a given stellar population, a SN with mass $> 20 \rm{M_{\odot}}$ can explode as early as 3 Myr, whereas SN with progenitor mass of $8 \rm{M_{\odot}}$ can explode as late as 40 Myr. An energy of $2 \times 10^{51}$ ergs per SN explosion is injected into the neighbouring cells.
    \item "\textbf{\texttt{hyper-sn}}": SN explosions happen at different times as in \texttt{smooth-sn}, with the injected energy extracted stochastically from a normal distribution centered at $1.2 \times 10^{51}$~ergs and ranging from $10^{50}$~ergs to $2 \times 10^{51}$~ergs (see also, \citealt{sukhbold2016,diaz2018,diaz2021}). Additionally, a metallicity dependent fraction of SN explodes as Hypernovae (HN), with an energy injection of $10^{52}$ ergs.
\end{itemize}

Dust in \texttt{SPICE} is not modelled as an active ingredient but rather through a direct dependence on the local gas-phase properties, i.e. its metallicity and ionization state. Following \citealt{Nickerson2018}, the dust number density is computed as
$n_{\rm d} = (Z/Z_\odot)  f_{\rm d} n_{\rm H}$, 
where $ Z $ represents the local metallicity, $ f_{\rm{d}} = 1 - x_{\rm{HII}} $ is the fraction of neutral hydrogen, and $ n\rm{_{H}} $ is the hydrogen number density. 
This relation ensures that the dust density depends on both chemical enrichment and the local ionization conditions.
Indeed, despite expectations of dust being typically destroyed in photo-ionized gas, observations have shown dust-to-ionized-gas ratios as high as 0.01 \citep{harper1971,contini2003,laursen2009}. Thus, the quantity $f_{\rm{d}}$ permits non-zero dust effects even in partially ionized environments.
The propagation of UV photons influences the gas via photo-ionization and photo-heating, but also through radiation pressure generated from both photo-ionization and through dust absorption. Infrared and optical photons interact with gas solely through radiation pressure on dust particles. 
Each photon group is characterized by specific dust absorption and scattering properties. When optical and UV photons are absorbed by dust, their energy is transferred to the infrared group, which later interacts with the gas through multi-scattering radiation pressure.

For a more comprehensive description of the simulations and the supernova feedback models discussed above, I refer the readers to B24. 

\section{Results}
\label{section:3}
In this Section, I present results from my analysis of how different SN feedback models affect the SFR and UVLF. 

\subsection{UV luminosity function}
\label{section:3.1}
\begin{figure}
\centering
    \includegraphics[width=60mm]{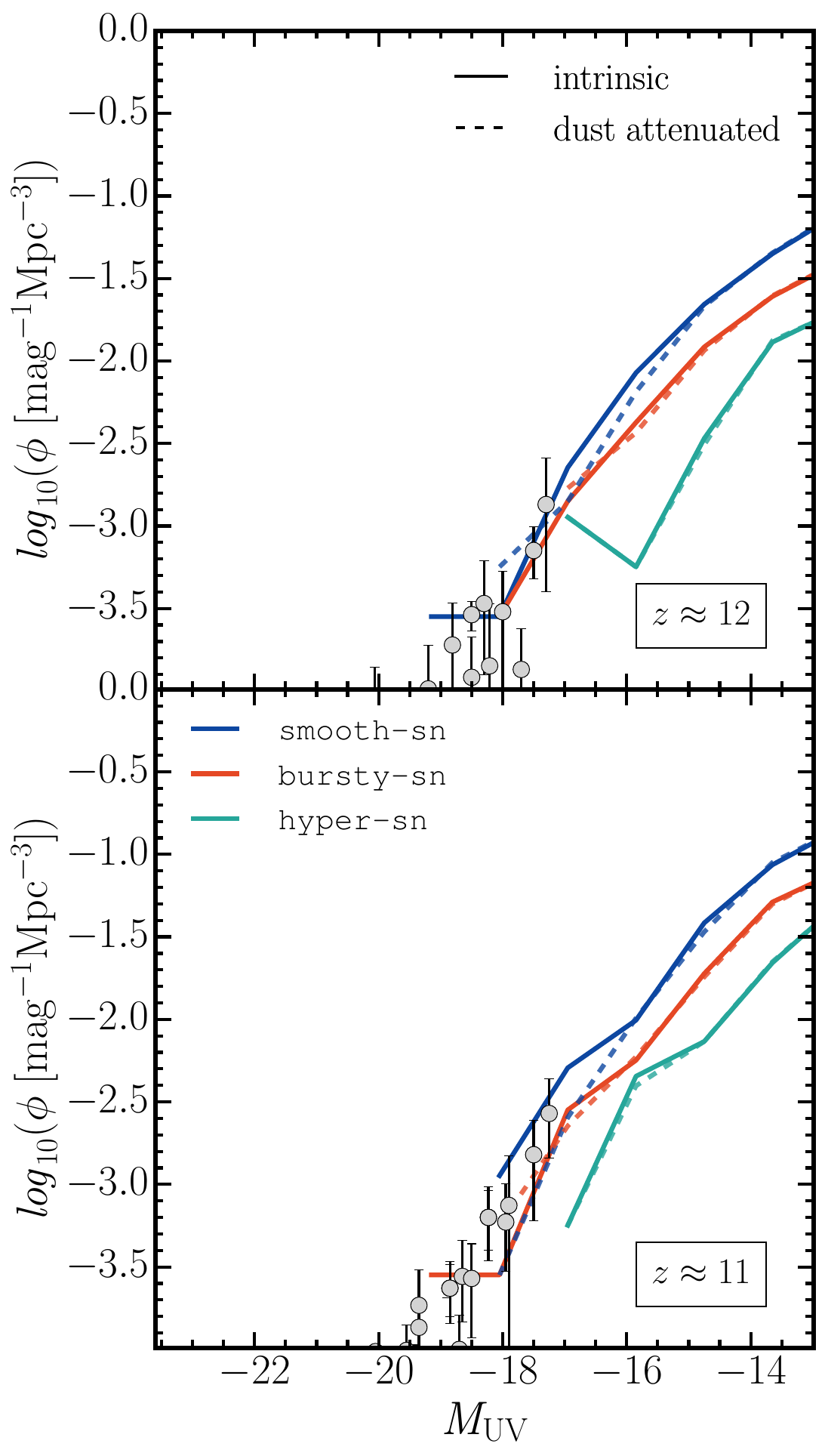}
    \includegraphics[width=61mm]{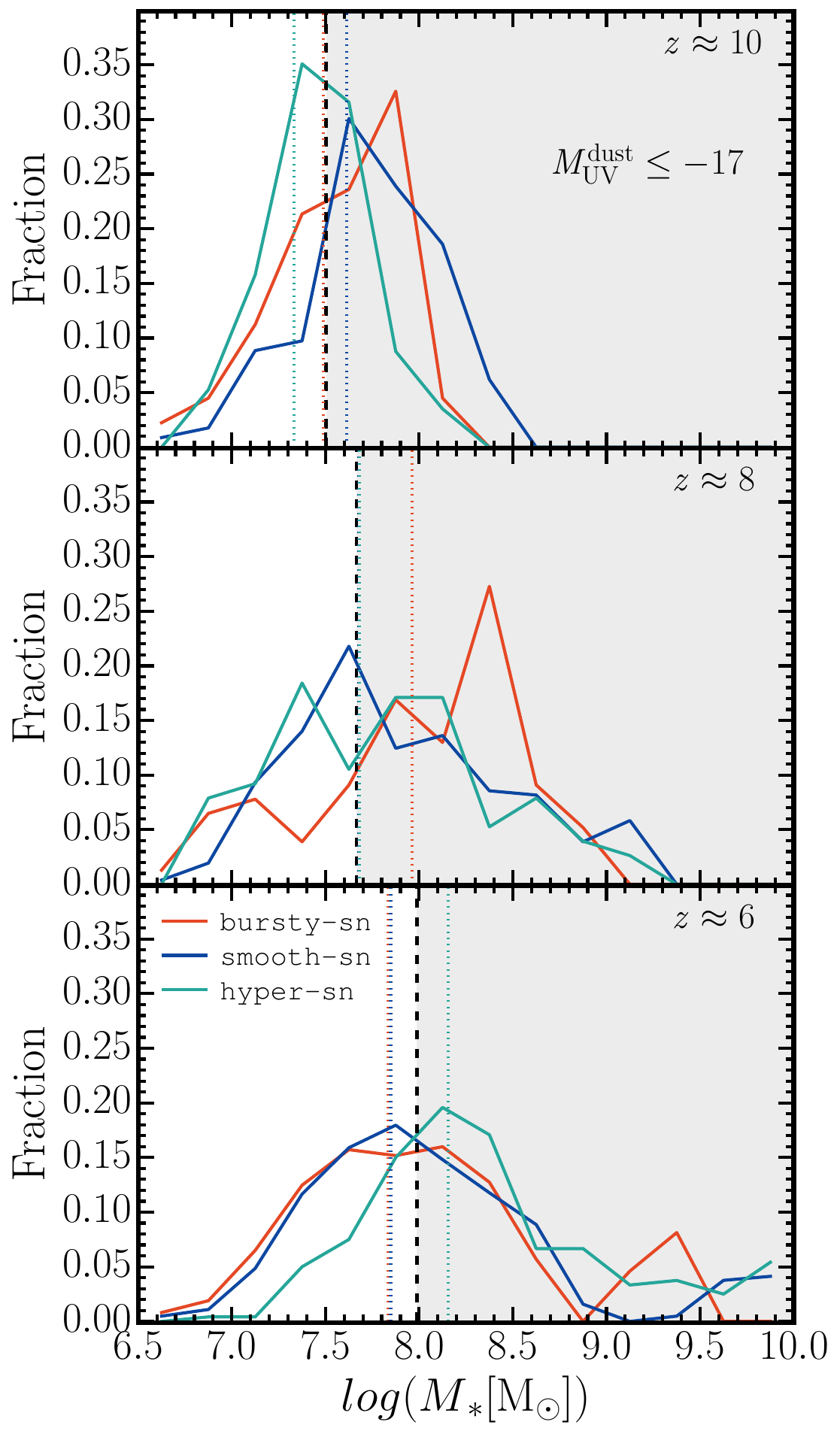}
  \caption{\textit{Left panel :}1500~$\si{\angstrom}$ luminosity functions at $z$ = 12 (\textit{top panel}) and 11 (\textit{bottom panel}) for three feedback models (in different colors). Solid and dashed curves refer to intrinsic and dust attenuated LFs, respectively. A compilation of observations from \texttt{HST} and \texttt{JWST} \citep{bouwens2015,harikane2022,naidu2022,adams2023,harikane2023,bouwens2023a,bouwens2023b,leung2023,donnan2023a,donnan2023b,perez2023,casey2024,robertson2024,mcleod2024,Whitler2025} is shown as gray data points.
  \textit{Right panel : }Distribution of stellar mass of objects with $\rm{\mathit{M}_{UV}^{dust}} \leq -17$ at  $z=10$ (\textit{top panel}), 8 (\textit{middle}) and 6 (\textit{bottom}). The corresponding median stellar masses are shown as vertical dotted lines, while the black dashed lines refer to the minimum stellar mass required for having such bright objects assuming the median SFE model of \citealt{mason2015,mason2023}.}
  \label{fig:uvlf_hoststellarmass}
\end{figure}

While in B24, the results at $z\leq10$ have been shown, in left panel of Figure \ref{fig:uvlf_hoststellarmass} I present the 1500~\AA\ LF for all feedback models  at $z=11$ and 12. 
The LFs have been derived from the stellar spectral energy distribution within a 10~$\si{\angstrom}$ bin centered around 1500~$\si{\angstrom}$ . I use the SED model of BPASSv2.2.1 \citep{elridge2017,stanway2018}, assuming the \citealt{chabrier2003} IMF (details are mentioned in B24). The solid and dashed lines in the figure indicate the intrinsic and dust-attenuated luminosity functions, respectively. 
To account for dust attenuation, I utilize the Monte Carlo radiative transfer code \texttt{RASCAS} \citep{michel2020}. In this approach, 100 rays are cast from each stellar particle within a halo out to its virial radius. The dust attenuation along each ray is computed following the model described in the previous section. 

As already noted in B24, also at these redshifts \texttt{smooth-sn} produces a higher intrinsic luminosity function (by 0.3-0.4 dex) at all magnitudes. 
Conversely, the \texttt{hyper-sn} model yields the lowest galaxy count across all magnitude bins. 
The \texttt{bursty-sn} model produces results which are typically in between those of the other models. 
The effect of dust attenuation (dashed curves) starts to be visible from an absolute magnitude of $M_{1500} \approx -15$, the value changing slightly depending on feedback model and redshift. 
When comparing the dust attenuated LFs to a compilation of observations from \texttt{HST} and \texttt{JWST} \citep{bouwens2015,harikane2022,naidu2022,adams2023,harikane2023,bouwens2023a,bouwens2023b,leung2023,donnan2023a,donnan2023b,perez2023,casey2024,robertson2024,mcleod2024,Whitler2025}, I notice a fairly good match in the range of magnitudes covered by \texttt{SPICE} (except for the \texttt{hyper-sn} model), which, given the limited box size, does not extend to the brightest observed galaxies. However, to assess whether the extrapolated UVLF aligns with observations, I parametrize and fit the curves using the commonly adopted \citealt{Schechter1976} function (see Appendix \ref{Appendix-uvlffitting}). For a more detailed discussion of the LF, I refer the reader to B24. 

In right panel of Figure \ref{fig:uvlf_hoststellarmass}, I further investigate the contribution to the bright end of the UVLF (with $\rm{\mathit{M}_{UV}^{dust}} \leq -17$) from objects with different stellar masses. To obtain a statistical sample, I have derived the distributions using all the simulation snapshots within a time interval of 100 Myr centered on each redshift shown. I observe that, for all feedback models, the stellar masses are always spread over a wide range, which extends to increasingly larger masses as redshift decreases, indicating that even objects with stellar masses as low as $10^{(6.5-7)} \rm{M_{\odot}}$ can contribute to the bright end of the LF. For all models the peak of the distribution drops with time along with the distribution becoming flatter and broader. Notably, though, all models exhibit a similar distribution, although the corresponding medians are  slightly different,  
with \texttt{smooth-sn}, \texttt{bursty-sn} and \texttt{hyper-sn} having the largest median value at $z\approx10$, 8 and 6, respectively. 
For a comparison to previous studies, I also show the stellar mass threshold required for having objects with $\rm{\mathit{M}_{UV}^{dust}} \leq -17$, obtained in the semi-analytic model of \citealt{mason2015,mason2023} (assuming the median SFE model), which has been used to explain the abundance of bright galaxies  detected by \texttt{JWST} at $z>10$.  
It is evident that their estimates are very similar to the median values obtained in the simulations. 

\begin{figure}
\centering
    \includegraphics[width=120mm]{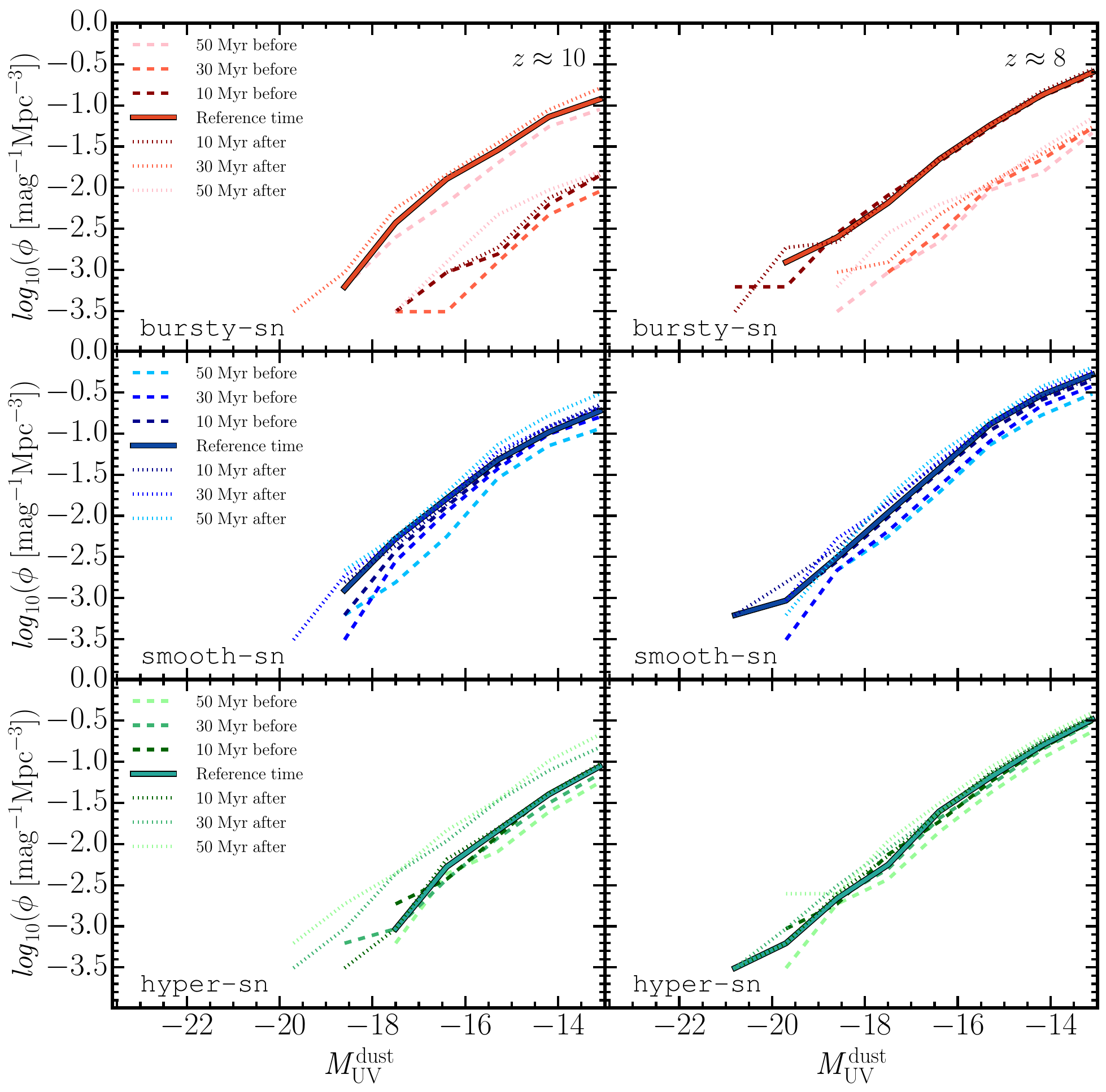}
  \caption{
Evolution of the dust-attenuated UVLF over time intervals of 10, 30, and 50 Myr (indicated by different color gradients, from dark to light), centered at $z \approx 10$ (\textit{left panel}) and 8 (\textit{right}), for three different feedback models: \texttt{bursty-sn} (\textit{top panels}), \texttt{smooth-sn} ({\it middle}), and \texttt{hyper-sn} ({\it bottom}). The dashed and dotted curves show the UVLF before and after the reference time. 
}
  \label{fig:uvlf_temporal_evol}
\end{figure}

To investigate the temporal evolution of the dust-attenuated UVLF on short timescales, in Figure \ref{fig:uvlf_temporal_evol} I present the UVLF in intervals of 10, 30 and 50 Myr around $z \approx 10$ and 8. 
While the general behaviour in the three models is similar, I find that the UVLF in the \texttt{bursty-sn} model experiences the most pronounced fluctuations, with a deviation of more than 1 dex. 
In comparison, the \texttt{smooth-sn} model exhibits a more gradual evolution of the UVLF, producing a steady increase in the number of galaxies. As already mentioned in B24, this suggests a smoother star formation process  due to the presence of gentler SN feedback. 
The \texttt{hyper-sn} model shows an intermediate behaviour at $z\approx10$, with variations up to 0.5 dex. At $z\approx8$, the variability in \texttt{hyper-sn} becomes slightly lower than in \texttt{smooth-sn}. 
In the following sections I will investigate more in detail the origin of such variability.

\subsection{Star formation rate variability}
\label{sfr_ind}

\begin{figure*}
\centering
        \includegraphics[width=110mm]{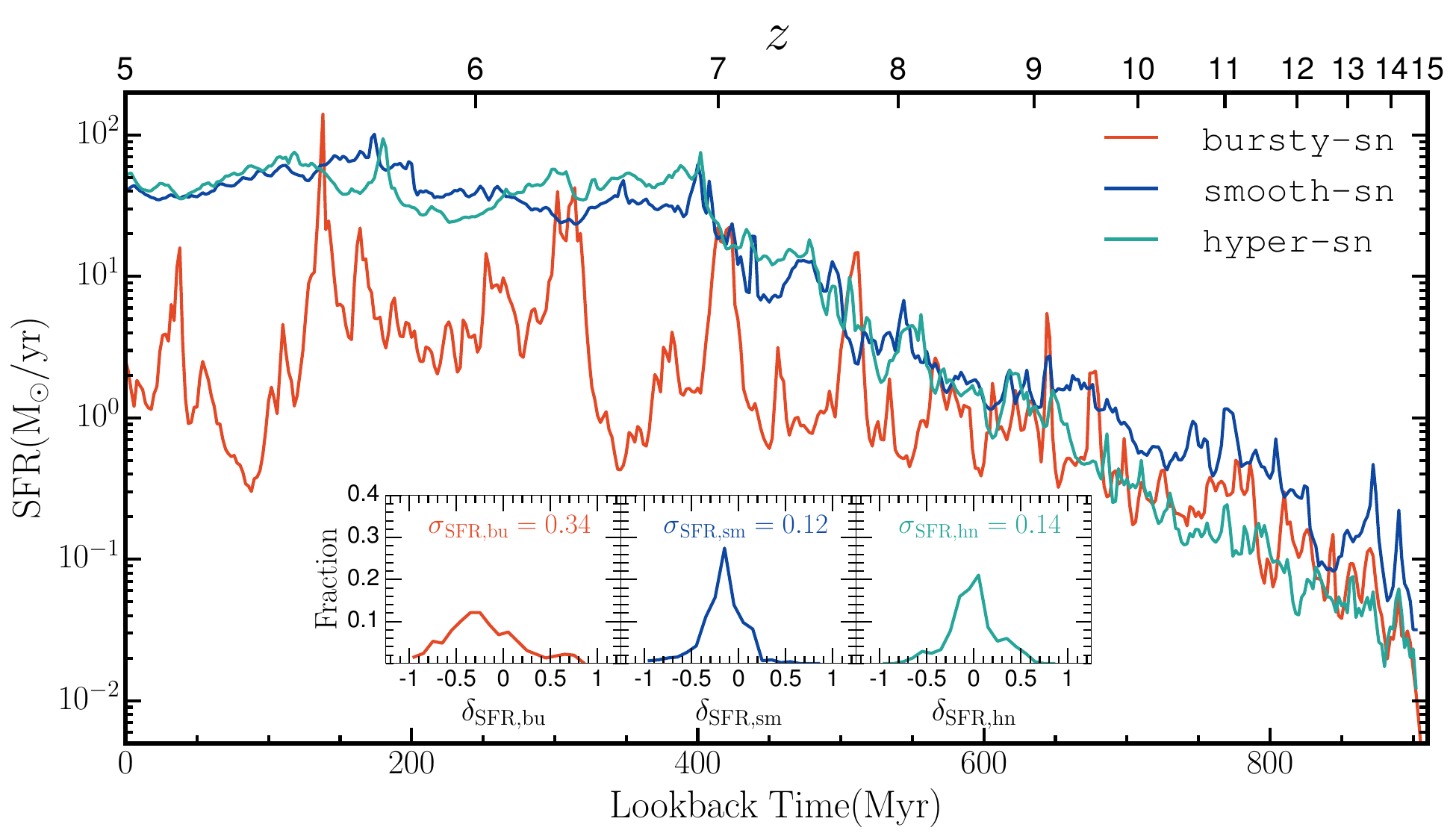}
  \caption{Temporal evolution of the SFR averaged over 2 Myr accounting for all stellar particles inside the virial radius of the most massive halo at $z = 5$, which has a total stellar mass of $2.1 \times 10^{9} \rm{\mathrm{M_{\odot}}}$, $1.9 \times 10^{10} \rm{\mathrm{M_{\odot}}}$ and $1.7 \times 10^{10} \rm{\mathrm{M_{\odot}}}$ for the \texttt{bursty-sn}, \texttt{smooth-sn} and \texttt{hyper-sn} model, respectively. Colors refer to the different SN feedback models. In the inset, I show the distributions of $\delta\rm{_{SFR,X}}$, together with the corresponding standard deviation, $\sigma\rm{_{SFR,X}}$ (see text for the details of the calculation),where `X' = `bu', `sm', `hn' for the \texttt{bursty-sn}, \texttt{smooth-sn} and \texttt{hyper-sn} model, respectively.
  }
  \label{fig:sfr_individual}

\end{figure*}

In this section, I explore how different SN feedback models impact the variability of the SFR. Figure \ref{fig:sfr_individual} shows the star formation history of the most massive halo from \texttt{SPICE} at $z = 5$ for the three models. 
I note that at this redshift the virial mass of the halo is $2.7 \times 10^{11} \rm{\mathrm{M_{\odot}}}$ 
and the total stellar mass is
$2.1 \times 10^{9} \rm{\mathrm{M_{\odot}}}$, $1.9 \times 10^{10} \rm{\mathrm{M_{\odot}}}$ and $1.7 \times 10^{10} \rm{\mathrm{M_{\odot}}}$ in the \texttt{bursty-sn}, \texttt{smooth-sn} and \texttt{hyper-sn} model, respectively. 
It is evident that overall \texttt{bursty-sn} produces the largest fluctuations in the SFR, with deviations of up to two orders of magnitude. In contrast, the \texttt{smooth-sn} model shows smaller fluctuations.
In \texttt{hyper-sn}, the SFR is the lowest among the models
during the first 200 Myr (i.e. for lookback times larger than 700 Myr), when the impact of HN explosions is the strongest. However, at later times the SFR is similar to that of \texttt{smooth-sn}.
The fluctuations in \texttt{hyper-sn} are comparable to those in \texttt{smooth-sn}.

For a more quantitative investigation of the SFR variability, I first compute the median star-formation main sequence in different halo mass bins and redshift intervals (corresponding to bins of 100 Myr\footnote{This has been calculated from the SFR values of all the halos present in each 100 Myr time intervals and then computing the median SFR as a function of halo mass. I have also performed the same analysis using bin sizes of 30 and 50 Myr, and found no qualitative differences in my conclusions.}), SFR$_{\rm median}$.
In the inset of Figure \ref{fig:sfr_individual} I show the variability of the SFR for the same halo, defined as $\rm{\mathit{\delta}_{SFR}}={\rm log}_{10}\rm{(SFR/SFR_{median})}$, which quantifies how the SFR of the halo deviates from the median SFR computed over the entire sample of halos in the same mass bin.
In the inset of Figure~\ref{fig:sfr_individual} I show the distribution of $\delta_{\rm SFR}$ for this halo during its entire lifetime, together with the corresponding standard deviation, $\sigma_{\rm SFR}$. As expected, \texttt{bursty-sn} shows the widest distribution, with $\sigma_{\rm SFR,bu}=0.34$, which is more than double the one of \texttt{smooth-sn} with $\sigma_{\rm SFR,sm}=0.12$. \texttt{hyper-sn} produces a $\delta_{\rm SFR}$ distribution which is similar to the one from \texttt{smooth-sn}, with a slightly higher standard deviation of $\sigma_{\rm SFR,hn} = 0.14$, reflecting the stronger fluctuations present at $z>9$. 

\begin{figure}
\centering
        \includegraphics[width=72mm]{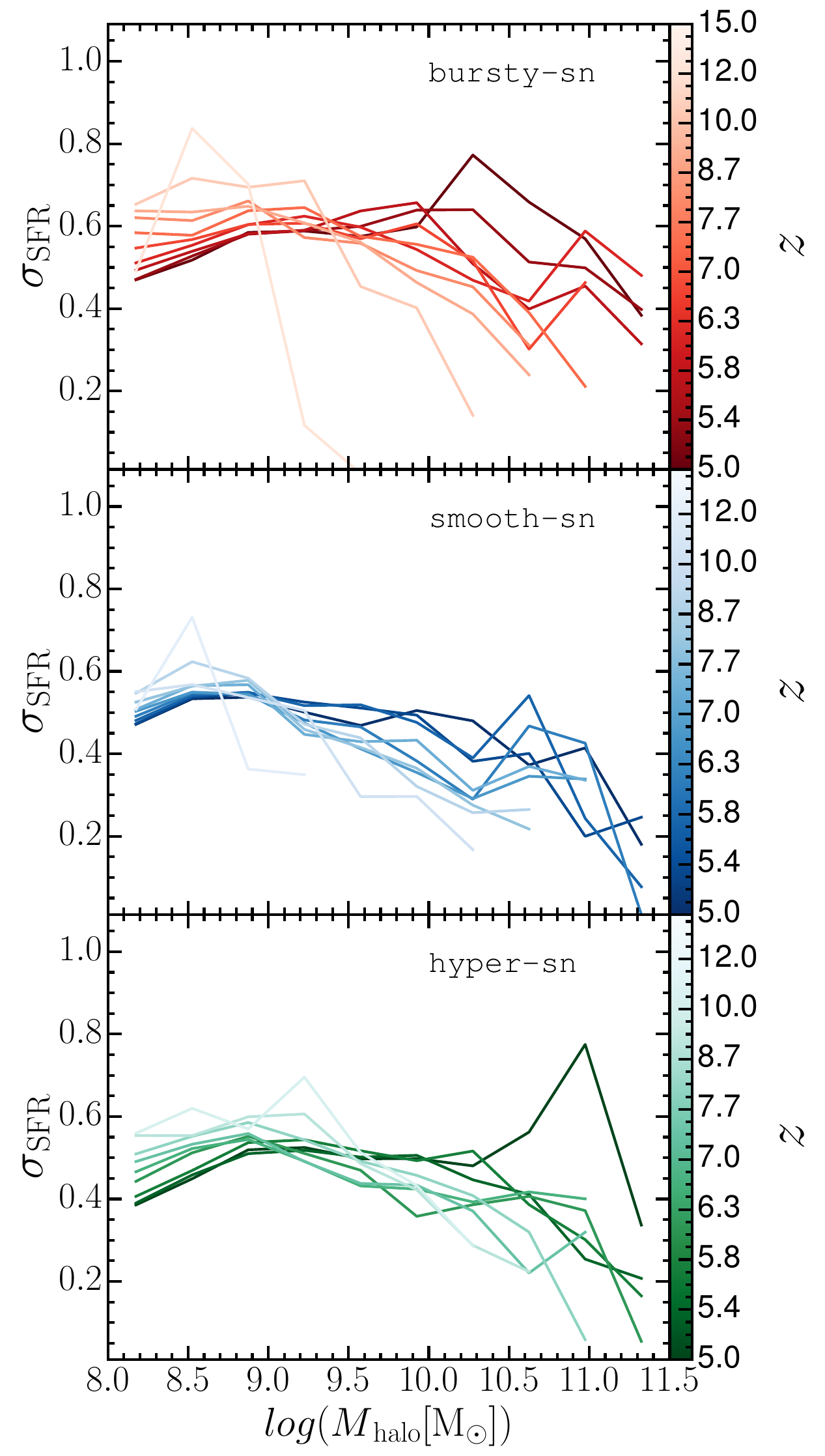}
        \includegraphics[width=70mm]{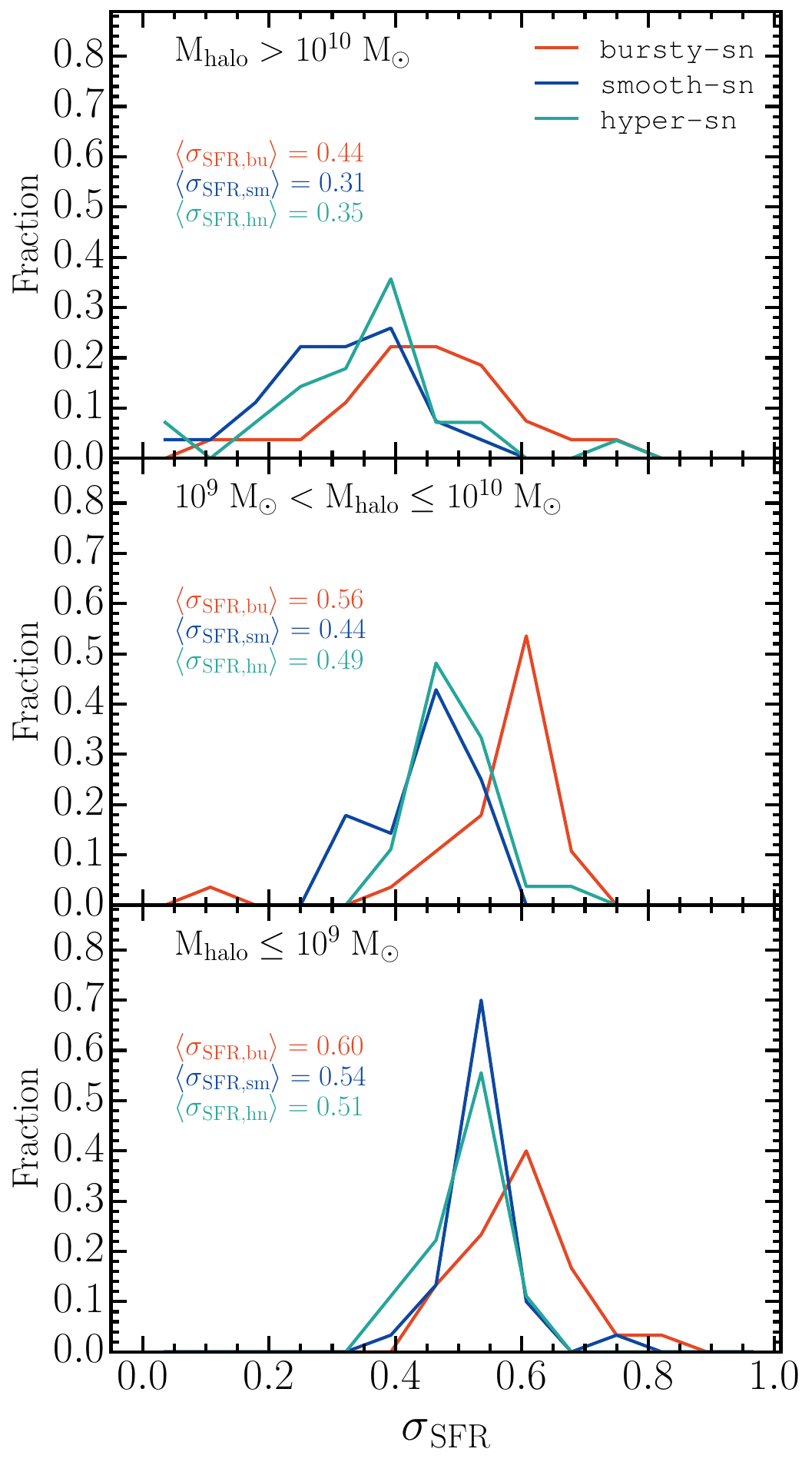}
  \caption{\textit{Left panel :}$\sigma\rm{_{SFR}}$ as a function of halo mass $M\rm{_{halo}}$ at different redshifts (indicated by the colorbar) for \texttt{bursty-sn} ({\it top panel}), \texttt{smooth-sn} ({\it middle}) and \texttt{hyper-sn} ({\it bottom}). \textit{Right panel :} Distribution of $\sigma\rm{_{SFR}}$ in different $M\rm{_{halo}}$ bins for the three SN feedback models, as indicated by the colors. Numbers refer to the corresponding average standard deviation. }
  \label{fig:sigmasfr_mhalo_bins_dist}
\end{figure}

In the left panel of Figure \ref{fig:sigmasfr_mhalo_bins_dist} I examine the mass dependence of $\sigma_{\rm SFR}$ at different redshifts.
Typically, smaller halos (${M_{\rm halo}} < 10^{9}$~ M$_\odot$) exhibit a higher degree of variability as compared to more massive ones, suggesting a stronger impact of SN feedback at lower masses.
For halos with ${M_{\rm halo}} < 10^{9}$~ M$_\odot$, all three models have a similar redshift-dependent trend, with the variability decreasing at lower redshifts. 
However, the \texttt{bursty-sn} model consistently shows the highest variability in most redshift bins, whereas \texttt{smooth-sn} and \texttt{hyper-sn} have a similar behaviour, with the latter showing a slightly higher variability. At $z<6$, all the models become comparable, although \texttt{hyper-sn} shows the lowest $\sigma_{\rm SFR}$ values.
For ${M_{\rm halo}} > 10^{10}$~M$_\odot$, the trend is reversed, i.e. the SFR variability increases as redshift decreases. In \texttt{bursty-sn},  $\sigma_{\rm SFR}$ remains consistently high and shows little dependence on halo mass at lower redshifts. The behavior of \texttt{hyper-sn}  mostly mirrors the one of \texttt{smooth-sn} except the peak at $z\approx5$ for halos with masses around $10^{10.5} \rm{M_\odot}$, which reaches the highest SFR variability among all models.

To get an overview of the statistical behaviour of $\sigma_{\rm SFR}$, in the right panel of Figure \ref{fig:sigmasfr_mhalo_bins_dist} I present the distribution of $\sigma_{\rm SFR}$ in three halo mass bins: $M_{\rm halo} > 10^{10} {\rm M}_\odot$, $10^{9} {\rm M}_\odot < M_{\rm halo} \leq 10^{10} {\rm M}_\odot$, and $M_{\rm halo} \leq 10^{9} {\rm M}_\odot$ for the entire redshift range covered in the left panel of Figure \ref{fig:sigmasfr_mhalo_bins_dist}. In the \texttt{bursty-sn} model, strong and regular fluctuations  dominate the overall population, resulting in a $\sigma\rm{_{SFR}}$ distribution skewed towards higher values at all masses in comparison to \texttt{smooth-sn}.
The \texttt{hyper-sn} model produces distributions that typically fall between those of the other two models, although they are much more similar to those of \texttt{smooth-sn}. In the highest mass bin, though, \texttt{hyper-sn} exhibits a $\sigma\rm{_{SFR}}$ even higher than in \texttt{bursty-sn}, which aligns with the pronounced peak observed in the left panel of Figure \ref{fig:sigmasfr_mhalo_bins_dist}. 
For all models, the $\sigma_{\rm SFR}$ distributions are wider in the highest mass bin, indicating a broader range of star formation variability. As I move to lower mass halos, the distributions become progressively narrower, indicating a smaller range of variability but centered around higher $\sigma_{\rm SFR}$ values. Additionally, as noted earlier in left panel of Figure \ref{fig:sigmasfr_mhalo_bins_dist}, the distributions for all models systematically shift towards higher $\sigma_{\rm SFR}$ values as halo mass decreases.

In the following section, I will explore how this mass and redshift-dependent SFR variability influences the variability of the UVLF.

\subsection{Variability of the  UV luminosity function}
\begin{figure}
\centering
        \includegraphics[width=72mm]{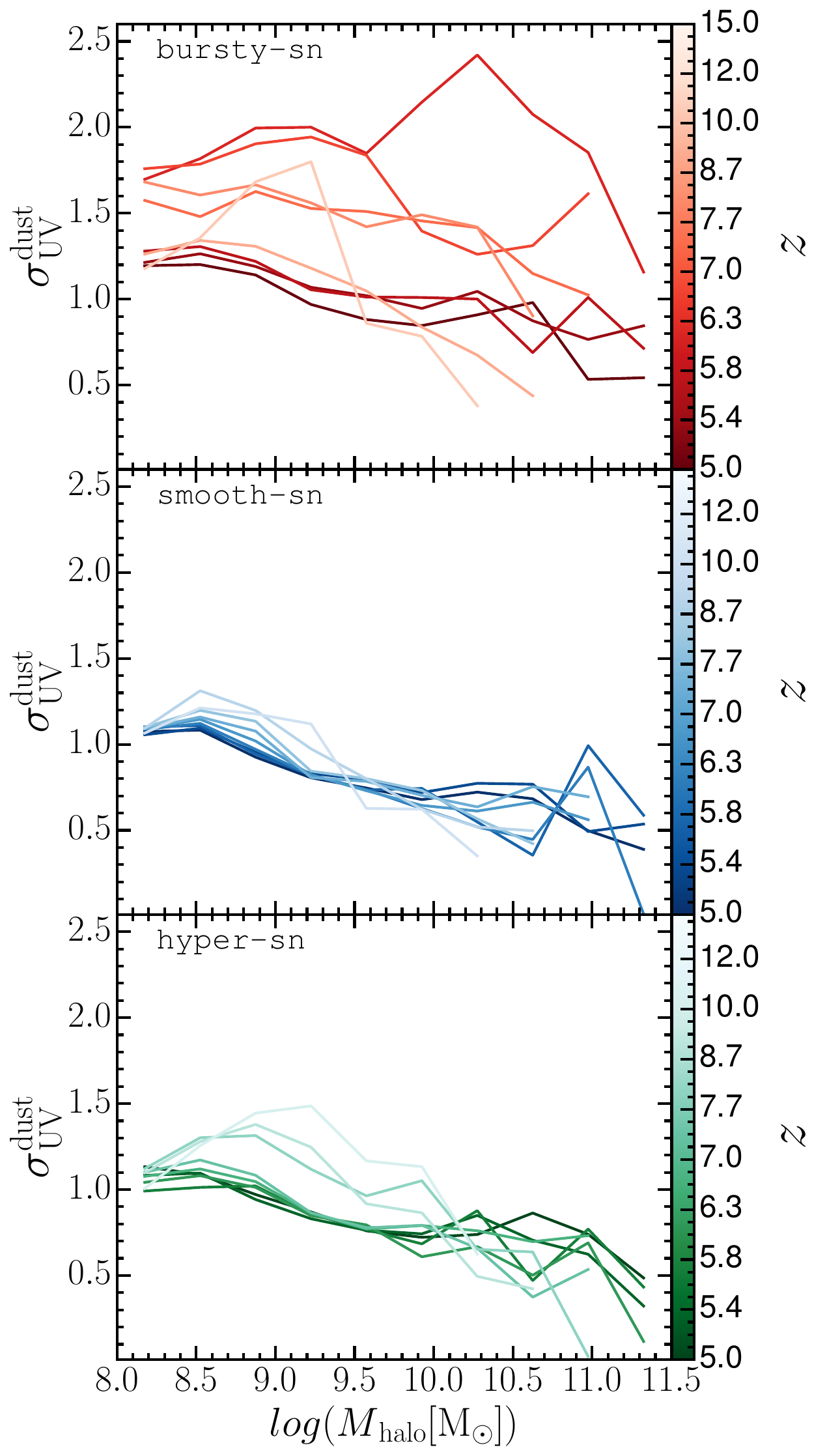}
        \includegraphics[width=70mm]{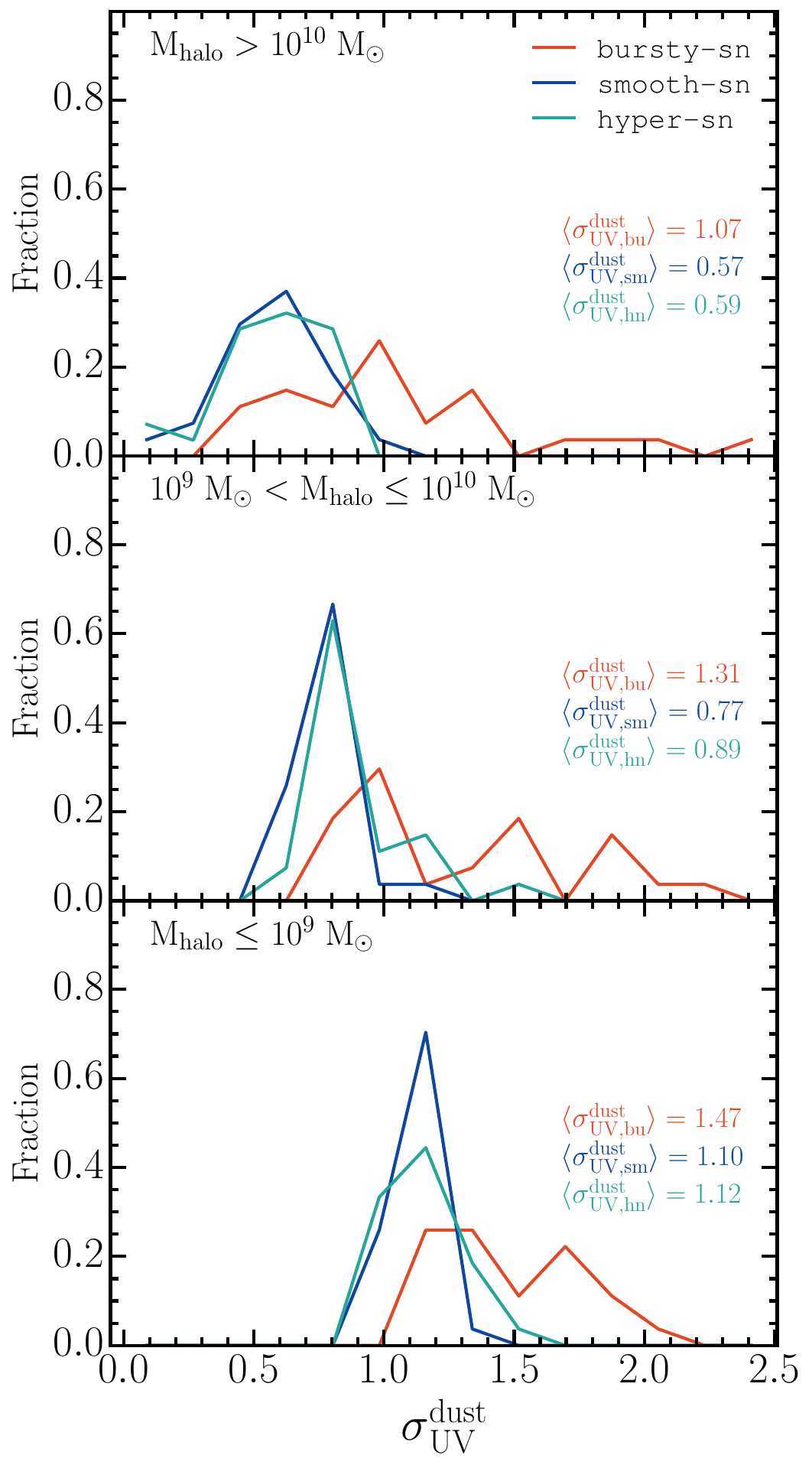}
  \caption{\textit{Left panel :}$\sigma\rm{_{UV}^{dust}}$ as a function of halo mass $M\rm{_{halo}}$ at different redshifts (indicated by the colorbar) for \texttt{bursty-sn} ({\it top panel}), \texttt{smooth-sn} ({\it middle}) and \texttt{hyper-sn} ({\it bottom}). \textit{Right panel :} Distribution of $\sigma\rm{_{UV}^{dust}}$ in different $M\rm{_{halo}}$ bins for the three SN feedback models, as indicated by the colors. Numbers refer to the corresponding average standard deviation.}
  \label{fig:sigmauv_mhalo_bins_dist}
\end{figure}

The variability in the UVLF is shaped by three primary factors: the accretion history of galaxies, feedback-regulated star formation histories, and dust attenuation (see \citealt{Shen2023}). The combined effect of non-linear feedback and accretion history on the SFR is encapsulated in the term $\sigma_{\mathrm{SFR}}$, which has been introduced in the previous section. Interstellar medium (ISM) properties \citep{heckman2001,ciardi2002,alexandroff2015}, feedback-driven dust destruction and dust distribution within galaxies \citep{Aoyama2018,Ocvirk2024,Esmarian2024,Zhao2024} further influence the escape of UV photons, affecting the observed UVLF.

To examine the variability of the UVLF in \texttt{SPICE}, I compute the scatter in UV luminosity around the median UVLF (see Section~\ref{section:3.1}).
For a consistent comparison with previous studies, I discuss the variability of the dust-attenuated UVLF ($\rm{\mathit{\sigma}_{UV}^{dust}}$), which generally shows a variability higher than the one of the intrinsic one \citep{Shen2023,pallottini_ferrera2023}.  In \texttt{SPICE}, the UVLF variability is influenced by both intrinsic flux fluctuations and dust attenuation, with the former playing a more dominant role. However, in approximately $25-35$$\%$ of cases (with the \texttt{smooth-sn} model showing the highest fraction), dust attenuation slightly reduces the overall variability of $\lesssim 4-5\%$ with respect to the intrinsic scatter.

Left panel of Figure \ref{fig:sigmauv_mhalo_bins_dist} shows the scatter in UV luminosity, $\sigma_{\rm UV}^{\rm dust}$, as a function of $M_{\rm halo}$ in the range $z=5-15$. At the highest redshifts, all models exhibit a similar dependence on halo mass, with smaller halos showing a larger UVLF variability, likely due to efficient SN feedback in this mass regime \citep{gelli2024}. As expected, \texttt{bursty-sn} has the highest variability among all models, with a UV luminosity scatter which starts to rise at $z\lesssim 10$ and reaches $\sigma_{\rm UV}^{\rm dust} \approx 2.5$ for $M_{\rm halo} \approx 10^{10.3}$~M$_{\odot}$ at $z \approx 6.5$. At lower redshift, $\sigma_{\rm UV}^{\rm dust}$ decreases again.
In comparison, for \texttt{smooth-sn} the UV luminosity scatter is always below $\sigma_{\rm UV}^{\rm dust}\approx 1.3$, whereas \texttt{hyper-sn} shows slightly higher variability, with a maximum value of $\sigma_{\rm UV}^{\rm dust} \approx 1.5$.
Indeed, \texttt{hyper-sn} exhibits a variability comparable to the one in \texttt{bursty-sn} until $z \approx 9$, while at lower redshift the variability decreases as a result of the reduced fraction of HN,
and becomes more similar to the \texttt{smooth-sn} one. The same behaviour had already been observed in Figure \ref{fig:sfr_individual} with respect to the SFR. At $z < 6$ all models have nearly identical variability levels, except for \texttt{bursty-sn} which shows larger fluctuations, with a maximum deviation up to 0.3-0.4 mag.

Similarly to Figure~\ref{fig:sigmasfr_mhalo_bins_dist}, in the right panel of Figure \ref{fig:sigmauv_mhalo_bins_dist} I show the distribution of $\rm{\mathit{\sigma}_{UV}^{dust}}$ in different halo mass bins. As found for the SFR, the curves are shifted towards higher $\rm{\mathit{\sigma}_{UV}^{dust}}$ for smaller halo masses, with \texttt{bursty-sn} having a broader distribution and the highest peak value overall. \texttt{smooth-sn} and \texttt{hyper-sn} show  distributions similar to each other, with the exception of the intermediate mass range (i.e. $10^{9} {\rm M}_\odot < M_{\rm halo} \leq 10^{10} {\rm M}_\odot$), where the latter is somewhat wider. In this mass regime, the average $\rm{\mathit{\sigma}_{UV}^{dust}}$ is about 0.89 for \texttt{hyper-sn} and 0.77 for \texttt{smooth-sn}, whereas \texttt{bursty-sn} has a higher average of 1.31.
For \texttt{bursty-sn}, the average $\rm{\mathit{\sigma}_{UV}^{dust}}$ increases with decreasing mass bin, ranging from 1.07 to 1.47, with a significant fraction of objects having $\rm{\mathit{\sigma}_{UV}^{dust}} > 1.5$. The maximum value reached in \texttt{smooth-sn} is $\sigma_{\rm UV}^{\rm dust} = 1.5$, with the average $\rm{\mathit{\sigma}_{UV}^{dust}}$ increasing from $0.57$ to $1.1$ from higher to lower halo mass bins.
The \texttt{hyper-sn} model falls in-between the other two with average $\rm{\mathit{\sigma}_{UV}^{dust}}$ values ranging from $0.59$ to $1.12$.

\begin{figure}
\centering
        \begin{center}
        \includegraphics[width=100mm]{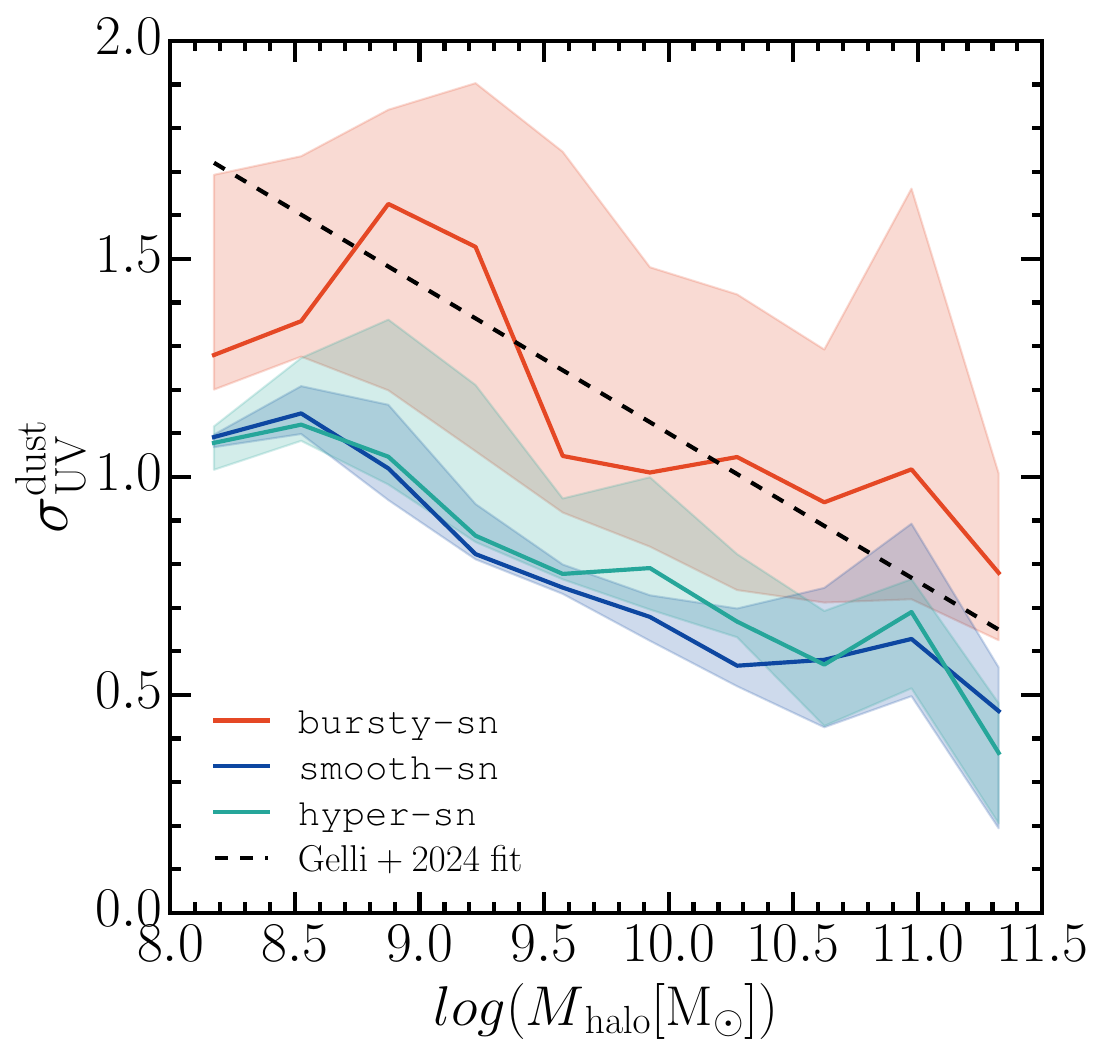}
  \caption{Dependence of $\sigma\rm{_{UV}^{dust}}$ on the DM halo mass $M\rm{_{halo}}$. The solid curves refer to the median value, while the shaded regions are the standard deviation. Colors refer to the different SN feedback models, while the black dashed curve represents the fit from \citealt{gelli2024}.  }
            
        \end{center}
  \label{fig:sigmauv_mhalo}
\end{figure}

In Figure \ref{fig:sigmauv_mhalo} I show the median $\sigma\rm{_{UV}^{dust}}$ as a function of $M\rm{_{halo}}$, where the former is calculated as the median value in each individual mass bin from all the curves corresponding to different redshift bins in Figure~\ref{fig:sigmauv_mhalo_bins_dist}. 
I find that the median curves in all models exhibit a similar slope, confirming that lower mass halos are more sensitive to feedback effects, producing more fluctuations compared to massive halos. This is primarily due to the shallower potential wells of lower mass halos, which facilitate repeated cycles of inflow, star formation, and outflow \citep{gelli2020,stern2021,furlanetto2022,legrand2022,gurvich2023,byrne2023,hopkins2023}. Among the models,  \texttt{bursty-sn}  consistently produces the highest values. Additionally, the scatter around the median differs among the models, and, as expected, the \texttt{bursty-sn} model shows the largest scatter because of the wider range of variability in UVLF, while \texttt{smooth-sn} has the smallest. 
I also note that the slope of my curves is consistent with the one from the analytical fit by \citealt{gelli2024} based on the results of the \texttt{FIRE-2} simulation at $z \approx 8$ \citep{sun2023}, although the amplitude is matched only by the \texttt{bursty-sn} model.

\begin{figure*}
\centering
        \includegraphics[width=120mm]{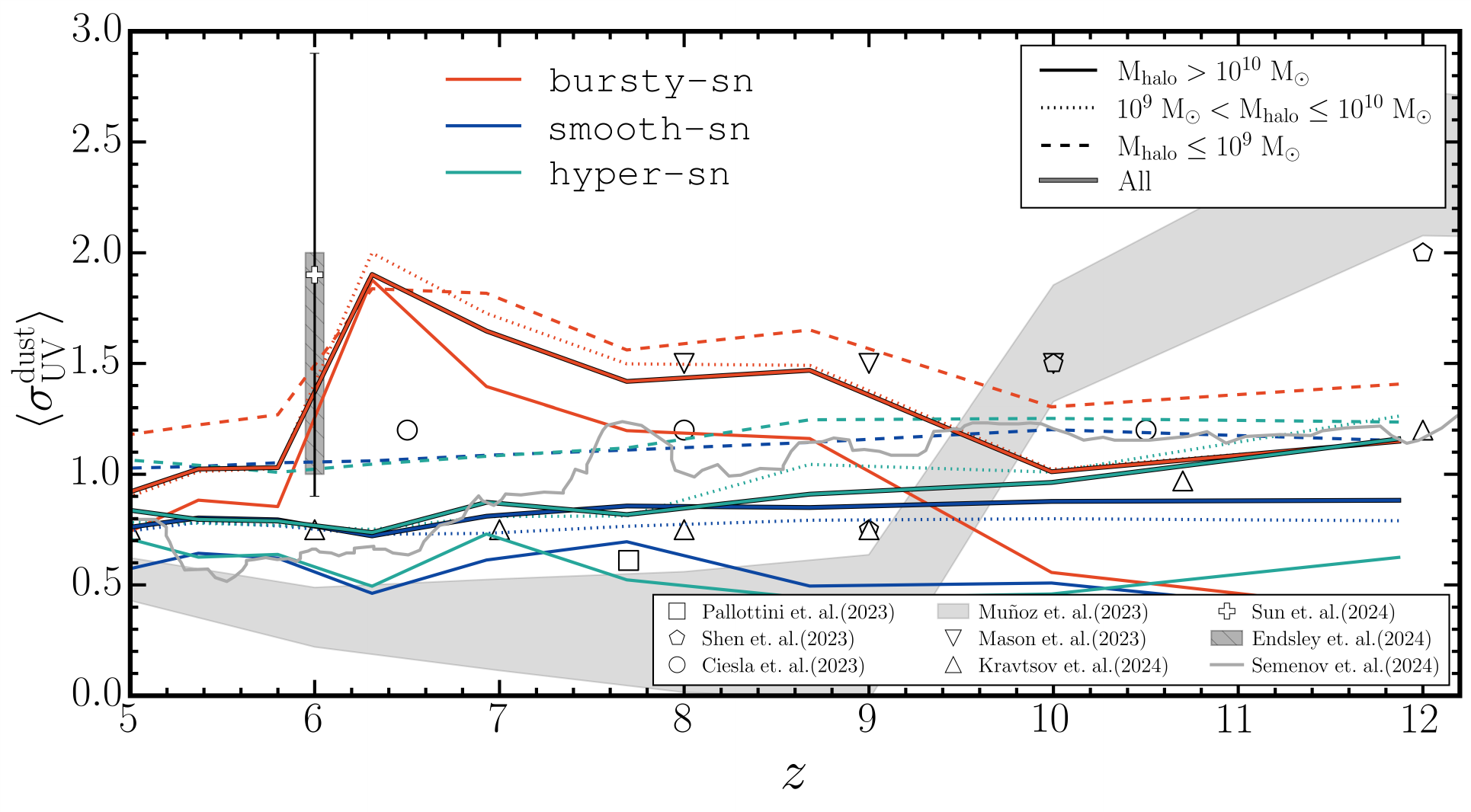}
  \caption{Redshift evolution of $\langle \sigma\rm{_{UV}^{dust}} \rangle$ for $M\rm{_{halo}}<10^9$~M$_\odot$ (dashed lines), $10^9$~M$_\odot< M\rm{_{halo}}<10^{10}$~M$_\odot$ (dotted) and  $M\rm{_{halo}}>10^{10}$~M$_\odot$ (solid). The curves for the entire sample are shown in solid curve with black edges. Colors refer to the three SN feedback models, while the values from observational and other theoretical studies \citep{pallottini_ferrera2023,Shen2023,ciesla2023,munoz2023,mason2023,kravtsov2024,sun2024,endsley2024,semenov2024} are shown in grey data points and shaded regions.
  }
  \label{fig:sigmauv_evol}
\end{figure*}

Finally, in Figure \ref{fig:sigmauv_evol} I show the redshift evolution of the average UV luminosity scatter $\langle \sigma_{\rm UV}^{\rm dust} \rangle$ in different halo mass bins (at all redshifts), as well as computed over the full sample.
I observe that in all models  $\langle \sigma_{\rm UV}^{\rm dust} \rangle$ is largest for smaller halos, consistently with Figure \ref{fig:sigmauv_mhalo_bins_dist}.
The curves for both the \texttt{smooth-sn} and \texttt{hyper-sn} models exhibit a clear and consistent trend across redshifts, with $\langle \sigma_{\rm UV}^{\rm dust} \rangle$ generally decreasing with decreasing redshift with an exception at the highest halo mass bin where the scatter in UV luminosity remains roughly constant. 
In contrast, across all masses and redshifts, the \texttt{bursty-sn} model consistently exhibits the highest $\langle \sigma_{\rm UV}^{\rm dust} \rangle$, indicating more pronounced fluctuations in UVLF compared to the other models. 
Similarly to the trend observed in Figure \ref{fig:sigmauv_mhalo_bins_dist}, also here I see that in the \texttt{bursty-sn} model, as the UVLF variability increases rapidly below $z\approx10$, has a peak of $\langle \sigma_{\rm UV}^{\rm dust} \rangle \approx 2$ at $z \approx 6.3$, and then it drops sharply. 
Since the halo assembly history is similar in all models, this behaviour is unlikely to be driven solely by halo mass evolution. For example, its origin may also be connected to galaxy morphology, gas accretion history, level of turbulence in the ISM, and the rise of a UVB, as around this time the reionization process begins to accelerate for the \texttt{bursty-sn} model. This will be addressed in more detail in a future study.
Notably, at $z \gtrsim 10$ for halos with $M_{\rm halo}>\rm{10^{9} M_\odot}$, \texttt{hyper-sn} produces the highest variability among all three models because of the impact of the stronger HN explosions dominating at that time. 

I compare the results from the \texttt{SPICE} simulations to values extracted from observational \citep{ciesla2023,endsley2024}, as well as theoretical \citep{munoz2023,mason2023,Shen2023,pallottini_ferrera2023,sun2023,sun2023a,sun2024,kravtsov2024,semenov2024} studies.
Employing a semi-empirical approach, \citealt{munoz2023} found a value of $\langle \sigma\rm{_{UV}^{\rm dust}} \rangle$ as high as 2.5 at $z\approx 12$, which drops to 0.8 at $z \leq 10$. While none of the models reproduces such large high-$z$ values, at lower redshift there is a better agreement, although in this case my $\langle \sigma\rm{_{UV}^{\rm dust}} \rangle$ are always slightly larger.
\citealt{mason2023} utilized a fully empirical framework to predict that a $\langle \sigma\rm{_{UV}^{\rm dust}} \rangle$ up to 1.5 in the range $8<z<10$ is necessary for models to align with existing observations. These numbers are similar to those predicted in the {\tt bursty-sn} model.
Following a similar methodology, \citealt{Shen2023} found a lower value of 0.75 at $z \approx 9$, which is instead consistent with those predicted by the {\tt smooth-sn} and {\tt hyper-sn} models. At $z \approx 10$ and 12, though, they predict values of 1.5 and 2.0, respectively, which are higher than those of my models.
Using a semi-analytical approach, \citealt{sun2024} (but see also \citealt{sun2023a,sun2023}, where bursty star formation in \texttt{FIRE-2} simulation has been used to explain high redshift \texttt{JWST} observations) found the most probable UVLF variability value to be $\approx1.9\pm 1.0$ at $z\approx6$. At the same redshift, \citealt{endsley2024} found a variability in the range 1-2, obtained by analyzing Lyman-break galaxies  assembled from ACS+NIRCam imaging in the GOODS and Abell 2744 fields. These values are generally reproduced in \texttt{bursty-sn}, as well as by halos with $M_{\rm halo}\leq\rm{10^{9} M_\odot}$ in the other models.
\citealt{ciesla2023}, using spectral energy distribution modeling with the \texttt{CIGALE} code, and assuming a non-parametric star formation history on the \texttt{JADES} public catalog, found $\langle \sigma\rm{_{UV}^{\rm dust}}\rangle \sim 1.2$ in the range $6.5 < z < 10.5$, which is fairly consistent with my {\tt bursty-sn} model. 
\citealt{pallottini_ferrera2023} analyzed 245 galaxies from the \texttt{SERRA} simulation suite at $z\approx 7.7$, finding a $\langle \sigma\rm{_{UV}^{\rm dust}} \rangle$ of 0.61. I note, though, that the method employed for this evaluation differs from that used in my study and other works. 
Finally, \citealt{semenov2024} investigated the importance of turbulent star formation by utilizing detailed modelling of cold turbulent ISM, star formation and feedback through zoom-in high resolution simulation of an early forming Milky Way analog, and they found that the variability decreases with time, similarly to the predictions of my models, with the exception of the $6<z<9$ range in \texttt{bursty-sn}. It is also interesting to note that the curves for the \texttt{smooth-sn} and \texttt{hyper-sn} models in the lowest mass bin aligns very well with \citealt{semenov2024} till $z\approx7.5$, whereas after that the latter shows much lower values. 

In general, I can conclude that the constraints from observational and other theoretical models can not specifically distinguish between the SN feedback models in \texttt{SPICE} due to their large scatter in $\sigma\rm{_{UV}^{\rm dust}}$ across different redshifts.

\section{Conclusions}
\label{section:4}

In this study I investigate the role played by SN feedback in driving the variability of the UVLF using the suite of radiation-hydrodynamic simulations \texttt{SPICE}.
This includes three models for SN feedback, which differ in terms of explosion energy and timing. 
The \texttt{bursty-sn} model, characterized by intense and episodic supernova explosions, shows the highest SFR variability, leading to significant fluctuations in UV luminosity. 
The \texttt{smooth-sn} model, in which energy injection happens more continuously and thus the effect of feedback is less disruptive, produces a higher and less variable SFR, resulting also in a lower UVLF variability. 
The \texttt{hyper-sn} model shows an intermediate trend, with higher variability at early times due to HN effects, and a behaviour similar to the one of \texttt{smooth-sn} at lower redshifts, when the fraction of HN events decreases. The main findings can be summarized as follows:

\begin{itemize}

    \item The good agreement between all models and the observed UV luminosity functions discussed in \citealt{bhagwat2024} is obtained even at the highest redshifts analyzed here, i.e. $z\approx 11$ and 12. 
    
    \item I find that not only the most massive objects contribute to the bright end of the LF, i.e. with dust corrected magnitude $\rm{\mathit{M}_{UV}^{dust}} \leq -17$, but also those with $\rm{\mathit{M}_{*} \sim 10^{(6.5-7)} M_{\odot}}$ can be a very bright object detected by \texttt{JWST}. The median stellar mass of these bright objects in my models is consistent with those derived by \citealt{mason2015,mason2023} at all redshifts.

    \item The nature of SN feedback has a significant impact on the temporal evolution of the UVLF, as I show for time intervals of 10, 30, and 50 Myr at $z\approx10$ and 8. The \texttt{bursty-sn} model produces the highest UVLF fluctuations, with a deviation of more than 1 dex. In comparison, the evolution of the UVLF in \texttt{smooth-sn} and \texttt{hyper-sn} is more gradual, with the latter having much higher fluctuations than the former at $z\approx10$. 

    \item The burstiness of star formation and UVLF is strongly influenced by the SN feedback, with the \texttt{bursty-sn} and \texttt{smooth-sn} models consistently exhibiting the highest and lowest variability, respectively. The disruptive nature of the \texttt{bursty-sn} model leads to both higher median values and larger scatter in $\sigma_{\rm UV}^{\rm dust}$, indicating stronger fluctuations in dust-obscured UV emission. In contrast, the \texttt{hyper-sn} model displays an intermediate behavior, balancing between the extremes of the other two models.

    \item All models show a similar dependence of the SFR and UVLF variability on the halo mass, with the extent of the fluctuations being higher for the lowest mass haloes, which are more susceptible to feedback effects. I find that the median $\sigma\rm{_{UV}^{dust}}$ as a function of $M\rm{_{halo}}$ in all models exhibit a similar slope, consistent with the one from the analytical fit from the \texttt{FIRE-2} simulation at $z \approx 8$ by \citealt{sun2023}, although the amplitude is matched only by the \texttt{bursty-sn} model.

    \item The redshift dependence of the average standard deviation $\langle \sigma_{\rm UV}^{\rm dust} \rangle$ is similar in  \texttt{smooth-sn} and \texttt{hyper-sn}, with a variability which remains almost constant, although in the latter model it is slightly higher. In contrast, the \texttt{bursty-sn} model shows an increase in $\langle \sigma_{\rm UV}^{\rm dust} \rangle$ from $z \approx 10$ to $\approx 6$, before declining. Additionally, the \texttt{bursty-sn} model consistently exhibits the highest values of $\langle \sigma_{\rm UV}^{\rm dust} \rangle$ across all masses and redshifts.

     \item I achieve the maximum UVLF variability as $\sigma_{\rm UV}^{\rm dust} \sim 2.5$ in the {\tt bursty-sn} model. Conversely, the smoother star formation history typical of the {\tt smooth-sn} model fails to induce a large variability, with values of $\sigma_{\rm UV}^{\rm dust}$ below 1.3. The {\tt hyper-sn} model lies is in between the other two, although its characteristics are more similar to those of the {\tt smooth-sn} one, with $\sigma_{\rm UV}^{\rm dust}$ extending up to 1.5.
\end{itemize}

This work emphasizes the critical role of SN feedback in shaping the variability of the UVLF. 
While my study suggests that SN feedback is the primary mechanism influencing the UVLF variability, the observed features arise from a complex interplay of multiple factors (i.e. UVB, gas accretion history, ISM turbulence) that requires further detailed investigation.
Here, though, this work provides new insights into how variations in feedback strength and timing impact the burstiness of star formation and UV emission, and I emphasize the importance of explaining how the variability depends on both mass and redshift, an aspect often overlooked in prior studies. This investigation also suggests the UVLF variability may alleviate the bright galaxy tension observed by \texttt{JWST} at high redshifts.
  
\chapter{The Influence of Ionizing Sources on Intergalactic Medium Properties}
\label{chap:chapter3}

\begin{flushright}
\begin{minipage}{0.5\textwidth}
\raggedleft
\textbf{\textit{``If I could wish for a thing, I would make neutrinos itchy for my enemies."}}\\[1ex]
\noindent\rule{0.5\textwidth}{0.4pt}\\[-0.2ex]
Géza Csörnyei
\end{minipage}
\end{flushright}

\begin{flushright}
\begin{minipage}{0.7\textwidth}
\raggedleft
\textbf{\textit{``Everything flows, no one can step into the same river twice, because everything flows and nothing stays."}}\\[1ex]
\noindent\rule{0.5\textwidth}{0.4pt}\\[-0.2ex]
Silvia Elizabeth Almada Monter
\end{minipage}
\end{flushright}


\textit{This work has been submitted for publication in the Monthly Notices of the Royal Astronomical Society \citep{Basu2025b}.}

\hspace{1cm}

The growth of primordial density fluctuations from gravitational instabilities gave rise to the formation of the first stars and galaxies, which started to emit  photons with enough energy to ionize the surrounding intergalactic medium (IGM) and initiate the last global phase transition of the universe, namely the Epoch of Reionization (EoR; \citealt{Loeb2001}). Observations of high-$z$ QSO absorption spectra \citep{Fan2003,Fan2006,Gunn1965,McGreer2015,bosman2018,Davies2018,Bosman2022}, Lyman-$\alpha$ emitters (LAEs; \citealt{Pentericci2014,Tilvi2014,Barros2017,Mason2018,Jung2022,Tang2023,Tang2024,nakane2024,napolitano2024,saxena2024,chen2024,ning2024}), information of the IGM thermal properties from QSO spectra \citep{Bolton2012,Raskutti2012,Gaikwad2020}, and of the Thompson scattering optical depth from Cosmic Microwave Background (CMB;  \citealt{Komatsu2011,planck2016,Pagano2020} 
indicate that reionization is an extended process, largely complete by $z\approx 5.5$ \citep{kulkarni2019,nasir2020}. 


This abundance of observations is accompanied by an increasingly sophisticated modeling of the reionization process, employing a variety of approaches, ranging from semi-analytic and semi-numeric models \citep{zhou2013,mesinger2007,santos2010,choudhury2018,hutter2021,choudhury2022}, to full radiation hydrodynamic simulations \citep{Baek2010,ocvirk2016,ockvirk2020, gnedin2014,finlator2018,Rosdahl2018,Trebitsch2021,kannan2022,garaldi2022,bhagwat2024}.
Although in most literature galaxies are thought to be the main drivers of reionization  \citep[e.g.][]{BeckerBolton2013,bouwens2015,robertson2015,Madau2017,dayal2020,hassan2018,Marius2020,Eide2020b,kannan2022}, it is still unclear which are the properties of those contributing the most (e.g. small vs. massive galaxies; \citealt{Naidu2020,Naidu2021,Matthee2022}). Additionally, the recent detection of a larger than expected number density of faint quasars at intermediate (i.e. $z\approx4$; \citealt{Giallongo2015,Giallongo2019,Boutsia2018}) and high (i.e. $z>6$;  \citealt{Weigel2015,McGreer2018,Parsha2018,Akiyama2018,harikane2023,maiolinoa2023,Maiolino2023,Goulding2023,Larson2023,Juodzbalis2023,greene2023}) redshift, have renewed the interest in quasars as possible strong contributors to the ionization budget \citep{grazian2024,madau2024}.

Whatever the main source of ionizing photons is, the physical properties of the IGM in terms of ionization and thermal state are influenced by a variety of sources, including the above mentioned stars and quasars, but also X-ray binaries (XRBs; \citealt{mirabel2011,fragos2013,fragosa2013,madaufragos2017,sazonov2017}), Bremsstrahlung emission from shock heated interstellar medium (ISM; \citealt{Chevalier1985,suchkov1994,strickland2000}), low energy cosmic rays \citep{nath1993,sazonov2015,leite2017,owen2019,bera2023,yokoyama2023,gessey2023,carvalho2024}, self-annihilation or decay of dark matter \citep{liu2016,cang2023} and plasma beam instabilities in TeV blazars \citep{chang2012,puchwein2012,lamberts2022}. 

As discussed for example in \citealt{Marius2018} and \citealt{Marius2020} (but see also, \citealt{Baek2010,Venkatesan2011,Grissom2014,enrico2019}), the more energetic sources are responsible for the partially ionized and warm gas found outside of the regions fully ionized by stars. Additionally, they can also increase the IGM temperature in comparison to the one obtained in star driven reionization scenarios. 
Their impact on studies of the 21cm signal from neutral hydrogen at high-$z$ has been widely recognized and investigated \citep{ross2017,ross2019,qingbo2020,qingbo2021,qingbo2022,kamran2022,cook2024,noble2024}.
Such high-energy sources may also impact the Lyman-$\alpha$ (Ly$\alpha$) forest by broadening thermal widths and lowering the IGM neutral fraction, thereby increasing the transmission flux.

Here, in this paper, I present a systematic study of the impact of different  source spectral energy distributions by post-processing \texttt{Sherwood}-type cosmological hydrodynamic simulation \citep{bolton2016} using the 3D multi-frequency radiative transfer code CRASH \citep[e.g.][]{ciardi2001,maselli2003,maselli2009,graziani2013,hariharan2017,glatzle2019,glatzle2022} which follows the formation and evolution of ionized hydrogen and helium along with the photoheating of IGM utilising different assumptions of the properties of ionizing sources. The chapter is organized as follows: Section~\ref{methodology} outlines the simulations, Section~\ref{results} presents my findings, and Section~\ref{discussion} concludes with a broader interpretation of the results and a summary.

\begin{figure*} 
\centering
    \includegraphics[width=60mm]{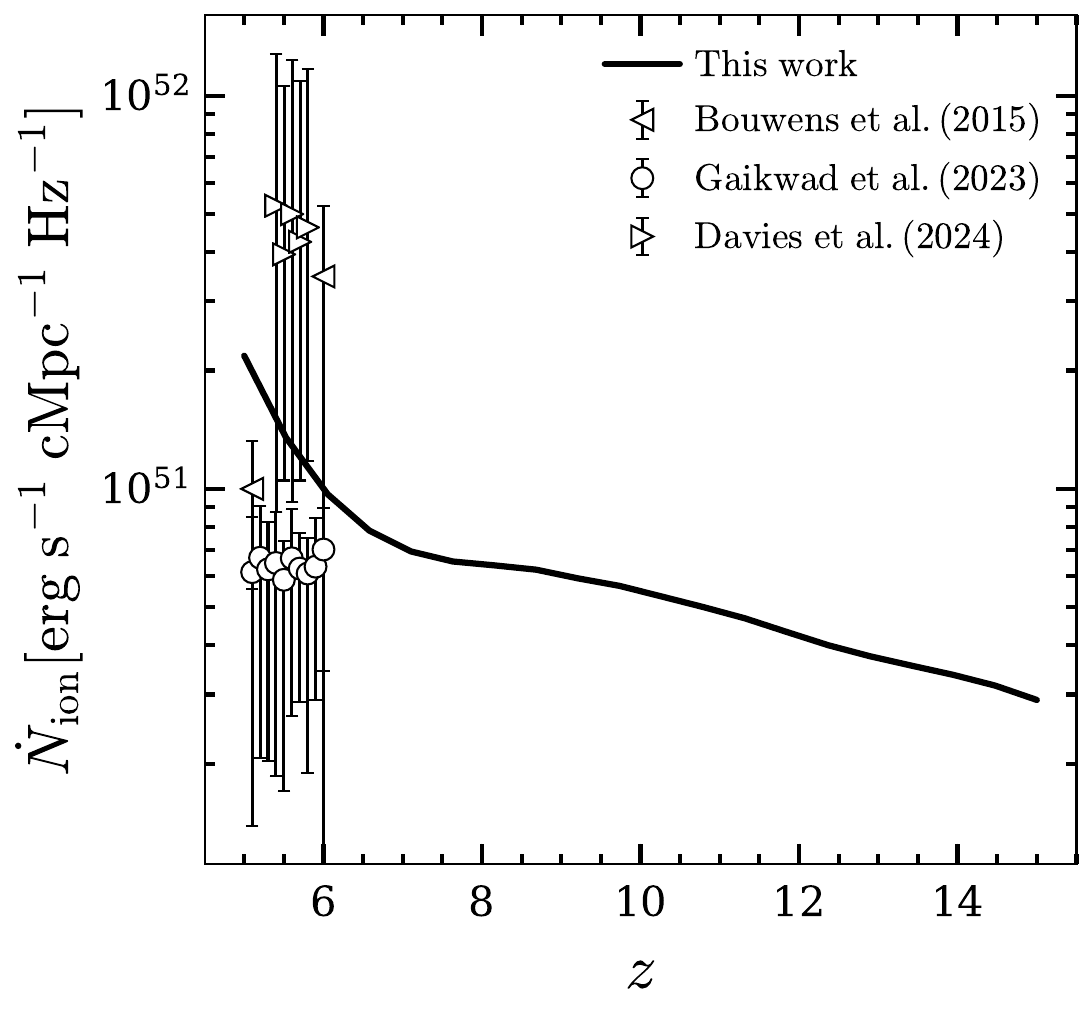} 
    \includegraphics[width=60mm]{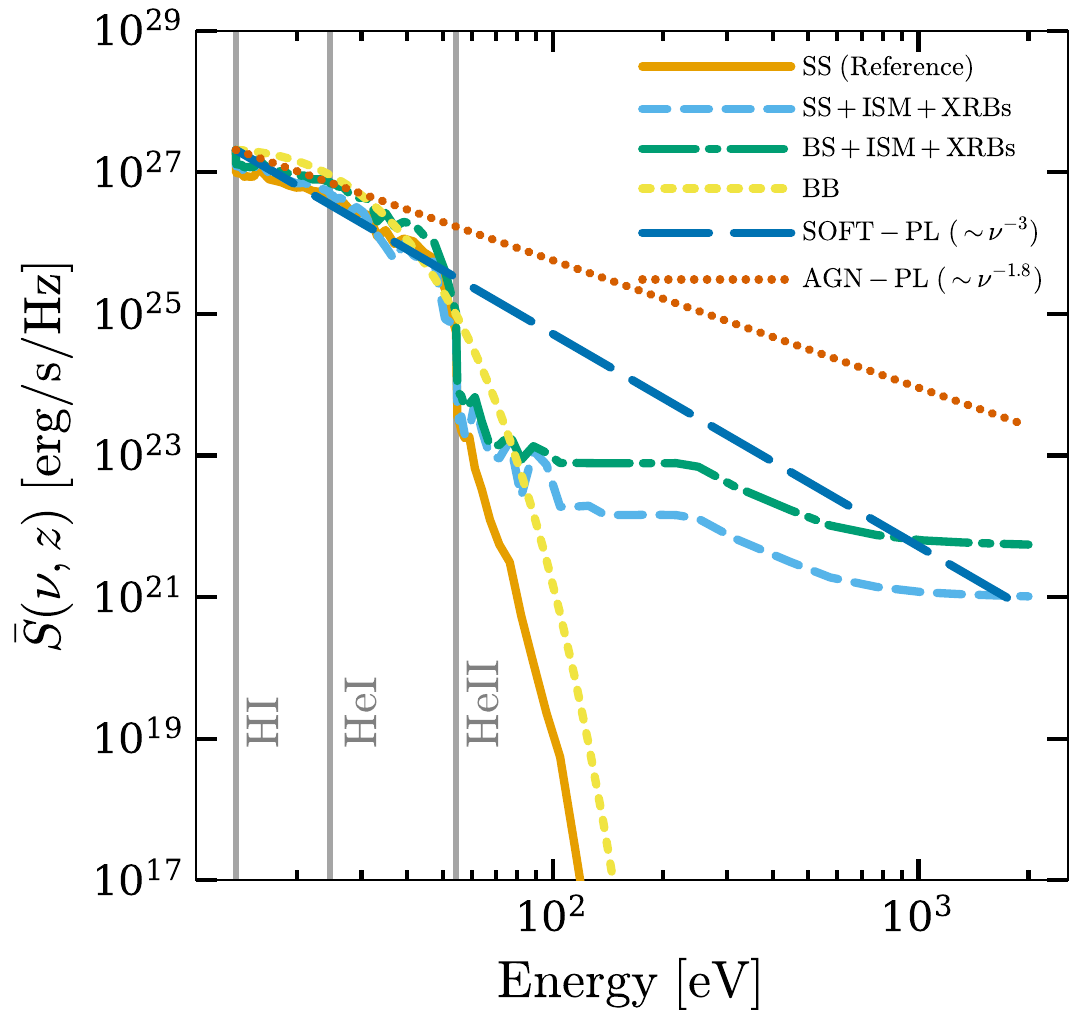}  
    \caption{ \textit{Left Panel:}Redshift evolution of the total ionizing photon production rate adopted in the study (black solid line) with a compilation of observational constraints \citep{becker2021,gaikwad2023,davies2024}. \textit{Right Panel:} averaged spectral energy distribution $\bar{S}(\nu,z)$ for: single stars (\texttt{SS}, solid orange curve); a combination of single stars, ISM and XRBs (\texttt{SS+ISM+XRBs}, dashed sky blue); a combination of binary stars, ISM and XRBs (\texttt{BS+ISM+XRBs}, dash-dotted green); a black-body spectrum with 50,000 K (\texttt{BB}, short-dashed yellow); a power-law spectrum with a slope of -3 (\texttt{SOFT-PL}, long-dashed blue), and a power-law spectrum with a slope of -1.8 (\texttt{AGN-PL}, dotted vermilion) The vertical gray lines indicate the ionization thresholds for neutral hydrogen (13.6 eV), neutral helium (24.6 eV) and singly ionized helium (54.4 eV).  
    }
    \label{fig:siminput}
\end{figure*}

\section{Methodology}
\label{methodology}
Here, I describe how I have post-processed the outputs of a hydrodynamical simulation (Section \ref{hydrosim}) with the multi-frequency radiative transfer (RT) code \texttt{CRASH} (Section \ref{crash}) to obtain a suite of simulations of reionization (Section \ref{section:2.3}).

\subsection{The cosmological hydrodynamic simulation}
\label{hydrosim}

To obtain the gas and source distribution and properties, I have used outputs at redshift 5.16 to 15 from a cosmological hydrodynamical simulation  performed in a comoving cubic box of size 20 $h^{-1}$ cMpc  with the parallel Tree-PM smoothed particle hydrodynamics (SPH) code \texttt{p-GADGET-3},   which is an updated version of \texttt{GADGET-2} \citep{Volker2001,gadget2}. The box contains $2\times1024^{3}$ gas and dark matter particles, each of mass $m\rm{_{gas}}=9.97 \times 10^{6}\, h^{-1} M_{\odot}$ and  $m\rm{_{DM}}=5.37 \times 10^{7}\, h^{-1} M_{\odot}$, respectively. Haloes are identified using the friends-of-friends algorithm with a linking length of $0.2$ times the mean inter-particle separation. The gravitational softening length for the simulation is set to 1/30th of the mean linear interparticle spacing, and star formation is incorporated using a simple and computationally efficient prescription which converts all gas particles with overdensity $\Delta > 10^{3}$ and temperature $T<10^{5} \rm{K}$ into collisionless stars \citep{Viel2004,viel2013}. Note that because of this simple treatment, the simulation does not self-consistently model star formation and feedback, but is perfectly appropriate to investigate IGM properties \citep{keating2016,kulkarni2016,mallik2024}. The model adopts a Planck 2013 consistent cosmology  \citep{planck2013} with $\rm{\Omega_{m}=0.308}$, $\rm{\Omega_{\Lambda}=0.692}$, $\rm{\Omega_{b}=0.0482}$, $\rm{\sigma_{8}=0.829}$, $\rm{\mathit{n}_{s}=0.963}$, $h=0.678$, where the symbols have their usual meaning.
Throughout the chapter , the gas is assumed to be of primordial composition with a hydrogen and helium number fraction of 0.92 and 0.08, respectively. 

\subsection{The radiative transfer code \texttt{CRASH}}
\label{crash}
The radiative transfer of ionizing photons through the IGM is implemented by post-processing the outputs of the hydrodynamic simulation using the 3D non-equilibrium and multi-frequency code \texttt{CRASH}  \citep[e.g.][]{ciardi2001,maselli2003,maselli2009,graziani2013,hariharan2017,glatzle2019,glatzle2022}. The code evaluates self-consistently the ionization states of hydrogen and helium, as well as the evolution of gas temperature. \texttt{CRASH} employs a Monte Carlo-based ray tracing approach, where the propagation of ionizing radiation is represented by multi-frequency photon packets that dynamically trace the spatial and temporal distribution of radiation within the simulation volume. The version of \texttt{CRASH} employed here has a comprehensive treatment of UV and soft X-ray photons, including X-ray ionization, heating, and secondary electron interactions \citep{graziani2018}. For an in-depth understanding of \texttt{CRASH}, I encourage readers to refer to the foundational papers.

\subsection{The radiative transfer simulation of reionization}
\label{section:2.3}

To post-process the simulation outputs with \texttt{CRASH}, I have gridded the gas density and temperature extracted from the snapshots onto $N_{\rm grid}^{3}$ cells.
More specifically, the RT is performed with $N_{\rm grid}=128$ (corresponding to a spatial resolution of 156 $h^{-1}\,\mathrm{ckpc}$) in the range $z=15-8$, while at $z<8$ the resolution is increased to $N_{\rm grid}=256$ (corresponding to 78 $h^{-1}\,\mathrm{ckpc}$) to ensure a better description of the Ly$\alpha$ forest related statistics. The gridded properties of the gas are obtained by assigning the particle data to a regular grid using the SPH kernel \citep{monaghan1992}, while the corresponding grid for the halo masses and positions is obtained by using the cloud-in-cell and friends-of-friends algorithms \citep{hockney1988,springel2001}. If more than one halo ends up in the same cell, their masses are combined. 
To reduce the large number of sources present in the simulation box and thus the computational time, I have adopted the source clustering technique developed by \citealt{Marius2020} with a luminosity limit of $0.1$ and maximum radius of $\sqrt{2}\times 2.01$ (see the original paper for more details). Every ionizing source emits $10^{4}$ photon packets which are lost once they exit the simulation box, i.e. no periodic boundary condition has been used. 

For each output $i$ at $z_i$ of the hydrodynamical simulation, the RT is followed for a time $\rm{\mathit{t}_{H}(\mathit{z}_{i+1}) - \mathit{t}_{H}(\mathit{z}_{i})}$, where $\rm{\mathit{t}_{H}(\mathit{z}_{i})}$ is the Hubble time corresponding to $z_{i}$. 
In order to account for the expansion of the Universe between the $i$-th and $(i+1)$-th snapshots, the gas number density is evolved as $n(\mathbf{x},z) = n(\mathbf{x}, z_{i})(1 + z)^{3}/(1 + z_{i})^{3}$, where $\mathbf{x} \equiv (x_{c}, y_{c}, z_{c})$ are the coordinates of each cell $c$. To avoid accounting twice for the UV background (which is already included in the hydrodynamic simulations), at $z\leq9$ all cells with $\rm{\mathit{T}<10^5 K}$ are assigned a temperature $\rm{\mathit{T}_{i+1}=\mathit{T}_{i}(1+\mathit{z}_{i+1}^{2})/(1+\mathit{z}_{i}^{2}}$), i.e. only the effect of the adiabatic cooling since the previous snapshot is taken into account in the evolution of the temperature, while the photoheating due to the UVB is discarded.  For the cells with $\rm{\mathit{T}\geq10^{5} K}$, the value of the temperature is not changed to take into account the heating from shocks, which is not captured by the RT (see also \citealt{maselli2003,whalen2008,Iliev2009,Cain2024} for additional details). 
The ionization fractions are initialized to their expected residual values from the recombination epoch, i.e.  $x_{\rm HII}=10^{-4}$ and  $x_{\rm HeII}=x_{\rm HeIII}=0$.

Traditionally, models of ionizing photon production rate (or emissivity) in reionization studies focus on the cumulative contribution of ionizing photons from evolving stellar populations, often incorporating time-dependent source formation histories to estimate the global photon budget. However, in this study, I adopt a different approach. While I still utilize an ionizing photon production rate-based framework, my focus is not on the integrated luminosity evolution over time, but rather on how variations in the spectral energy distribution (SED) of the sources influence IGM and hence Ly$\alpha$ forest observables. By isolating the impact of SED shape, assuming a fixed ionizing photon production rate model, I aim to determine whether different assumptions about the source spectra can significantly alter interpretations of the Ly$\alpha$ forest. To this end, the source ionizing photon production rate is evaluated according to the prescription introduced by \citealt{Ciardi2012}, namely, I have assumed a total comoving hydrogen ionizing ionizing photon production rate at each redshift, $\epsilon\rm{_{tot}(\mathit{z}_i)}$, and distributed it in terms of rate of ionizing photons (${\dot{N}_{\text{ion}}}$) among all the halos  proportionally to their mass. I display my adopted redshift evolution of ${\dot{N}_{\text{ion}}}$ in the left panel of Figure \ref{fig:siminput}. This approach assures by design that the  ionizing photon production rate is broadly consistent with $z= 5-6$ observations \citep{becker2021,gaikwad2023,davies2024}.
Additionally, this approach does not require any assumption on the escape fraction of ionizing photons, which is a very uncertain quantity (\citealt{vanzella2016,vanzella2018,matthee2018,naidu2018,steidel2018,fletcher2019}; for a review see \citealt{wood2000,Pratika2018}). Differently from \citealt{Ciardi2012}, though, rather than having a purely linear relation between the ionizing ionizing photon production rate and the halo mass, I have incorporated Gaussian noise scatter with mean $\mu=0$ and standard deviation $\sigma=0.01$ in the log-log scale, conserving the total ionizing ionizing photon production rate (as done in \citealt{enrico2019,Basu2024}).

The SED of the sources has been derived following \citealt{Marius2018}, \citealt{Marius2020} and \citealt{qingbo2022}, which include the contribution from single (SS) and binary (BS) stars, XRBs and thermal Bremsstrahlung from supernova-heated ISM. While I refer the reader to the original papers for more details, here I note that stars dominate the total SED budget at energy $\rm{\mathit{h}_p \nu \leq 60}$ eV, while the ISM contribution is dominant above that into hard UV and the soft X-rays (i.e.in the range $\rm{\mathit{h}_p \nu \in [60,500]}$ eV),  and the XRBs start to contribute  at $\rm{\mathit{h}_p \nu > 500}$ eV (see Figure 2 in \citealt{Marius2018} and Figure 1 in \citealt{qingbo2022} for a visual inspection of individual SEDs of different types of ionizing sources). 
For all RT simulations, the spectra of ionizing sources extend to a maximum frequency of $\rm{\sim 2 \ keV}$ and are discretized into 82 frequency bins, with  bins more densely spaced around the ionization thresolds of hydrogen (13.6 eV) and helium (24.6 eV and 54.4 eV). I run three  simulations: one with ionizing sources consisting solely of single stars (\texttt{``SS"}), another adding also the contribution from the ISM and XRBs (\texttt{``SS+ISM+XRBs"}), and a third that includes binary stars along with the ISM and XRBs (\texttt{``BS+ISM+XRBs"}). For comparison, I also perform three additional simulations using simplified SED prescriptions commonly adopted in the literature. These include a blackbody spectrum with $T$ = 50,000 K (\texttt{``BB"}), a galaxy-like power-law SED with a slope of -3 (\texttt{``SOFT-PL"}), and an AGN-type power-law SED with a slope of -1.8 (\texttt{``AGN-PL"}). For simplicity, following \citealt{Marius2018} and \citealt{Marius2020}, I adopt the globally averaged SEDs (averaging over the spectral shapes for all sources) for the first three simulations. For the last three simulations, I assume a constant model for all sources. The resulting SEDs for all six simulations are shown in the right panel of Figure \ref{fig:siminput}.

\begin{figure*}
\centering
    \includegraphics[scale=0.355]{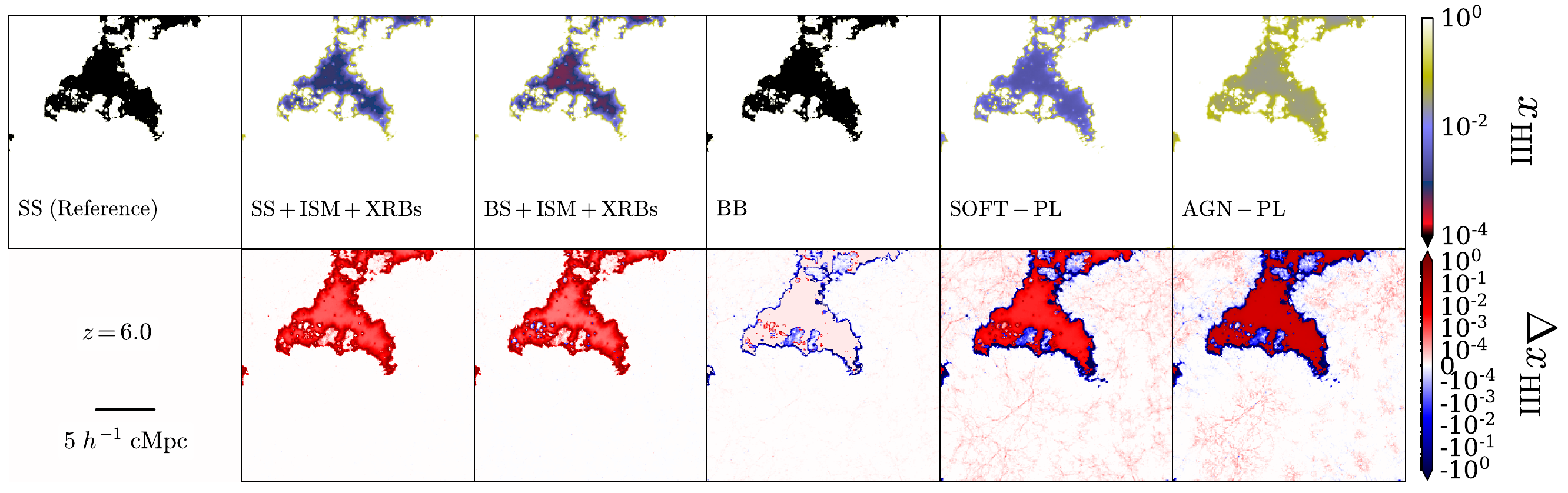}     
    \includegraphics[scale=0.355]{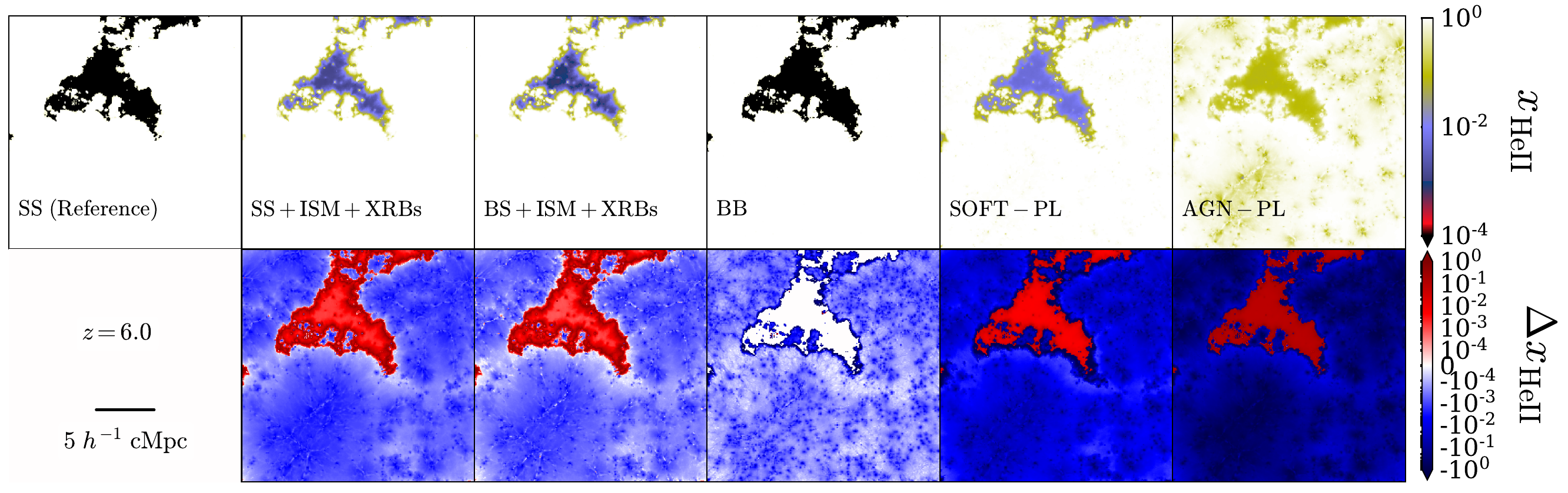}      
    \includegraphics[scale=0.355]{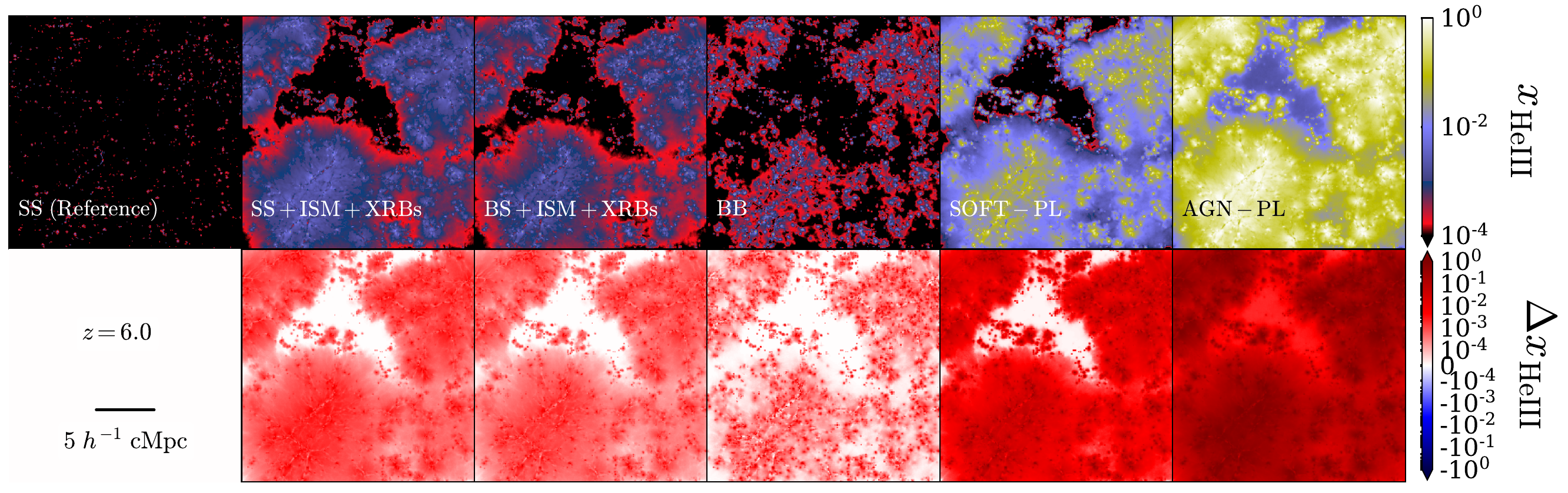}   
    \includegraphics[scale=0.355]{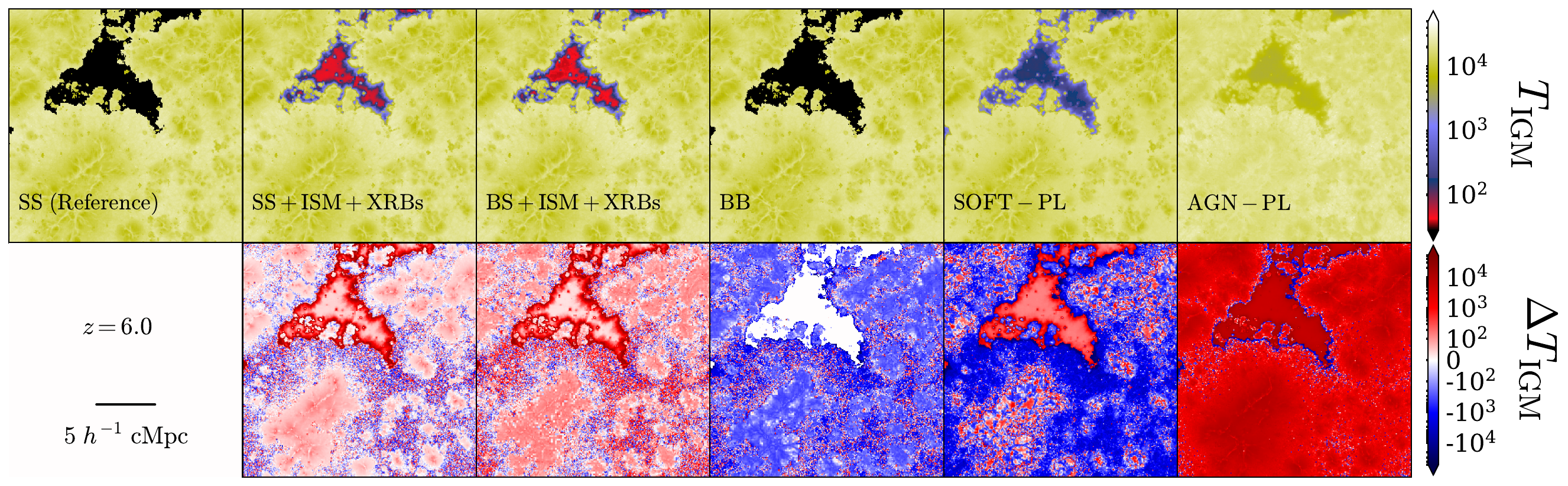}  
    \caption{From top to bottom, slice maps of the H~{\sc ii}, He~{\sc ii}, He~{\sc iii} fractions  and IGM temperature at $z = 6$ for different combinations of SED models: single stars  ({\tt SS}, first column);  single stars, ISM and XRBs ({\tt SS+ISM+XRBs}, second); binary stars, ISM and XRBs ({\tt BS+ISM+XRBs}, third); blackbody ({\tt BB}, fourth); galaxy power law ({\tt SOFT-PL}, fifth) and AGN power law ({\tt AGN-PL}, sixth). The lower sets of rows show the difference with respect to the simulations with single stars only. The maps are $20 h^{-1} \rm{cMpc}$ wide and $78 h^{-1} \rm{ckpc}$ thick.}
    \label{fig:slice_map}
\end{figure*}

\section{Results}
\label{results}

The central question I seek to explore is: \textit{what is the impact of SED models on the properties of the IGM as revealed by the Ly$\alpha$ forest?} To address this, I will first examine the qualitative differences introduced by different SEDs of sources and the resulting photoionization rate, then delve into the quantitative details of reionization and thermal histories. Finally, I will analyze their influence on the observable features associated with the Ly$\alpha$ forest.

\subsection{Qualitative overview of IGM properties}
\label{slice_compare}

For a qualitative discussion of the results, I present the maps of the ionization and thermal state of the IGM at $z = 6$ for all simulations in Figure \ref{fig:slice_map}, together with maps showing the difference relative to \texttt{SS}.

In the \texttt{SS} model, the gas is mostly either fully ionized and hot ($x_{\rm HII} \approx x_{\rm HeII} \approx 1$, $T_{\rm IGM} \gtrsim 10^4$ K) or nearly neutral and cold ($x_{\rm HII} \approx x_{\rm HeII} \lesssim 10^{-4}$, $T_{\rm IGM} \sim 10$ K). 
In contrast, the higher-energy photons emitted by the ISM and XRBs in the \texttt{SS+ISM+XRBs} model introduce extra heating and partial ionization outside of the fully ionized regions, raising the gas temperature by few hundred degrees. 
This is evident in the difference maps of H~{\sc ii}, He~{\sc ii}, and $\rm{\mathit{T}_{IGM}}$, where red pixels highlight regions with increased ionization and heating. He~{\sc ii} behaves similarly to H~{\sc ii}, with the exception that within fully ionized regions, He~{\sc ii} is partially converted into He~{\sc iii}, as it is clearly visible from both maps and difference maps. The impact of He~{\sc iii} is also seen as an increment in the temperature of the ionized regions. I also note that the noise-like pattern in the $\rm{\mathit{T}_{IGM}}$ difference maps within fully ionized regions results from Monte Carlo noise, as obtaining exactly the same temperature between simulations with different ionizing sources is statistically unlikely. However, this is not the case for the ionization fraction once a cell is fully ionized. 
The inclusion of binary stars in the \texttt{BS+ISM+XRBs} simulation displays a very similar ionization and thermal structure of the IGM as \texttt{SS+ISM+XRBs}. However many of the higher energy photons are used in lifting the level of partial ionization surrounding the fully ionized regions, resulting in fewer photons available to penetrate further into the neutral IGM, leading to reduced partial ionization in more distant regions. The \texttt{BS+ISM+XRBs} model results in a slightly lower He~{\sc iii} fraction than \texttt{SS+ISM+XRBs} because a significant fraction
of ionizing photons is consumed in expanding the ionized bubbles, while the total ionizing photon budget per time step remains fixed.

While comparing with the simplified SED prescriptions commonly adopted in literature for reionization studies, I notice certain differences with respect to my reference \texttt{SS} model. The extent of the fully ionized regions (ie. $x_{\rm HII} \gtrsim 0.99$) in the \texttt{BB} model is slightly smaller than in the reference \texttt{SS} one, as seen from the dark blue pixels surrounding the ionized gas in the difference maps of $x_{\rm HII}$ and $x_{\rm HeII}$. This is primarily due to a slight difference in the numbers of H~{\sc i} ionizing photons, otherwise, the maps in the \texttt{BB} model are qualitatively same as the \texttt{SS} model. However, The amount of doubly ionized helium is higher in \texttt{BB}, since this model produces more photons at and slightly beyond the He~{\sc ii} ionization energy compared to the \texttt{SS} model. However, having less IGM temperature (denoted as blue pixels in the temperature difference maps) is a combined effect of having lower number of photons around the H~{\sc i} ionization energy (since the total ionizing photon production rate is constant in all the models, model with lower number of ionizing photons in the higher frequency regime will have more H~{\sc i} ionizing photons) and higher He~{\sc ii} regions. For similar reason, in the \texttt{SOFT-PL} model, the extent of fully ionized regions is even smaller.
The \texttt{SOFT-PL} model also has a higher fraction of gas partially ionized by the higher energy tail of the SED, with many cells reaching $x_{\rm HII} \sim 10^{-2}$. The temperature maps further reflect this, showing increased heating in the partially ionized regions and inside ionized bubbles, similar but more pronounced feature as \texttt{BB} model. However, in the dense areas, additional helium double ionization heating become the dominant factor in raising the temperature. The difference maps in $x_{\rm HII}$ reveal filamentary structures with positive $\Delta x_{\rm HII}$, representing regions which has lower recombination rate due to boosted temperature. 
In the \texttt{AGN-PL} model, all the effects discussed above are even more pronounced. The higher number of high-energy photons leads to stronger partial ionization of H~{\sc i} and He~{\sc i}, with many cells reaching $x_{\rm HII}$ and $x_{\rm HeII} \sim 10^{-1}$. These energetic photons also drive more helium double ionization, creating large regions of fully ionized helium. As a consequence, the overall temperature is significantly higher, and the temperature difference maps appear mostly red, reflecting widespread heating across the slice. 

While these differences highlight the qualitative impact of various SED model assumptions, a more insightful approach is to perform a quantitative analysis. In the following sections, I will focus on this aspect.

\subsection{H~{\sc i} photoionization rate}
\label{photoionizationrate}
\begin{figure}   
\centering
\includegraphics[width=80mm]{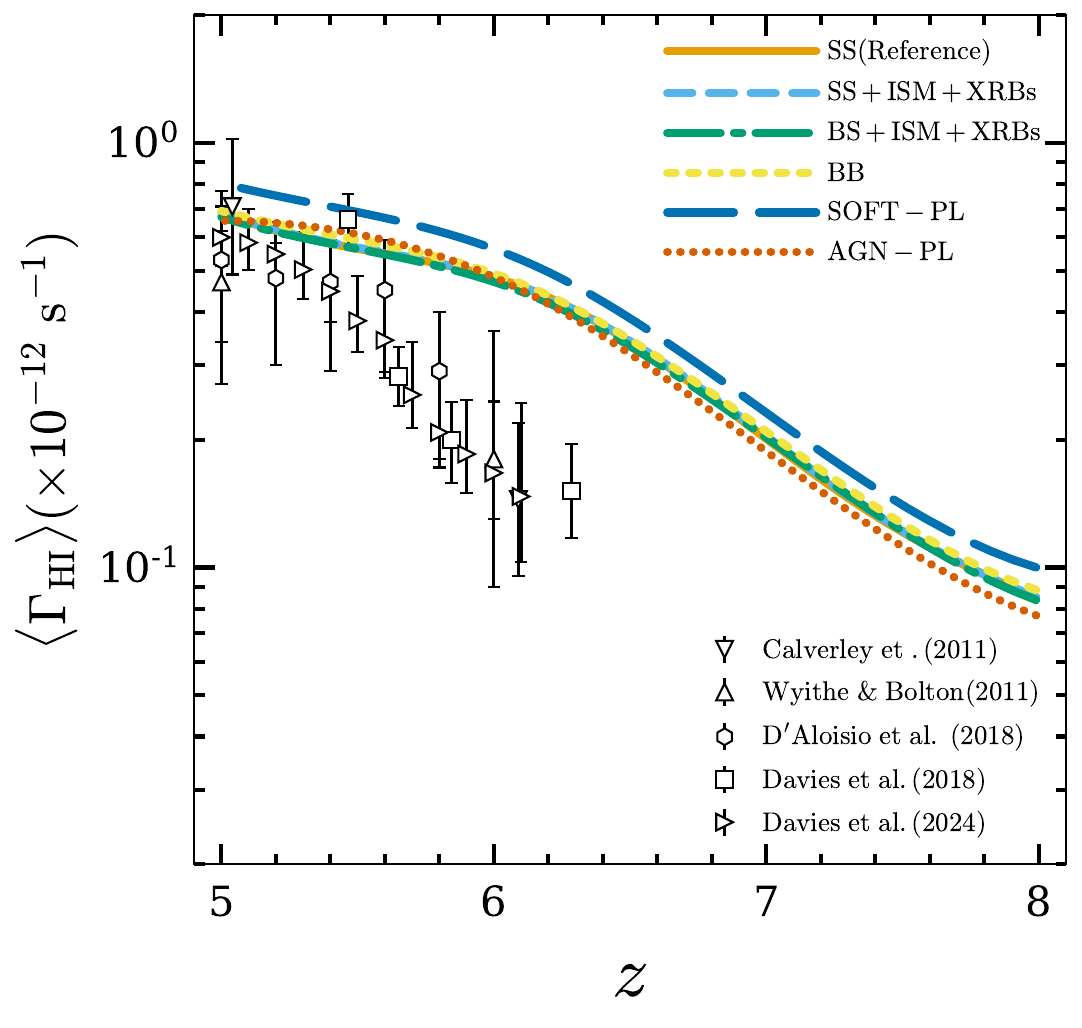}   
    \caption{The resulting volume averaged H~{\sc i} photoionization rate for all SED models: \texttt{SS} (solid orange curve), \texttt{SS+ISM+XRBs} (dashed sky blue), \texttt{BS+ISM+XRBs} (dash-dotted green), \texttt{BB} (short-dashed yellow), \texttt{SOFT-PL} (long-dashed blue) and \texttt{AGN-PL} (dotted vermilion). The symbols denote a compilation of observational constraints from the literature \citep{Calverley2011,wyithe2011,daloisio2018,Davies2018, davies2024}.}
    \label{fig:gammahi}
\end{figure}
In Figure \ref{fig:gammahi}, I present the volume averaged H~{\sc i} photoionization rate ($\rm{\langle \mathit{\Gamma}_{HI} \rangle}$) extracted from my simulations at $z \lesssim 8$, where I achieve higher spatial resolution for RT calculations (as discussed in Section \ref{section:2.3}). 
Since the photoionization rate is not a direct output of my simulations, I estimate it indirectly by assuming ionization equilibrium in each simulation cell (following the approach as done in \citealt{Ciardi2012}). Under this assumption, the rate can be expressed as:
\begin{equation}
\Gamma_{\mathrm{HI}} = \alpha_{\mathrm{HII}}(T) \frac{n_e n_{\mathrm{HII}}}{n_{\mathrm{HI}}} - \gamma_{e\mathrm{HI}}(T)n_e,
\end{equation}
where $\alpha_{\mathrm{HII}}(T)$ and $\gamma_{e\mathrm{HI}}(T)$ are the temperature-dependent hydrogen recombination and collisional ionization rates (in units of cm$^3$\,s$^{-1}$), respectively. All other quantities have their standard physical meanings. This equilibrium-based estimate is generally valid for most of the cells in my simulation volume after reionization. 
The results are shown for all simulations, alongside observational constraints from \citealt{Calverley2011}, \citealt{wyithe2011}, \citealt{daloisio2018}, \citealt{Davies2018} and \citealt{davies2024}. Note that, I have calculated  $\rm{\mathit{\Gamma}_{HI}}$ taking into account only the ionized cells (with $x_{\rm HII} \gtrsim 0.99$), although I notice that computing $\rm{\langle \mathit{\Gamma}_{HI} \rangle}$ from all cells in the simulation volume shows marginal differences (i.e. deviation within 5 $\%$ at $z\lesssim7$). All simulations follow a similar overall trend, with $\rm{\langle \mathit{\Gamma}_{HI} \rangle}$ increasing by approximately an order of magnitude from $z = 8$ to $z = 5$. At $z \sim 6$, the evolution of $\rm{\langle \mathit{\Gamma}_{HI} \rangle}$ slows down and eventually flattens, marking the phase when the majority of the simulation volume becomes ionized. 

Although most of the simulations yield photoionization rates that are systematically higher than the observations, there are distinct, albeit minor, differences between the models. When comparing the different models to the \texttt{SS} reference simulation, I observe that despite the general trend being similar, differences emerge. Notably, the \texttt{SOFT-PL} model overall yields the largest $\rm{\langle \mathit{\Gamma}_{HI} \rangle}$ value, higher by a factor of $\sim 1.15$ compared to \texttt{SS}. This is primarily due to lesser IGM temperature which essentially expresses in higher recombination rate and hence the H~{\sc i} photoionization rate. Meanwhile, models with other SEDs, i.e. \texttt{BB}, \texttt{SS+ISM+XRBs}, and \texttt{BS+ISM+XRBs} exhibit similar trends, with \texttt{BB} producing the highest $\rm{\langle \mathit{\Gamma}_{HI} \rangle}$ among them because of same reason.
The behaviour of the \texttt{AGN-PL} model is particularly interesting, as at the highest redshifts it exhibits the lowest $\rm{\langle \mathit{\Gamma}_{HI} \rangle}$ value, but at $z<6$ it becomes higher than all the other models (except for \texttt{SOFT-PL}). This is primarily because of a combined effect of lesser recombination rate, excess amount of electrons due to pronounced double ionization of helium as well as higher collisional ionization rate due to higher temperature boosting. In the following sections, I conduct a detailed analysis of IGM ionization and heating, comparing various simulations, and then explore the impact on the characteristics of the Ly$\alpha$ forest. 

\begin{figure}
\centering
\includegraphics[width=90mm]{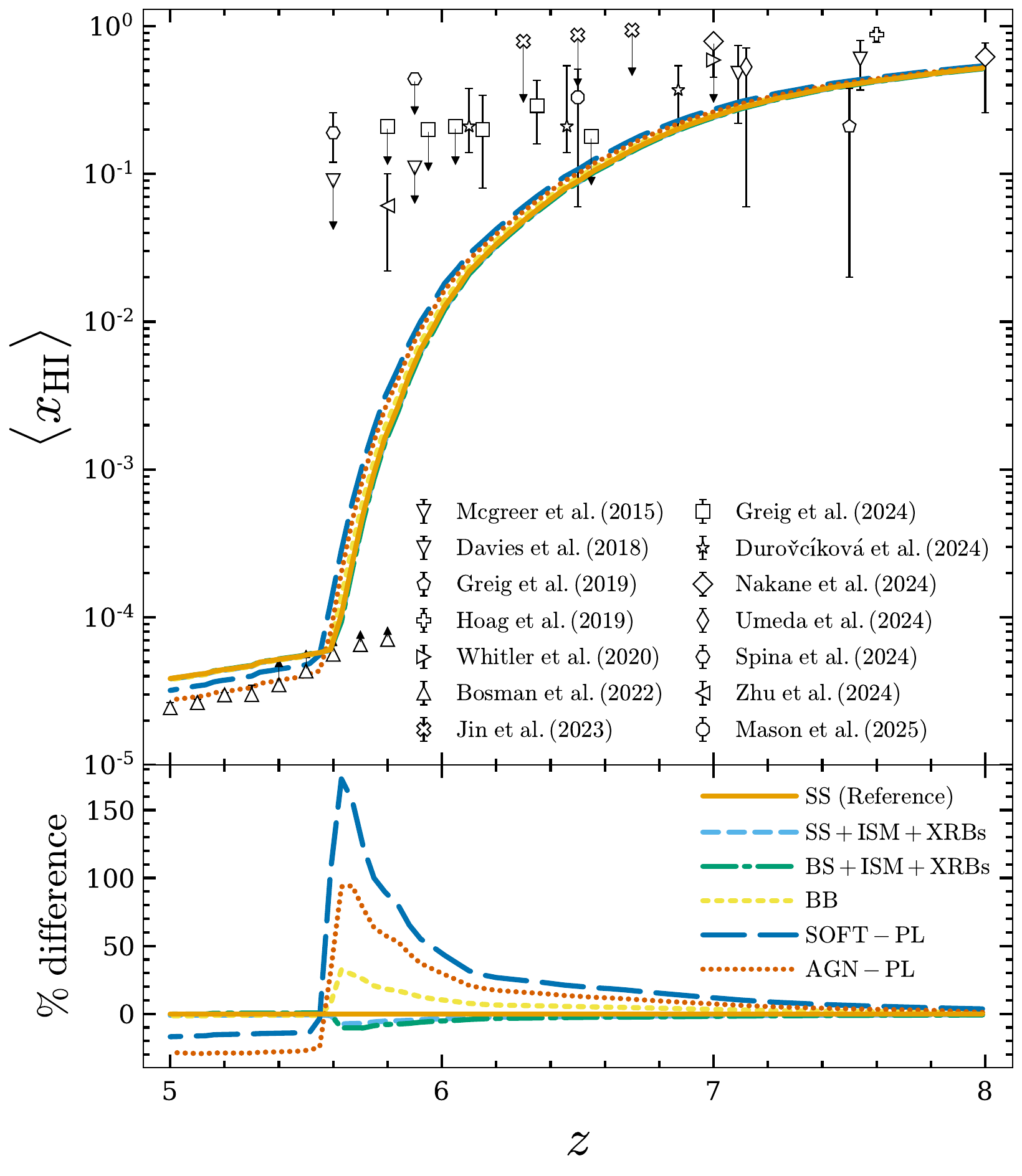}   
    \caption{{\it Top panel:} Redshift evolution of the volume averaged $\mathit{x}\rm{_{HI}}$ for all SED models: \texttt{SS} (solid orange curve), \texttt{SS+ISM+XRBs} (dashed sky blue), \texttt{BS+ISM+XRBs} (dash-dotted green), \texttt{BB} (short-dashed yellow), \texttt{SOFT-PL} (long-dashed blue) and \texttt{AGN-PL} (dotted vermilion). The symbols denote a compilation of observational constraints from the literature \citep{McGreer2015,greig2019,hoag2019,Whitler2020,Bosman2022,Jin2023,Greig2024,durovcicova2024,nakane2024,umeda2024,spina2024,zhu2024,Mason2025}. {\it Bottom panel:} Fractional difference with respect to the \texttt{SS} simulation.}
    \label{fig:XHIevol}
\end{figure}

\subsection{Reionization history}
\label{sec:reion}
To discuss the overall effect induced by different SED models, in Figure \ref{fig:XHIevol} I display the redshift evolution of the volume averaged H~{\sc i} fraction for all simulations, together with the fractional difference with respect to \texttt{SS}, and a compilation of observational results \citep{McGreer2015,greig2019,hoag2019,Whitler2020,Bosman2022,Jin2023,Greig2024,durovcicova2024,nakane2024,umeda2024,spina2024,zhu2024,Mason2025}. With the exception of \texttt{AGN-PL}, all simulations produce a neutral hydrogen fraction which is slightly higher than observational constraints at $z \lesssim 5.5$, where my primary focus lies. The curves are able to reproduce most data points at higher redshift, where though there is less consistency among different observations.  

As discussed earlier, the inclusion of binary stars or more energetic sources leads to increased partial ionization and heating, resulting in a lower H~{\sc i} fraction. The most significant differences appear at $z = 5.6$, where the \texttt{SS+ISM+XRBs} and \texttt{BS+ISM+XRBs} models show deviations of approximately 7.5$\%$ and 10$\%$, respectively. In contrast, the \texttt{BB} model maintains a higher H~{\sc i} fraction up to $z = 5.6$, indicating a slightly delayed reionization (by about $\Delta z \sim 0.05$) compared to the \texttt{SS} model. This delay is even more pronounced in the \texttt{SOFT-PL} and \texttt{AGN-PL} models, which sustain higher H~{\sc i} fractions over a similar redshift range. This occurs because, as these models contain more high-energy photons, they produce fewer H~{\sc i}-ionizing photons, given that the total number of ionizing photons is kept constant (as discussed in Section \ref{slice_compare}). The \texttt{SOFT-PL} model shows the largest deviation, with a fractional difference exceeding 170$\%$ at $z = 5.6$, while the deviation for the \texttt{AGN-PL} model reaches about 100$\%$.
As reionization progresses and $\langle x_{\rm HI} \rangle$ falls below $10^{-4}$, the differences between the models diminish and eventually vanish. However, during the final stages of reionization, the \texttt{SOFT-PL} and \texttt{AGN-PL} models predict a Universe that is more ionized than in the \texttt{SS} model, with deviations of approximately 15$\%$ and 30$\%$, respectively—primarily due to additional heating and its impact on the photoionization rate, as discussed in Section \ref{photoionizationrate}.

\begin{figure}
\centering
\includegraphics[width=90mm]{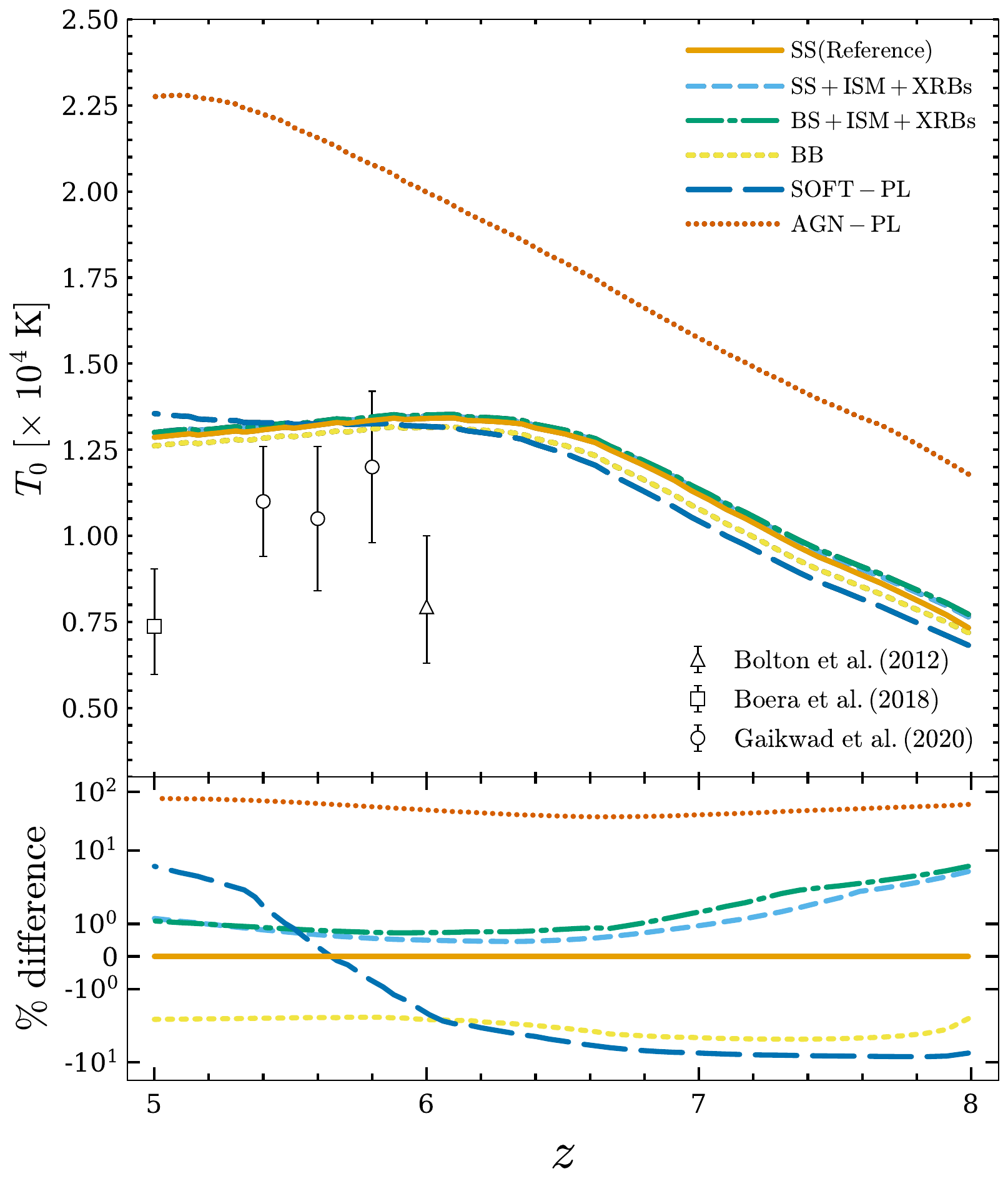}   
    \caption{{\it Top panel:} Redshift evolution of the IGM temperature at mean density ($\mathit{T}\rm{_{0}}$) for \texttt{SS} (solid orange curve), \texttt{SS+ISM+XRBs} (dashed sky blue), \texttt{BS+ISM+XRBs} (dash-dotted green), \texttt{BB} (short-dashed yellow), \texttt{SOFT-PL} (long-dashed blue) and \texttt{AGN-PL} (dotted vermilion). The symbols denote a compilation of observational constraints from the literature \citep{Bolton2012,boera2019,Gaikwad2020}. {\it Bottom panel:} Fractional difference with respect to the \texttt{SS} simulation.}
    \label{fig:tempevol}
\end{figure}

\subsection{Thermal history}
\label{thermal_history}
The ionization of hydrogen and helium injects energy into the IGM through photo-heating, raising its temperature, and consequently affecting the dynamics of gas, its cooling rate, and its accretion onto galaxies. In Figure \ref{fig:tempevol}, I show the redshift evolution of the IGM temperature at mean density, calculated by selecting all the cells in the simulation volume with gas density within $\pm1\%$ of the mean density.  
As expected, with the exception of \texttt{AGN-PL}, $ T_{0} $ rises with decreasing redshift, reaching a peak at $ z= 6 $ when about 99\% of the simulation volume is ionized. Subsequently, the rate of heat injection from photo-heating diminishes, leading to a plateau followed by a gradual decline, which is much shallower, though, than the one  suggested by a compilation of observational data \citep{Bolton2012,boera2019,Gaikwad2020}. The similar behaviour has been observed in the context of helium reionization by \citealt{Basu2024}. However, the observations are dominated by systematics, rather than an issue with the IGM thermal evolution. I have tested that the shallower evolution observed is not due to additional heating associated to the increasing number of cells experiencing hydrodynamical feedback effects (with cells having IGM temperature $> 10^{5}$ K), which remain challenging to disentangle or accurately correct for while post processing with the RT code. In Appendix \ref{thermal_history_with_Tcuts}, I explore this issue quantitatively by computing the thermal history taking into account the cells below some temperature thresholds.
Interestingly, the \texttt{SOFT-PL} and \texttt{AGN-PL} models deviate from this trend, showing no significant decline in $T_0$. While the \texttt{SOFT-PL} model exhibits a brief saturation at $z = 6$, the temperature continues to rise at lower redshifts, primarily due to the effects of helium double ionization. A similar trend is evident in the \texttt{AGN-PL} model, though with a significantly higher amplitude, consistently exceeding the temperatures of other models by more than 7500 K due to the abundant energetic photons populating this SED. Interestingly, though, while the evolution of the  H~{\sc i} fraction is mostly consistent with observations, in particular at $z<5.5$, the IGM temperature is always substantially higher than the data points.  

Both \texttt{SS+ISM+XRBs} and \texttt{BS+ISM+XRBs} exhibit temperatures higher than \texttt{SS} due to additional heating from more energetic photons, with binary stars contributing even more significantly. The maximum temperature difference reaches approximately $5\%$ at $z = 8$, but gradually decreases to about $1\%$ at lower redshifts, as reionization progresses.  
The \texttt{BB} model consistently shows lower temperatures, with a nearly constant deviation of $\sim 4\%$ for $z \lesssim 6$ as discussed in Section \ref{slice_compare}. Differently, \texttt{SOFT-PL}  initially produces temperatures lower than the \texttt{SS} model, reaching a maximum negative deviation of $\sim 10\%$ at $z = 8$. 
However, below $z = 6$, with most of the H~{\sc i} and He~{\sc i} in the simulation volume getting ionized, this deviation gradually decreases and eventually turns positive, reaching approximately $8\%$ at $z = 5$ due to more pronounced effect from double helium reionization. 
As already mentioned, the \texttt{AGN-PL} model is characterized by larger values of $T_0$ at all redshifts, with deviations of $\sim 100\%$.

In the following section, I will investigate if the differences noticed in IGM temperature and ionization state affect the Ly$\alpha$ forest properties.

\subsection{Ly$\alpha$ forest statistics}
\label{lyalpha_forest}
To investigate Ly$\alpha$ forest properties, I have generated synthetic spectra by extracting 16384 sightlines at each RT snapshot, each spanning the full box length parallel to the $z$-direction. For each pixel $i$ in a sightline I evaluate the normalized transmitted flux $F(i) = \exp[-\tau(i)]$, where
\begin{equation}
    \tau(i) = \frac{c \sigma_{\alpha} \delta R}{\sqrt{\pi}} \sum_{j=1}^{N} \frac{n\mathrm{_{HI}}(j)}{b\mathrm{_{HI}}(j)} \ \tilde{V}(i,j)
\end{equation}
is the Ly$\alpha$ optical depth. Here $N$ is the number of pixels in a sightline, $\sigma_{\alpha}$ is the Ly$\alpha$ scattering cross-section, $c$ is the speed of light, $\delta R$ is the size of a pixel in proper distance units, $n_\mathrm{{HI}}(j)$ is the H~{\sc i} number density at the position of pixel $j$, $\mathrm{\mathit{b}_{HI}(\mathit{j}) = (2 \mathit{k}_{B} \mathit{T}_{IGM} / \mathit{m}_{H})^{1/2}}$, $\mathrm{\mathit{k}_{B}}$ is the Boltzmann constant, $\mathrm{\mathit{m}_{H}}$ is the hydrogen mass, and $\tilde{V}$ is the convoluted Voigt profile approximation provided by \citealt{tepper2006}. The latter depends on the IGM temperature and the peculiar velocity. Throughout this chapter , I ignore the contamination from other lines into the wavelength window of the Ly$\alpha$ forest as this is expected to have a negligible
impact on my results \citep{hellsten1998,yang2022}.

\begin{figure}
\centering
    \includegraphics[width=130mm]{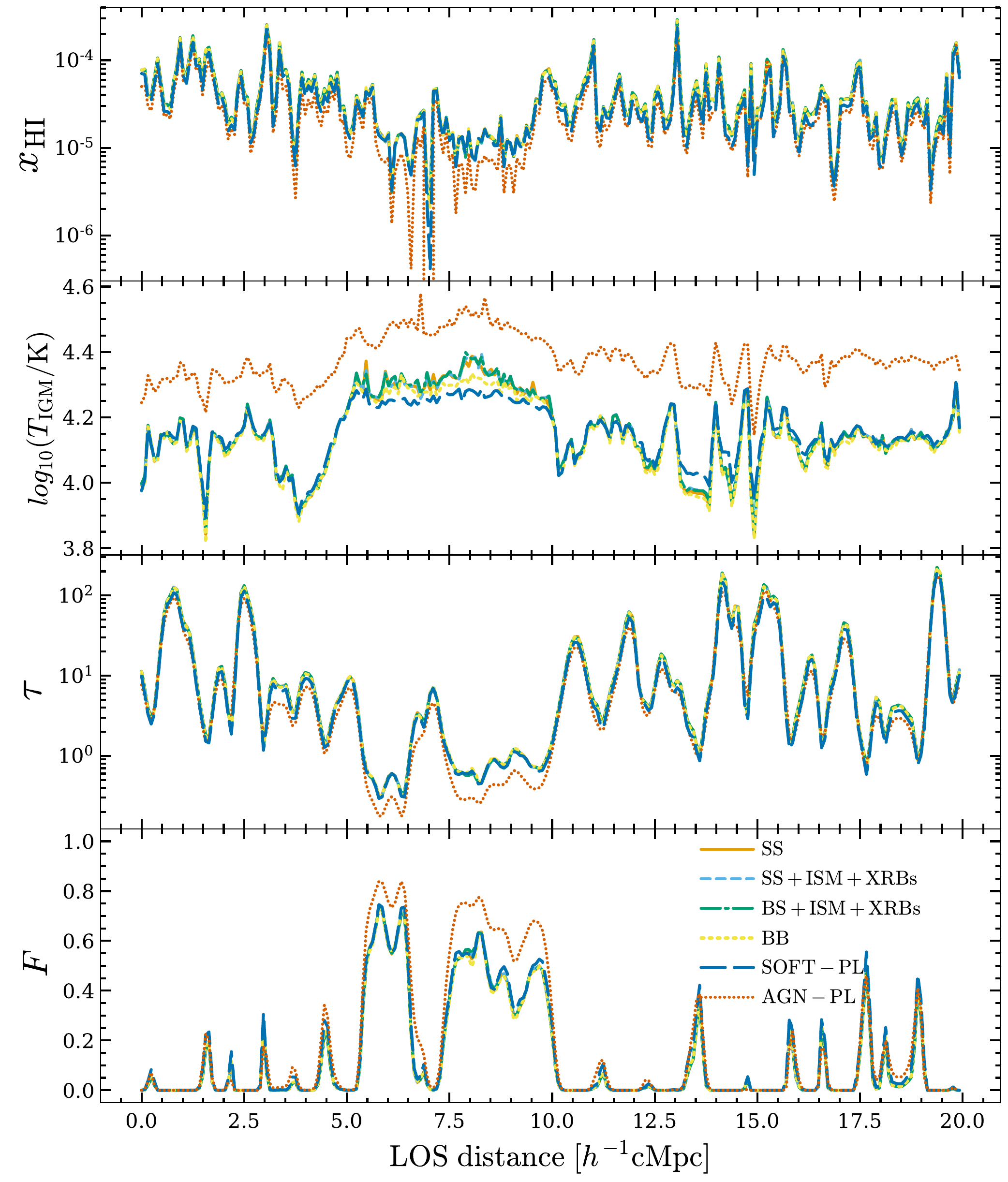}
    \caption{IGM properties along a reference line of sight at \textit{z} = 5 for \texttt{SS} (solid orange curve), \texttt{SS+ISM+XRBs} (dashed sky blue), \texttt{BS+ISM+XRBs} (dash-dotted green), \texttt{BB} (short-dashed yellow), \texttt{SOFT-PL} (long-dashed blue) and \texttt{AGN-PL} (dotted vermilion). From top to bottom the panels refer to the H~{\sc i} fraction, IGM temperature, Ly$\alpha$ optical depth, and normalized transmitted flux. }
    \label{fig:los}
\end{figure}

\subsubsection{Individual lines of sight}
\label{los}

As a reference, in Figure \ref{fig:los} I display the neutral hydrogen fraction ($\rm{\mathit{x}_{HI}}$), the IGM temperature ($\rm{\mathit{T}_{IGM}}$), the Ly$\alpha$ optical depth ($\rm{\tau}$), and the Ly$\alpha$ transmitted flux ($F$) along a single line of sight (LOS) at $z=5$. For clarity, I omit the pixel index when referring to the values of these physical quantities.

The abundance of H~{\sc i} remains remarkably consistent across all simulations, with only slight variations. Notably, a marginally lower value is observed in the {\tt AGN-PL} model, as illustrated also in Figure~\ref{fig:XHIevol}. 
The ionized regions are characterized by an IGM temperature $T_{\rm IGM} \sim 10^4$~K, with  {\tt AGN-PL} exhibiting the highest value of $\sim 10^{4.4}$~K, as discussed in previous sections. In all models the highest temperatures are reached in regions which have been recently ionized. 
The behaviour observed in the H~{\sc i} fraction is directly reflected in the corresponding Ly$\alpha$ optical depth, which has typical values below 5 in regions of higher ionization. It is in these areas that I also observe a larger transmission of Ly$\alpha$ photons. When comparing different SED models, the {\tt AGN-PL} consistently exhibits the highest transmission, suggesting that its additional heating affects the width and shape of transmission features as a result of the lower recombination rate due to the higher temperature.
Instead, the other models do not show substantial variations in the transmitted flux, as the differences in optical depth introduce only subtle fluctuations primarily driven by the additional heating. This slightly enhances the transmission by reducing small-scale variations in optical depth. 

For a better comparison, I have computed the mean transmission flux ($\rm{\bar{\mathit{F}}}$) for the entire simulation volume in each SED models at $z=5$. As I find out, most of the SED models i.e. {\tt SS}, {\tt SS+ISM+XRBs}, {\tt BS+ISM+XRBs} and {\tt BB}, show almost similar $\rm{\bar{\mathit{F}}}$ value as 0.14. The slightly higher IGM temperature in the {\tt SOFT-PL} model displays slightly higher $\rm{\bar{\mathit{F}}}$, say, 0.16. This difference gets more pronounced in the {\tt AGN-PL}, having about $\sim 30 \%$ increase in the mean transmission flux compared to {\tt SS} model, with a value of 0.18. 

While the characteristics discussed above are common to the majority of lines of sight, to provide a more quantitative comparison, in the next sections I calculate quantities which more accurately capture the statistical behaviour of the Ly$\alpha$ forest.

\begin{figure}   
\centering
\includegraphics[width=100mm]{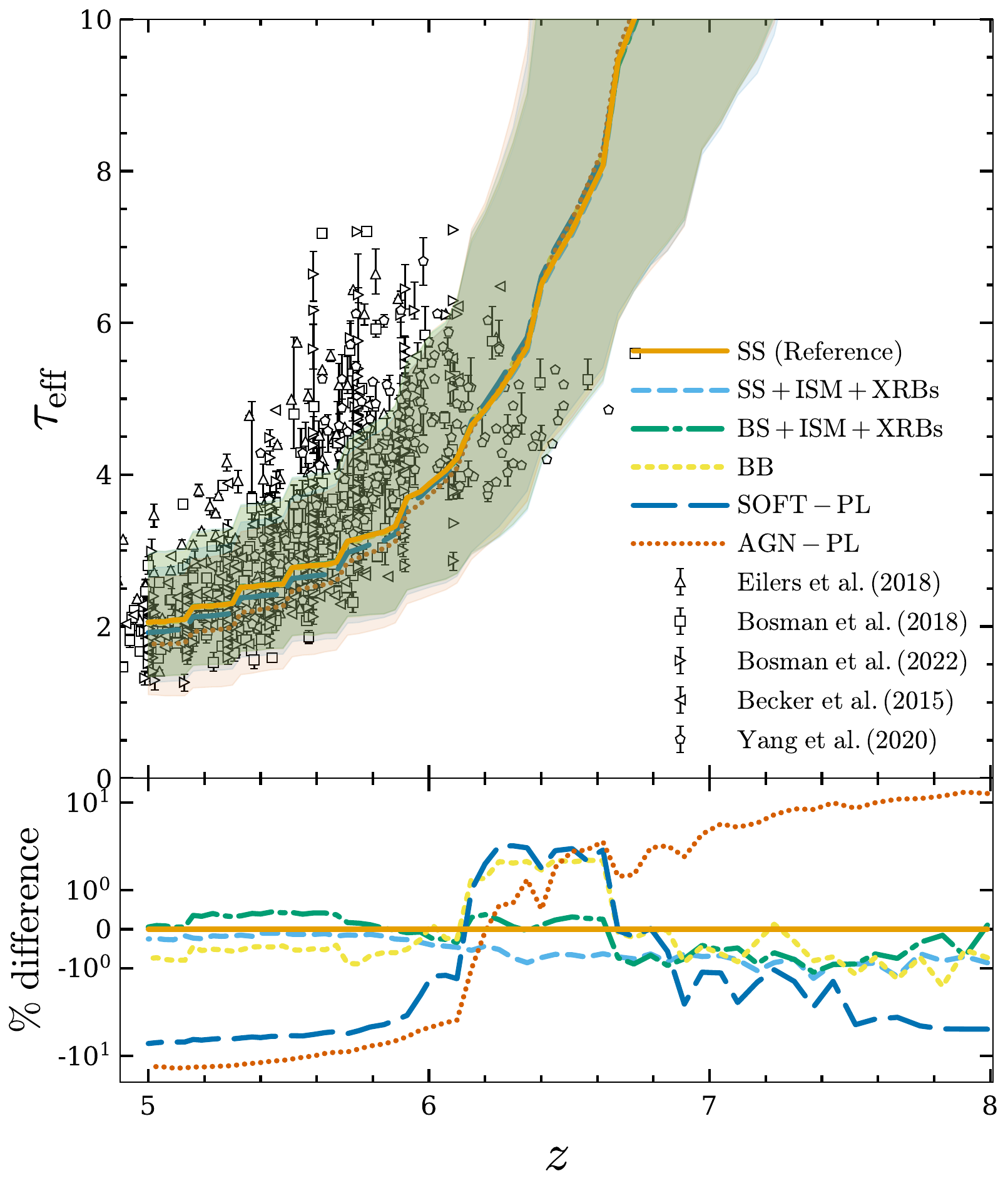}
    \caption{{\it Top panel:} Redshift evolution of the median Ly$\alpha$ effective optical depth for \texttt{SS} (solid orange curve), \texttt{SS+ISM+XRBs} (dashed sky blue), \texttt{BS+ISM+XRBs} (dash-dotted green), \texttt{BB} (short-dashed yellow), \texttt{SOFT-PL} (long-dashed blue) and \texttt{AGN-PL} (dotted vermilion). The shaded regions indicate 95$\%$ confidence intervals. The symbols denote a compilation of observational constraints from the literature \citep{Becker2015,eilers2018,yang2020,bosman2018,Bosman2022}. {\it Bottom panel:} Fractional difference with respect to the \texttt{SS} simulation. }
    \label{fig:taueff_evol}
\end{figure}

\begin{figure}   
\centering
\includegraphics[width=100mm]{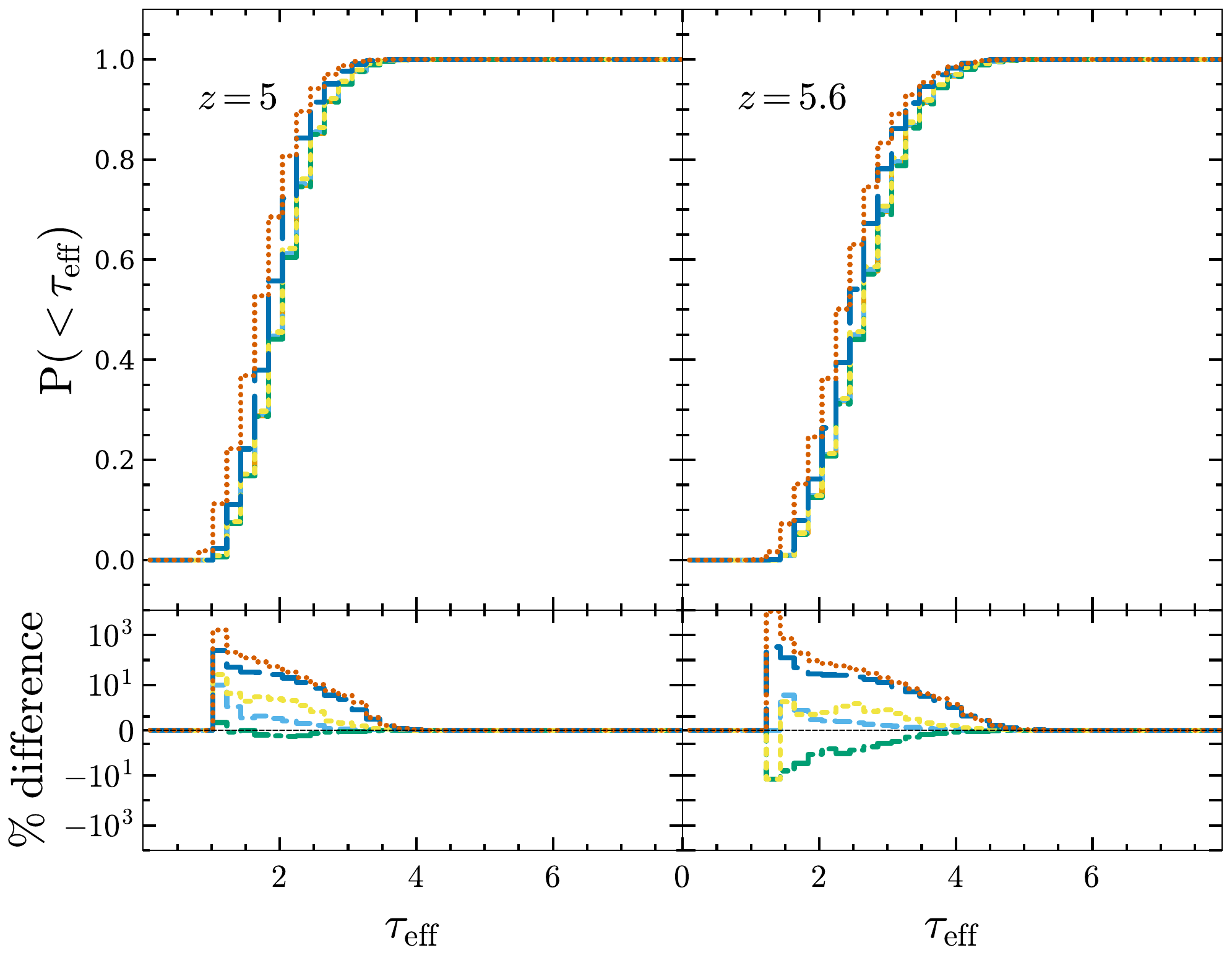}
    \caption{{\it Top panels:} Cumulative distribution of Ly$\alpha$ effective optical depth as $z=5.0$ and 5.6 for \texttt{SS} (solid orange curve), \texttt{SS+ISM+XRBs} (dashed sky blue), \texttt{BS+ISM+XRBs} (dash-dotted green), \texttt{BB} (short-dashed yellow), \texttt{SOFT-PL} (long-dashed blue) and \texttt{AGN-PL} (dotted vermilion). {\it Bottom panels:} Fractional difference with respect to the \texttt{SS} simulation. }
    \label{fig:taueff_cdf}
\end{figure}

\subsubsection{Effective optical depth}

The first statistical quantity I discuss is the Ly$\alpha$ effective optical depth ($\tau_{\rm eff}$), a widely used characterization of the Ly$\alpha$ forest. To calculate this, I combine multiple synthetic spectra to obtain a length of 50$\rm{\mathit{h}^{-1} cMpc}$, traditionally adopted to represent observational data (e.g. \citealt{Bosman2022}), resulting in over $\sim 5500$ sightlines at each redshift. The effective optical depth is then computed as:
\begin{equation}
\tau_{\rm{eff}} = - \ln (\langle F \rangle),
\label{columndensity}
\end{equation}
where $\langle F \rangle$ is the mean transmitted flux of all the pixels in each new sightline. Figure \ref{fig:taueff_evol} displays the redshift evolution of the median $\tau_{\rm eff}$, along with the central $95 \%$ of the data, as well as a compilation of observational constraints from \citealt{Becker2015}, \citealt{eilers2018}, \citealt{bosman2018}, \citealt{yang2020} and \citealt{Bosman2022}. It is important to note that the input ionizing photon production rate in the \texttt{SS} simulation is calibrated to match the observational estimates of $\langle \rm F \rangle$ at $z \lesssim 5.5$ \citep{Becker2015,eilers2018,bosman2018,Bosman2022}. As the ionizing photon production rate is same in all simulations, this ensures that the reference model is consistent with empirical data, while allowing me to examine deviations in the other models. All simulations produce a very similar behaviour, with $\tau_{\rm eff}$ decreasing with the progress of reionization, and slightly flattening at $z\lesssim6$. It is also interesting to see that all models are consistent with the majority of observational data within the scatter.

As can be better appreciated in the bottom panel of the figure, for $z \gtrsim 6$, the differences between models, with the exception of {\tt AGN-PL} and {\tt SOFT-PL}, are modest, with a maximum deviation from the {\tt SS} model of $\sim 1\%$, which becomes $<0.5\%$ at $z\sim5$ for {\tt SS+ISM+XRBs} and {\tt BS+ISM+XRBs}. This suggests that the SEDs of BS, ISM and XRBs do not significantly alter the effective optical depth. The deviations remain at the 1\% level also for a black-body spectrum. 
On the other hand, both {\tt SOFT-PL} and {\tt AGN-PL} at $z=5$ produce an effective optical depth which is $\sim 10\%$ lower than the {\tt SS} one, highlighting the impact of the larger abundance of more energetic photons resulting in hotter gas temperatures, which then lower the neutral fraction through the reduced recombination rate, which in the end produce a higher transmission of Ly$\alpha$ photons in comparison to the other models. 

To further examine the differences among the SED models in terms of $\tau_{\rm eff}$, Figure \ref{fig:taueff_cdf} presents the cumulative distribution functions (CDFs) at three redshifts: $z = 6$, $5.5$, and $5$. I emphasize that my simulated optical depths are not calibrated to reproduce any observed CDF or mean transmission flux. As expected, the CDF systematically shifts toward lower optical depths as reionization progresses and more regions become ionized. When comparing different SED models to the \texttt{SS} reference, I find negligible deviations at both the high-$\tau_{\rm eff}$ end ($\tau_{\rm eff} > 6, 5, 4$ for $z = 6, 5.5,$ and $5$, respectively, where the cumulative probability is already unity) and the low-$\tau_{\rm eff}$ end, where the cumulative probability remains near zero. The differences are most apparent at intermediate $\tau_{\rm eff}$ values. Most models deviate from \texttt{SS} by less than $5\%$, whereas \texttt{SOFT-PL} and \texttt{AGN-PL} exhibit much larger discrepancies, reaching up to $\sim 1000\%$, with the latter showing the strongest effect.

Overall, aside from \texttt{SOFT-PL} and \texttt{AGN-PL}, all other models yield very similar Ly$\alpha$ averaged quantity, particularly in terms of the median and scatter of $\tau_{\rm eff}$. Nonetheless, it remains interesting to investigate how these models influence small- and large-scale fluctuations, which I address in the next section.

\subsubsection{1-D flux power spectra}
\label{lya_ps}

\begin{figure}  
\centering
    \includegraphics[width=100mm]{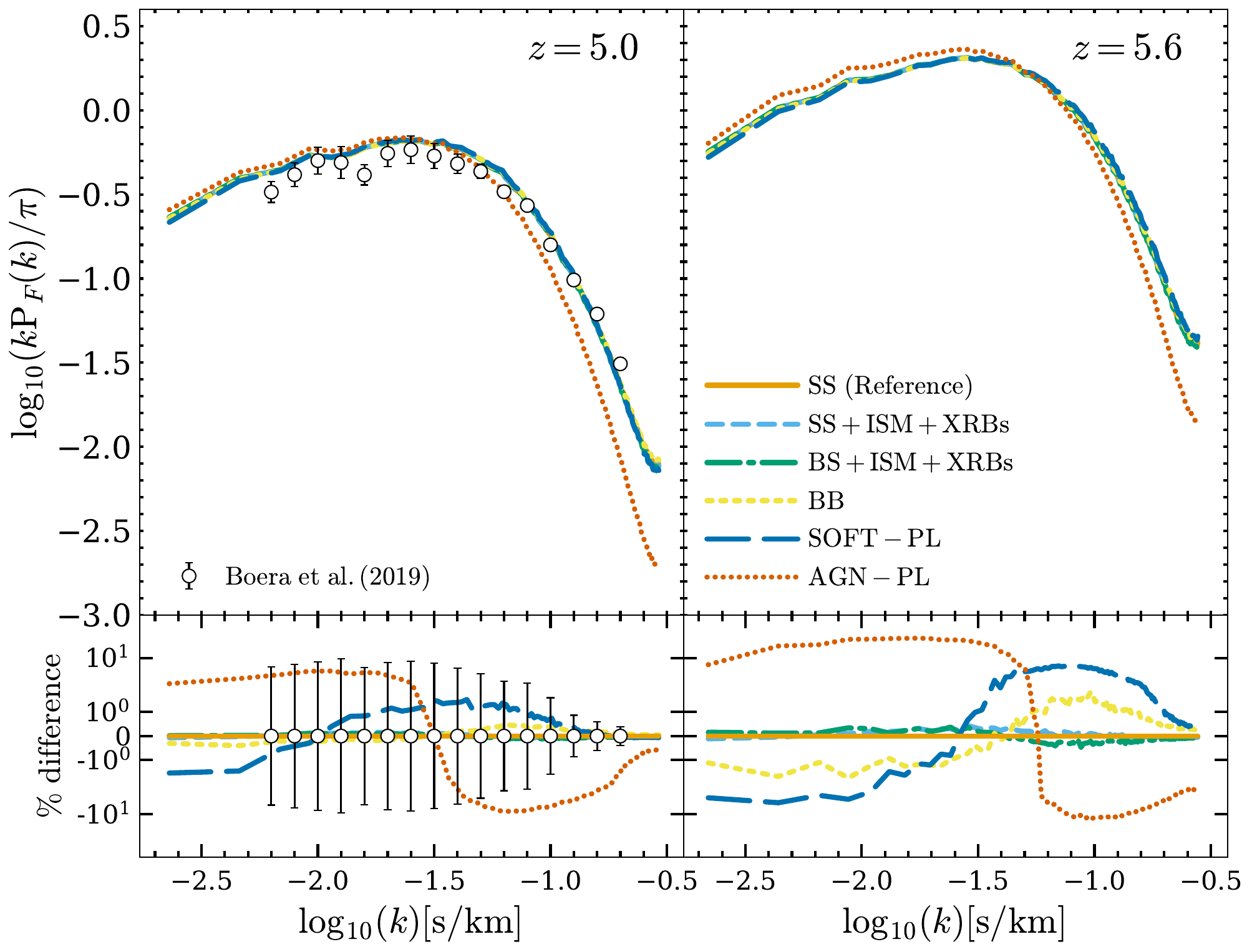}
    \caption{{\it Top panels:} 1-D dimensionless Ly$\alpha$ forest transmission flux power spectra at $z=5$ (\textit{left panel}) and 5.6 (\textit{right}) for \texttt{SS} (solid orange curve), \texttt{SS+ISM+XRBs} (dashed sky blue), \texttt{BS+ISM+XRBs} (dash-dotted green), \texttt{BB} (short-dashed yellow), \texttt{SOFT-PL} (long-dashed blue) and \texttt{AGN-PL} (dotted vermilion). The symbols denote a compilation of observational constraints from the literature \citep{boera2019}.
    {\it Bottom panel:} Fractional difference with respect to  the \texttt{SS} simulation along with the uncertainties of the observational constraints.}
   \label{fig:ps_w_scaling}
\end{figure}

Here now, I calculate the 1-D power spectrum (PS) of Ly$\alpha$ transmission flux using all sightlines from my simulation, and plot it in Figure~\ref{fig:ps_w_scaling} at $z=5$ and $z=5.6$. The latter redshift is chosen because it is when the largest differences  in $\langle \rm{\mathit{x}_{HI} \rangle}$ emerge, and at the same time there are enough regions of high transmission to render my exploration of their fluctuations more statistically significant. Note that, I have rescaled the mean transmission flux in all the simulations to match with \citealt{Bosman2022}.  The overall shape of the power spectra is similar at both redshifts, increasing from $k \sim 10^{-0.5} \ \mathrm{s \ km^{-1}}$ to $k \sim 10^{-1.6} \ \mathrm{s \ km^{-1}}$, which marks the typical scale of fluctuations in the Ly$\alpha$ forest, and then decreasing at larger scales. However, the power at $z=5$ is lower than at $z=5.6$ because by then the IGM is mostly ionized, making the Ly$\alpha$ transmission more uniform. 
The suppression of power at the highest $k$-values comes mostly from  the thermal broadening kernel.  

At \( z = 5.6 \), models incorporating more energetic sources, \texttt{SS+ISM+XRBs} and \texttt{BS+ISM+XRBs}, exhibit only a slight deviation from the baseline \texttt{SS} model, showing a modest \( 0.5\% \) increase in the flux power spectrum at \( k \sim 10^{-1.6} \, \mathrm{s \, km^{-1}} \). Other SED models, such as \texttt{BB} and \texttt{SOFT-PL}, display more pronounced differences. These two models share a similar trend: an enhancement in power at small spatial scales and a suppression at larger scales compared to the \texttt{SS} model, with deviations reaching up to \( 8\% \) in the case of \texttt{SOFT-PL}. The decrease in power at small scales arises from the combined effects of ionization, recombination, and photoheating in dense regions, which amplify fluctuations in those environments. In contrast, at larger scales, the partial ionization and associated photoheating lead to reduced fluctuations in the power spectrum for the \texttt{BB} and \texttt{SOFT-PL} models. 
Interestingly, the \texttt{AGN-PL} model stands out as at low $k$ it has about $10\%$ higher fluctuations than \texttt{SS} due to a more pronounced impact from the inhomogeneous reionization and heating process (see also, \citealt{Molaro2022}), while at higher $k$, the deviation in power drops before peaking again in the negative side at a $10\%$ deviation, driven by additional heating from helium double ionization. As the UV background becomes more homogeneous by $z=5$, fluctuations diminish, and differences in comparison to \texttt{SS} reach a maximum of $2\%$, mainly influenced by more pronounced double ionization of helium affecting the thermal Doppler broadening because of the increased IGM temperatures. By this stage, the deviations for \texttt{AGN-PL} remains at a higher value of $\sim 10\%$.

I note that all models are consistent with observational data from \citealt{boera2019} at $z=5$, with the exception of \texttt{AGN-PL} at $k \gtrsim 10^{-1.2} \mathrm{s \, km^{-1}}$. Such small differences with respect to my reference \texttt{SS} model remain within the observational uncertainties at $z=5$, with the exception of the extreme \texttt{AGN-PL} model. While current observations make it difficult to isolate the impact of different SEDs on the flux power spectrum, surveys with next-generation facilities like the Extremely Large Telescope (\texttt{E-ELT}, see \citealt{Dodorico2024}) and upgraded Dark Energy Spectroscopic Instrument (\texttt{DESI}, see \citealt{Karacayli2020}) could reduce the error bars to $1\%$–$2\%$, enabling a more precise assessment of how to distinguish the contribution from sources with different SEDs, and help in understanding if the assumptions on the sources SED in the theoretical models used to interpret Ly$\alpha$ forest data might induce any bias.

\subsubsection{Source-IGM connection}
\label{gal-igm}

\begin{figure*}   
\includegraphics[width=53mm]{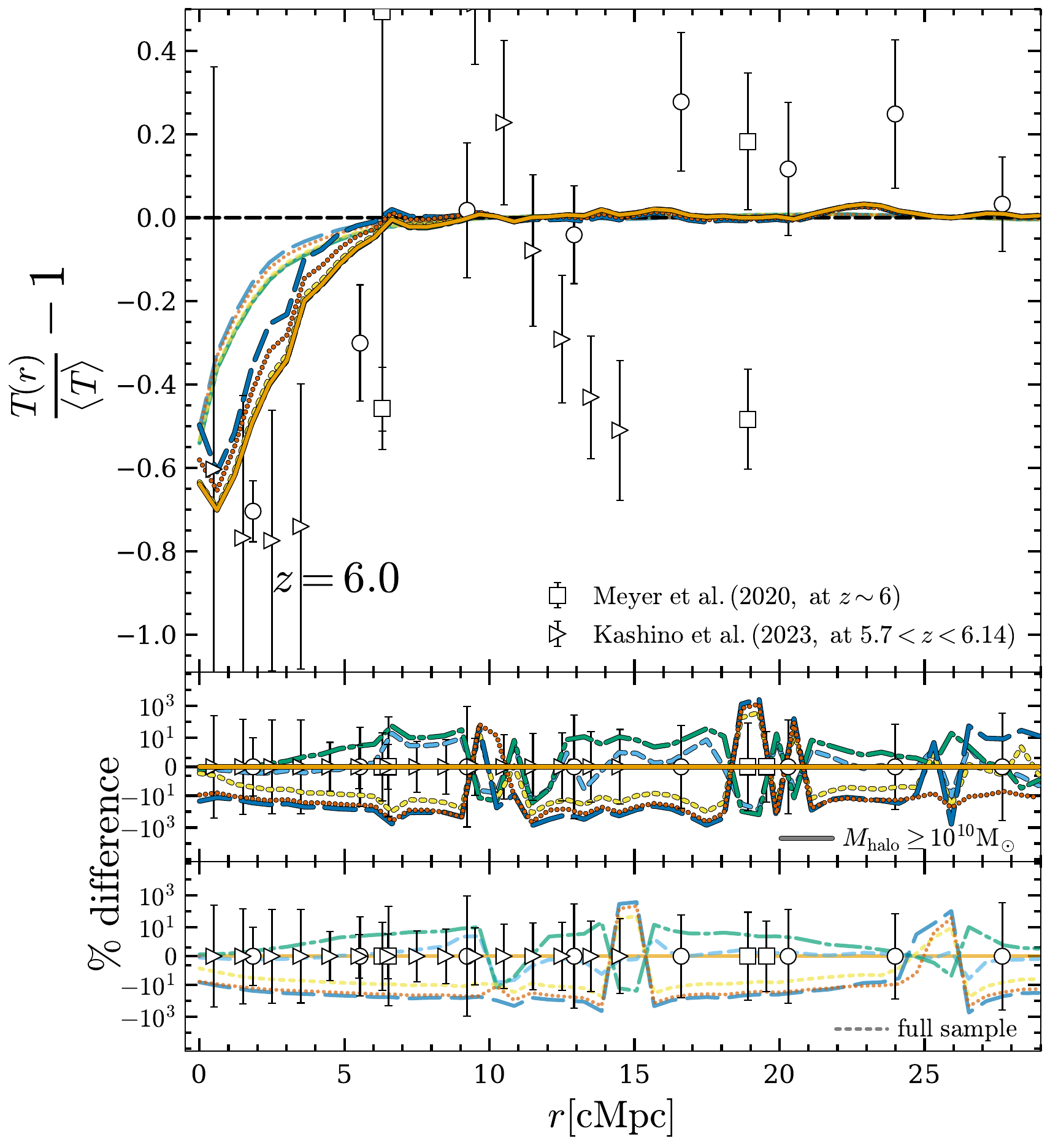}
\includegraphics[width=53mm]{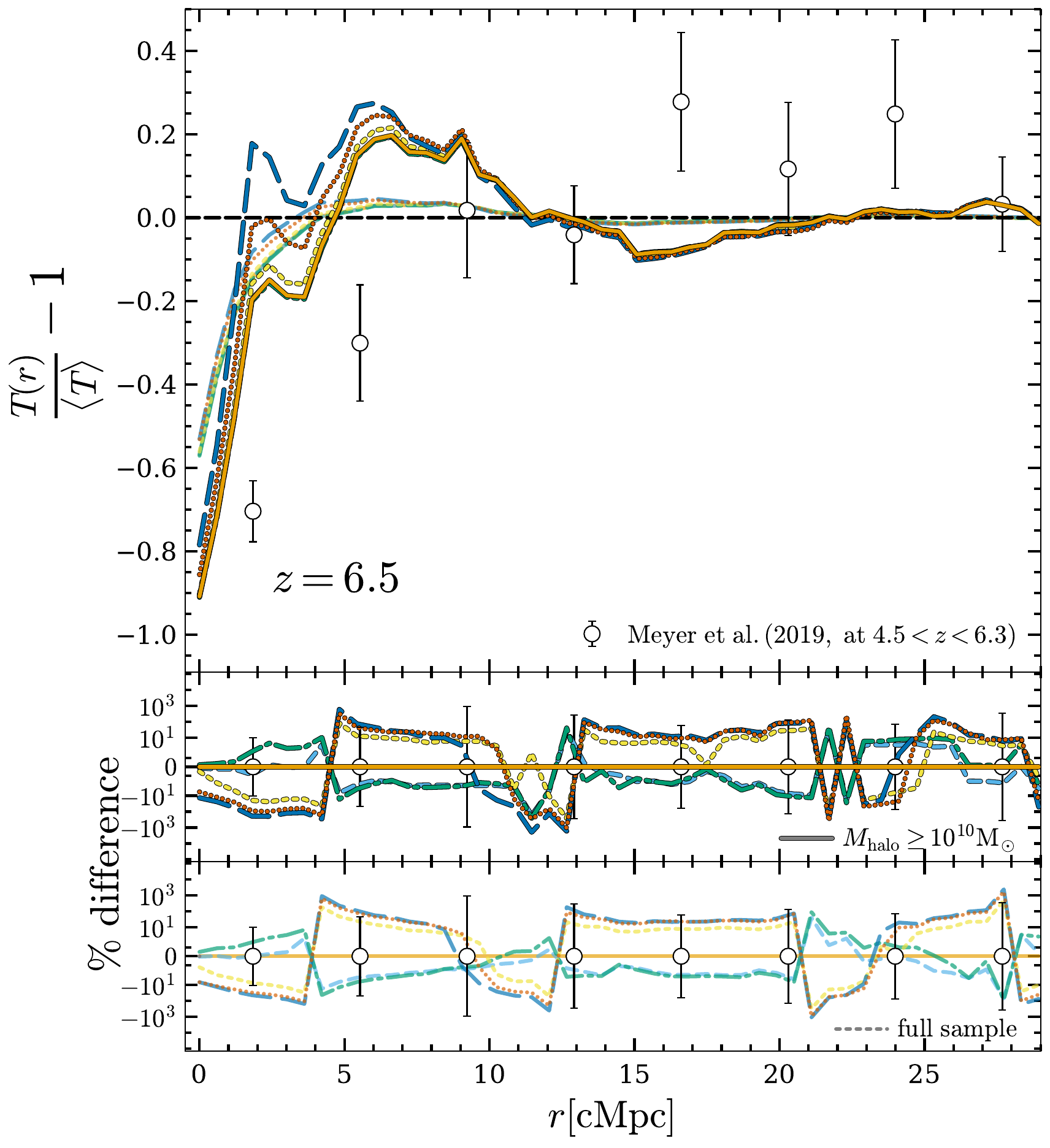}
\includegraphics[width=53mm]{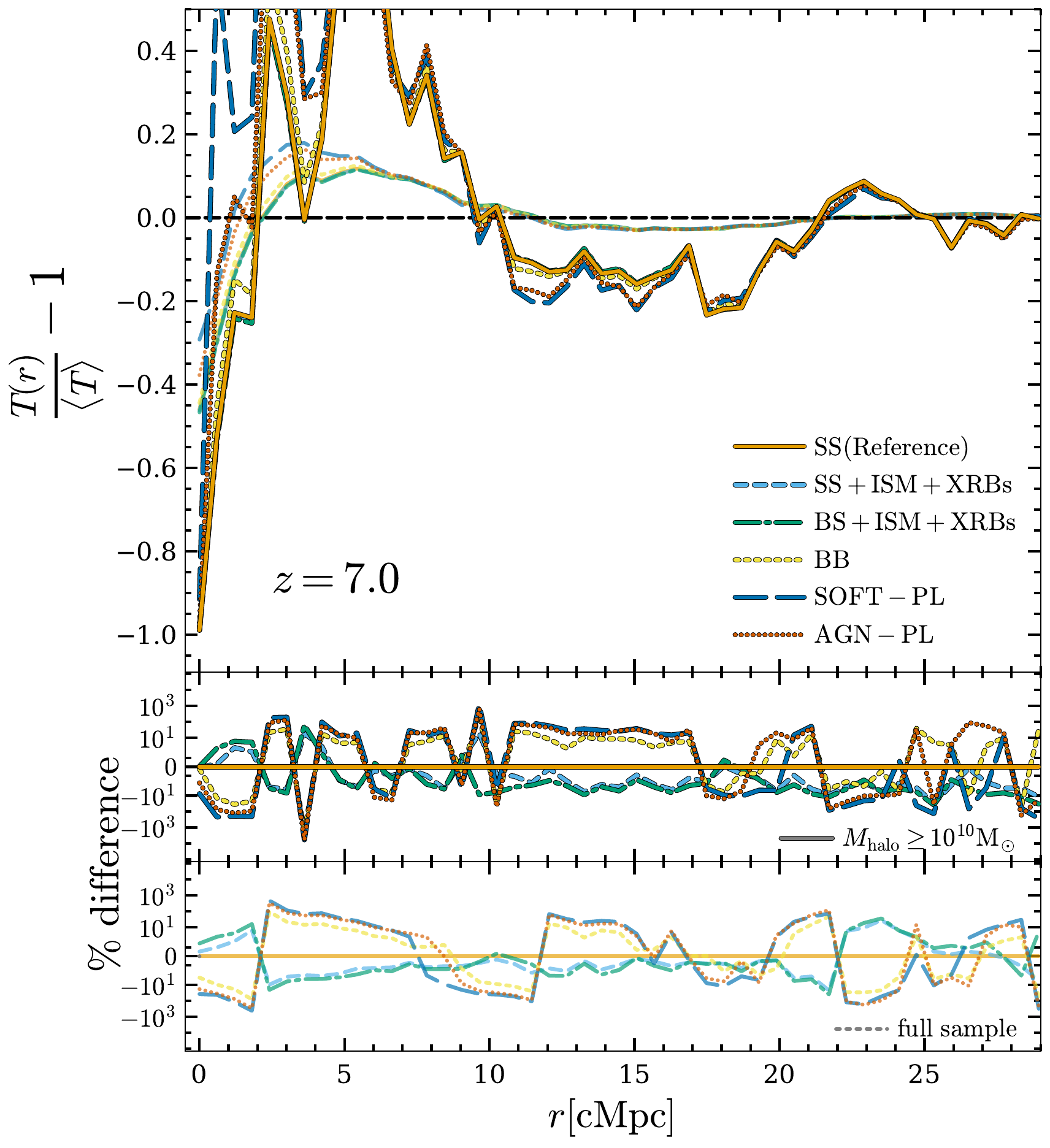}
    \caption{{\it Top panels:} Excess Ly$\alpha$ forest transmission at $z=6$ (left panel), 6.5 (middle) and 7 (right) for \texttt{SS} (solid orange curve), \texttt{SS+ISM+XRBs} (dashed sky blue), \texttt{BS+ISM+XRBs} (dash-dotted green), \texttt{BB} (short-dashed yellow), \texttt{SOFT-PL} (long-dashed blue) and \texttt{AGN-PL} (dotted vermilion). 
    The transparent (opaque) curves are representing the similar curves but considering all the sample of haloes (haloes of $\rm{\mathit{M}_{halo} \geq 10^{10} M_{\odot}}$) The symbols denote a compilation of observational constraints from the literature \citep{meyer2019,meyer2020,kashino2023}. 
    {\it Bottom panel:} Fractional difference with respect to  \texttt{SS} simulation along with the uncertainties of the observational constraints for the sample of haloes of $\rm{\mathit{M}_{halo} \geq 10^{10} M_{\odot}}$ (upper part) and full sample of haloes (lower part).}
    \label{fig:excess_transmission}
\end{figure*}

To investigate more in detail the local effect of different SED models, 
I examine how the average IGM transmissivity ($T$) (i.e. the ratio between the transmitted flux and the inferred continuum flux at the same location) varies with distance $r$ from nearby sources. This approach, first proposed by \citealt{kakiichi2018} and subsequently adopted also by \citealt{meyer2019}, \citealt{meyer2020} and \citealt{kashino2023}, provides a complementary way to probe the relation between sources and their surrounding IGM.
A key finding from these works is that the Ly$\alpha$ transmitted flux shows a broad peak at intermediate distances (i.e. 10 $\rm{cMpc}/h$ $\lesssim$ $r$ $\lesssim$ 30 $\rm{cMpc}/h$) from ionizing sources, which can be interpreted as due to a proximity effect, suggesting that galaxies enhance ionization in their immediate surroundings, leading to increased Ly$\alpha$ transmissivity. While observational studies have made significant progress in this area \citep{koki2025,Kashino2025}, theoretical investigations remain limited, with the notable exception of \citealt{garaldi2022} and \citealt{garaldi2024}, who examined this phenomenon using high-resolution radiation-hydrodynamics simulations. For more insights, very recently \citealt{conaboy2025} examined this connection between high-redshift galaxies and Ly$\alpha$ transmission utilising models of \texttt{Sherwood-relics} simulation.

Following the efforts made by \citealt{garaldi2022}, to create a synthetic version of observations, in Figure \ref{fig:excess_transmission} I present the excess IGM transmissivity relative to its average across all $r$, expressed as $ T(r) / \langle T \rangle - 1 $, at $ z = 6 $, 6.5, and 7. The general trend of the curves is mostly the same at all redshifts - in the closest regime of the sources (at low radius) they show a negative value near  -1 expressing an anti-correlation between source distribution and Ly$\alpha$ transmission. Then towards higher radius, they reach a peak indicating positive correlation, and finally they converge to zero at the largest distances from the sources, where the transmission is the same as the average one. 
Our simulations reveal an excess transmissivity at $ r \leq 10 $ cMpc. As redshift increases, I observe that the amplitude of the excess transmission increases and shifts to larger distances. This behaviour arises due to three key factors, namely that  at earlier times there are: (i) lower overdensities, which limit the suppression of Ly$\alpha$ flux near sources due to weaker hydrogen recombination; (ii) smaller ionized bubbles, which restrict the distance at which flux is enhanced; and (iii) a rapid decline in the average transmissivity $ \langle T \rangle $, while highly ionized regions continue to produce localized transmission spikes. These factors collectively shape the evolution of Ly$\alpha$ transmissivity and its dependence on source proximity over cosmic time. At $r \geq 20 \rm{cMpc}$, all models converge as excess the transmission approaches zero, since the local transmission becomes the same as the global mean. 

In the bottom panels of Figure \ref{fig:excess_transmission}, I show the fractional difference in excess transmission relative to the \texttt{SS} model. At $z = 6$, in the regions closest to the sources ($r \lesssim 9~\rm{cMpc}$), the \texttt{AGN-PL}, \texttt{SOFT-PL}, and \texttt{BB} models exhibit the strongest excess transmission, peaking at small distances before gradually declining outward. In contrast, the \texttt{BS+ISM+XRBs} model yields the lowest excess transmission, while the \texttt{SS+ISM+XRBs} model transitions from negative deviations at $ r \lesssim 6.5~\rm{cMpc}$ to positive deviations at larger radii. At $z = 6.5$, the transition to positive excess transmission shifts inward to $r \lesssim 4~\rm{cMpc}$, and by $z = 7$ it contracts further to $r \leq 2~\rm{cMpc}$. In the regions with positive excess transmission, the \texttt{AGN-PL}, \texttt{SOFT-PL}, and \texttt{BB} models consistently produce the largest deviations, showing a peak at intermediate distances that is most prominent at $z = 6.5$ and $7$, while the overall excess transmission decreases by $z = 6$. Meanwhile, the \texttt{BS+ISM+XRBs} and \texttt{SS+ISM+XRBs} models remain slightly negative, with deviations $\sim 10\%$ lower than \texttt{SS}.

I also repeat this calculation for a subset of haloes with $\rm{\mathit{M}_{halo}\geq 10^{10} M{\odot}}$ to investigate the effect of focusing on more massive systems. The trends across SED models remain broadly similar to the full halo sample, but the amplitude of the excess transmissivity is higher, and the curves are noticeably noisier due to the small number of haloes in this mass range, particularly at $z=7$.

In the same figure, I also show the available observational measurements of excess Ly$\alpha$ transmission for comparison. These include the C~{\sc iv}-selected galaxies from \citealt{meyer2019}, the Ly$\alpha$ emitters (LAEs) and Lyman-break galaxies (LBGs) from \citealt{meyer2020}, and the [O~{\sc iii}] emitters from \citealt{kashino2023}. The different galaxy selection methods naturally probe distinct environments, which contributes to the diversity in their measured profiles. At $z = 6$, my simulations show excess transmission at smaller distances than the \citealt{meyer2019} and \citealt{meyer2020} measurements but are more consistent with the shape reported by \citealt{kashino2023}. Beyond $r \gtrsim 20~\rm{cMpc}$, my predictions are in broad agreement with most observational points, except for \citealt{kashino2023}, whose measurements exhibit a sharp decline already beyond $r \gtrsim 12~\rm{cMpc}$. At $z = 6.5$, my models tend to overpredict the excess transmission within $r \lesssim 10~\rm{cMpc}$ compared to \citealt{meyer2019}, while the agreement improves at larger radii. By $z = 7$, the measurements are sparse, but my predicted profiles remain broadly consistent within the large uncertainties.
The observational error bars are generally larger than the model-to-model deviations, especially at $z = 7$, which limits the ability to distinguish between SED scenarios. However, in some cases—most notably for the \texttt{AGN-PL}, \texttt{SOFT-PL}, and \texttt{BB} models at $z = 6$ and $6.5$—the predicted deviations exceed the observational uncertainties at intermediate distances ($r \sim 5$–15 $\rm{cMpc}$), suggesting that these models could be distinguishable with current or near-future data. Even at the largest radii, where the deviations decline, some instances remain comparable to observational error bars, highlighting that the excess Ly$\alpha$ transmissivity around galaxies can provide a promising test of different ionizing source models as more precise measurements become available.

\section{Conclusion and Discussion}
\label{discussion}
This chapter investigates the influence of various sources SED modelling, including X-ray binaries, Bremsstrahlung emission from shock heated ISM, and binary stars, on the IGM properties during the final stages of the epoch of reionization, as probed by the Ly$\alpha$ forest. I perform a comparative study including also more idealized, commonly adopted, SEDs, such as power-law and  blackbody spectra. Using simulations obtained by post-processing outputs of a \texttt{Sherwood}-type hydrodynamic simulation \citep{bolton2016} with the 3D radiative transfer code \texttt{CRASH} \citep{ciardi2001,maselli2003,maselli2009,graziani2013,hariharan2017,glatzle2019,glatzle2022}, I explored the impact of different SEDs on the ionization and thermal state of the IGM, as well as their   imprint on Ly$\alpha$ forest observables. My major findings are:
\begin{itemize}
    \item While the broad morphology of the fully ionized H\,{\sc ii} regions is similar across all SED models, the details of the ionization and thermal structure differ. Models including harder spectral components (e.g. ISM, XRBs, AGN-like SEDs) produce more extended partially ionized regions, enhanced He\,{\sc iii} fractions, and higher surrounding IGM temperatures, compared to the models dominated by stellar emission (single or binary stars, or blackbody-like spectra).

    \item The evolution of the volume-averaged neutral hydrogen fraction is very similar across all SED models. Small differences arise at $z \gtrsim 6$, where models including harder spectral components (e.g. ISM, XRBs) show a slightly lower $\langle x_{\rm HI} \rangle$ due to enhanced partial ionization and heating, while power law models remain marginally more neutral. By the end of reionization ($z$ $\lesssim 5.5$), the \texttt{AGN-PL} and \texttt{SOFT-PL} models predict a universe more ionized than the other models.
    
    \item The IGM temperature at mean density increases as reionization progresses, peaking around $z$ $\approx 6$ for most models. The \texttt{SS+ISM+XRBs} and \texttt{BS+ISM+XRBs} configurations show a modest increase in temperature compared to \texttt{SS}, driven by additional photoheating. Differently, the \texttt{BB} model results in slightly cooler temperatures. The \texttt{SOFT-PL} model has a behaviour similar to the \texttt{BB} one, but towards the end of reionization the gas becomes hotter than all the previous models because of  helium double ionization heating. The \texttt{AGN-PL} model consistently produces much higher temperatures, exceeding \texttt{SS} by  more than 7500 K at $z$ = 5, due to the large fraction of high-energy photons in its SED which are very efficient in fully ionizing helium. This scenario, though, is inconsistent with observational constraints.

    \item All models reproduce the general decline with redshift of the Ly$\alpha$ effective optical depth ($\tau_{\rm eff}$) obtained from observational data. Differences among models at  $z$ $\lesssim$ 6 are generally within 1$\%$, with the exception of \texttt{AGN-PL} and \texttt{SOFT-PL} which produce $\tau_{\rm eff}$ values up to 10$\%$ lower than those of the reference \texttt{SS} model due to their higher ionization and heating efficiency.
    
    \item The 1D power spectrum of Ly$\alpha$ transmission flux exhibits differences similar to the other quantities, with most models showing deviations of $\lesssim$2$\%$ from \texttt{SS} mainly localized at intermediate scales due to thermal broadening. Differently, \texttt{AGN-PL} produces up to 10$\%$ higher power on large scales due to spatial ionization variations, and 10$\%$ lower power on small scales from stronger heating. While subtle, these signatures could be observable with future high-precision spectroscopic surveys (i.e. \texttt{E-ELT} or \texttt{DESI}).

    \item By analyzing Ly$\alpha$ transmission as a function of distance from sources, I observe enhanced transmissivity (proximity effect) within $\sim$10 cMpc of massive halos. This effect is strongest at earlier times ($z $$\sim$ 7), when ionized regions are smaller and the IGM is more neutral. Hard-spectrum sources like \texttt{AGN-PL} and \texttt{SOFT-PL} produce the strongest proximity effects, while stellar models, particularly BS+ISM+XRBs, show more modest enhancements. These results align with recent observational studies and suggest that proximity zone statistics could distinguish between different source populations in the reionization era.
\end{itemize}

Our findings emphasize that the choice of SED modeling is critical for interpreting Ly$\alpha$ forest measurements during the late stages of reionization. Adopting simplified spectra risks underestimating the contribution of high energy sources, which subtly alter $\tau_{\rm eff}$, flux power, and local transmissivity, potentially biasing constraints on the thermal and ionization history of the IGM. The differences across models, while often modest in global statistics, are most pronounced in proximity zone behavior and the small- to intermediate-scale flux power spectrum, which represent the most promising observables to disentangle source populations. With the advent of next-generation spectroscopic surveys such as \texttt{E-ELT} and future \texttt{DESI} campaigns, improved measurements of these statistics will allow me to test the contribution of X-ray binaries, shock-heated ISM, and other high-energy sources. Utilising such data with physically motivated SED models will be essential to robustly connect Ly$\alpha$ forest observations to the sources driving the end of reionization and to refine my understanding of the IGM’s thermal evolution.

  \chapter{Impacts of Quasars in Helium Reionization}
\label{chap:chapter4}

\begin{flushright}
\begin{minipage}{0.5\textwidth}
\raggedleft
\textbf{\textit{``Rage, rage against the dying light. Do not go gentle into the good night."}}\\[1ex]
\noindent\rule{0.5\textwidth}{0.4pt}\\[-0.2ex]
Nikita
\end{minipage}
\end{flushright}

\begin{flushright}
\begin{minipage}{0.7\textwidth}
\raggedleft
\textbf{\textit{``Travel expands the mind; whether it’s around the globe or through the stars is simply a choice."}}\\[1ex]
\noindent\rule{0.5\textwidth}{0.4pt}\\[-0.2ex]
Abinaya Swaruba Rajamuthukumar
\end{minipage}
\end{flushright}

\textit{This work has been published in the Monthly Notices of the Royal Astronomical Society, Volume 532, Issue 1, July 2024, Pages 841–858}\citep{Basu2024}.

\hspace{1 cm}

Reliably constraining helium reionization is very challenging. The most direct avenue for observing this transformation is via the He~{\sc ii} Lyman-$\alpha$ forest in the spectra of high-$z$ quasars.
For instance, \citealt{Dixon2009} interpreted the rapid decrease in the He~{\sc ii} optical depth at \textit{z} $\sim$ 2.7 as indicative of the end of helium reionization. 
Moreover, the fluctuations in opacity observed along multiple lines of sight offer valuable insights into the concluding stages of this epoch \citep{Anderson1999,Heap2000,Smette2002,Reimers2005}. \citealt{Furlanetto2011} suggested that the large fluctuations observed at \textit{z} $\sim$ 2.8 were more likely the product of ongoing reionization. The fact that the process of helium reionization is extended is confirmed by sightline-to-sightline variation of optical depth in more recent observations \citep{worseck2016,worseck2019,makan2021,makan2022}, but these are limited by the low number of sightlines due to the intrinsic difficulties of observing the He~{\sc ii} Lyman-$\alpha$ forest. In the near future,  \texttt{WEAVE} \footnote{The WHT Enhanced Area Velocity Explorer, \url{https://ingconfluence.ing.iac.es/confluence//display/WEAV}} is expected to more than double the number of sightlines at $z>2$ over several thousand square degrees\footnote{Although the wavelength coverage of WEAVE does not include the He~{\sc ii} Lyman-$\alpha$ line, this survey can}, serving as a target selection of clean sightlines where the He~{\sc ii} Lyman-$\alpha$ forest can be detected by subsequent observations. 
Additionally, WEAVE-QSO \citep{pieri2016} will constrain the IGM temperature at $z \sim 3$, providing new tighter constraints on helium reionization. 
Overall, despite the difficulties, a consensus has emerged that helium reionization occurred in the redshift range $2.7 \lesssim z \lesssim 4.5$ \citep{Miralda1993,Giroux1995,Croft1997,Fardal1998,Schaye2000,Theuns2002,Riccoti2000,Bolton2006,Bolton2009,Meiksin2012}.

A detailed understanding of helium reionization is crucial not only to test the structure formation framework, but also to assess its impact on galaxies and the IGM. In fact, the excess energy from the He~{\sc ii} to He~{\sc iii} transition significantly heats up the IGM \citep{hui&gnedin1997,Schaye2000,McQuinn2009,Garzili2012}. 
However, modelling helium reionization is computationally very challenging. It requires to simultaneously capture the quasars clustering properties on scales of hundreds of Mpc and the galaxy-scale gas physics. To cope with such requirements, many numerical simulations of this epoch focus on scales $\le 100$ $\rm{cMpc}$ \citep{Sokasian2002,Meiksin2012,Compostella2013,compostella2014}, at the cost of missing the large ionized bubbles expected during helium reionization and failing to include the rarest sources. Alternatively, several studies employ various combinations of analytic and semi-analytic models \citep{Tom2000,Gleser2005,upton2020}, that however result in less solid predictions (e.g. concerning the morphology of ionized helium bubbles, heating of the IGM, He-ionizing background). 

Until now, the modeling of He~{\sc ii} reionization has typically either relayed on the simulated (instead of observed) QSOs population, or on selecting the observed QLF at a specific redshift and extrapolating it based on some emissivity evolution \citep{Compostella2013,compostella2014,Garaldi2019}. Here, I explore the prediction of the recent smoothly-time-evolving QLF from \citealt{Shen2020} on helium reionization, its morphology and the properties of the $2<z<4$ IGM. To this end, I post-process the large-scale hydrodynamical simulation \texttt{TNG300} \citep[e.g.][]{pillepich2018,nelson2018} using the 3D multi-frequency radiative transfer code \texttt{CRASH} \citep[e.g.][]{ciardi2001,glatzle2022} which follows the formation and evolution of He~{\sc iii} regions produced by a population of quasars extracted from the \citealt{Shen2020} QLF. I forward model these simulations to produce synthetic Lyman-$\alpha$ forest spectra and compare them to available data. Finally, I investigate the impact of the recent \textit{James Webb Space Telescope} (\texttt{JWST}) observations of a large number of active galactic nuclei (AGNs) at $z>5$ \citep[e.g.][]{Fudamoto2022,harikane2023,maiolinoa2023,Maiolino2023,Goulding2023,Larson2023,Juodzbalis2023,greene2023}. I introduce the simulations in Section \ref{method}, whereas my results are discussed in Section \ref{results}. I summarize the conclusion and the future prospects in Section \ref{conclusions}.

\section{Methodology}
\label{method}
In order to provide a faithful picture of helium reionization, I have combined the outputs of a hydrodynamical simulation (section \ref{tng300}) with a multi-frequency radiative transfer code (section \ref{crash}). The latter is sourced by a population of quasars following the QLF from \citealt[][in particular their Model 2]{Shen2020}, that are placed in the simulations volume as described in Section \ref{tng+crash}. This procedure ensures that my results are as genuine predictions of the observed QLF as possible.

\subsection{The \texttt{TNG300} hydrodynamical simulation}
\label{tng300}

In this work I have used the \texttt{TNG300} simulation, which is a part of the \texttt{Illustris TNG project} \citep{volker2018,naiman2018,marinacci2018,pillepich2018,nelson2018}, to model the formation and evolution of structures in the Universe. It has been performed with the \texttt{AREPO} code \citep{springel2010}, which is used to solve the idealized magneto-hydrodynamicals equations \citep{pakmor2011} describing the non-gravitational interactions of baryonic matter, as well as the gravitational interaction of all matter. The simulation employs the recent \texttt{TNG} galaxy formation model \citep{weinberger2017,pillepich2018}, and star formation is incorporated by converting gas cells into star particles above a density threshold of $\textit{n}\rm{_{H}}$ $\rm{\sim 0.1 cm^{-3}}$, following the Kennicutt-Schmidt relation \citep{springel2003}. Stellar populations are self-consistently evolved, and inject metals, energy and mass into the  interstellar medium (ISM) throughout their lifetime, including their supernova (SN) explosions. AGN feedback has two-modes: a more efficient kinetic channel at low Eddington ratio (`radio mode') and a less efficient thermal channel at high Eddington ratio (`quasar mode') \citep{weinberger2018}.

\texttt{TNG300} has been run in a comoving box of length $L_\mathrm{box} = 205 \,h^{-1} {\rm cMpc}$, with (initially) $2\times2500^{3}$ gas and dark matter (DM) particles. The average gas particle mass is $\bar{m}\rm{_{gas}}=7.44 \times 10^{6}\,M_{\odot}$, while the DM particle mass is constant and amounts to $m\rm{_{DM}}=3.98 \times 10^{7}\,M_{\odot}$. Haloes are identified on-the-fly using a friends-of-friends algorithm with a linking length of $0.2$ times the mean inter-particle separation. \texttt{TNG300} adopts a \citep{planck2016}-consistent cosmology with $\Omega_{m}=0.3089$, $\Omega_{\Lambda}=0.6911$, $\Omega_{b}=0.0486$, $h=0.6774$, $\sigma_{8}=0.8159$ and $n_{s}=0.9667$, where the symbols have their usual meaning.

In particular, from \texttt{TNG300} I have used 19 outputs covering the redshift range $5.53 \leq z \leq 2.32$. For the additional simulations presented in Section \ref{jwst}, I employ additional outputs at $z = 6$ and $z=5.85$. These serve as basis for the radiative transfer, that I describe next.

\subsection{The \texttt{CRASH} radiative transfer code}
\label{crash}
I have implemented the radiative transfer (RT) of ionizing photons through the IGM by post-processing the outputs from the hydrodynamical simulation with the code \texttt{CRASH} \citep{ciardi2001,maselli2003,maselli2009,Maselli2005,partl2011},
which self-consistently calculates the evolution of the hydrogen and helium ionization state and the gas temperature. \texttt{CRASH} uses a Monte-Carlo-based ray tracing scheme, where the ionizing radiation and its time varying distribution in space is represented by multi-frequency photon packets travelling through the simulation volume. The latest version of CRASH features a self-consistent treatment of UV and soft X-ray photons, in which X-ray ionization, heating as well as detailed secondary electron physics \citep{graziani2013,graziani2018} and dust absorption \citep{glatzle2019,glatzle2022} are included. I refer the reader to the original papers for more details on \texttt{CRASH}. 

\texttt{CRASH} performs the RT on grids of gas density and temperature. In my setup, these correspond to snapshots from the \texttt{TNG300} simulation (see Section \ref{tng300}). In order to account for the expansion of the Universe between the $i$-th and $(i+1)$-th snapshots, the gas number density is evolved as $n(\mathbf{x},z) = n(\mathbf{x}, z_{i})(1 + z)^{3}/(1 + z_{i})^{3}$, where $\mathbf{x} \equiv (x_{c}, y_{c}, z_{c})$ are the coordinates of each cell c. 

Since I are only interested in helium reionization, I follow radiation covering the energy range $h_p \nu \in [54.4 \ \rm{eV}, 2 \ \rm{keV}]$, where $h_p$ is the Planck constant, and I fix the IGM H~{\sc I} and He~{\sc II} ionization fractions to $x_{\rm HI} = 10^{-4}$ and $x_{\rm HeI}=0$. This corresponds to a fully-completed hydrogen reionization.

I generate five RT outputs at regular time intervals in between each pair of adjacent \texttt{TNG300} snapshots. 

\begin{figure*}
\centering
    \includegraphics[width=120mm]{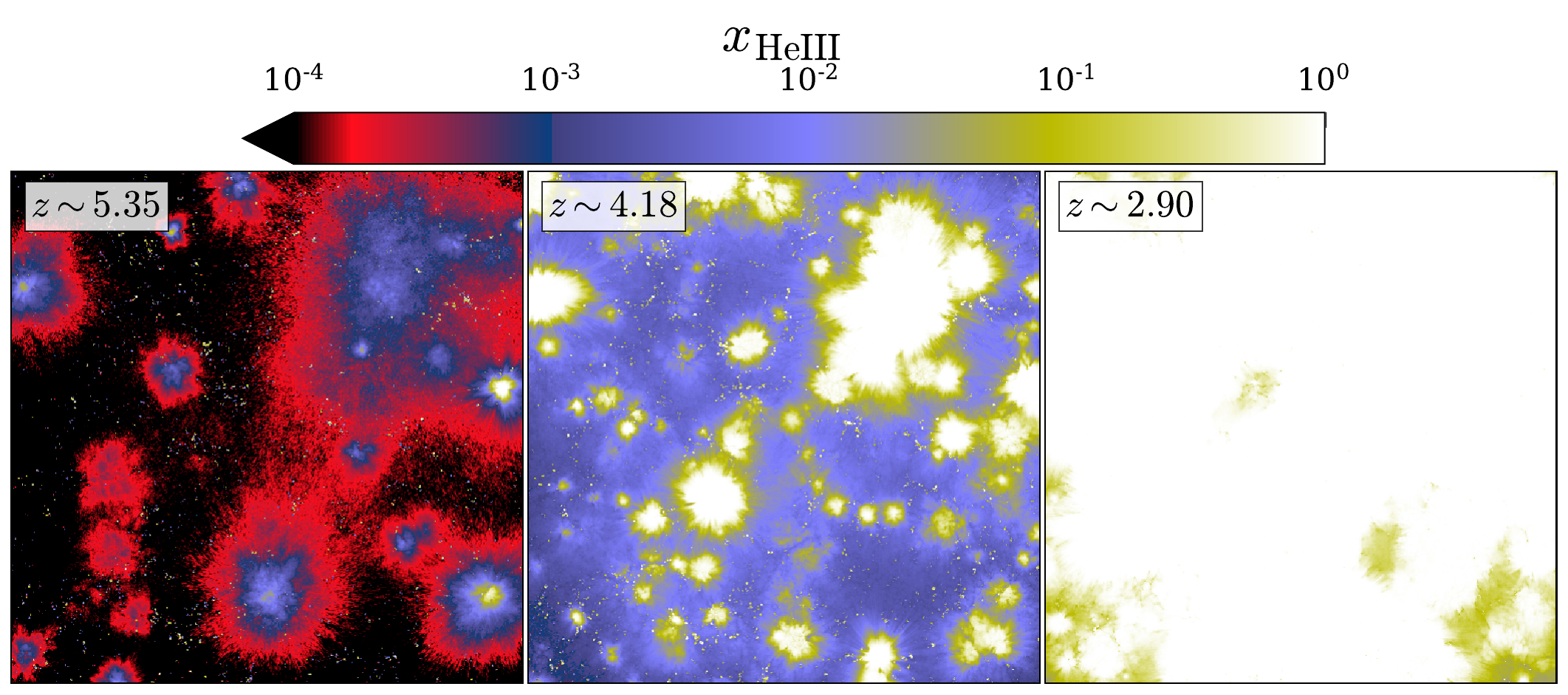}
    \includegraphics[width=120mm]{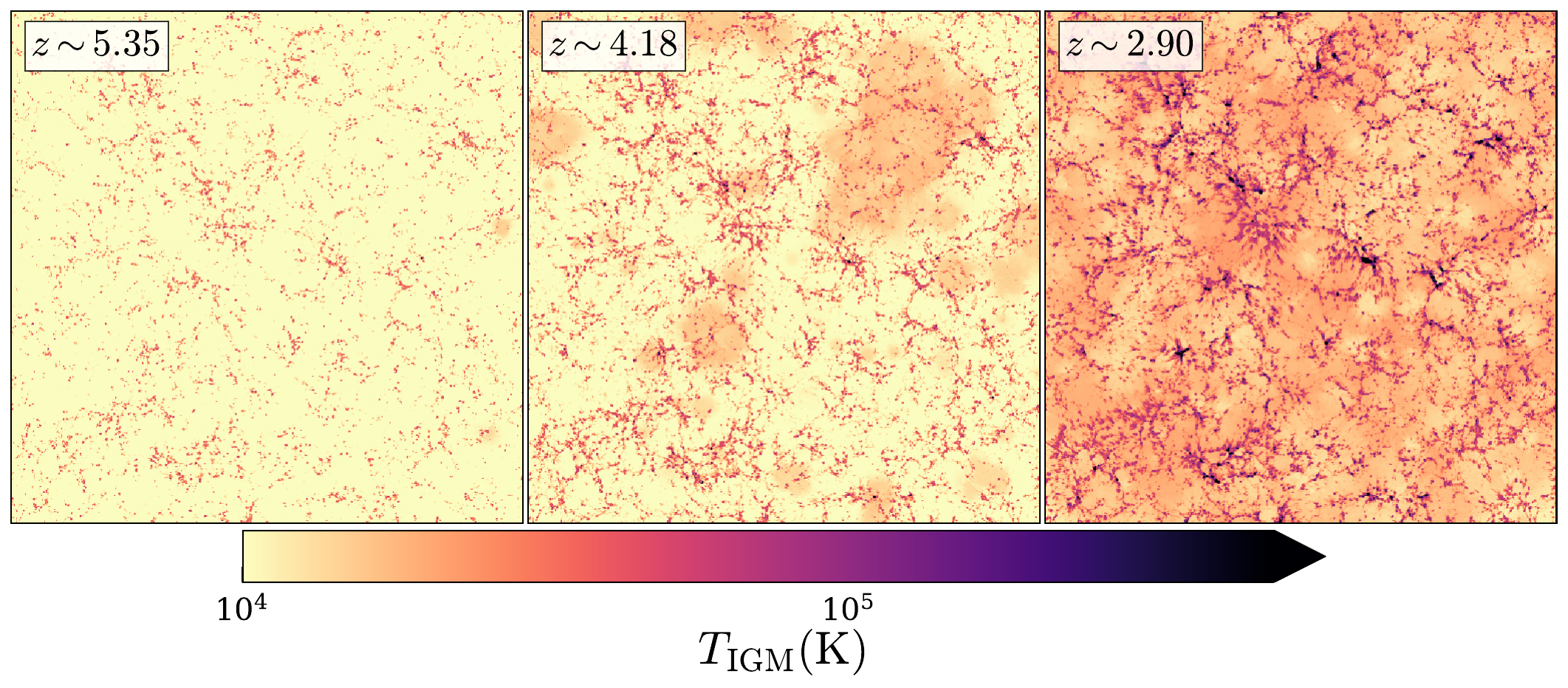}
    
    \caption{Slice across the simulation box of \texttt{N512-ph1e6} at \textit{z} =5.35, 4.18, and 2.90 (from left to right). Top and bottom panels show the He~{\sc iii} fraction and the gas temperature, respectively. The maps are $205 h^{-1} \rm{cMpc}$ wide and $400 h^{-1}$ ckpc thick.}
    \label{fig:reion_history}
\end{figure*}

\subsection{Combining \texttt{CRASH} and \texttt{TNG300}}
\label{tng+crash}

\subsubsection{Initial Conditions and Outputs}
\label{ics}
To incorporate the hydrodynamical simulation outputs into \texttt{CRASH}, the gas density and temperature extracted from the snapshots need to be deposited onto 3D uniform grids. I adopt a smoothing length for each Voronoi gas cell equal to the radius of the sphere enclosing its 64 nearest neighbours, and use the usual SPH cubic-spline kernel for its volume distribution. I initialize the He~{\sc iii} fraction to $x_{\rm HeIII}=10^{-4}$. 
To account for a fully-completed hydrogen reionization in the initial conditions, I set a gas temperature floor of $10^4$~K (which is approximately the average of the two available observations \citep{walther2019,Gaikwad2020} of IGM temperature at mean density ($T_{0}$) at $z \sim 5.4$, see figure \ref{fig:temp_evol} for a reference). This is consistent with my assumption of having  H~{\sc ii} fraction equal to unity in all cells.
Additionally, I assume that the fraction of H~{\sc i} and He~{\sc i} remain fixed for the entire range of redshifts.

\subsubsection{Spectral Energy Distribution}
\label{sed}
To derive the SED of quasars, I employ the model developed by \citealt{Marius2018, Marius2020}. In the range 13.6 $\rm{eV}$--200 $\rm{eV}$, this model explicitly calculates the average over a sample of 108,104 QSO SEDs from \citealt{Krawczyk2013} in the interval  $0.064 < z < 5.46$. Beyond 200 $\rm{eV}$, the spectral shape is modeled as a power law with an index of -1. I assume no evolution of the SED with redshift (refer to Section 2 in \citealt{Marius2018} for more details). For all RT simulations, the spectra of ionizing sources extend to a maximum frequency of $\sim 2 \ \rm{keV}$, covering the UV and soft X-ray regime. The discretization of the source spectrum is finer around the ionization thresholds of He~{\sc ii} ($\sim$ 54.4 $\rm{eV}$).

\subsubsection{Quasar distribution}
\label{qso_assignment}

In order to ensure that the simulated helium reionization is a genuine prediction of the QLF from \citealt[][Model 2]{Shen2020}, I disregard the black holes within \texttt{TNG300}, since their location and properties critically depend on the black hole seeding prescription, accretion physics and feedback model implemented in the simulation. Instead, I follow a more agnostic approach and populate the simulated haloes with quasars following the prescription described below. 

While the QLF offers predictions for the number of quasars of a given luminosity, it lacks information on their spatial distribution. Thus, to associate quasars to halos I adopt an abundance matching approach which establishes a one-to-one correspondence between the quasar bolometric luminosity ($L\rm{_{bol}}$) and the mass of DM haloes ($M\rm{_{DM}}$).
In practice, this amounts to placing the $i$-th brightest quasar at the center of the $i$-th most massive halo in the simulation. I refer to this approach as `direct'. 
In order to account for randomness in the $L\rm{_{bol}}$ -- $M\rm{_{DM}}$ correspondence, I design a slightly modified approach. In this `fiducial' method I split the simulated halo mass function and the QLF into $N_\mathrm{bins} = 20$ equal bins in logarithmic space. I then place the quasars as follows:
(i) I first select a quasar from the QLF at a given redshift, accounting for the finite simulation volume; 
(ii) then I identify the bin of the QLF in which the quasar resides (e.g. the $i$-th brightest bin of QLF); (iii) I choose a random halo from the corresponding $i$-th most massive bin of the halo mass function;
(iv) I place the quasar at the centre of such halo;
(v) finally, I introduce a 1\% random noise on the $L\rm{_{bol}}$ of the quasar;
(vi) I repeat this step for all the quasars in the simulation volume.
Note that the QLF computed self-consistently from the \texttt{TNG300} simulation is not in agreement with the \citealt{Shen2020} model. Therefore, I opted to re-assign QSO luminosities as described in order to ensure my results are as genuine predictions of the observed QLF as possible.
Additionally, I did not incorporate any correction to the simulated AGN feedback, which is likely to leave an imprint on the properties of nearby gas. The reason is that the impact of feedback is highly non-linear and coupled to other simulated processes, and it is therefore impossible to remove that directly in post-processing. However, if I were hypothetically able to undo the heat injection associated with AGN feedback, it would likely improve the comparison with observations (specifically, by raising the effective optical depths in the low-$z$ regime) which I discuss later in the section \ref{forest_tau_eff}. 

In both approaches, the quasar-halo pairing is repeated for each snapshot of the hydrodynamical simulation, and haloes that hosted a QSO in the previous snapshot are temporarily excluded from the procedure. This approximately accounts for the quasar duty cycle, since the interval between two snapshots is comparable (although somewhat larger) than the expected quasar lifetime \citep{Morey2021,Khrykin2021,Soltinsky2023}. In the remainder of the paper we show results from our fiducial approach. In Appendix \ref{appendix:BH_to_halo} we demonstrate that these two approaches give nearly identical results, with only $\mathcal{O}(10\%)$ differences in the initial stages of helium reionization, as reionization fronts takes more time to escape the largest haloes because of their higher densities.

\setlength{\tabcolsep}{5pt}
\begin{table}
\centering
    \begin{tabular}{lccccc}
        \hline
        Simulation Name   &$N\rm{_{grid}}$ &$N_{\gamma}$ &PBC  &QSO assignment  &$z_{\rm{final}}$\\
        \hline
        \hline
        \texttt{N256-ph1e5}   &$256^{3}$   &$10^{5}$  &No  &fiducial &2.32\\
        \texttt{N256-ph5e5}   &$256^{3}$   &$5 \times 10^{5}$  &No  &fiducial &2.32\\
        \texttt{N256-ph1e6}   &$256^{3}$   &$10^{6}$  &No  &fiducial &2.58\\
        \texttt{N256-ph1e5-PBC}   &$256^{3}$   &$10^{5}$  &Yes  &fiducial &2.32\\
        \hline
        \texttt{N512-ph1e5}   &$512^{3}$   &$10^{5}$  &No  &fiducial &2.32\\
        \textbf{\texttt{N512-ph5e5}}   &$\boldsymbol{512^{3}}$   &$\boldsymbol{5 \times 10^{5}}$  &\textbf{No}  &\textbf{fiducial} &\textbf{2.32}\\
        \texttt{N512-ph1e6}   &$512^{3}$   &$10^{6}$  &No  &fiducial &2.73\\
        \texttt{N512-ph1e5-PBC}   &$512^{3}$   &$10^{5}$  &Yes  &fiducial &3.71\\
        \texttt{N512-ph5e5-DIR}   &$512^{3}$   &$5 \times 10^{5}$  &No  &direct &2.44\\
        \hline
        \texttt{N768-ph5e5}   &$768^{3}$   &$5 \times 10^{5}$  &No  &fiducial &2.73\\
        \texttt{N768-ph1e6}   &$768^{3}$   &$10^{6}$  &No  &fiducial &3.82\\
        \hline
    \end{tabular}
    \caption{List of simulations and associated parameters. From left to right: name of the simulation, grid size ($N\rm{_{grid}}$), number of photon packets emitted per source per timestep ($N_{\gamma}$), presence of periodic boundary conditions (PBC), the method used to assign quasars to the DM haloes and the final redshift until which the simulation has run ($z_{\rm{final}}$). The reference simulation is shown in bold.}
    \label{tab:table-sim}
\end{table}

\subsubsection{Simulation Setup}
\label{sim_setup}

I have run a suite of simulations with different resolutions for the gas and radiation fields. The former is quantified as the dimension of the Cartesian grid ($N_{\rm{grid}}$) used for the deposition of the \texttt{TNG300} particle data. The latter is controlled by the number of photon packets emitted by each source for every output of the hydrodynamical simulation\footnote{The RT time step is computed as the ratio of the time between two hydrodynamical outputs and $N_{\gamma}$, so that at each RT time step, every source emits one packet.}. I have explored $N_{\rm{grid}} = 256^{3}$, $512^{3}$ and $768^{3}$, corresponding to spatial resolutions of $\delta x = L_\mathrm{box} / N_{\rm{grid}}^{1/3} \approx 800$, $400$ and $267$ $h^{-1}\,\mathrm{ckpc}$, respectively, and $N_{\gamma} = 10^{5}$, $5 \times 10^{5}$ and $10^{6}$.
I refer the reader to Appendix \ref{appendix:convergence_Ngamma} and \ref{appendix:convergence_Ngrid} for a quantitative discussion of the convergence tests performed with respect to photon packet sampling and grid dimension, respectively. Details of the simulations have been summarized in Table \ref{tab:table-sim}. Notice that computational constraints prevented us from running the highest-resolution boxes to low redshift, and therefore they are used only for convergence tests. I designate simulation \texttt{N512-ph5e5} as my reference and present only results from this run unless stated otherwise.

In all simulations, the photon packets are lost once they leave the simulation box, i.e. no periodic boundary conditions (PBC) have been applied. This choice is dictated by the otherwise steep rise in the computational cost once the vast majority of the volume is ionized, since in the presence of PBC photons can cross the simulation box multiple times before being absorbed. I have checked that this does not significantly affect my results (see Appendix \ref{appendix:convergence_PBC}). Nevertheless, I take the conservative approach of removing in my analysis all cells within $5 \times \delta x$ from each side of the box.

\section{Results}
\label{results}
The main question I want to address here is: \textit{What are the implications for helium reionization of the recent QLF constraints?}
In order to answer it, in the following I discuss the outcome of my fiducial simulation and compare it with available data. Then, I explore the implications of the high AGN number density reported by recent \texttt{JWST} observations.

\subsection{Reionization history}
\label{reion_history}
For a visual examination of the progress of helium reionization, I show in figure \ref{fig:reion_history} the He~{\sc iii} fraction (top panels) and IGM temperature (bottom panels) at \textit{z} = 5.35, 4.18 and 2.90 in a slice extracted from simulation \texttt{N512-ph1e6}. From the figure, it is evident how He~{\sc ii} reionization proceeds in an inhomogeneous manner, with ionized regions initially forming around quasars, evolving rapidly, and eventually merging \citep[e.g.][]{Compostella2013,laplante2017}. The partially-ionized gas surrounding ionized regions primarily results from soft X-ray and hard UV photons, as already discussed in prior research \citep{koki2017,Marius2018,Marius2020}. The bottom panels show how the ionization of He~{\sc ii} coincides with heating of the IGM.

\begin{figure}
\centering
    \includegraphics[width=120mm]{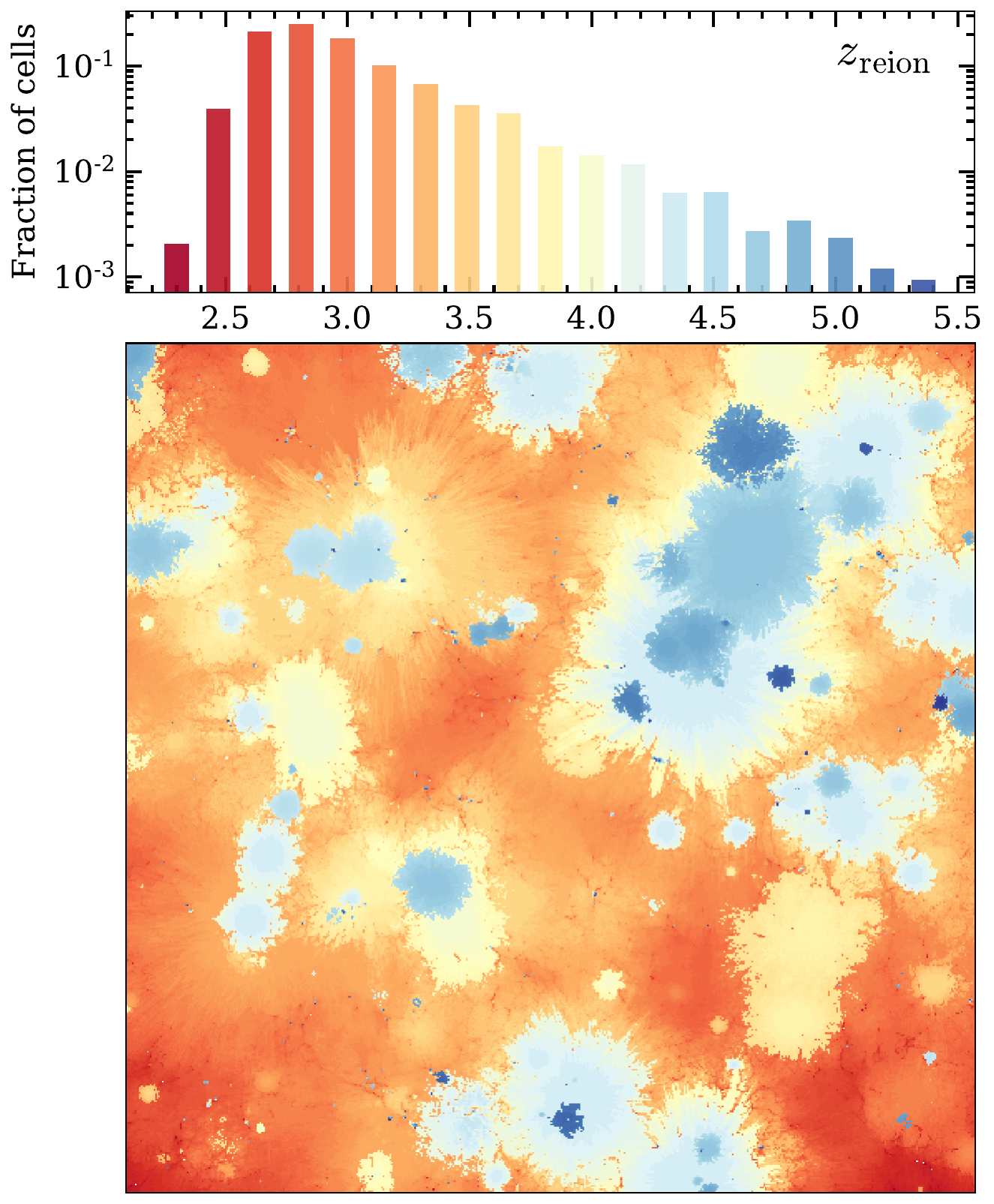}
    
    \caption{Fraction of cells (top panel) in the entire simulation volume that underwent reionization at a given redshift ($z\rm{_{reion}}$, see text for the definition). The color associated to each individual bin is used in the bottom panel to color code  cells in a slice of \texttt{N512-ph1e6} (the same slice as figure \ref{fig:reion_history}).}
    \label{fig:reion_redshift}
\end{figure}

\begin{figure}
\centering
    \includegraphics[width=100mm]{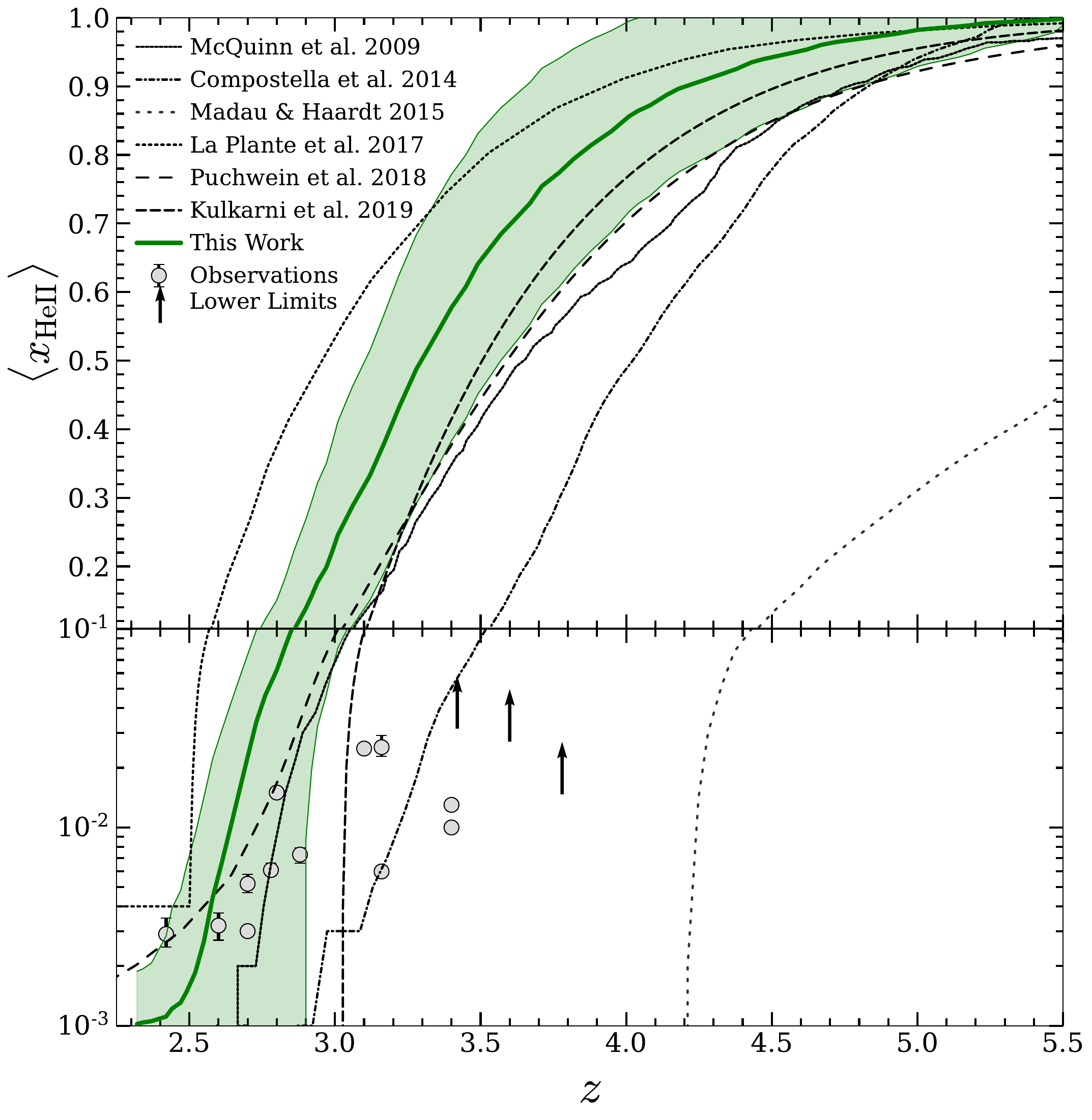}
    
    \caption{Redshift evolution of the volume averaged He~{\sc ii} fraction for  \texttt{N512-ph5e5} (solid green line). Other theoretical models \citep{McQuinn2009,compostella2014,madau2015,laplante2017,puchwein2019,kulkarni2019} are shown in black thin lines with different linestyles as mentioned in the legend. The symbols denote a collection of observational constraints from \citealt{worseck2016}, \citealt{davies2017}, \citealt{worseck2019}, \citealt{makan2021}, and \citealt{makan2022}. The shaded region displays the central 68$\%$ of the data.}
    \label{fig:xheii}
\end{figure}

To further demonstrate the highly-inhomogeneous nature of helium reionization, I show in figure \ref{fig:reion_redshift} the redshift $z\rm{_{reion}}$ at which each cell in the same slice as figure \ref{fig:reion_history} is reionized. I define $z\rm{_{reion}}$ as the redshift at which the local He~{\sc iii} fraction exceeds 0.99 for the first time. Consistently with the anticipated inside-out scenario, the process initially occurs in proximity to the quasars. As the process unfolds, the majority of cells in the simulation volume experience ionization at lower redshifts, specifically $z < 3$, revealing a prominent peak around $z \sim 2.7$ in the histogram of $z\rm{_{reion}}$. This is reflected in the redshift evolution of the volume-averaged He~{\sc ii} fraction, that is shown in figure \ref{fig:xheii} for my fiducial run. Symbols denote observational constraints from \cite{worseck2016}, \cite{davies2017}, \cite{worseck2019}, \cite{makan2021} and \cite{makan2022}.
my model completes reionization too late to accommodate the reported He~{\sc ii} fraction at $3 \lesssim z \lesssim 3.5$. However, an observational determination of this quantity is notoriously difficult, especially at higher redshift, due to the small number of sightlines suitable for such measurement. 
Additionally, the local nature of such constraints  combined with the inhomogeneous nature of reionization enables multiple values to coexist at the same time. Finally, the steeply-declining sensitivity of my probes to large neutral fractions biases observational results. my reionization history, though, appears delayed also compared to other theoretical models \citep{madau2015, laplante2017, puchwein2019, McQuinn2009, compostella2014, kulkarni2019}, with the exception of \citealt{laplante2017}, which employ a fully coupled radiation-hydrodynamical approach and model quasar activity as a lightbulb with luminosity-dependent lifetimes reproducing QLF constraints from SDSS and \texttt{COSMOS}\footnote{The Cosmic Evolution Survey, \url{https://cosmos.astro.caltech.edu/}}\citep{Masters2012,Mcgreer2013,Ross2013}.
It is essential to note that all theoretical models differ in their methodologies and modeling of source properties, with many lacking a RT scheme, so that a detailed comparison is not straightforward and outside the scope of this work. 

Once helium reionization is completed, the residual He~{\sc ii} fraction in my model is somewhat lower than the one inferred from observations. This is confirmed by the analysis performed in Section \ref{forest_tau_eff}. This discrepancy is indicative of the fact that my simulations do not fully capture the surviving sinks of radiation in the post-helium-reionization Universe. 
This is due to the fact that the resolution of the \texttt{TNG300} simulation is not high enough to capture the Lyman limit systems scales, as well as that by gridding the density field I smooth  out the largest gas overdensities that can self-shield from radiation and therefore maintain He~{\sc ii} reservoirs until late time. 

\begin{figure}
\centering
    \includegraphics[width=100mm]{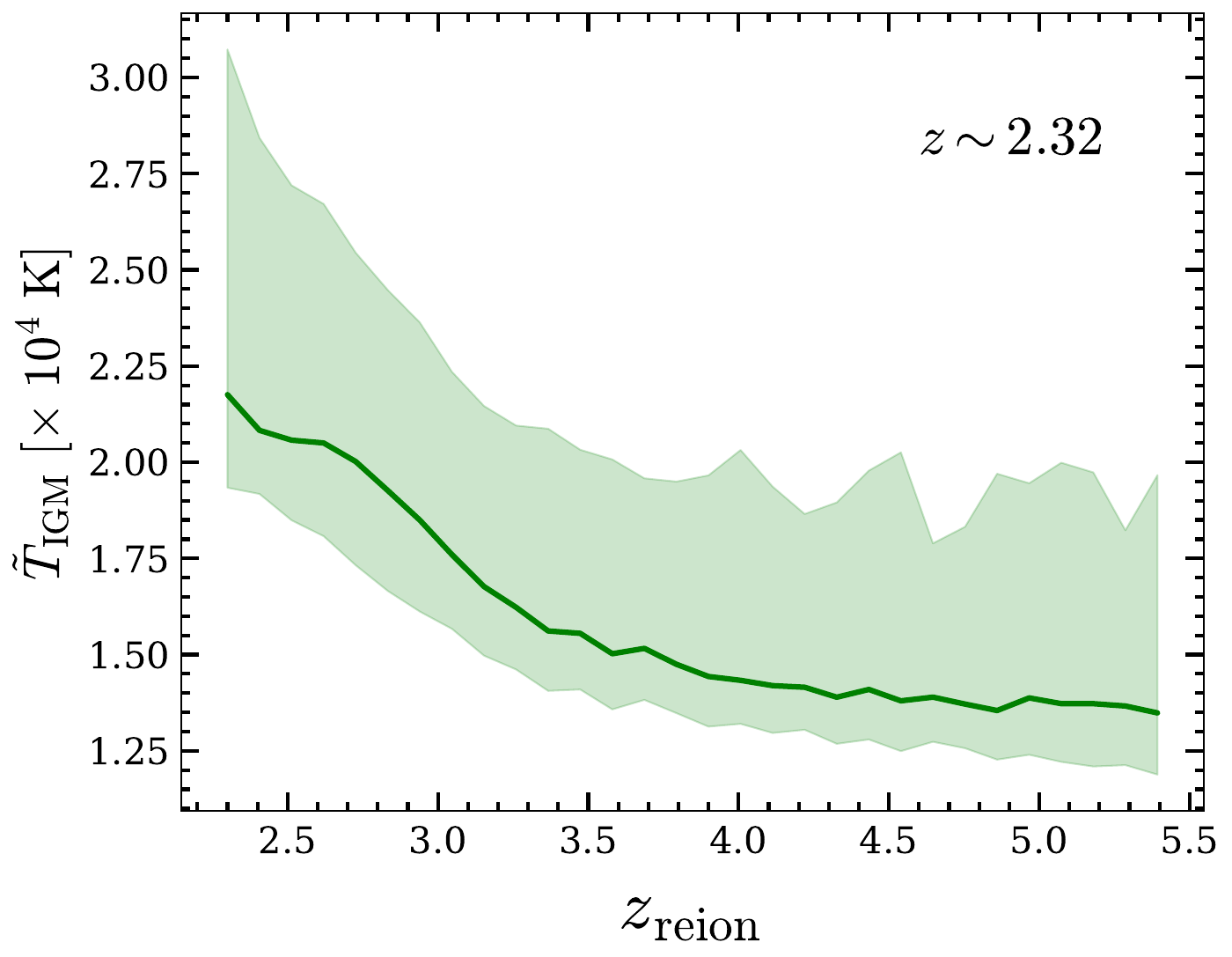}
    
    \caption{Correlation between the median IGM temperature (solid line) and the reionization redshift at $z\sim2.32$.  Only cells from \texttt{N512-ph5e5} with $T$ $\leq 40,000$~K are considered. The shaded region displays the central $68\%$ of the data.}
    \label{fig:reion_redshift_temp}
\end{figure}

Finally, I demonstrate the impact of helium reionization onto the IGM properties in figure \ref{fig:reion_redshift_temp}, where I show the dependence of the median IGM temperature ($\tilde{T}\rm{_{IGM}}$) at $z=2.32$  as a function of $z\rm{_{reion}}$ for my reference simulation \texttt{N512-ph5e5}. Note that here I only considered cells with $ T \leq 40,000 \ \rm{K}$ filter out cells affected by feedback precesses other than photo-ionization, which typically have higher temperatures. From the figure it is clear that an early He~{\sc ii} reionization results in a longer period of adiabatic cooling and thus a lower IGM temperature at the specific redshift investigated (the same trend is found at other redshifts).

\begin{figure}
\centering
    \includegraphics[width=100mm]{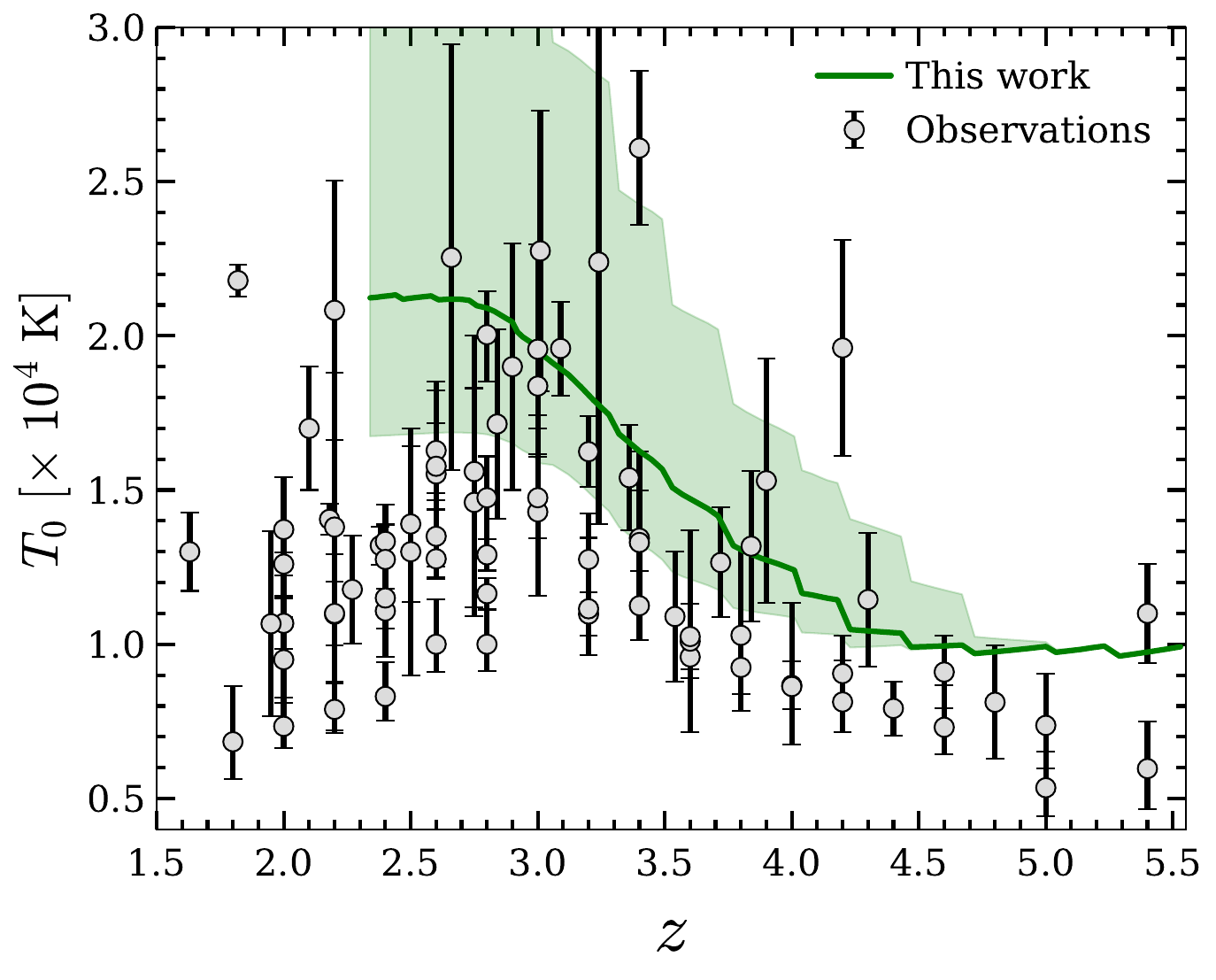}
    \caption{Redshift evolution of the median IGM temperature at mean density (solid line) for \texttt{N512-ph5e5}. The symbols denote a collection of observational constraints from \citealt{Schaye2000}, \citealt{Bolton2010}, \citealt{Lidz2010}, \citealt{becker2011}, \citealt{Bolton2012}, \citealt{Garzili2012}, \citealt{bolton2014}, \citealt{boera2014}, \citealt{rorai2018}, \citealt{hiss2018}, \citealt{boera2019}, \citealt{walther2019}, \citealt{Gaikwad2020}, and \citealt{gaikwad2021}. The shaded region displays the central $68\%$ of the data.}
    \label{fig:temp_evol}
\end{figure}

\subsection{Thermal history}
\label{thermal_history}
The ionization of helium injects energy in the IGM through photo-heating, rising its temperature. The evolution of IGM temperature is not only a test of reionization models, but also the main physical mechanism linking helium reionization to the evolution of structures, since it  hanges the dynamics of gas, its cooling rate and its accretion onto galaxies. I compute the IGM temperature at mean density by selecting all the cells in the simulation volume having gas density within $1\%$ of mean gas density. In figure \ref{fig:temp_evol}, I show the redshift evolution of the median of this quantity (solid curve) alongside the central 68$\%$ of the data (shaded region). As anticipated, $T_{0}$ increases rapidly in the first half of helium reionization through photo-heating. Once the majority of the simulation volume has been reionized at $z \sim 3$, the rate of heat injection due to photo-heating diminishes, resulting in a flattening at $z \sim 2.7$, when $\gtrsim 99\%$ of cells are ionized and the He~{\sc ii} reionization process is basically complete. 

In figure \ref{fig:temp_evol} I also plot a collection of observational data by \citealt[computed using Lyman-$\alpha$ line width distribution]{Schaye2000}, \citealt[from analysing QSO proximity zone]{Bolton2010}, \citealt[using Morlet wave filter analysis]{Lidz2010}, \citealt[using curvature statistics]{becker2011}, \citealt[from Doppler line width of Lyman-$\alpha$ absorption]{Bolton2012}, \citealt[via wavelet filtering analysis]{Garzili2012}, \citealt[from line-width distributions]{bolton2014}, \citealt[from curvature statistics]{boera2014}, \citealt[from distribution of H~{\sc i} column density and Doppler parameters]{rorai2018}, \citealt[from distribution of Doppler parameters]{hiss2018}, \citealt[via Lyman-$\alpha$ forest flux power spectra]{boera2019}, \citealt[via Lyman-$\alpha$ forest flux power spectra]{walther2019} and \citealt[via flux power spectra, Doppler parameter distribution, wavelet and curvature statistics]{Gaikwad2020,gaikwad2021}.
Despite a large scatter among them, these observations yield a coherent picture in which the IGM undergoes a phase of rapid heating at $3 \lesssim z \lesssim 4$, and then rapidly cools down. 
In the initial phase and until $z \sim 4$, my fiducial simulation predicts somewhat higher temperature compared to observations. This mismatch is a consequence of introducing a temperature floor at $10^{4}$ K (see Section~\ref{ics}). This does not affect my conclusions, since my entire analysis relies on later times, where the self-consistently calculated IGM temperature significantly exceeds this value.

Interestingly, The IGM temperature at mean density does not decrease at all at $z \lesssim 3$, but rather stays constant. There are multiple possible reasons for this. First, if the simulated heating provided by feedback processes during structure formation is too large, it might artificially compensate for the adiabatic cooling of the IGM, maintaining its temperature approximately constant. I have checked in my simulation that this plays only a minor role by excluding from the computation of $T_0$ all cells affected by feedback in {\tt TNG300} and finding that this only changes the IGM temperature at mean density by approximately 5$\%$. Alternatively, the quasar SED employed might be responsible for this offset. In fact, if my SED overestimates the number of photons with $E_\gamma \gg 54.4$ eV, it might artificially boost the IGM photo-heating. However, I remind the reader that such SEDs are based on a  compilation of observations covering the redshift range investigated (see section~\ref{sed}). Finally, it might be the case that the simulated end of helium reionization is not rapid enough, producing a broad peak, of which I miss the late-time part. This explanation is in broad agreement with the somewhat late end of helium reionization in my model discussed above. 

It should be noted that the majority of the observational constraints on $T_{0}$ are obtained by calibrating observable quantities with simulations, therefore introducing model-dependencies in the inferred physical quantities. Additionally, many of the simulations employed for this task assume a spatially-uniform time-varying UV-background, thus missing the effect of a fluctuating He~{\sc ii} ionizing background. Nevertheless, the evidence for a peak at $z\sim3$ appears strong, at least from a qualitative perspective.

Interestingly, the flattening point in the simulated curve aligns well with the peak in $T_{0}$ for the majority of observations, indicating that my predicted end of the helium ionization process is similar to the observed one. However, the shallower evolution of $T_{0}$ around this ending phase (i.e. turn-over point) suggests a more extended period of helium reionization. I will explore this in more detail in the next sections by extracting synthetic He~{\sc ii} Lyman-$\alpha$ forest properties.

\subsection{He~{\sc ii} Lyman-$\alpha$ forest}
\label{forest}
To directly compare my model with observations of He~{\sc ii} Lyman-$\alpha$ forest, I have generated synthetic spectra by extracting $16384$ sightlines at each RT snapshot, each spanning the full box length in the z direction.
For each pixel $i$ in a sightline I evaluate
the normalized transmission flux $F(i) = \exp[-\tau_{\rm{HeII}}(i)]$, where
\begin{equation}
    \tau_{\rm{HeII}}(i) = \frac{c \sigma_{\alpha} \delta R}{\sqrt{\pi}} \sum_{j=1}^{N} n\mathrm{_{HeII}}(j) \ \tilde{V}(i,j),
\end{equation}
is the He~{\sc ii} Lyman-$\alpha$ optical depth. 
Here $N$ is the number of pixels in a the sightline, $\sigma_{\alpha}$ is the Lyman-$\alpha$ scattering cross-section, $c$ is the speed of light, $\delta R$ is the size of a pixel in proper distance units, $n_\mathrm{{HeII}}(j)$ is the He~{\sc ii} number density at the position of pixel $j$ and $\tilde{V}$ is the convoluted Voigt profile approximation provided by \citealt{tepper2006}. The latter depends on the IGM temperature and the peculiar velocity. Throughout this chapter, I ignore the contamination from other lines into the wavelength window of the He~{\sc ii} Lyman-$\alpha$ forest. This is not expected to have any impact on my results. In fact, such contamination is a major obstacle to observations but the data actually employed to constrain the helium reionization are typically free of this issue, making my choice reasonable.

\begin{figure*}
\centering
    \includegraphics[width=120mm]{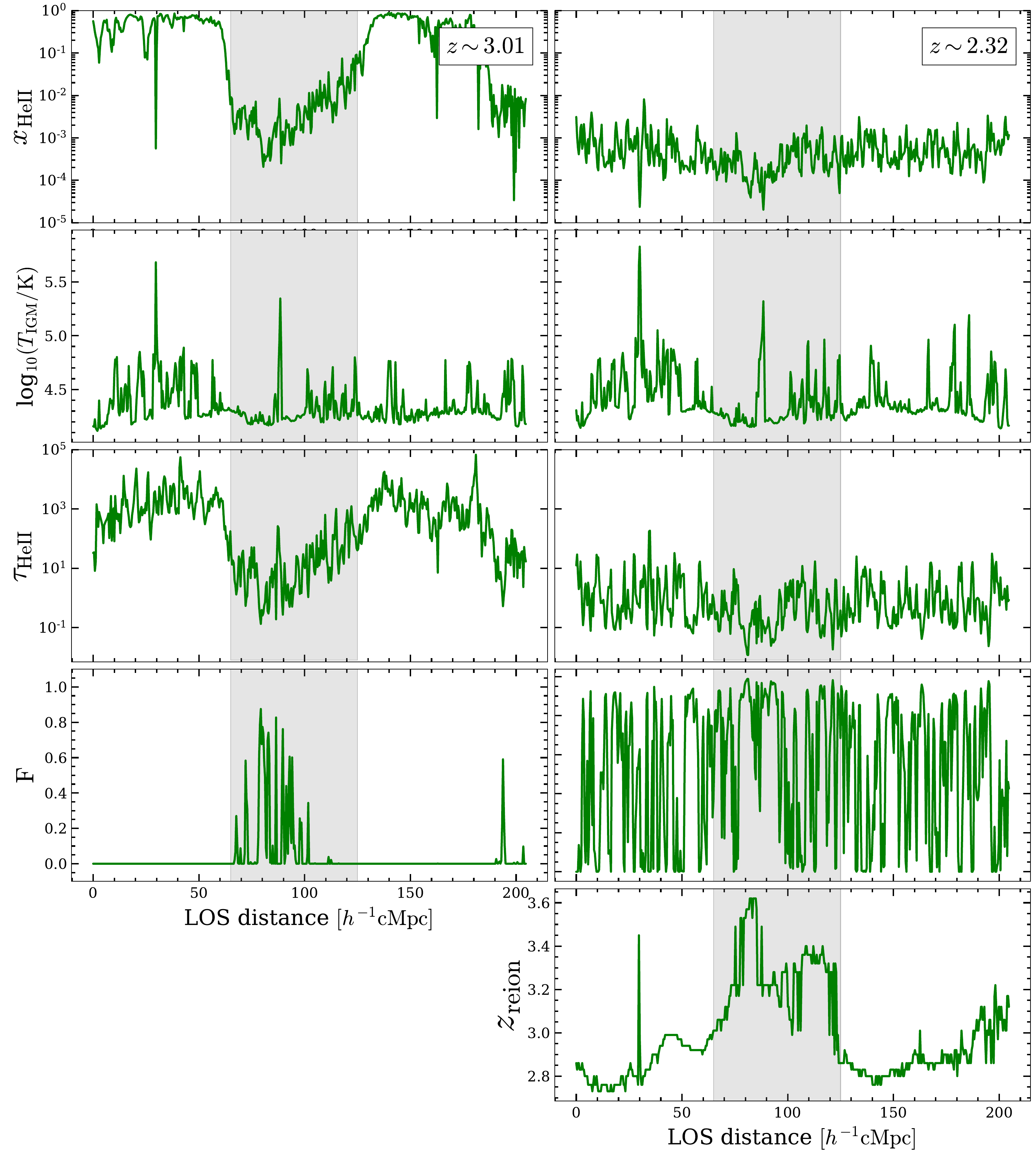}
    
    \caption{Line of sight IGM properties at \textit{z} = 3.01 and 2.32 extracted from my reference simulation \texttt{N512-ph5e5}. The fmy panels from top to bottom display the He~{\sc ii} fraction, the IGM temperature, the He~{\sc ii} optical depth, and the normalized transmission flux. The local He~{\sc ii} reionization redshift along the line of sight extracted at $z=2.32$ is shown in the right bottom panel. The shaded region highlights cells that undergo an early reionization.}
    \label{fig:los}
\end{figure*}

\begin{figure}
\centering
    \includegraphics[width=90mm]{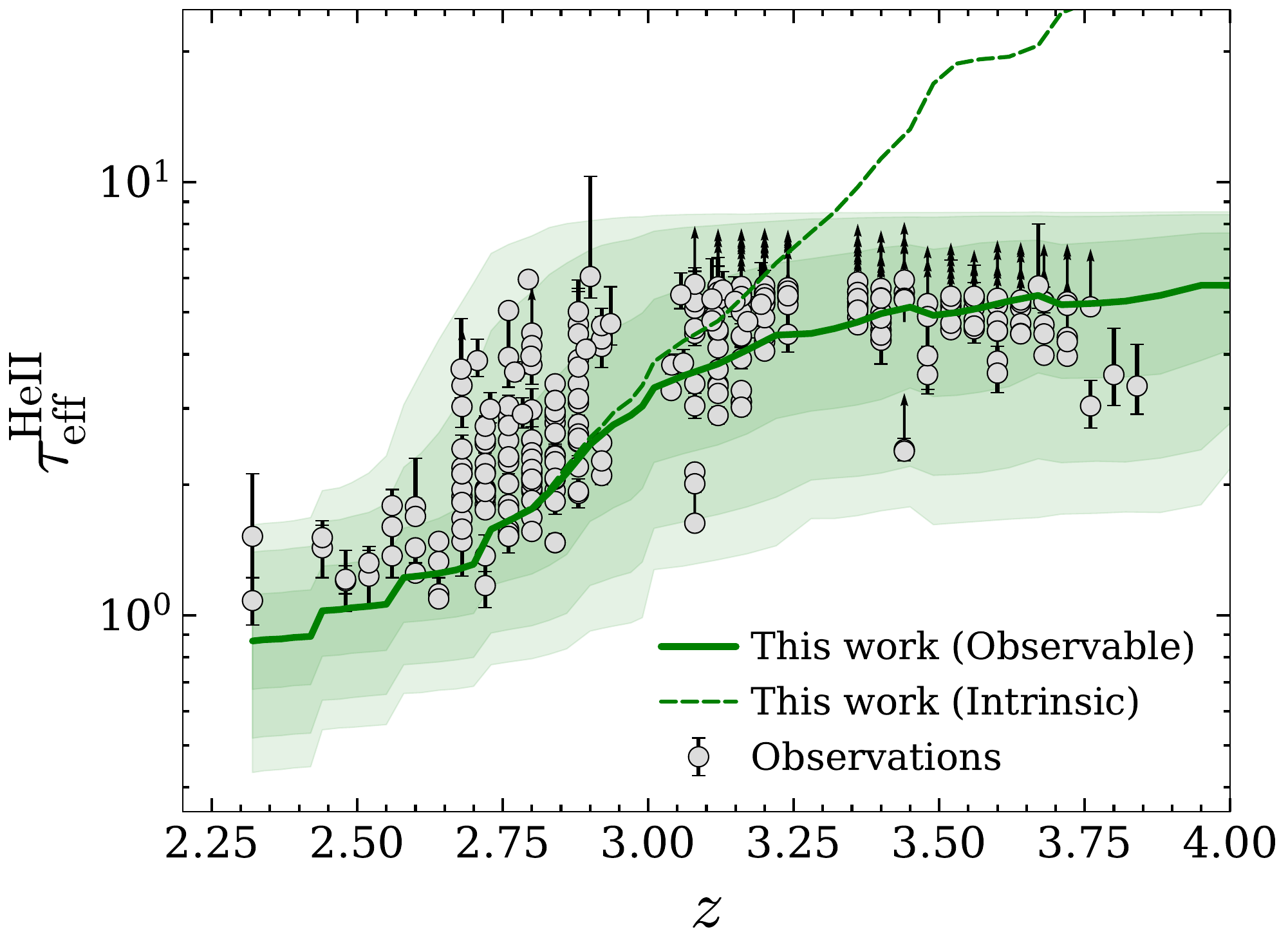}
    
  \caption{Redshift evolution of the median He~{\sc ii} effective optical depth computed from the synthetic spectra in my reference simulation \texttt{N512-ph5e5} (solid green curve). The shaded regions indicate 68\%, 95\%, and 99\% confidence intervals. Observational estimates from \citealt{worseck2016}, \citealt{worseck2019}, \citealt{makan2021} and \citealt{makan2022} are shown as symbols. 
  For a consistent comparison with observations, in the evaluation of the median I include only spectral chunks with an effective optical depth below 8.56 (i.e. `observable', see text for more details), while the dashed line refers to the median calculated for the `intrinsic' distribution.}
    \label{fig:taueff}
\end{figure}

\subsubsection{Individual line of sight}
\label{forest_1los}
To demonstrate the results of my forward modeling procedure, I show in figure \ref{fig:los} (from top to bottom) the IGM ionization state, gas temperature, transmitted flux and optical depth along one individual line of sight at $z \sim 3.01$ (left column) and 2.32 (right column). For the sake of clarity, here (and for all the following plots) I omit the $i$ when referring to values of the physical quantities associated to a single pixel. As anticipated, at $z \sim 3.01$  $x_{\textrm{HeII}}$ (top row) is notably higher than at $z \sim 2.32$, when most helium is fully ionized. Correspondingly, as redshift decreases, $T_{\rm IGM}$ (second row) rises alongside He~{\sc iii} fractions, leading to lower values of $\tau_{\rm HeII}$ (third row) and more areas of transmitted radiation (bottom). Notice that the IGM temperature is set by the combined effect of photo-heating and other feedback processes associated to structure formation (such as shock heating, AGN feedback, etc.). Therefore it can locally reach values significantly higher than those expected from pure photo-heating and, simultaneously, drive localised collisional ionization of He~{\sc ii} into He~{\sc iii}. This can be seen in the narrow temperature spikes in the figure. I also show the local He~{\sc ii} reionization redshift ($z\rm{_{reion}}$) extracted at the end of the simulation (i.e. at $z\sim2.32$). As discussed in section \ref{reion_history}, regions undergoing an earlier reionization (highlighted by the shaded region)
are characterized by a slightly lower temperature than the rest.

\subsubsection{Effective optical depth}
\label{forest_tau_eff}
 
To facilitate direct comparison with observations, I evaluate the He~{\sc ii} effective optical depth $(\tau_{\rm{eff}}^{\rm{HeII}})$, a widely used characterization of the Lyman-$\alpha$ forest. To achieve this, I partition each synthetic spectrum into chunks of length 50 $ h^{-1}{\rm cMpc}$ (corresponding to $\Delta z \approx 0.04$ at the relevant redshifts) following \citealt{worseck2016}, resulting in a sample of over 60,000 chunks at each redshift. The He{\sc ii} effective optical depth is then computed as:
\begin{equation}
\tau_{\rm{eff}}^{\rm{HeII}} = - \ln (\langle F \rangle),
\label{columndensity}
\end{equation}
where $\langle F \rangle$ is the mean transmitted flux of all the pixels in each chunk. In order to approximately mimic the fact that observations are not able to distinguish values of the optical depth above a certain threshold $\tau_{\rm{eff,th}}^{\rm{HeII}}$, I consider observable those spectral chunks with $\tau_{\rm{eff}}^{\rm{HeII}}<\tau_{\rm{eff,th}}^{\rm{HeII}}$. I set $\tau_{\rm{eff,th}}^{\rm{HeII}}=8.56$ as representative of the typical maximum optical depth detected in the observations at $z>3$ \citep{worseck2016,worseck2019,makan2021,makan2022}. 

Figure \ref{fig:taueff} displays the redshift evolution of the median observable He~{\sc ii} effective optical depth $\tau_{\rm{eff}}^{\rm{HeII}}$ (solid line) along with the central 68\%, 95\% and 99\% of the data (shaded regions of increasing transparency), as well as aformentioned observational data points. 
As a reference, I also show the median of the intrinsic distribution (i.e. the one without any optical depth threshold) using a dashed line. The difference between the intrinsic and the observable distribution is crucial for investigating the highest-redshift observations available, that are only able to sample the low-$\tau_{\rm{eff,th}}^{\rm{HeII}}$ part of the intrinsic distribution. I also predict that at $z\lesssim 3$ observations are able to effectively sample the entire distribution. 

In fact, while the intrinsic median $\tau_{\rm{eff,th}}^{\rm{HeII}}$ keeps increasing with increasing redshift, the observable one flattens out at $z \gtrsim 3.5$, with the bulk of effective optical depths becoming less and less accessible to observations, which are only sensitive to the most transparent regions of the IGM.
Following the completion of the reionization process at $z \sim 2.7$, the effective optical depth shows a residual non-zero value, due to residual He~{\sc ii} in self-shielded systems and recombinations in the IGM. The mild time evolution seen in my simulations develops as a consequence of (i) the evolving thermal state of the IGM mainly through adiabatic cooling and heat injection due to structure formation, and (ii) the reduced recombination rate stemming from the expansion of the Universe lowering the density in the IGM. The noticeable scatter around the median curve reflects the patchy nature of the He~{\sc ii} reionization process, and indeed it decreases once this process is over. 
I find an agreement with most data points within the $95\%$ confidence intervals, demonstrating the effectiveness of my simulation in reproducing the observed behavior of He~{\sc ii} effective optical depth. This success is indicative of the fact that the QLF from \citealt{Shen2020} provides a reasonable helium reionization history and topology, when combined with a accurate hydrodynamical and RT simulation. 

\begin{figure}
    \centering
    \includegraphics[width=80mm]{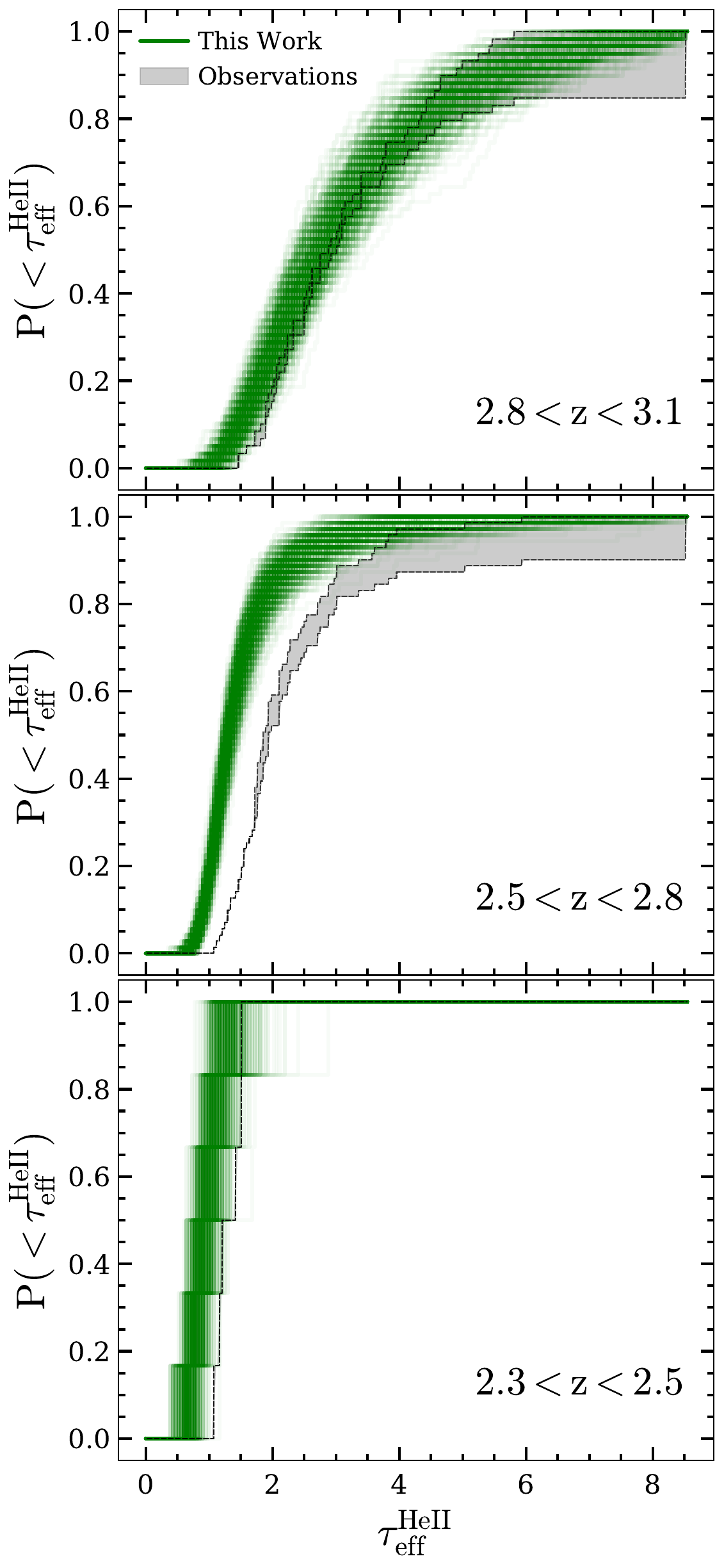}
    
    \caption{Cumulative distribution function of He~{\sc ii} effective optical depth in different redshift bins, averaged over spectral chunks of length $50 h^{-1} {\rm cMpc}$ extracted from simulation \texttt{N512-ph5e5}. All the curves are generated by sampling the true distribution in the simulation with the same number of observational data points in respective redshift bins (see text for more details). The grey shaded regions are observational limits adopting the "optimistic" and "pessimistic" approach used by \citealt{bosman2018}.}
    \label{fig:cum_taueff}
\end{figure}

When looking at the details, however, I find some discrepancies, which are explored in the following. From figure \ref{fig:taueff}, it can be seen that at $z \lesssim 3$ most of the observed effective optical depths lie above the median value derived from my simulations. In order to better compare the distribution of effective optical depths at fixed redshift, I compute its cumulative distribution function (CDF) in three redshift bins spanning the range $z$ = 2.3 - 3.1. 
In order to provide a fair comparison with observations, for each redshift bin investigated I create 500 realization of the predicted optical depth distribution. I do so by randomly selecting a number of synthetic $\tau_{\rm{eff}}^{\rm{HeII}}$ equal to the number of observed values within the bin. I show these realization in figure \ref{fig:cum_taueff} as thin green lines, along with the observed CDF (obtained employing the same observations shown in figure \ref{fig:taueff}). For the latter, I adopt the "optimistic/pessimistic" approach of \citealt{bosman2018} to deal with pixels having flux below the detection threshold, thus resulting in a range of possible observed values (gray shading).  It should be noted that my simulated optical depths are not calibrated to match any observed CDF nor mean transmitted flux. 
Despite this, in the highest redshift bin (corresponding to an approximately 90\% ionised IGM) the simulated CDF reproduces well the observed one. As reionization progresses, the CDF slope becomes as expected progressively steeper. For $z\lesssim 2.8$, though, the simulated CDF is systematically shifted to lower optical depths in comparison to observations, although the shape is still well reproduced. In the lowest redshift bin, the CDF gets saturated at $\tau_{\rm{eff}}^{\rm{HeII}} \sim 1.5$, confirming that the last stage of the reionization process has been reached. Here as well, the simulated CDF is somewhat shifted to lower optical depths. These discrepancies can be revealing of two different phenomena, namely (i) that in my model helium reionization is completed slightly too early with respect to the observed data and (ii) that the limited resolution of my simulations prevents us from fully resolving the residual sinks of radiation in the post-reionization Universe that would increase the IGM effective optical depth. From the discussion in Section \ref{reion_history} and the resolution test I have performed I conclude that these differences are probably due to a combination of both effects.
Unfortunately, the resolution needed to properly capture the small scale Lyman-limit systems would make my simulations prohibitively expensive. Therefore, for a more accurate modeling of the final phases of the process, a sub-grid prescription is required (e.g. \citealt{Mao2020,Bianco2021,Cain2021}),
which I plan to include in future investigations.
\subsubsection{Characterisation of transmission regions}
\label{forest_spikes}
\begin{figure}
\centering
    \includegraphics[width=100mm]{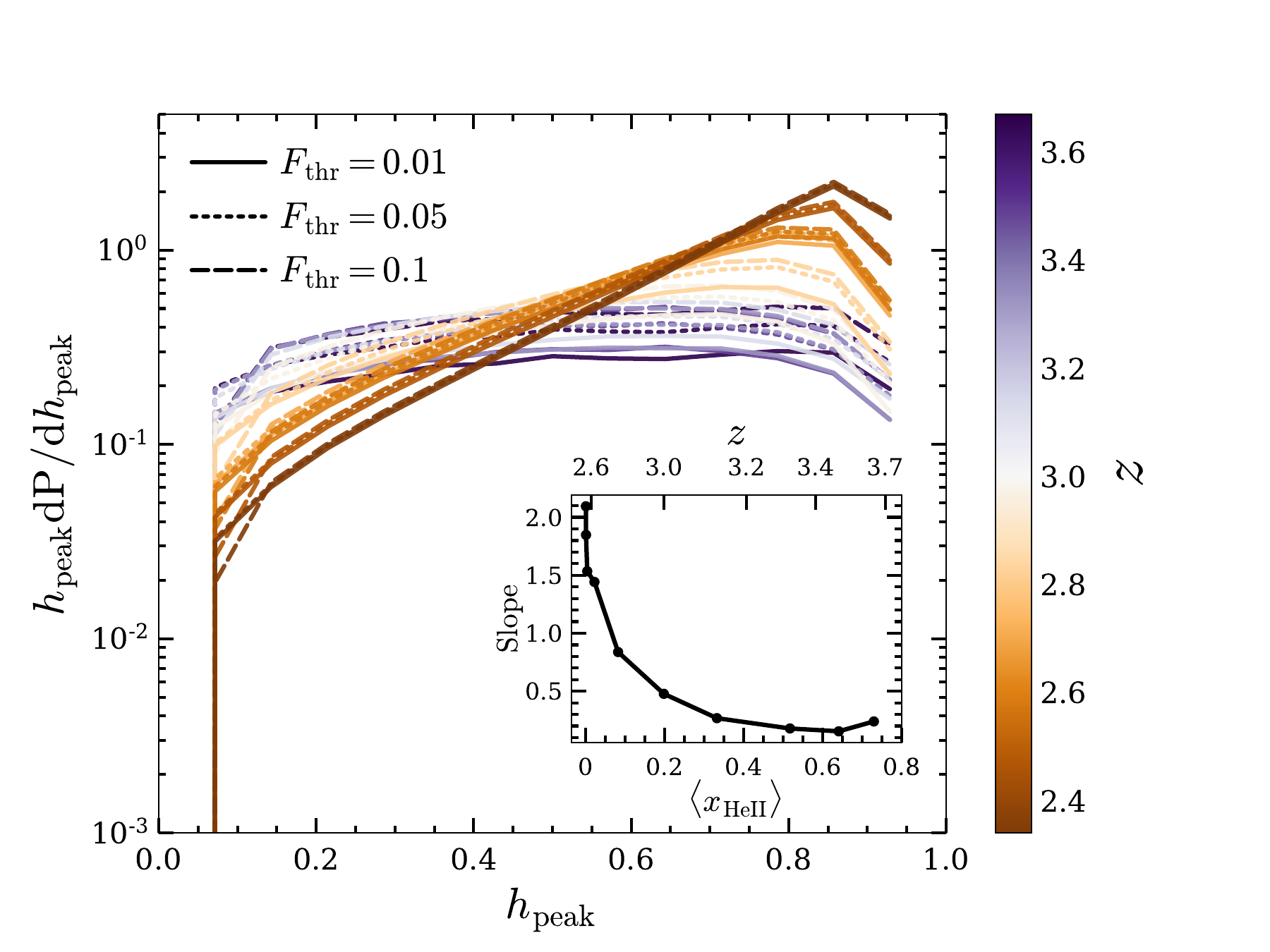}

    \caption{Distribution of  peak heights ($h\mathrm{_{peak}}$) computed over the sample of He~{\sc ii} Lyman-$\alpha$ forest spectra extracted from simulation \texttt{N512-ph5e5} at different redshifts (indicated by the color bar). Different line styles refer to different values of the adopted flux threshold, $F_{\rm thr}$. In the inset, the relation between the slope of the distribution (see text for more details) and $\langle x_\mathrm{HeII}\rangle$ for $F_{\rm thr}=0.01$ is shown as black solid curve.}
    \label{fig:transmitted_region_prop_hpeak}
\end{figure}

\begin{figure}
\centering
    \includegraphics[width=100mm]{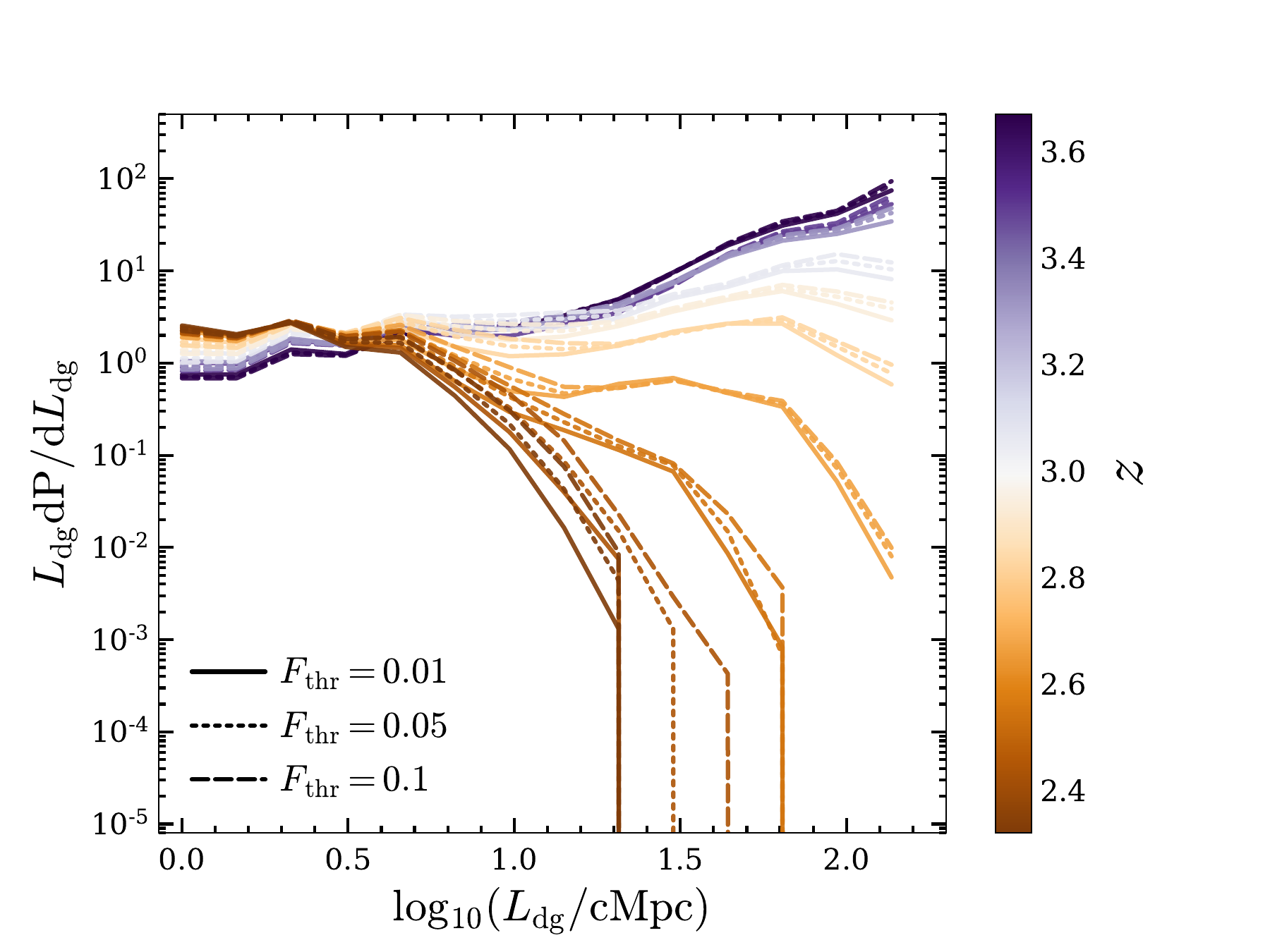}

    \caption{Distribution of  dark gaps lengths ($L\mathrm{_{dg}}$) computed over the sample of He~{\sc ii} Lyman-$\alpha$ forest spectra extracted from simulation \texttt{N512-ph5e5} at different redshifts (indicated by the color bar). Different line styles refer to different values of the adopted flux threshold, $F_{\rm thr}$.}
    \label{fig:transmitted_region_prop_Ldark}
\end{figure}

\begin{figure}
\centering
    \includegraphics[width=90mm]{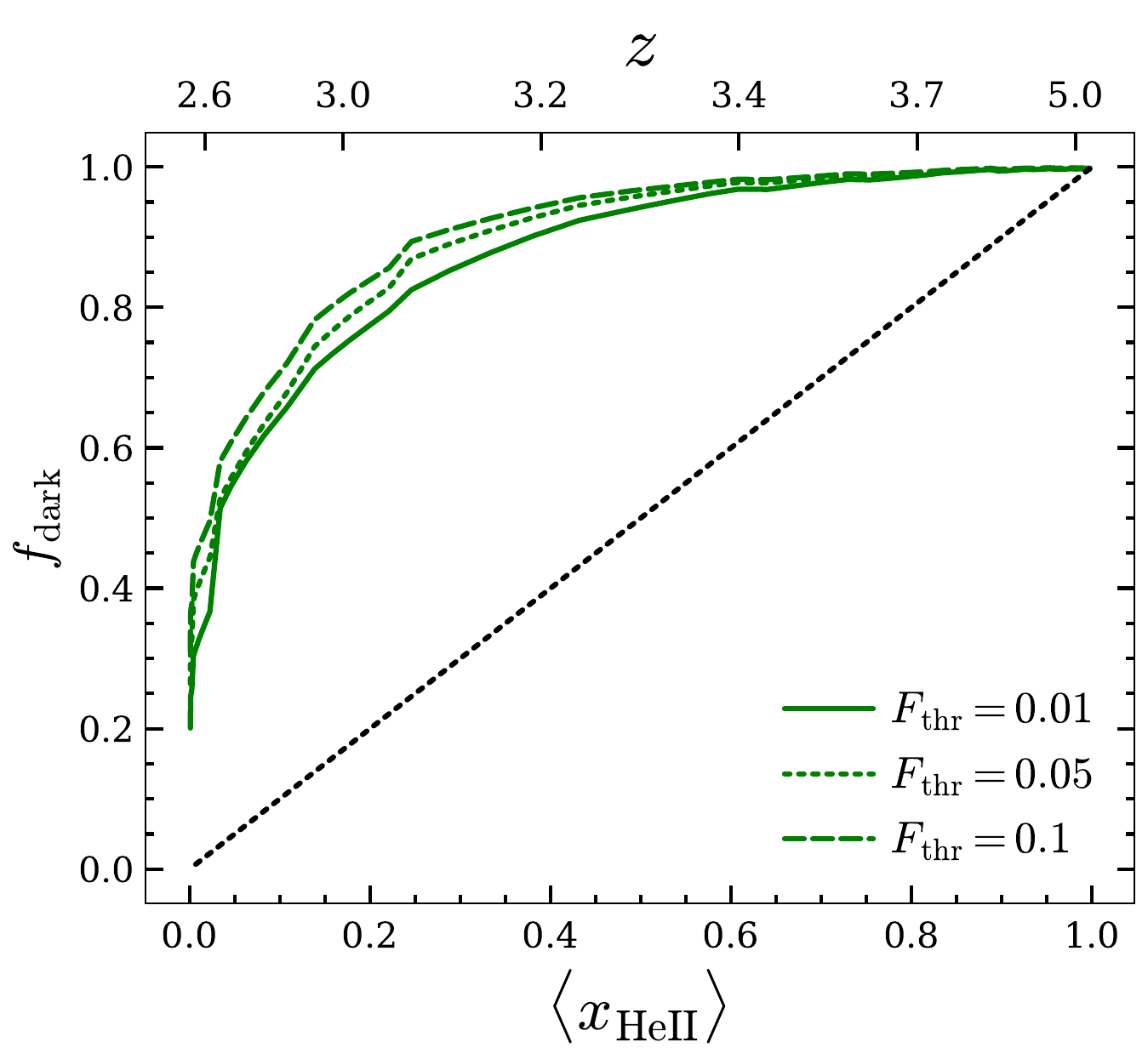}
    
    \caption{Evolution of the dark pixel fraction ($f_\mathrm{dark}$) as a function of the volume-averaged He~{\sc ii} fraction ($\langle x_\mathrm{HeII}\rangle$) in synthetic spectra of the He~{\sc ii} Lyman-$\alpha$ forest extracted from my fiducial simulation. Different line styles (green curves) refer to different values of the adopted flux threshold, $F_{\rm thr}$, in the definition of dark pixels. The black dotted curve shows one-to-one correspondence between these two quantities. In the top axis, the corresponding redshifts are displayed.}
    \label{fig:darkfraction_evol}
\end{figure}
Here I move to a characterization of individual features in the He~{\sc ii} Lyman-$\alpha$ forest. These carry information on the sources of reionization \citep[e.g. ][]{Garaldi2019, Gaikwad2020}, but are also much more dependent on the simulation resolution than the average quantities discussed so far. I note that the resolution requirements in the case of helium reionization are less stringent than in the case of hydrogen reionization, since the larger bias of the sources implies that the ionized regions are typically much larger, and so are the features in the forest. Therefore, the resolution necessary to capture their existence, if not their details, is lower. In order to maintain a conservative approach, in the following I only investigate quantities that I have found to be resilient to changes in resolution. For instance, I found that the transmission peak height is more robust than its width against resolution changes, since the former only depends on the most ionized regions (which are typically close to a very bright source and therefore fully ionized regardless of the resolution), while the latter depends on the details of the ionized region edges, which are much more sensitive to the employed resolution. Consequently, I elect to investigate only the former. 
Nevertheless, the results that follow need to be taken as qualitative more than quantitative, since an improved spatial resolution will still affect their details. 

I start by identifying transmission regions in the synthetic spectra. I follow the procedure developed in \citealt{Garaldi2019b,garaldi2022}, adapted from \citealt{Gnedin2017}, and consider a pixel as part of a transmission spike if its normalised flux is higher than a threshold value, $F\mathrm{_{thr}}$. I then characterize the spike through its height, $h_\mathrm{_{peak}}$, defined as the maximum transmitted normalized flux among all pixels associated to it. 
In figure \ref{fig:transmitted_region_prop_hpeak} (main panel) I show the probability density of $h\mathrm{_{peak}}$ for different redshift and $F\mathrm{_{thr}}$. I observe a strong dependence on $z$, with an almost flat curve at $z>3$, indicating a prevalence of small transmission spikes over large ones (notice that the vertical axis is multiplied by $h\mathrm{_{peak}}$). This can easily be understood as at this redshift ionized regions are still relatively small around the first quasars active in the simulated volume, so part of the transmitted flux is absorbed while traveling through the nearby neutral regions. As redshift decreases, more sources turn on and the ionized regions grow in size. This is reflected in a tilt of the peak height distribution towards larger values. Interestingly, this change in the slope does not break the linearity of the relation (for the chosen variables and axis scaling). This allows us to derive an empirical relation between the slope of this distribution and the volume-averaged He~{\sc ii} fraction in the simulation. 
To obtain the former I perform a least-square fit of $\log(h_\mathrm{_{peak}} \mathrm{d}P/\mathrm{d}h_\mathrm{_{peak}})$ in the range $0.2 \leq h\mathrm{_{peak}} \leq 0.8$. I show the co-evolution of these two quantities in the inset of figure \ref{fig:transmitted_region_prop_hpeak}. When $\langle x_\mathrm{HeII}\rangle $ is well above 40$\%$ (or $z>3$), the slope is $\sim $ 0, consistently with the flat distribution discussed above. Towards lower redshift, instead, I observe a steep rise of the slope, confirming the sensitivity of this probe to the last phases of helium reionization with percolation of ionized bubbles. 
Finally, I note that the rightmost bin of the $h_\mathrm{_{peak}}$ distribution shows a decline. This is due to the fact that in my simulations the IGM is not yet sufficiently ionized to allow a complete transmission of the incoming flux. The qualitative behavimy described is independent of the value adopted for $F\mathrm{_{thr}}$, although there are some minor quantitative differences at the highest redshifts. 

To complement the peak height distribution analysis, I compute the distribution of the absorbed regions in the spectrum using the dark gap (DG) statistics \citep{paschos2005,Fan2006,Gallerani2006,Gallerani2008}, where a DG is defined as a continuous region with normalised transmission flux below $F\mathrm{_{thr}}$, and it is characterized by its length, $L\mathrm{_{dg}}$. In figure \ref{fig:transmitted_region_prop_Ldark} I show the distribution of $L\mathrm{_{dg}}$ for various redshifts and $F\mathrm{_{thr}}$ values. 
It should be noted that $L\mathrm{_{dg}}$ is limited by construction to the length of each synthetic spectra, which in turn is constrained by the simulation box length (since I do not employ periodic boundary conditions). 
The distribution of $L\mathrm{_{dg}}$ exhibits a clear trend throughout the entire redshift range investigated. The occurrence of the longest gaps ($L\mathrm{_{dg}} \gtrsim 10\, \rm{cMpc}$) monotonically decreases with cosmic time, as a consequence of the increasing number of ionized regions towards the lower redshifts. This drop is significantly more rapid between $z\sim2.7 \lesssim z \lesssim 3$. Conversely, the occurrence of shorter gaps increases with the development of helium reionization. The distribution of $L\mathrm{_{dg}}$ is also practically insensitive to the adopted threshold value. I have confirmed that this statistics is fairly independent of the simulation resolution in the redshift range explored. 

Finally, I compress the information provided by the dark gaps distribution into a single number by computing the fraction of pixels ($f\mathrm{_{dark}}$) in all spectra at a a given redshift that have normalized flux below a threshold value $F\mathrm{_{thr}}$. 
In the context of hydrogen reionization, this quantity is often employed to derive model-independent upper limits on the neutral fraction as $x_\mathrm{HI} \leq f_\mathrm{dark, HI}$ \citep{mesinger2010,McGreer2011,McGreer2015}. 
In figure \ref{fig:darkfraction_evol} I show the co-evolution of this dark fraction with volume-averaged He~{\sc ii} fraction in my fiducial simulation. 
A distinct drop of $f\mathrm{_{dark}}$ is visible at $z\sim3.3$, where $\sim$ 50$\%$ of Helium is fully ionized. As expected, $f\mathrm{_{dark}}$ increases with increasing $F\mathrm{_{thr}}$. 
The fact that throughout my simulation $f\mathrm{_{dark}}$ is higher than $\langle x_\mathrm{HeII}\rangle$ (i.e. all green curves are well above the black dotted curve) confirms that this approach works also in the context of helium reionization. 

In general, with enough data available, all these statistics characterizing transmission regions can be used to constrain the timing of He~{\sc ii} reionization, as well as the source properties, as discussed e.g. by \citealt{garaldi2022}.

\begin{figure} 
\centering
    \includegraphics[width=100mm]{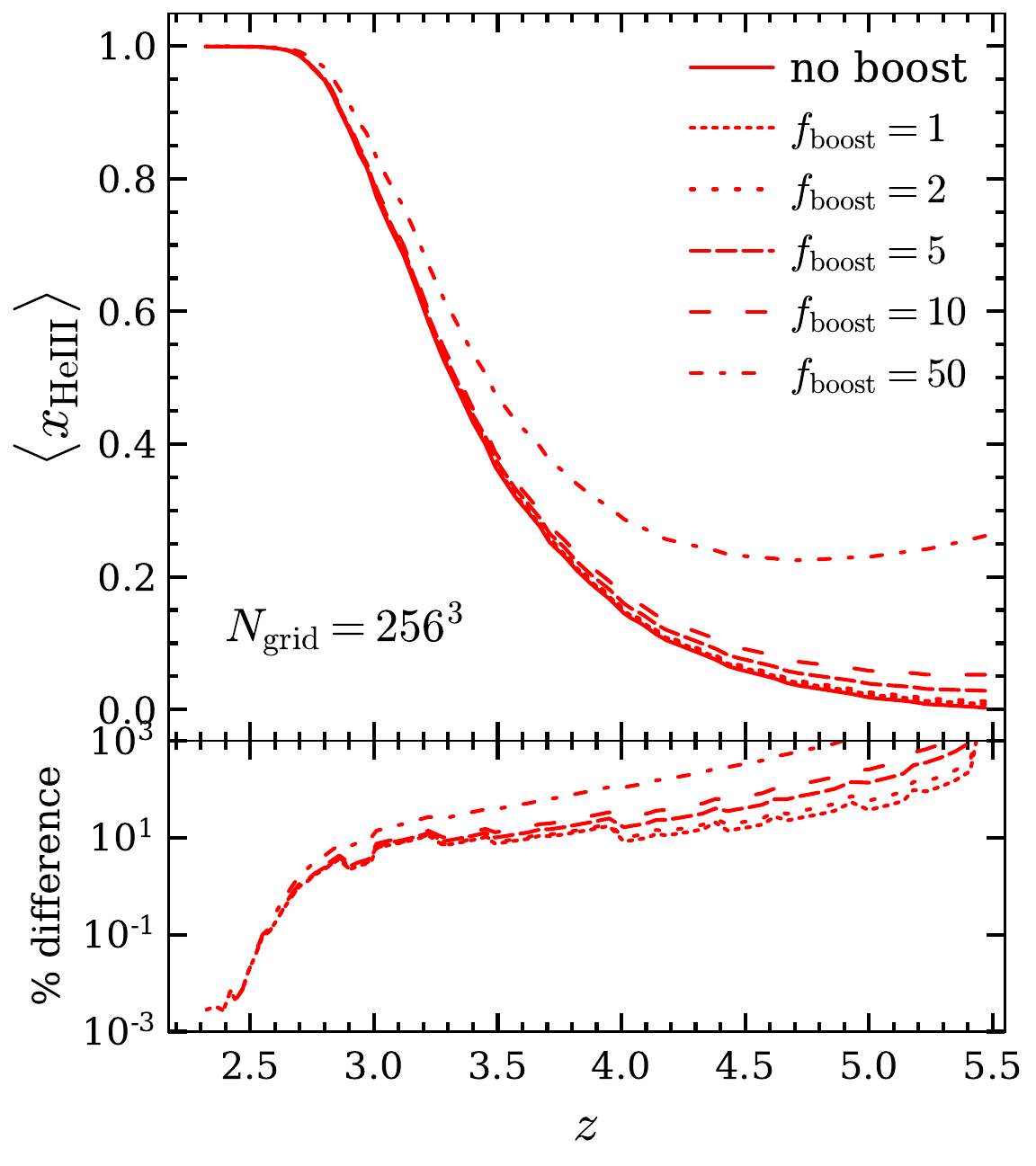}
    \caption{{\it Top panel}: evolution of the volume-averaged He~{\sc ii} fraction for simulations featuring a boosted QLF at $z>5.53$. The boosting factor is reported in the legend. Notice that I assume a step-like transition at $z=5.53$ between the boosted and fiducial QLF.
    The solid curve ($f_\mathrm{boost}=1$) correspond to \texttt{N256-ph1e5}, i.e. a lower-resolution version of my fiducial simulation, which is started at $z_\mathrm{in}=5.53$ with the intergalactic helium singly ionised. {\it Bottom panel}: relative differences of each run with respect to the fiducial \texttt{N256-ph1e5}.}
    \label{fig:jwst}
\end{figure}

\subsection{Impact of \texttt{JWST} detection of $z>5$ QSOs}
\label{jwst}

The \texttt{JWST} has recently detected numerous AGNs at $z > 5$, although in relatively small survey areas. If confirmed to be representative of the entire population, these would imply a significantly boosted QLF (up to approximately a factor of 30 in the faint end of the QLFs) with respect to pre-\texttt{JWST} estimations \citep[e.g.][]{harikane2023,maiolinoa2023,Maiolino2023,Goulding2023,Larson2023,Juodzbalis2023,greene2023}.
Under the assumption that the characteristics of such objects are similar to those of lower redshift ones (in particular, their UV spectrum and their near-unity escape fraction of ionizing photons, but see \citealt{Christiani2016}), such population could have an impact on both hydrogen and helium reionization. Here I assess their contribution to the latter. I caution, however, that my approach is relatively simple, and as such its results should be taken as qualitative. 

Given the large differences in the QLF derived by different studies, and in order to minimize the number of parameters in my study while maintaining flexibility, I choose to model the population of $z>5$ QSOs by boosting my fiducial QLF \citep[i.e Model 2 of ][]{Shen2020} by a factor $f_\mathrm{boost} $\footnote{Note that the $f_\mathrm{boost}$ mentioned here differs from the one used in \citealt{pacucii2023}, where it is employed to quantify a boost in black hole masses observed by \texttt{JWST} to account for non-detectability. Effectively, this will cause a uniform shift in the QLF towards more massive luminous bins while I instead introduce a boost across all luminosity bins.} = 1, 2, 5, 10, 50  at $z>5.53$, while maintaining its value at lower redshift, where observations of the QLF are more robust. I then run simulations identical in setup to those discussed so far, but starting from $z_\mathrm{init} = 6$. To reduce the computational requirement, these additional simulations are run only with $N_{\rm grid}=256^3$.

Figure \ref{fig:jwst} shows the simulated helium reionization histories (top panel) and their relative difference with respect to my fiducial QLF (bottom) for the different $f_\mathrm{boost}$ simulated. 
I also show a run labeled `no boost', which I take as reference in the bottom panel. This is a lower-resolution version (to match the other runs discussed in this section) of my fiducial simulation. I use this to show the impact of different initial redshift for the simulation, since `no boost' starts at $z_\mathrm{init} = 5.53$ (like my fiducial run and other runs previously discussed), while the simulation labeled `$f_\mathrm{boost} = 1$' begins at $z_\mathrm{init}=6$. From an inspection of the bottom panel it can be seen that such difference is generally negligible with respect to modification to the QLF. 
For most $f_\mathrm{boost}$ values, the He~{\sc ii} fraction at $z=5.53$ is only mildly affected and the difference between the various runs is quickly erased once helium reionization is ongoing. The exception to this is my most extreme model ($f_\mathrm{boost} = 50$), that maintains some small difference all the way to the completion of helium reionization. Although not shown here for the sake of brevity, I find similar qualitative and quantitative results with respect to the He~{\sc ii} effective optical depth and associated statistics.
The findings unequivocally indicate that an increased (up to an order of magnitude) number of quasars at $z\gtrsim5$ (or more precisely a higher emissivity from quasars at such earlier times), as suggested by \texttt{JWST}, does not necessarily impact the ionization and thermal state of the IGM during the ending phase of this epoch. Should the number density of quasars be larger, their characteristics (chiefly, the ionizing photons escape fraction) be different than lower-$z$ QSOs, or should the boost persist to later times, my results would have to be revised through a tailored, more accurate model.

\section{Discussion and Conclusions}
\label{conclusions}

Recent observational advancements have pushed my ability to probe helium reionization past its tail end. At the same time, new observations-based models of the QLF have become available. Most notably, some of these models enforce a smooth evolution through time of the QLF, enabling us to bypass the need to assume an ionising emissivity evolution to rescale the observed QLF at a given redshift, as often done in the past \citep{Compostella2013, compostella2014, Garaldi2019b}. In this chapter, I present the first simulations of helium reionization directly implementing one such QLF \citep[namely the Model 2 from][]{Shen2020}. I carry out a thorough exploration of its predictions concerning helium reionization, including a comparison to the latest observations and a set of predictions for the features of the He~{\sc ii} Lyman-$\alpha$ forest that can be measured in current and future data. To this end, I combine the \citealt{Shen2020} QLF with the \texttt{TNG300} cosmological hydrodynamical simulation and the 3D radiative transfer code \texttt{CRASH} (see section \ref{tng+crash}). Another distinguishing feature of my simulations is the inclusion of observationally-derived quasar spectra (see section \ref{sed}). This makes my prediction entirely independent from the (often very approximate) physics of black hole seeding, formation and evolution implemented in large-volume numerical simulations. Employing this setup, I simulate the evolution of the IGM from $z_\mathrm{init} = 5.53$ to the completion of helium reionization and forward model my runs to provide faithful predictions of the He~{\sc ii} Lyman-$\alpha$ forest. Finally, inspired by recent \texttt{JWST} observations, I develop a simple numerical model to assess the impact of high-redshift QSOs on helium reionization. For that, I assume that (i) the QLF is unchanged at $z\leq 5.53$ with a step-like transition, (ii) that the large number of high-redshift AGNs observed results in a rigid shift of my fiducial QLF towards larger number densities, and (iii) that the spectrum of high-redshift quasars is identical to their low-redshift counterparts. While simplistic, this approach captures the main features relevant for helium reionization, and therefore my results are expected to be qualitatively robust.

The main results can be summarized as follows: 
\begin{itemize}
    \item I predict a later helium reionization history with respect to the majority of models available in the literature. While many details of the modeling differ, the largest impact likely comes from the different QLF employed.
    
    \item My fiducial simulation predicts an IGM temperature at mean density which is somewhat higher than most of the observational data points. This is likely due to the fact that the latter are obtained by calibrating with respect to simulations employing a uniform UV background, thus missing the impact of inhomogeneous helium reionization.

    \item The simulated IGM temperature at mean density remains constant even after completing the reionization process, although the point of flattening of the curve aligns well with the observational turn-over point.
    
    \item The observed and simulated He~{\sc ii} effective optical depth are in good agreement once observational limitations are properly taken into account. They also show a comparable amount of scatter throughout the simulated history. Therefore, I expect my model to properly capture the bias of the sources responsible for helium reionization and the resulting patchiness of such process. 

   \item I find residual differences between the simulated and observed distribution of He~{\sc ii} effective optical depth at fixed redshift, in particular at $z \le 2.8$. This can be ascribed to an early end of reionization in my simulations. 

    \item I present a comprehensive analysis of transmission regions and dark gaps in simulated He~{\sc ii} Lyman-$\alpha$ forest spectra. I demonstrate how the shape of their distribution is linked to the underlying IGM ionization state, opening up the possibility to use such measure to provide additional constraints from observed quasar spectra. 

    \item I find that, in my simplified approach to estimate the impact on helium reionization of the large number of AGNs observed by \texttt{JWST}, their effect is negligible unless the QLF is boosted by more than an order of magnitude at $z\gtrsim5.5$. 

\end{itemize}

In summary, I have presented a detailed characterization of helium reionization in light of the latest constraints on the QLF and the latest measurements of the IGM optical depth, combining a state-of-the-art hydrodynamical simulation with accurate radiative transfer. my results show an overall convergence of predicted and observed properties in the redshift range $2.3 \le z\le 3.5$. 
Nevertheless, the discrepancies I notice between my fiducial simulation and observations are likely indicative of a lack of self-consistency between current observations of helium reionization and of the QLF itself. Unfortunately, my simulations are unable at present to point toward one of these two as culprit. Additionally, it is possible that assumptions in my modeling contribute (or even generate) the discrepancies discussed before. Among these, I would like to highlight the spectral energy distributions of QSOs and the quasar ionizing escape fraction. The latter, in particular, is virtually always assumed to be unity (in simulations like the one presented in this work), but it has been shown to vary significantly \citep{Christiani2016}. Testing and overcoming these assumptions are as essential as improving observational datasets (in particular concerning helium reionization) in order to assess whether my understanding of quasar assembly and evolution is consistent with that of intergalactic helium reionization. I plan to follow this route in a future dedicated study.
Investigations like this demonstrate the power of combining different domains of knowledge to achieve a deeper understanding of structure formation throughout cosmic history.

  \chapter{Probing Helium reionization through the $^{3}\mathrm{He}^{+}$ transition line}
\label{chap:chapter5}

\begin{flushright}
\begin{minipage}{0.7\textwidth}
\raggedleft
\textbf{\textit{``Everyone has one, but don't be one to each other."}}\\[1ex]
\noindent\rule{0.5\textwidth}{0.4pt}\\[-0.2ex]
Joanne Tan
\end{minipage}
\end{flushright}

\begin{flushright}
\begin{minipage}{0.7\textwidth}
\raggedleft
\textbf{\textit{``Where we go from there is a choice I leave to you."}}\\[1ex]
\noindent\rule{0.5\textwidth}{0.4pt}\\[-0.2ex]
Akash Vani (\textit{inspired from `The Matrix'})
\end{minipage}
\end{flushright}

\textit{This work has been submitted for publication in the Monthly Notices of the Royal Astronomical Society \citep{Basu2025c}.}

\hspace{1cm}

At redshift $z \approx 1100$, hydrogen recombined as the Universe expanded and cooled sufficiently to allow electrons to bind with protons. At earlier times ($z \approx 6000$), helium experienced its first recombination (He~{\sc iii} $\to$ He~{\sc ii}), i.e. the one pertaining its inner electron, which is bound with an energy of 54.4 eV \citep{Switzer2008b}. As the Universe continued to cool, singly ionized helium (He~{\sc ii}) recombined into neutral helium (He~{\sc i}) at $z\approx 1800$ \citep{Switzer2008a}. 
With time, the neutral atoms formed during these recombination epochs were gradually reionized by photons emitted from various emerging astrophysical sources.
As the first ionization potential of helium is similar to the one of hydrogen, He~{\sc i} is expected to be ionized into He~{\sc ii} at the same time of hydrogen, i.e. by $z \sim 5.5$ \citep{Fan2006,Becker2015,bosman2018,Bosman2022}. However, the reionization of singly ionized helium (He~{\sc ii} $\to$ He~{\sc iii}) occurs at a later time, requiring the energetic photons emitted by quasars \citep{Compostella2013,upton2016,daloisio2017,Mitra2018,Garaldi2019,Basu2024}. 

Following hydrogen recombination, the IGM comprised approximately $24\%$ helium by number, predominantly in the form of $^4\mathrm{He}$, with a much smaller abundance of $^3\mathrm{He}$—about one part in $10^5$ relative to $^4\mathrm{He}$ \citep{kneller2004}. The $^3\mathrm{He}$ isotope is particularly significant due to its nonzero magnetic dipole moment, which enables hyperfine splitting in the hydrogen-like $^3\mathrm{He}^+$ ion, resulting in a rest-frame transition at 8.67 GHz (3.5-cm), analogous to the well-known 21-cm line from neutral hydrogen \citep{field1959,madau1997,shaver1999,tozzi2000,ciardi2003,furnaletto2006,zaroubi2013,acharya2024}. While much of the effort to explore the high-redshift IGM has focused on the 21-cm line, the $^3\mathrm{He}^+$ hyperfine transition offers a unique window into key astrophysical processes in the early Universe \citep{furnaletto2006, bagla2009, McQuinn2009, takeuchi2014, vasiliev2019}. Since the evolution of \textsc{H ii} and He~{\sc ii} during the Epoch of Reionization (EoR) is primarily driven by stellar sources, the resulting signals at 21-cm and 3.5-cm are expected to be anti-correlated. Because of that, as proposed by \citealt{bagla2009}, beyond tracing the high-$z$ evolution of the helium component in the IGM, the 3.5-cm signal could also provide constraints independent from those of the 21-cm signal. Additionally, this line can help to constrain the properties of ionizing sources and serve as a powerful mean for detecting quasars \citep{khullar2020}. It also has a potential to probe large scale filamentary structures as discussed by \citealt{takeuchi2014}.

Although the abundance of He~{\sc ii} is lower than the one of \textsc{H ii}, the $^3\mathrm{He}^+$ hyperfine transition has several advantages over the 21-cm line, as (i) the foreground contamination is smaller at higher rest-frame frequency, and (ii) its spontaneous decay rate is $\sim 680$ times larger, enhancing the signal strength. However, the detectability of this signal is limited by the sensitivity of most current telescopes. While there are several radio telescopes which are operating in this relevant frequency range, the Square Kilometre Array (\texttt{SKA}; specifically in the Phase 2, using \texttt{SKA-mid}) is the most promising one to detect this signal. 

While \citealt{bagla2009} estimated the signal strength using semi-analytic methods, more detailed numerical simulations are necessary to capture the full complexity of He~{\sc ii} evolution, including spatial inhomogeneities, radiative transfer effects, and feedback mechanisms that semi-analytic models may not fully incorporate. \citealt{takeuchi2014} provided a precise estimation of the ionization state of $^3\mathrm{He}^+$ and studied the physical processes that influence its spin temperatures over a wide redshift range ($z \sim 0-8$). However, their work did not include radiative transfer effects, which are crucial for a more accurate representation of how ionizing radiation propagates through the IGM. 
The investigation by \citealt{khullar2020} addressed some of these limitations by using radiative transfer simulations (\citealt{Marius2020}) to explore the possibility of detecting the $^3\mathrm{He}^+$ signal, considering various ionizing sources, including stars, X-ray binaries, accreting black holes, and shock-heated gas in the interstellar medium. Additionally, they also studied a high-$z$ quasar to investigate its environment (see also \citealt{koki2017}). However, a more detailed investigation at lower redshifts, where helium reionization is ongoing, is crucial for assessing the feasibility of detecting the 3.5cm signal over a wider redshift range. Since the impact of quasars on reionization is redshift-dependent, focusing on this later stage provides new insights into how the evolving He~{\sc ii} ionizing radiation field influences the IGM.

Recent observational work has sought to constrain the prospects of helium hyperfine signal detection. \citealt{Trott2024} presented the first limits on the power spectrum  of the $^3\mathrm{He}^+$ hyperfine transition at $z = 3{-}4$ on spatial scales of 30 arcminutes, using 190 hours of archival interferometric data from the Australia Telescope Compact Array (ATCA). Although noise and residual radio frequency interference limited their sensitivity, the study demonstrated the feasibility of detecting the helium signal with current telescopes. It also emphasized the potential of next-generation instruments, such as \texttt{SKA}, which, with their higher sensitivity, could yield cosmologically relevant signals and offer deeper insights into helium reionization.

In this study, I discuss the expected $^3\mathrm{He}^+$ hyperfine transition line  at $z < 5$, estimated from the latest cosmological simulations of helium reionization \citep[][hereafter AB24]{Basu2024}. I also perform a quantitative comparison between the simulated $^3\mathrm{He}^+$ signal and the first observational constraints from \citealt{Trott2024}. The simulation incorporates recent constraints on the quasar luminosity function (QLF) and successfully reproduces a majority of helium reionization observations, including the He~{\sc ii} Lyman-$\alpha$ forest. Specifically, I introduce the simulation methodology in Section \ref{method}, present the results in Section \ref{results}, and conclude with a discussion in Section \ref{conclusion}. Throughout this chapter , I adopt a flat $\Lambda \rm{CDM}$ cosmology, consistent with \cite{planck2016}, using the following parameters: $\Omega_{\rm m} = 0.3089$, $\Omega_{\Lambda} = 0.6911$, $\Omega_{\rm b} = 0.0486$, $h = 0.6774$, $\sigma_{8} = 0.8159$, and $n_{\rm s} = 0.9667$, where the symbols have their usual meanings.

\begin{figure*}
\centering
    \includegraphics[width=160mm]{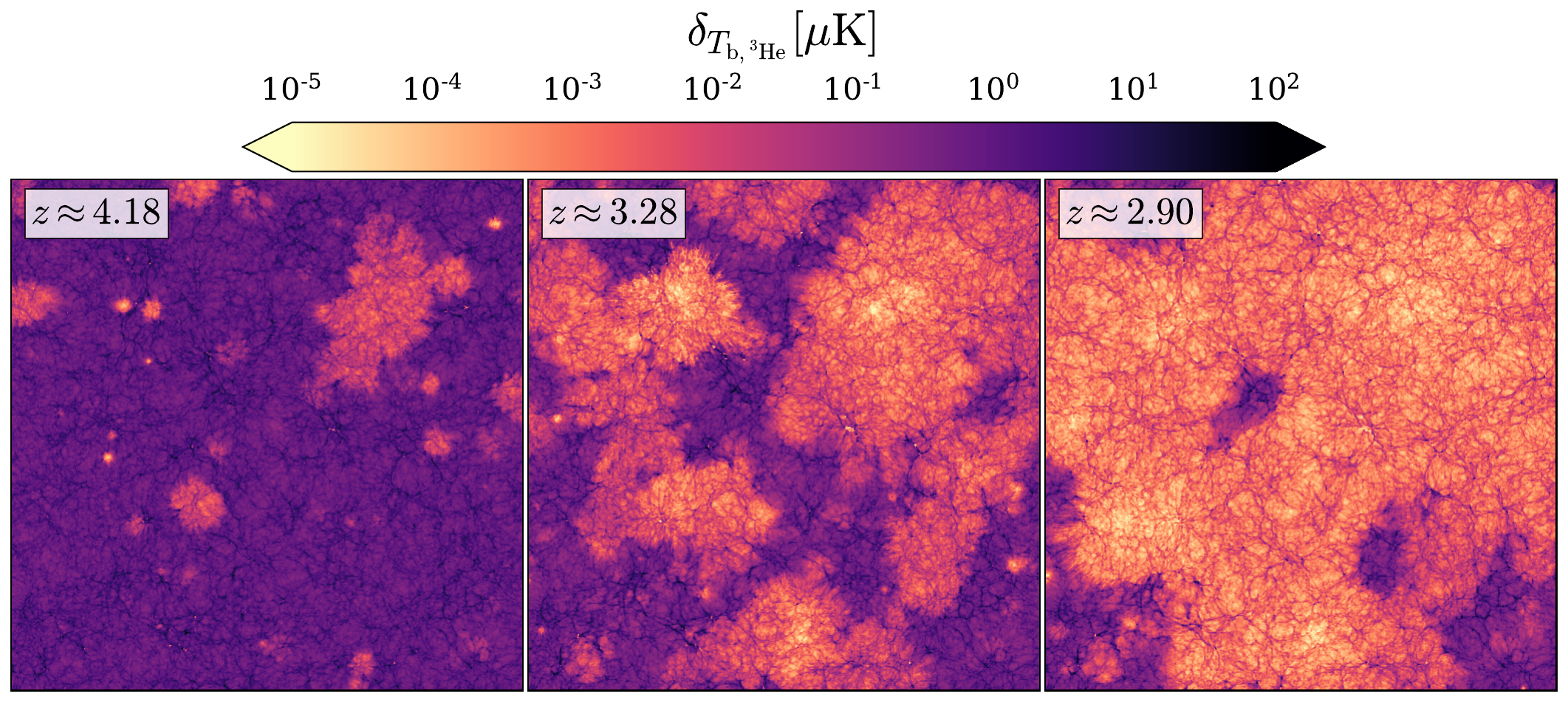}
    \caption{Slice map of $^3\rm{He}^+$ differential brightness temperature across the simulation box at $z\approx4.18$, 3.28 and 2.90, from left to right.  The maps are 205$\rm{\mathit{h}^{-1} cMpc}$ wide and 400$\rm{\mathit{h}^{-1} ckpc}$ thick.}
    \label{deltatb_map}
\end{figure*}

\section{Methods}
\label{method}

\subsection{Simulation of helium reionization}

In the following, I briefly describe the simulation used to estimate the 3.5~cm signal, while I refer the reader to AB24 for more details. 

The radiative transfer (RT) simulation has been performed by post-processing the \texttt{TNG300} hydrodynamical simulation, which is part of the \texttt{Illustris TNG} project \citep{volker2018,naiman2018,marinacci2018,pillepich2018,nelson2018}. This simulation was performed using the \texttt{AREPO} code \citep{springel2010}, which solves the idealized magneto-hydrodynamical equations \citep{pakmor2011} governing the non-gravitational interactions of baryonic matter, as well as the gravitational interactions of all matter. \texttt{TNG300} employs the latest \texttt{TNG} galaxy formation model \citep{weinberger2017,pillepich2018}, with star formation implemented by converting gas cells into star particles above a density threshold of $\textit{n}\rm{_{H}} \sim 0.1 \ \rm{cm^{-3}}$, following the Kennicutt-Schmidt relation \citep{springel2003}. The simulation is run in a comoving box of length $L_\mathrm{box} = 205 \,h^{-1} \, \mathrm{cMpc}$, with (initially) $2 \times 2500^{3}$ gas and dark matter particles. The average gas particle mass is $\bar{m}\rm{_{gas}} = 7.44 \times 10^{6} \, M_{\odot}$, while the dark matter particle mass is fixed at $m\rm{_{DM}} = 3.98 \times 10^{7} \, M_{\odot}$. Haloes are identified on-the-fly using a friends-of-friends algorithm with a linking length of $0.2$ times the mean inter-particle separation.  I utilized 19 outputs covering the redshift range $5.53 \leq z \leq 2.32$.

The radiative transfer of ionizing photons through the IGM has been implemented with the \texttt{CRASH} code \citep[e.g.][]{ciardi2001,maselli2003,maselli2009,Maselli2005,partl2011}, which computes self-consistently the evolution of hydrogen and helium ionization states, as well as gas temperature. \texttt{CRASH} employs a Monte Carlo-based ray-tracing scheme, where ionizing radiation and its spatial and temporal variations are represented by multi-frequency photon packets propagating through the simulation volume. The latest version of \texttt{CRASH} includes UV and soft X-ray photons, accounting for X-ray ionization, heating, detailed secondary electron physics \citep{graziani2013,graziani2018}, and dust absorption \citep{glatzle2019,glatzle2022}. For more details, I refer the reader to the original \texttt{CRASH} papers. The RT is performed on grids of gas density and temperature extracted from \texttt{TNG300} snapshots and tracks radiation with energy $h_{\rm P} \nu \in [54.4 \ \mathrm{eV}, 2 \ \mathrm{keV}]$, assuming fully ionized hydrogen (i.e. $x_{\rm HI} = 10^{-4}$) and fully singly ionized helium (i.e. $x_{\rm HeI} = 0$ and $x_{\rm HeIII} = 10^{-4}$). 

As sources of ionizing radiation I consider quasars, which are assigned to halos in the simulation according to a modified abundance-matching approach (see AB24 for more details) to reproduce the QLF of \citealt{Shen2020}.
The quasar SED is the same adopted in \citealt{Marius2018} and \citealt{Marius2020}, which is obtained by averaging over 108,104 SEDs observed in the range $0.064 < z < 5.46$ \citep{Krawczyk2013} at $h_{\rm P} \nu <200$~eV, while it follows a power law with index $-1$ at higher energies (see Section 2 of \citealt{Marius2018}).  

\subsection{The 3.5cm signal}

The simulation described in the previous section provides the spatial and temporal distributions of the gas number density, $\rm{\mathit{n}_{gas}}$, and temperature, $\rm{\mathit{T}_{gas}}$, as well as of the fractions of H~{\sc ii}, He~{\sc ii}, and He~{\sc iii} (denoted as $\rm{\mathit{x}_{HII}}$, $\rm{\mathit{x}_{HeII}}$, and $\rm{\mathit{x}_{HeIII}}$, respectively). 
The brightness temperature associated with the hyperfine transition of $^{3}\mathrm{He}^{+}$, $\rm{\mathit{\delta T}_{b,^{3}He}}$, is evaluated in each cell of the simulated volume as (see eq.~61 of \citealt{furnaletto2006}):
\begin{align}
    \rm{\mathit{\delta T}_{b,^{3}He}} \approx & \ 0.5106 \ \rm{\mathit{x}_{HeII} \ (1+\delta) \ (1 - \frac{\mathit{T}_{CMB}}{\mathit{T}_{s}})} \nonumber \\
    & \times \ \rm{(\frac{[^{3}He/H]}{10^{-5}}) \ (\frac{\Omega_{b}\mathit{h}^{2}}{0.0223}) \sqrt{\frac{\Omega_{m}}{0.24}} \ (1+\mathit{z})^{1/2} \ \mu K}, 
    \label{deltatb_formula}
\end{align}
where $\delta = (\mathit{n}_{\mathrm{gas}} - \bar{\mathit{n}}_{\mathrm{gas}})/\bar{\mathit{n}}_{\mathrm{gas}}$ is the gas overdensity, with $\bar{\mathit{n}}_{\mathrm{gas}}$  mean gas number density within the simulation box. $\mathit{T}_{\mathrm{CMB}} = 2.725(1+\mathit{z}) \, \mathrm{K}$ denotes the CMB temperature at redshift $z$, $\rm{\mathit{T}_{s}}$ is the spin temperature, and the relative abundance of $^{3}\mathrm{He}$ to hydrogen, $[\mathrm{^{3}He/H}] \approx 10^{-5}$, is determined by Big Bang nucleosynthesis. 
I adopt the common assumptions $\frac{\mathit{H}(z)}{(1+\mathit{z})(dv/dr)} \sim 1$ and $\rm{\mathit{T}_{s} \sim \mathit{T}_{\mathrm{gas}}}$.

The power spectrum (PS) of the differential brightness temperature is defined as:
\begin{align}
    \label{ps-equ}
    \rm{\mathit{P}(k) = \langle \ \mathit{\delta T}_{b,^{3}\rm{He}} (k) \mathit{\delta T}_{b,^{3}\rm{He}}(k)^{*} \rangle},
\end{align}
where $\rm{\mathit{\delta T}_{b,^{3}\rm{He}} (k)}$ is the Fourier transform of the differential brightness temperature field, and $\rm{\mathit{\delta T}_{b,^{3}\rm{He}} (k)^{*}}$ is its complex conjugate. 
In the following, I will express the results in terms of the dimensionless power spectrum, defined as:
\begin{align}
    \rm{\Delta_{3.5cm}^{2} = \frac{\mathit{k}^{3}}{2 \pi^{2}} \mathit{P}(k)}.
\end{align}

\begin{figure}
\centering
    \includegraphics[width=100mm]{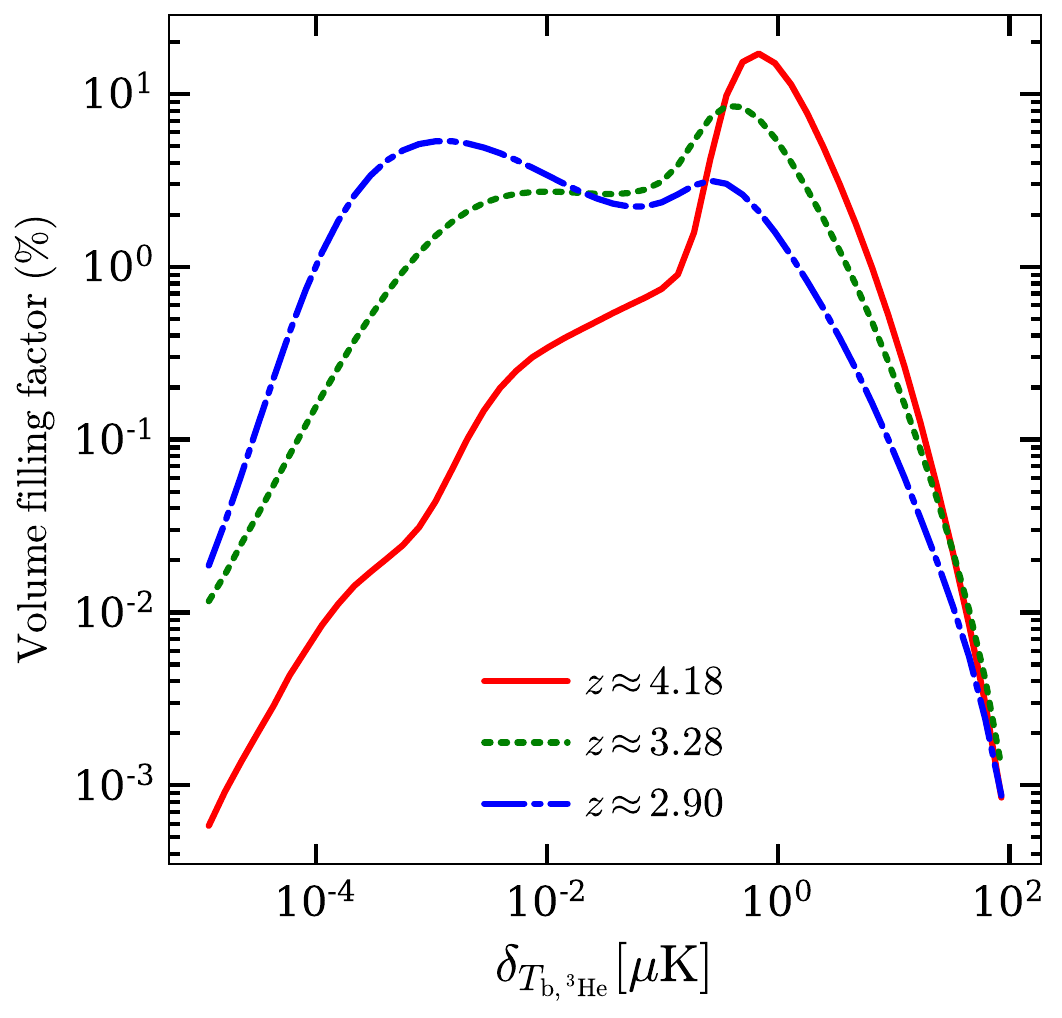}
    \caption{Volume filling factor of differential brightness temperature of the hyperfine transition of $^{3}\rm{He}^{+}$ at $z\approx4.18$ (solid red curve), 3.28 (dotted green) and 2.90 (dash-dotted blue).}
    \label{deltatb_dist}
\end{figure}

\section{Results}
\label{results}

In this section, I discuss the features and potential observables associated to the 3.5cm transition line extracted from the helium reionization simulations introduced in AB24.

\subsection{Overview}

In Figure \ref{deltatb_map} I present maps of $\delta T_{\mathrm{b},^{3}\mathrm{He}}$ for a slice of the simulation box at   $z \approx 4.18$, $3.28$, and $2.90$, corresponding to the redshifts
of the observations by \citealt{Trott2024}. At $z \approx 4.18$, most of the helium in the simulation volume is in the form of He\,\textsc{ii}, resulting in  values of $\delta T_{\mathrm{b},^{3}\mathrm{He}}$ typically exceeding $10\,\mu\mathrm{K}$. However, in regions that have been fully ionized to He\,\textsc{iii} by the ionizing radiation emitted by quasars, $\delta T_{\mathrm{b},^{3}\mathrm{He}}$ drops to $\approx 10^{-4}\,\mu\mathrm{K}$. At $z \approx 3.28$ more gas has low values of $\delta T_{\mathrm{b},^{3}\mathrm{He}}$, reflecting the expansion of He\,\textsc{iii} regions throughout the volume. By $z \approx 2.90$, nearly the entire slice of the simulation box exhibits low $\delta T_{\mathrm{b},^{3}\mathrm{He}}$ values, marking the near-completion of helium reionization. 

For a more quantitative analysis, Figure~\ref{deltatb_dist} shows the volume filling factor of $\delta T_{\mathrm{b},^{3}\mathrm{He}}$ at the same redshifts as the slice maps in Figure~\ref{deltatb_map}. At all redshifts, the volume filling factor shows a double-peaked structure. The peak at higher $\delta T_{\mathrm{b},^{3}\mathrm{He}}$ values corresponds to regions where helium is still singly ionized, while the lower peak is linked to areas where helium has been fully ionized to He\,\textsc{iii}. At $z \approx 4.18$, the fraction is dominated by a strong peak around $1\,\mu\mathrm{K}$, which is over two orders of magnitude higher than the secondary peak at $\delta T_{\mathrm{b},^{3}\mathrm{He}} \sim 10^{-3}\,\mu\mathrm{K}$. At this redshift, there is a third peak towards an even lower value, at $\delta T_{\mathrm{b},^{3}\mathrm{He}} \sim 10^{-4}\,\mu\mathrm{K}$, but with a very low volume filling factor. By $z \approx 3.28$, the volume filling factor at $\delta T_{\mathrm{b},^{3}\mathrm{He}} \lesssim 0.1\,\mu\mathrm{K}$ increases significantly due to ongoing helium reionization, while the contribution from higher emission regions declines. At $z \approx 2.90$, the peak prominence shifts entirely to lower $\delta T_{\mathrm{b},^{3}\mathrm{He}}$ values, with the dominant peak centered around $4 \times 10^{-4}\,\mu\mathrm{K}$, marking the completion of helium reionization.

\begin{figure}
\centering
    \includegraphics[width=100mm]{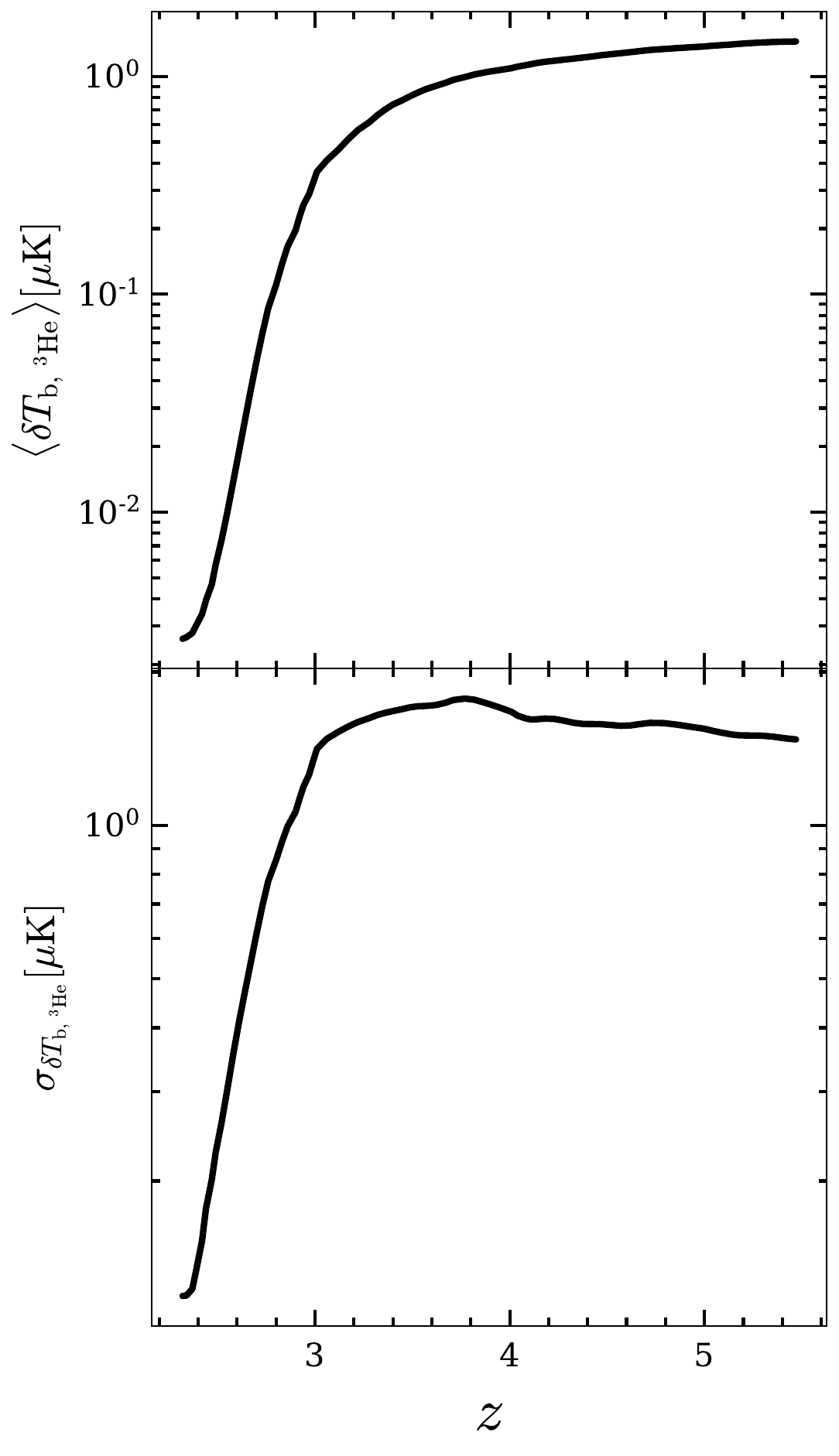}
    \caption{Redshift evolution of the volume averaged differential brightness temperature (\textit{top panel}) and standard deviation (\textit{bottom}) of the hyperfine transition of $^{3}\rm{He}^{+}$.}
    \label{3he_evolution}
\end{figure}

In the top panel of Figure \ref{3he_evolution} I show the volume-averaged differential brightness temperature, ie. $\langle \delta T_{\rm b,\, ^3He} \rangle$. Throughout the entire redshift range $\langle \delta T_{\rm b,\, ^3He} \rangle$ remains positive because $T{\rm _S}$ is always higher than the CMB temperature (see Equation \ref{deltatb_formula}). 
The redshift evolution of $\langle \delta T_{\rm b,\, ^3He} \rangle$ closely follows the one of the He~{\sc ii} fraction (see Figure 3 in AB24), with a nearly constant value until $z\approx4$, when only $\sim$15\% of He~{\sc ii} has been ionized. At lower redshift the average signal starts to decline rapidly, reaching $\approx 2.5 \times 10^{-2}$ by $z\approx 2.3$. This drop is linked to the completion of helium reionization.  

To track how fluctuations in the global 3.5cm signal evolve over time, in the bottom panel of the Figure I examine the evolution of the standard deviation of $\delta T_{\rm b,^{3}\rm{He}}$, i.e. $\sigma_{\delta T_{\rm b,\, ^3He}}$. Initially, the fluctuations increase slightly from 1.4 $\mu$K at $z\approx 5.5$ to  1.7 $\mu$K by $z\approx 3$. By this time, about 80\% of He~{\sc ii} in the simulation volume has been fully ionized into He~{\sc iii}. After $z\approx 3$, the fluctuations start to decrease because the reionization process has smoothed out most variations in the abundance of He~{\sc ii} in the IGM. This transition occurs at a redshift when the temperature of the IGM at mean density begins to saturate, as observed in AB24. Once reionization is complete, the remaining fluctuations are induced by fluctuations in the gas density.

In the following sections, I will focus on potentially observable quantities related to the 3.5cm transition line and their trends with redshifts.

\subsection{Power spectra of 3.5cm signal}

\begin{figure}
\centering
    \includegraphics[width=90mm]{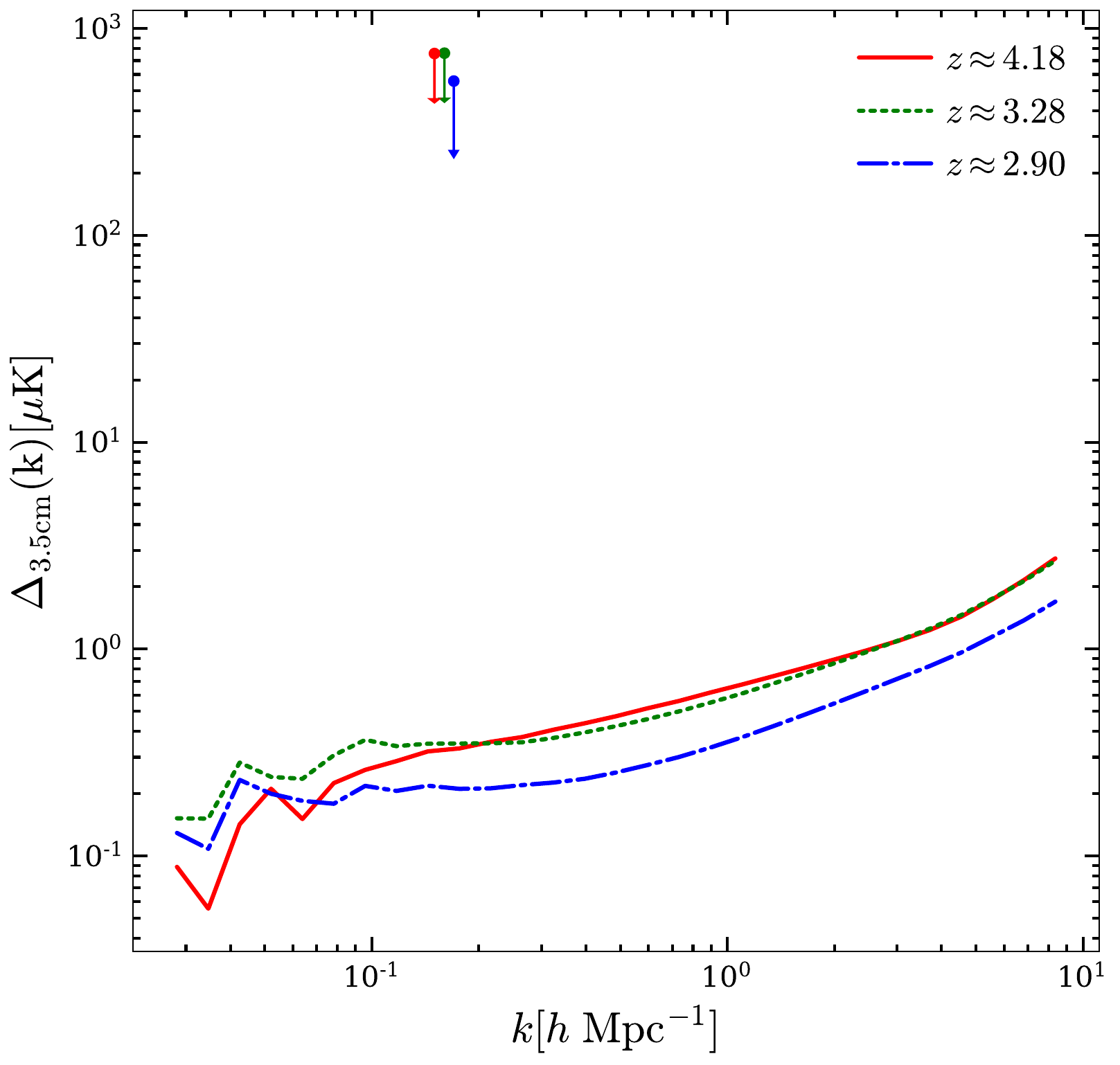}
    \caption{Power spectra of the 3.5cm signal at $z\approx4.18$ (solid red curve), 3.28 (dotted green) and 2.90 (dash-dotted blue). Observational upper limits from \citealt{Trott2024} at the same redshifts are shown as circles.}
    \label{ps}
\end{figure}

\begin{figure}
\centering
    \includegraphics[width=100mm]{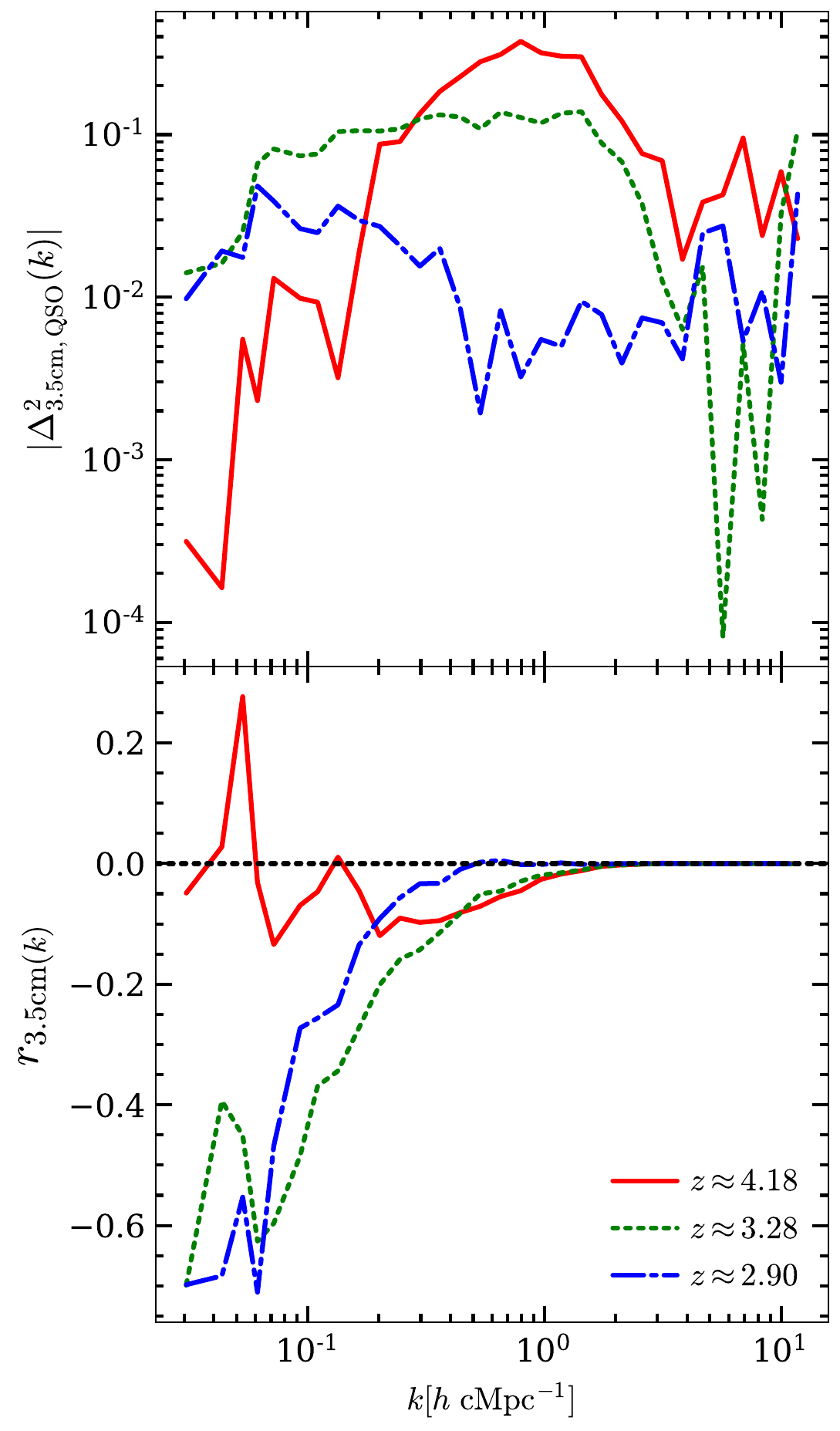}
    \caption{\textit{Top panel:} Simulated 3.5cm - QSO dimensionless cross-power spectra at $z\approx4.18$ (solid red curve), 3.28 (dotted green) and 2.90 (dash-dotted blue). \textit{Bottom:}  Cross-correlation coefficient between these two fields at the same redshifts. The black dotted curve denotes no correlation  as a reference.}
    \label{he3_qso_crossPS}
\end{figure}


Similarly to the 21cm line, the power spectrum of the signal is expected to be the first detectable quantity. Thus, in Figure \ref{ps} I present the dimensionless power spectra as $\rm{\Delta_{3.5cm}}(\mathit{k})$. All curves increase towards higher $k$ values, indicating stronger fluctuations at smaller spatial scales. At $z\gtrsim 5$, $\rm{\Delta_{3.5cm}}(\mathit{k})$ is mainly dominated by the over-density of gas matter and the inhomogeneous temperatures from the hydrodynamic simulations.
As reionization proceeds, the impact of quasars emerges in a scale dependent way. 
From $z \approx 4.18$ to $3.28$, the amplitude of the fluctuations at $\rm{k \gtrsim 0.2 \, \mathit{h} \, Mpc^{-1}}$ (corresponding to spatial scales of $\leq 50 \, \rm{\mathit{h}^{-1} \, Mpc}$) remains largely unchanged, while it drops by $z\approx2.90$. Essentially, the quasars radiation increases the gas temperature in their vicinity, reducing the fluctuations of $\rm{\mathit{\delta T}_{b,^{3}\rm{He}}}$, as well as the amplitude of  $\rm{\Delta_{3.5cm}}(\mathit{k})$ slightly. On the larger spatial scales, the effect of the inhomogeneous heating becomes more pronounced (see also, \citealt{pritchard2007}), increasing the amplitude of $\rm{\Delta_{3.5cm}}(\mathit{k})$ from $z \approx 4.18$ to $3.28$ (when, $\rm{\mathit{x}_{HeII}}$ drops from $85\%$ to less than $50\%$, see also Figure 3 in AB24). By $z \approx 2.90$, $\rm{\Delta_{3.5cm}}(\mathit{k})$ decreases further since the majority of the simulation volume is fully ionized, and fluctuations are reduced at all scales.

In the Figure I also include the first observational upper limits on the 3.5cm brightness temperature fluctuation from \citealt{Trott2024}, which remain consistently about 3-4 orders of magnitude higher than my theoretical predictions. This is primarily attributed to the adopted simple model of foreground emission in \citealt{Trott2024} and to the sensitivity limitations of the telescope.

\subsection{Cross-correlation between the 3.5cm signal and quasars}

As initially discussed by \citealt{lidz2009}, cross-correlation between galaxies and 21cm emission promises to be an excellent probe of the EoR, as it is sensitive to the size and filling factor of H~{\sc ii} regions, and it would alleviate the effect of observational systematics. 
This topic has been extensively investigated in recent years \citep{wiersma2013,vrbanec2016,hutter2018,vrbanec2020,hutter2023,laplante2023,hartman2025}. 
A similar approach could be applied to the 3.5cm signal, as it could provide additional information on the timing and morphology of helium reionization. 
To explore this, I compute the cross-power spectrum between the differential brightness temperature of the \(^3\mathrm{He}^+\) signal and the spatial distribution of quasars in my simulations. The quasar density field can be characterized by the density contrast \(\delta_{\text{QSO}}(\mathbf{x}) = (n_{\text{QSO}}(\mathbf{x})/{\langle n_{\text{QSO}} \rangle}) - 1\),  where \(n_{\text{QSO}}(\mathbf{x})\) is the local number density of quasars, and \(\langle n_{\text{QSO}} \rangle\) is its spatial average. Following the methodology of \citealt{lidz2009}, I calculate the cross-power spectrum as a function of wave number \(k\). The top panel of Figure~\ref{he3_qso_crossPS} shows the resulting cross-power spectrum, while the bottom panel presents the corresponding cross-correlation coefficient, defined as \(r_{\rm 3.5cm,\text{QSO}}(k) = P_{\rm 3.5cm,\text{QSO}}(k)/{\sqrt{P_{\rm 3.5cm}(k) P_{\text{QSO}}(k)}}\), where \(P_{\rm 3.5cm,\text{QSO}}(k)\) is the cross-power spectrum, and \(P_{\rm 3.5cm}(k)\) and \(P_{\text{QSO}}(k)\) are the auto-power spectra of the \(^3\mathrm{He}^+\) signal and quasar density field, respectively. 

At $z \approx 4.18$, the cross-power spectrum peaks at $k \sim 0.8 \, \mathrm{h\, cMpc^{-1}}$, while the cross-correlation coefficient indicates a positive correlation between quasars and the 3.5cm signal on the largest spatial scales ($k \lesssim 0.1 \, \mathrm{h\, cMpc^{-1}}$). At this stage, quasars are still rare and have only just begun ionizing their local environments. Since they preferentially reside in overdense regions, which contain more matter and thus more He\,\textsc{ii}, these regions exhibit stronger 3.5cm emission prior to being ionized. The resulting positive correlation on large scales does not arise from direct quasar emission, but rather from the fact that both the quasars and the 3.5cm signal trace the same large-scale overdense regions in the IGM that are rich in He\,\textsc{ii}.  

As ionized bubbles begin to grow around these quasars, the 3.5cm emission in their immediate surroundings starts to decrease. This lowers the amplitude of the cross-power spectrum at intermediate and small scales (larger $k$), even though the cross-correlation coefficient at these scales gradually transitions toward negative values. Physically, the anti-correlation strengthens because regions hosting quasars now correspond to He\,\textsc{iii} bubbles, where the 3.5cm signal is suppressed, while the emission is maintained in the more distant, mostly underdense regions that lack quasars. By $z \approx 3.28$, the cross-power spectrum amplitude has dropped, and its peak has shifted toward larger spatial scales, reflecting the typical size of He\,\textsc{iii} bubbles. At this stage, the cross-correlation coefficient is negative up to $k \sim 1 \, \mathrm{h\, cMpc^{-1}}$, consistent with the bubble scale. By $z \approx 2.9$, when most of the volume is ionized, the cross-power spectrum continues to decline at all scales, and the anti-correlation shifts to lower $k$. On the smallest scales (high $k$), the cross-correlation coefficient approaches zero, as the remaining 3.5cm emission becomes uncorrelated with the sparse quasar distribution. Throughout this evolution, the curves remain noisier than standard 21cm--galaxy cross-power spectra due to the much lower quasar number density. Nevertheless, these trends demonstrate that the 3.5cm--quasar cross-power spectrum provides a sensitive probe of He\,\textsc{iii} bubble growth and the morphology of the helium reionization process.

\begin{figure}
\centering
    \includegraphics[width=140mm]{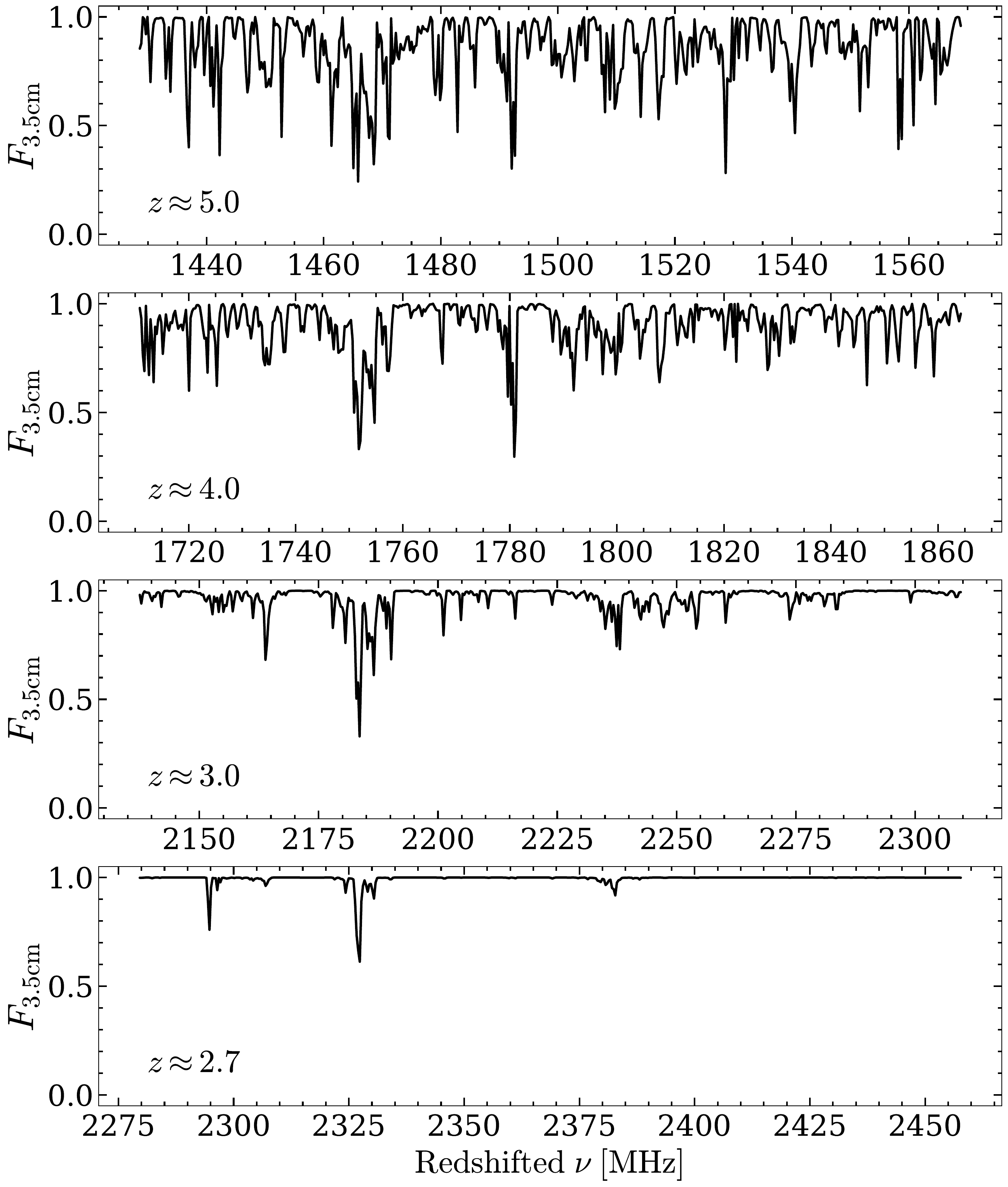}
    \caption{Line of sight 3.5cm transmission flux extracted from the fiducial simulation in AB24 at $z\approx 5.0$, 4.0, 3.0 and 2.7 (from top to bottom).}
    \label{3_5cm_absorption}
\end{figure}

\subsection{The 3.5cm forest}
\label{3_5cm_forest}
Analogously to the 21cm forest (as explored e.g. in \citealt{carilli2002,furlanetto2006,xu2011,ciardi2013,ciardi2015,Soltinsky2023,soltinsky2025}), here I investigate the absorption feature of He~{\sc ii} against a background source. It can essentially be a unique probe of the heating process and the thermal history of the late stage of helium reionization, as the signal is strongly dependent on the IGM temperature and ionization state. The photons emitted by a radio-loud source at redshift \( z_{\rm s} \) with frequencies \( \nu > \nu_{3.5\,\mathrm{cm}} \) will be removed from the source spectrum with a probability \( 1 - e^{-\tau_{3.5\,\mathrm{cm}}} \), due to absorption by \( ^3\mathrm{He}^+ \) along the line of sight (LOS) at redshift \( z = \frac{\nu_{3.5\,\mathrm{cm}}}{\nu(1 + z_{\rm s})} - 1 \). The optical depth \( \tau_{3.5\,\mathrm{cm}} \) for this hyperfine transition can be written as (following \citealt{furlanetto2006}): 
\[
\tau_{3.5\mathrm{cm}}(z) = \frac{3 h_p c^3 A_{3.5\,\mathrm{cm}}}{32\pi k_B \nu_{3.5\,\mathrm{cm}}^2}  \frac{x_{\mathrm{HeII}} n_{\mathrm{He}}}{T_s (1 + z) \left( \frac{dv_\parallel}{dr_\parallel} \right)},
\]
where \( \nu_{3.5\,\mathrm{cm}} \approx 8.67\,\mathrm{GHz} \) is the rest-frame frequency of the \( ^3\mathrm{He}^+ \) hyperfine transition, \( A_{3.5\,\mathrm{cm}} \approx 1.95 \times 10^{-12}\,\mathrm{s}^{-1} \) is the Einstein coefficient for spontaneous emission, \( x_{\mathrm{HeII}} \) is the fraction of singly ionized helium, \( n_{\mathrm{He}} \) is the total helium number density, \( T_{\rm s} \) is the spin temperature, and \( dv_\parallel / dr_\parallel \) is the proper velocity gradient along the LOS (including Hubble flow and peculiar velocities). The other constants and variables have their standard physical meanings. The optical depth for the \( ^3\mathrm{He}^+ \) transition may then be calculated in discrete form for pixel \( i \) as:
\[
\tau_{\rm 3.5cm},i = \frac{3 h_{\rm p} c^3 A_{\rm 3.5cm}}{32 \pi^{3/2} \nu_{\rm 3.5cm}^2 k_{\rm B}} \frac{\delta v}{H(z)} \sum_{j=1}^N \frac{n_{\mathrm{HeII},j}}{b_j T_{{\rm s},j}} \exp\left[ -\frac{(v_{H,i} - u_j)^2}{b_j^2} \right],
\]
with Hubble velocity \( v_{H,i} \) and velocity width \( \delta v \). Here, \( b_j = \sqrt{2k_{\rm B} T_{{\rm K},j}/m_{\mathrm{He}}} \) is the Doppler parameter determined by the kinetic temperature \( T_{\rm K} \) of the gas. The term \( u_{\rm j} = v_{\rm H,j} + v_{\mathrm{\rm pec},j} \) includes both the Hubble velocity and the peculiar velocity of the gas. 
In Figure \ref{3_5cm_absorption}, I show the redshift evolution of the transmission flux, $F_{\rm 3.5cm}={\rm exp}[-\tau_{\rm 3.5cm}]$. For
the sake of clarity, here (and for all the following plots) I omit the $i$ when referring to values of the physical quantities associated to a single pixel. As anticipated, the strong absorption features at $z\approx 5.0$ are slowly disappearing towards lower redshift as more helium is getting doubly ionized. Despite this, even at $z\approx2.7$, when more than $99\%$ of He~{\sc ii} is fully ionized, few strong absorption features tracing the higher density singly ionized gas are visible. 

Preliminary modeling based on simulated mock spectra including instrumental effects and noise suggests that detecting the 3.5cm forest with instruments like \texttt{SKA-mid} is promising. This offers a new opportunity to probe the ionization state and small-scale structure of the IGM during and after He~\textsc{ii} reionization. I plan to investigate this observational prospect in a follow up work.

\section{Discussion and Conclusions}
\label{conclusion}

In this study, I investigated the $^{3}\mathrm{He}^{+}$ hyperfine transition line at 8.67 GHz (3.5cm) as a novel probe of helium reionization in the IGM, focusing particularly on the redshift range $z < 5$. Using the most recent simulations of helium reionization \citep{Basu2024} run by post processing outputs from the hydrodynamic simulation \texttt{TNG300} \citep{volker2018,naiman2018,marinacci2018,pillepich2018,nelson2018} with the radiative transfer code \texttt{CRASH} \citep{ciardi2001,maselli2003,maselli2009,Maselli2005,partl2011}, I examined the spatial and temporal evolution of the 3.5cm signal and its connection to the ionization state of helium driven by quasars. I also discussed relevant probes related to this signal to constrain the helium reionization. My main findings are as follows:

\begin{itemize}
    \item The spatial distribution of the differential brightness temperature, $\delta T_{\rm b,\, ^3He}$, shows a clear transition from He\,\textsc{ii}-dominated gas at $z \approx 4$ to a He\,\textsc{iii}-dominated one at $z \approx 2.90$. The signal intensity drops from values $\gtrsim 1\,\mu$K to $\lesssim 0.1\,\mu$K as reionization progresses.

    \item  The volume-averaged $\delta T_{\rm b,\, ^3He}$ remains relatively constant until $z \sim 4$, after which it declines sharply, mirroring the evolution of the He\,\textsc{ii} fraction. The standard deviation increases slightly during the early stages of reionization and then decreases as the IGM becomes more (uniformly) ionized.

    \item The 3.5cm power spectrum, \( \Delta_{3.5cm}(k) \), increases with \( k \), indicating stronger small-scale fluctuations. Initially, these are driven by gas density and temperature variations. The simulated power spectra are 3–4 orders of magnitude below the current observational upper limits set by \citealt{Trott2024}. This highlights the challenge of detecting the signal with current instruments.

    \item The 3.5cm-quasar cross-power spectrum evolves from a large-scale positive correlation at $z \approx 4.18$, peaking at $k \sim 0.8\,h\,\mathrm{cMpc^{-1}}$, to a strong anti-correlation by $z \approx 3.28$ as He\,\textsc{iii} bubbles grow and suppress 3.5cm emission around quasars. The peak of the spectrum shifts to larger spatial scales, reflecting typical bubble sizes, and by $z \approx 2.9$ the signal weakens further, with the correlation approaching zero on small scales.

    \item I find that the 3.5cm forest exhibits strong absorption features at $z \approx 5$, which gradually weaken toward $z \approx 2.7$ as helium becomes doubly ionized, with detectable transmission fluctuations persisting even when over $99\%$ of He\,\textsc{ii} is ionized.

\end{itemize}

Overall this study highlights the utility of the $^{3}\mathrm{He}^{+}$ hyperfine transition line as a probe to explore the latest stages of helium reionization driven by quasars. Although current observational limits lie well above the theoretical predictions, upcoming improvements in radio instrumentation are expected to narrow this gap, similar to the progress seen in 21cm studies.
To obtain more reliable insights, future progress will rely on improved instrumental sensitivity to achieve the high signal-to-noise ratios required for detecting the $^{3}\mathrm{He}^{+}$ signal. This work will help exploring instrumental noise and foreground modeling in simulations to improve our understanding of observational constraints. The enhanced sensitivity of the \texttt{SKA}, particularly \texttt{SKA-mid}, will be pivotal in improving the detectability of the $^3\mathrm{He}^+$ hyperfine signal and advancing my understanding of the final stages of helium reionization.

  \chapter{Summary and Outlook}

\begin{flushright}
\begin{minipage}{0.5\textwidth}
\raggedleft
\textbf{\textit{``The world was born not with the Big Bang, but when man began asking questions."}}\\[1ex]
\noindent\rule{0.5\textwidth}{0.4pt}\\[-0.5ex]
Mimasha Pandit
\end{minipage}
\end{flushright}

\begin{flushright}
\begin{minipage}{0.7\textwidth}
\raggedleft
\textbf{\textit{``Two roads diverged in a wood, and I took the one less traveled by, and that has made all the difference."}}\\[1ex]
\noindent\rule{0.5\textwidth}{0.4pt}\\[-0.5ex]
Soumik Goswami
\end{minipage}
\end{flushright}

\section{Summary of the Thesis}

This thesis examines fundamental astrophysical processes that govern the evolution of the Universe's ionization and temperature. It presents four interconnected studies that investigate how various models of ionizing sources affect observable phenomena and influence the properties of early galaxies. By integrating both theoretical models and observational data, this work explores the variability of the UVLF of early galaxies observed by \texttt{JWST}, the influence of SED models on the Lyman-$\alpha$ forest, the role of quasars in helium reionization, and the potential of the $\mathrm{^3He^+}$ transition line as a cosmic probe. These studies offer vital insights into the process of cosmic reionization, the formation of the first galaxies, and provide strategies for future observational efforts. Here I summarize the key findings of this thesis.

\subsubsection{Impact of source characteristics on high-$z$ galactic observables}

In the study, described in Chapter \ref{chap:chapter2}, I investigate the variability of the UVLF at $z > 5$ using the \texttt{SPICE} suite of cosmological, radiation-hydrodynamic simulations, which include three distinct SN feedback models: \texttt{bursty-sn}, \texttt{smooth-sn}, and \texttt{hyper-sn}. 
The \texttt{bursty-sn} model, driven by intense and episodic SN explosions, produces the highest fluctuations in the SFR. Conversely, the \texttt{smooth-sn} model, characterized by gentler SN feedback, results in minimal SFR variability. The \texttt{hyper-sn} model, featuring a more realistic prescription that incorporates HN explosions, exhibits intermediate variability, closely aligning with the \texttt{smooth-sn} trend at lower redshifts.
These fluctuations in SFR significantly affect the $\rm{\textit{M}_{UV} - \textit{M}_{halo}}$ relation, a proxy for UVLF variability. Among the models, \texttt{bursty-sn} produces the highest UVLF variability, with a maximum value of 2.5. In contrast, the \texttt{smooth-sn} and \texttt{hyper-sn} models show substantially lower variability, with maximum values of 1.3 and 1.5, respectively.
However, in all cases, UVLF variability strongly correlates with host halo mass, with lower-mass halos showing greater variability due to more effective SN feedback in their shallower gravitational wells. The \texttt{bursty-sn} model, though, results in higher amplitudes.
Variability decreases in lower mass haloes with decreasing redshift for all feedback models.
In this study underscores I explore the critical role of SN feedback in shaping the UVLF, and highlights the mass and redshift dependence of its variability, suggesting that UVLF variability may alleviate the bright galaxy tension observed by \texttt{JWST} at high redshifts.

\subsubsection{Impact of SED modelling on IGM properties}

In the study described in Chapter \ref{chap:chapter3}, I investigate the influence of the SEDs of ionizing sources on the thermal and ionization state of the IGM and the resulting Lyman-$\alpha$ forest statistics during the Epoch of Reionization. Cosmological hydrodynamical simulations from the \textsc{Sherwood} suite are post-processed using the multi-frequency radiative transfer code \textsc{crash}, exploring six SED models that span single and binary stellar populations, XRBs, ISM bremsstrahlung, blackbody sources, and power-law AGN-like spectra. Unlike conventional approaches that vary the emissivity, this analysis fixes the total hydrogen ionizing emissivity to isolate the impact of SED shape. Results show that harder SEDs, such as AGN-PL, significantly enhance IGM photoheating, produce extended He\,\textsc{iii} regions, and lower the effective Lyman-$\alpha$ optical depth due to broader thermal Doppler broadening. In contrast, models with softer spectra (e.g. SS-only) yield cooler IGM temperatures and more compact ionized regions. The analysis includes comparisons of H\,\textsc{i} photoionization rates, reionization and thermal histories, and multiple Lyman-$\alpha$ forest observables: effective optical depth, transmission flux probability distributions, 1D flux power spectra, and the spatial dependence of Lyman-$\alpha$ transmissivity around galaxies. While all models align with current observational constraints at $z \lesssim 6$, statistically significant variations emerge, particularly in high-transmission tails and small-scale flux fluctuations. The impact of SED variations is also evident in the enhanced Lyman-$\alpha$ transmissivity near galaxies, with AGN-PL and SOFT-PL models showing the strongest proximity effects. These findings emphasize that assumptions about ionizing SEDs introduce non-negligible biases in the interpretation of Lyman-$\alpha$ forest data and highlight the potential of next-generation surveys (e.g. \texttt{ELT}, \texttt{DESI}) to constrain the nature of reionization-era sources through high-precision Lyman-$\alpha$ forest measurements.

\subsubsection{Impact of QSOs on Helium Reionization}

In the study described in Chapter \ref{chap:chapter4}, I examine how quasars influence Helium reionization. Recent models of the QLF based on large observational compilations exhibit a smooth evolution with time, unlike earlier versions. This development eliminates the need to assume an ionizing emissivity evolution when simulating helium reionization with observations-based QLFs, thereby providing more robust constraints. One such QLF is combined with a cosmological hydrodynamical simulation and 3D multi-frequency radiative transfer. The resulting reionization history is consistently delayed compared to most models available in the literature. The predicted intergalactic medium temperature exceeds observational estimates at $z \lesssim 3$. Forward modeling of the He~{\sc ii} Lyman-$\alpha$ forest shows that the model produces an extended helium reionization and reproduces the bulk of the observed effective optical depth distribution. However, it over-ionizes the Universe at $z \lesssim 2.8$, a consequence of small-scale Lyman Limit Systems not being resolved. Transmission regions and dark gaps in He~{\sc ii} Lyman-$\alpha$ forest sightlines are characterized in detail. Their sensitivity to the timing and progression of helium reionization is quantified, providing a new avenue for future observational investigations of this epoch. Finally, the implications of the large population of active galactic nuclei detected at $z \gtrsim 5$ by \texttt{JWST} are explored. Results indicate that such a population has negligible impact on observables at $z \leq 4$, except in the most extreme scenarios. This suggests that the observed abundance of high-redshift AGNs does not significantly influence helium reionization.

\subsubsection{Probing Helium Reionization with 3.5 cm hyperfine line}

In the study described in Chapter \ref{chap:chapter5}, I explore the helium reionization through the 8.67 GHz (3.5 cm) hyperfine transition of singly ionized helium-3 (\(^3\text{He}^+\)), using advanced cosmological simulations post-processed with the radiative transfer code CRASH. Focusing on redshifts \( z < 5 \), a regime where quasars dominate the ionizing background, the evolution of the \(^3\text{He}^+\) signal is traced as helium transitions from He~{\sc II} to He~{\sc III}. The differential brightness temperature \( \delta T_b,3\text{He} \) exhibits a significant decline, from values greater than 1 $\mu$K at \( z \approx 4.2 \) to less than 0.1 $\mu$K by \( z \approx 2.9 \), reflecting the progression of reionization. This is also observed in the volume-averaged signal, which remains relatively constant until \( z \sim 4 \), after which it drops rapidly as the He~{\sc II} fraction decreases. The standard deviation peaks at approximately 1.7 $\mu$K around \( z \approx 3 \) before falling off at lower redshifts. The power spectrum of brightness temperature fluctuations, \( \Delta ³\text{He}^2(k) \), reveals an increase in small-scale power (for \( k \gtrsim 0.2 \, h \, \text{Mpc}^{-1} \)) driven by quasar-induced heating, which diminishes as reionization completes. These theoretical predictions remain 3–4 orders of magnitude below current observational upper limits, underscoring the challenge of detection with present facilities. Cross-correlating the 3.5 cm signal with quasar positions shows a shift from weak positive to strong negative correlation (with \( r \lesssim -0.4 \) at intermediate scales), tracing the spatial imprint of quasar-driven ionization fronts. The results demonstrate that the \(^3\text{He}^+\) hyperfine line is a sensitive and complementary tracer to the 21-cm signal, offering valuable insights into the late stages of reionization, particularly with future instruments such as the \texttt{SKA}.

\section{Outlook}

This thesis opens up several exciting directions for future research. One important area to explore, building on the work in Chapter \ref{chap:chapter2}, is how variations in the UVLF might affect galaxy morphology. Bursty star formation could explain many features we see in galaxies at high redshifts. Recently, researchers have become more interested in understanding how galactic discs form and when galaxies switch from being dispersion dominated to rotationally supported. By studying how UV fluctuations, caused by stellar feedback, affect the ratio of velocity dispersion to rotation (V/$\sigma$), we could learn more about how early stellar feedback influences galaxy formation. To gain a clearer picture, it will be important to separate out the different factors involved, such as the UV background, gas accretion, and turbulence in the ISM caused by supernovae, as each of these plays a role in shaping galaxies at high redshifts.

In addition, based on the findings in Chapter \ref{chap:chapter3}, it would be valuable to further investigate how SED modeling, especially the effects of X-ray photons, affects galactic properties, particularly through Lyman-$\alpha$ emission. Harder X-ray photons can add extra heating to the gas, which could change the Lyman-$\alpha$ transmission profile, making this a key area for further study. Understanding how this heating affects the way light passes through the gas could help us interpret the state of the gas and how it interacts with radiation. Another important area to look at is how metal lines might affect the Lyman-$\alpha$ forest in quasar spectra. This contamination could interfere with our understanding of the IGM and its evolution, so studying it could improve our analysis of early Universe gas.

The work in Chapters \ref{chap:chapter4} and \ref{chap:chapter5} provides useful information about the properties of the IGM at lower redshifts. These insights could help us better understand the signatures of helium reionization, a key event in cosmic history, through different types of observations. It would be interesting to look at temperature fluctuations in the IGM during helium reionization to help us figure out the timing and details of this process. In addition, refining models of high-redshift quasars, including the study of `little red dots', could give us a better understanding of how quasars influenced the timing of helium reionization. Finally, future observational studies of \(^3\text{He}^+\) could also play a crucial role in advancing our understanding of helium reionization. Using advanced techniques, such as exploring the 3.5 cm absorption features on the spectra from high redshift objects, bright in radio signal (i.e. radio-loud quasars or afterglow from gamma-ray bursts), could allow us to examine the distribution and properties of He~{\sc ii} tracing the spatial inhomogeneity of the low redshift Universe.  These studies can also provide key insights into how quasar radiation influences local proximity zones and its evolution over time. Such approaches could help constrain the timeline of helium reionization with greater precision. Together, these future research directions hold great promise for improving our understanding of the early Universe and the role of stellar feedback and reionization in shaping the cosmos.

  \begin{appendix}

\appendix
\chapter{Comparison between different methods for computing $\sigma\rm{_{UV}}$}
\label{Appendix-uvlfcompare}

\begin{figure}
\centering
        \includegraphics[width=75mm]{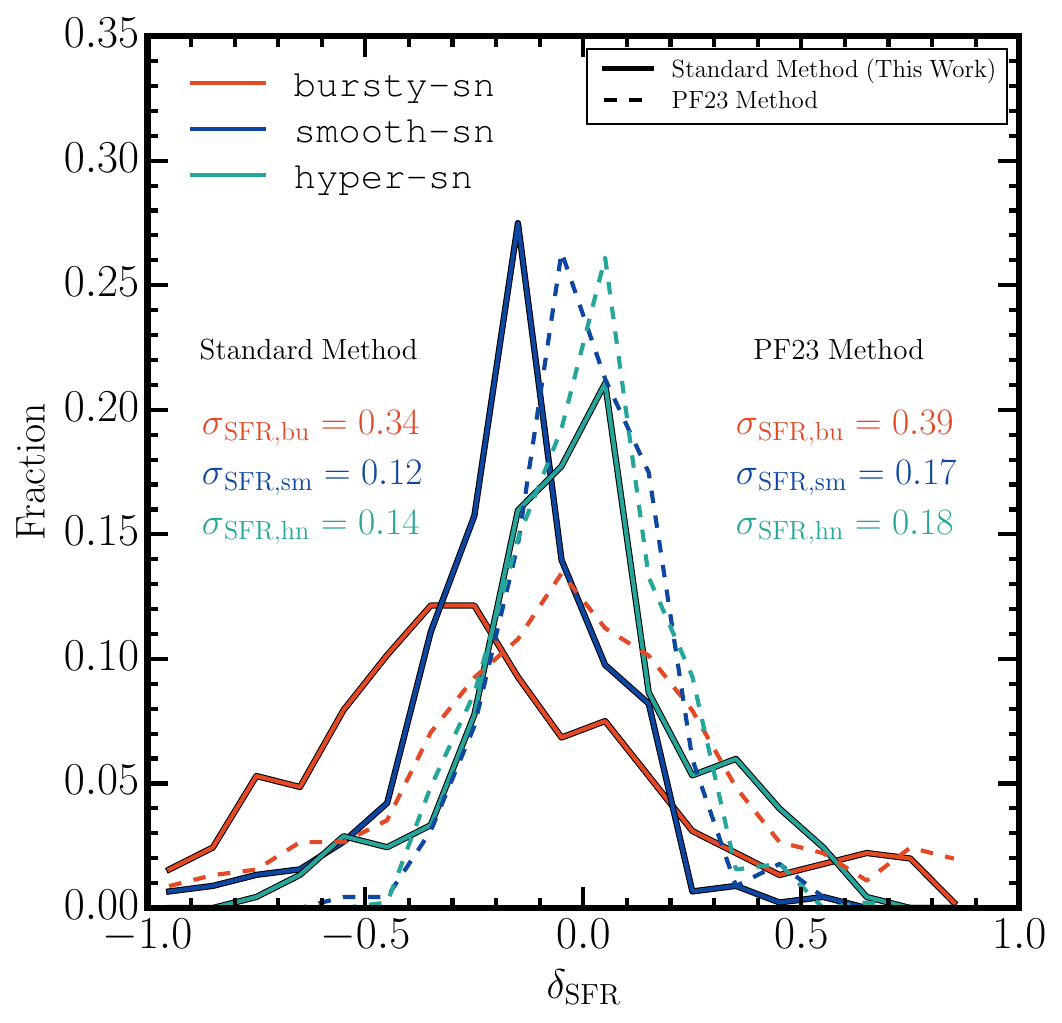}
        \includegraphics[width=75mm]{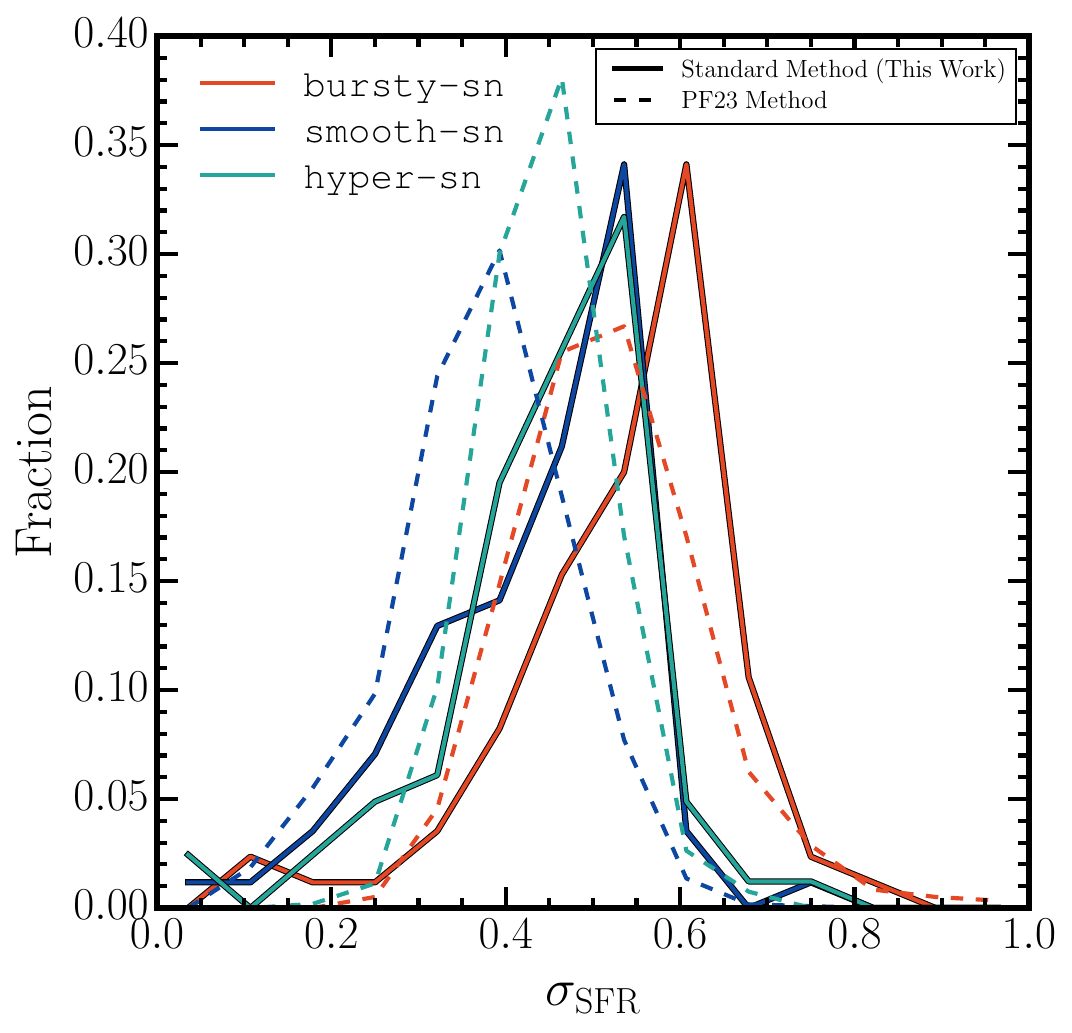}
  \caption{\textit{Left Panel:} Distribution of $\delta_{\rm{SFR}}$ computed from the temporal evolution of the most massive halo at $z=5$ as shown in Figure \ref{fig:sfr_individual}. Colors refer to different SN feedback models. The solid and dashed curves represent the results obtained using the standard method and those from the method proposed by \citet{pallottini_ferrera2023}. Numbers refer to the corresponding standard deviation. \textit{Right Panel:} Distribution of $\sigma_{\rm{SFR}}$ computed over the entire sample of halos present at $z=5$.}
  \label{fig:deltasfr_compare_sigmasfr_compare}
\end{figure}

In this section, I discuss how my major findings, discussed in Chapter \ref{chap:chapter2} are affected by adopting the method proposed by \citet{pallottini_ferrera2023} (hereafter referred to as \texttt{PF23}) in the calculation of the variability. I begin by analyzing the temporal evolution of the SFR for the most massive halo at $z=5$  illustrated in Figure \ref{fig:sfr_individual}. Each curve is fitted with a third-order polynomial on a log-linear scale (i.e. $\text{SFR}_{\text{fit}}$), following the technique used by \texttt{PF23}. The stochastic variability of the SFR is then quantified by computing the residuals, defined as $\delta_{\rm{SFR}} = \log_{10}(\text{SFR}/\text{SFR}_{\text{fit}})$, from which I derive the standard deviation, $\rm{\mathit{\sigma}_{SFR}}$. The results of this approach are compared with the standard method presented in Section \ref{sfr_ind} and summarized in the left panel of Figure \ref{fig:deltasfr_compare_sigmasfr_compare}. I observe that the $\delta_{\rm{SFR}}$ distribution using the \texttt{PF23} method is slightly shifted toward higher values across all models compared to the standard method. This shift is reflected in the slightly higher values of $\rm{\mathit{\sigma}_{SFR}}$, with 0.39 for \texttt{bursty-sn}, 0.17 for \texttt{smooth-sn}, and 0.19 for \texttt{hyper-sn}, each exceeding the standard method's results by approximately $\Delta(\mathit{\sigma}_{\rm{SFR}}) \sim 0.04-0.05$. To examine the behavior of the entire halo population at $z=5$, right panel of Figure \ref{fig:deltasfr_compare_sigmasfr_compare} also presents the distribution of $\sigma_{\mathrm{SFR}}$ calculated using both methods across all SN feedback models. Interestingly, I observe that the \texttt{PF23} method yields a distribution of $\sigma_{\mathrm{SFR}}$ values that are slightly lower across all models when compared to the standard method. This finding contrasts with the trend seen in the left panel of Figure \ref{fig:deltasfr_compare_sigmasfr_compare}. Specifically, I note that the peak of the $\sigma_{\mathrm{SFR}}$ distribution shifts by approximately $0.15 - 0.2$.

The key difference lies in the \texttt{PF23} method’s focus on the temporal evolution of individual halos up to the redshift of interest, which emphasizes fluctuations around the median evaluated for the same halo, differently from other methods which calculate the median from the full sample of halos.  This distinction leads to a fundamentally different interpretation of the underlying variability being measured, with the \texttt{PF23} method offering a more detailed view of individual halo evolution over time.

\chapter{Fitting and Extrapolation of UVLF}
\label{Appendix-uvlffitting}

To investigate whether the \texttt{SPICE} UVLF, incorporating scatter, can effectively reproduce the observed luminosity functions (discussed in Chapter \ref{chap:chapter2}), I fit each UVLF with a \citet{Schechter1976} function\footnote{The functional form is given by : \\ 
$\rm{\Phi(\mathit{M}_{UV}) = 0.4 \ln(10) \, \phi^* \, \left[ 10^{0.4(\mathit{M}_{UV}^* - \mathit{M}_{UV})(\alpha + 1)} \right] 
\exp\left[ -10^{0.4(\mathit{M}_{UV}^* - \mathit{M}_{UV})} \right]}$, where $\phi^*$ is the normalization factor, $\rm{\mathit{M}_{UV}^*}$ is the characteristic magnitude and $\alpha$ is the faint-end slope.}
. In Table \ref{Table-1}, I show the fitted parameters for $z=5$, 6, 7, 9, 10, 12.
I note that for $\rm{\mathit{M}_{UV}\lesssim -19}$, \texttt{SPICE} lacks a sufficient statistical sample of halos for a robust fitting. As a result, the Schechter function fails to provide an optimal fit, leading to progressively higher deviations from the SPICE LFs towards lower $\rm{\mathit{M}_{UV}}$ values relative to the fitted characteristic magnitude ($\rm{\mathit{M}_{UV}^{*}}$). Consequently, the reduced $\rm{\chi^{2}}$ values are notably low in some cases where the deviations are larger, particularly at higher redshifts.

\begin{figure}
\centering
        \includegraphics[width=100mm]{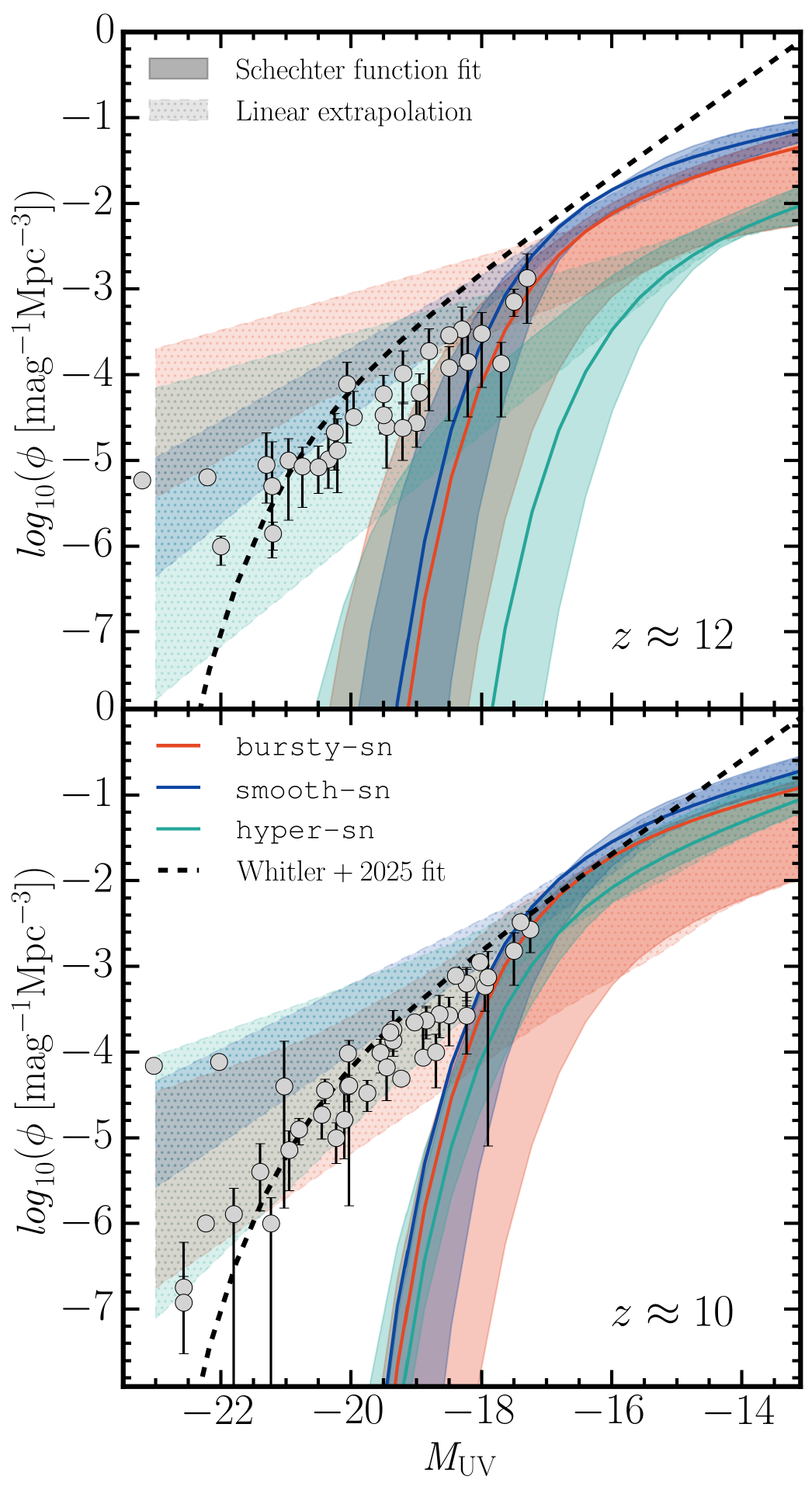}
  \caption{UVLF at $z = 12$ (\textit{top panel}) and 10 (\textit{bottom}) obtained after fitting with a \citet{Schechter1976} function (solid lines). Shaded regions represent the 2-$\sigma$ scatter calculated from the evolution of the UVLF over time intervals of 10, 30, and 50 Myr around the reference redshifts. The hatched shaded regions indicate a direct extrapolations of the UVLFs. Colors refer to different SN feedback models. The black dashed curve indicates the result from \citet{Whitler2025} fitted to their observational F115W dropout sample assuming a \citet{Schechter1976} function, while a compilation of observations from \texttt{HST} and \texttt{JWST} \citep{bouwens2015,harikane2022,naidu2022,adams2023,harikane2023,bouwens2023a,bouwens2023b,leung2023,donnan2023a,donnan2023b,perez2023,casey2024,robertson2024,mcleod2024,Whitler2025} is shown as gray data points.}
  \label{fig:UVLF_fit}
\end{figure}

As a reference, in Figure \ref{fig:UVLF_fit} I plot the fitted curves at $z=10$ and 12 as solid lines, while the shaded regions represent the 2-$\sigma$ scatter obtained by performing the fits for the curves corresponding to 10, 30, and 50 Myr before and after the reference redshift.
As mentioned above, the fit at the brightest end of the LF is not robust, and indeed all models fall below observational data, although it has previously been noticed \citep{Bowler2020,Donnan2024,Whitler2025} that high-$z$ galaxies with $\rm{\mathit{M}_{UV}\lesssim -22}$ are in excess of the exponential decline of the Schechter function. 
At both redshifts, the \texttt{bursty-sn} and \texttt{smooth-sn} models perform similarly in reproducing observations within the range $-19 \lesssim \rm{\mathit{M}_{UV}} \lesssim -17$, with the latter exhibiting slightly smaller scatter. Both models align more closely with the data toward the brighter end, up to $\rm{\mathit{M}_{UV}} \approx -20$, at $z \approx 12$. Beyond this point, they systematically underpredict the data, with deviations becoming progressively larger at lower $\rm{\mathit{M}_{UV}}$. The \texttt{hyper-sn} model, while displaying a greater scatter in comparison to \texttt{smooth-sn} at $z \approx 12$, does not achieve better agreement with the data, instead showing a significant offset from observations. At $z \approx 10$, however, its predictions remain comparable to those of other models, with the lowest scatter observed.
 
To provide additional information, I also include the 2-$\sigma$ scatter obtained by linearly extrapolating the \texttt{SPICE} UVLFs corresponding to 10,
30, and 50 Myr before and after the reference redshift beyond the $\rm{\mathit{M}_{UV}}$ range available in \texttt{SPICE}, without applying any fitting procedure. These curves align more closely with observations across the entire $\mathit{M}_{\rm UV}$ range, with \texttt{hyper-sn} offering the best overall agreement at both redshifts. The \texttt{bursty-sn} model tends to overpredict the counts at $z \approx 12$, but closely matches most observational constraints at $z \approx 10$. Conversely, \texttt{smooth-sn} performs the worst overall, but shows better agreement at $z \approx 12$ than at $z \approx 10$.

\setlength{\tabcolsep}{1.8pt}
\begin{table*}
    \begin{tabular}{c|ccc|ccc|ccc|cccc}
      \hline
      $z$ & \multicolumn{3}{c|}{$\phi^{*}(\times 10^{-3})$} & \multicolumn{3}{c|}{$\rm{\mathit{M}_{UV}^*}$} & \multicolumn{3}{c|}{$\alpha$} & \multicolumn{3}{c}{$\rm{\chi_{\nu}^2}$} \\
      & Bursty & Smooth & Hyper & Bursty & Smooth & Hyper & Bursty & Smooth & Hyper &  Bursty & Smooth & Hyper & \\
      \hline
      \hline
      5 & $75.95$ & $256.11$ & $23.36$ & $-17.32$ & $-16.84$ & $-17.81$ & $-1.53$ & -1.40 & -1.67 & $0.40 $& 2.68 & 0.39 \\
      7 & $44.19$ & $155.57$ & $63.89$ & $-17.12$ & $-16.62$ & $-16.93$ & $-1.61$ & -1.45 & -1.59 & $0.45 $& 1.92 & 0.78 \\
      9 & $0.18$ & $136.72$ & $34.28$ & $-19.96$ & $-15.93$ & $-16.53$ & $-1.89$ & -1.39 & -1.65 & $0.06 $& 0.33 & 0.12 \\
      10 & $30.84$ & $67.20$ & $12.23$ & $-16.51$ & $-16.14$ & $-16.51$ & $-1.48$ & -1.48 & -1.67 & $0.21 $& 0.15 & 0.06 \\
      12 & $14.18$ & $21.13$ & $4.06$ & $-16.38$ & $-16.66$ & $-15.22$ & $-1.43$ & -1.42 & -1.55 & $0.02 $& 0.02 & 0.01 \\
      \hline
    \end{tabular}
    \caption{Fitted parameters of the UVLF functional forms at various redshifts, modeled using the \citet{Schechter1976} function. The parameters include the overall normalization ($\phi^{*}$), the characteristic magnitude ($\rm{\mathit{M}_{UV}^*}$), and the faint-end slope ($\alpha$). The reduced chi-square ($\chi^2_{\nu}$) values assess the goodness of fit.}
    \label{Table-1}
\end{table*}

\end{appendix}

\chapter{Impact of hydrodynamical feedback on $\rm{\mathit{T}_{0}}$}
\label{thermal_history_with_Tcuts}

One of the key limitations of studying the EoR using post-processed cosmological hydrodynamical simulations with RT codes is the difficulty in accurately accounting for hydrodynamical effects, particularly those influencing the temperature of simulation cells. Disentangling or correctly interpreting cells with IGM temperatures exceeding $10^{5}$ K is especially challenging. This complexity can introduce uncertainties when analyzing IGM properties, such as those of the Ly$\alpha$ forest, which are highly sensitive to temperature. At lower redshifts near the end of reionization, these effects can artificially boost the mean IGM temperature. 

\begin{figure}   
\centering
    \includegraphics[width=100mm]{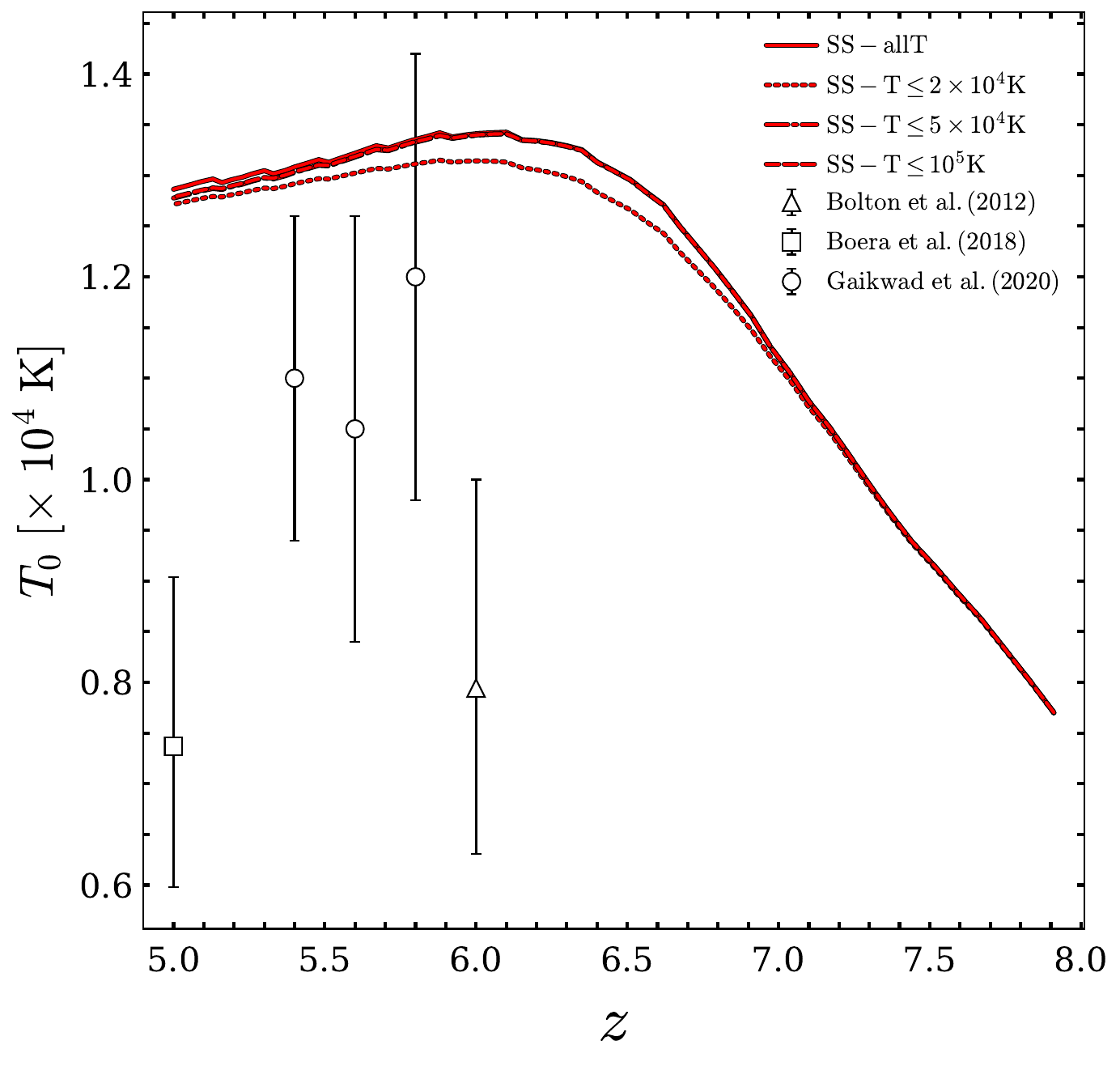}
    \caption{Redshift evolution of the IGM temperature at mean density ($\mathit{T}\rm{_{0}}$) computed for different temperature thresolds as discussed in the main text. The symbols denote a compilation of observational constraints from the literature \citep{Bolton2012,boera2019,Gaikwad2020}.}
   \label{fig:T0_w_cuts}
\end{figure}

To explore this, we systematically assess the impact of excluding high temperature cells affected by hydrodynamical feedback on to IGM thermal history. Specifically, we compute the temperature at mean density, $\rm{\mathit{T}_{0}}$, for the \texttt{SS} model by applying a temperature threshold (i.e. 20,000 K, 50,000 K and $10^{5}$ K) and including only those cells below the specified limit. We then compare the results to observational constraints from the literature \citep{Bolton2012,boera2019,Gaikwad2020}. This has been shown in Figure \ref{fig:T0_w_cuts}. Our findings show that all the $\rm{\mathit{T}_{0}}$ curves follow a broadly similar trend, with some notable differences. When applying a temperature cutoff at 20,000 K, the resulting $\rm{\mathit{T}_{0}}$ is approximately 500 K lower than the fiducial value obtained by including all cells. For higher threshold values, the deviations from the fiducial curve are minimal. At the lowest redshift considered, $z = 5$, a small difference remains; however, the evolution of $\rm{\mathit{T}_{0}}$ below $z = 6$ appears largely saturated relative to the observational data. This highlights potential inconsistencies between the observations and the simulated results.

\chapter{Convergence tests}
\label{appendix:convergence}

In this section, I describe a series of convergence tests I performed for the simulations presented in Chapter \ref{chap:chapter4}. Starting from my fiducial run, I systematically vary individual numerical parameters and assess their impact on the simulated IGM properties. I refer the readers to Table \ref{tab:table-sim} for a detailed overview of all simulation parameters.

\subsection{Convergence with photon sampling}
\label{appendix:convergence_Ngamma}

\begin{figure}
\centering
\includegraphics[width=100mm]{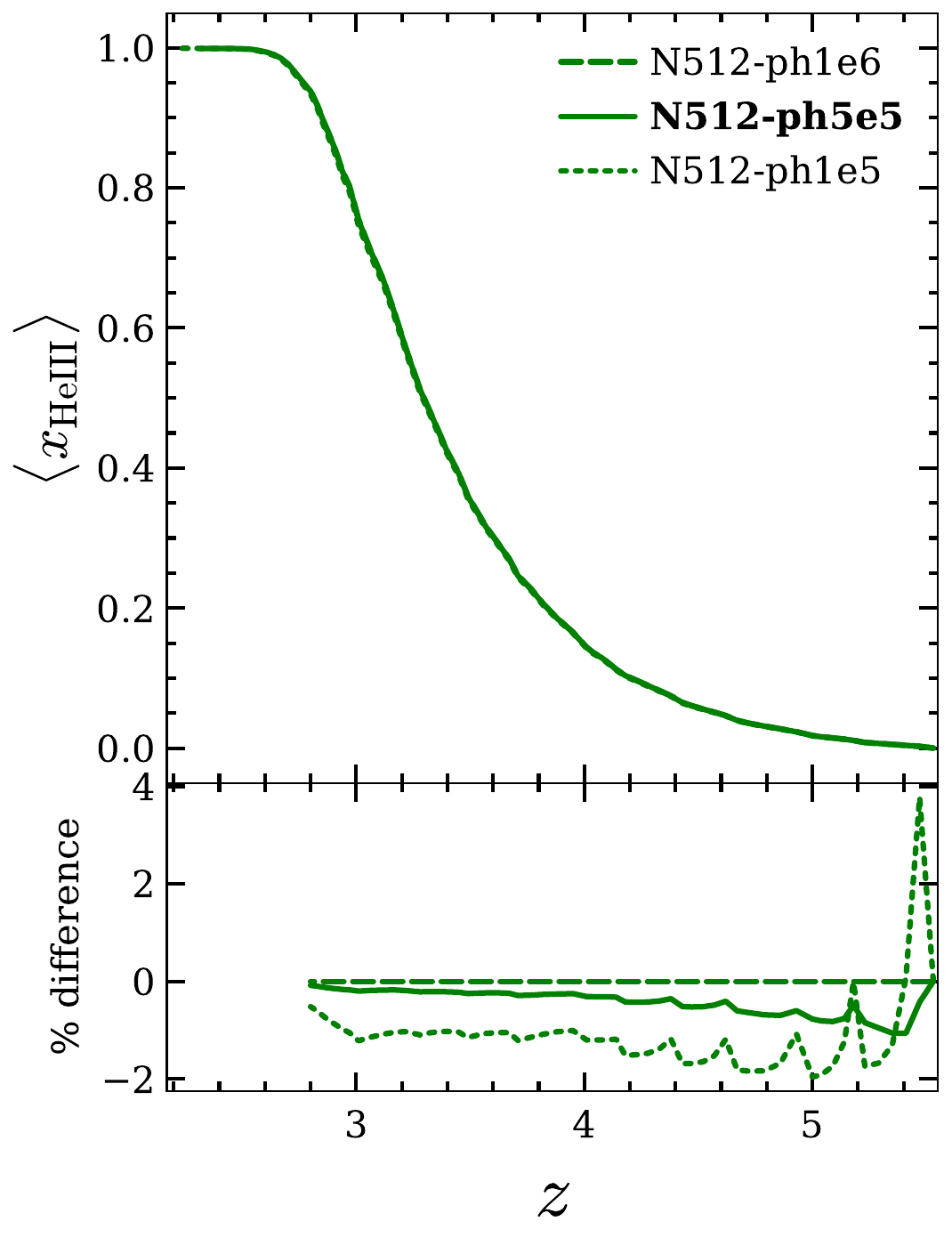}
\caption{{\it Top panel}: evolution of the volume-averaged He~{\sc iii} fraction in simulations with  $N_{\gamma}$ = $10^6$ (dashed curve), $5 \times 10^5$ (solid) and $10^5$ (dotted). {\it Bottom panel}: relative differences with respect to the highest resolution simulation \texttt{N512-ph1e6}. The fiducial simulation is marked in bold font.}
\label{fig:photon_pckt_convergence}
\end{figure}

I start by investigating the numerical convergence with respect to the number of photon packets emitted by each source at each timestep of the simulation ($N_{\gamma}$). I explore $N_{\gamma}$ = $10^5$, $5 \times 10^5$ and $10^6$, corresponding to the runs labeled as \texttt{N512-ph1e5}, \texttt{N512-ph5e5} and \texttt{N512-ph1e6}, respectively. I present the  evolution of volume-averaged He~{\sc iii} fraction in these three models in the top panel of figure \ref{fig:photon_pckt_convergence}. My reference simulation \texttt{N512-ph5e5} (solid line) shows an excellent convergence (within $2\%$) with the highest resolution simulation (\texttt{N512-ph1e6}, dashed line) in the entire simulated redshift range, with a relative difference (bottom panel) below 1\% towards the end of reionization. The lowest resolution simulation \texttt{N512-ph1e5} (dotted line) is also converged within 3\% throughout the entire reionization history. 

\begin{figure}
\centering
\includegraphics[width=100mm]{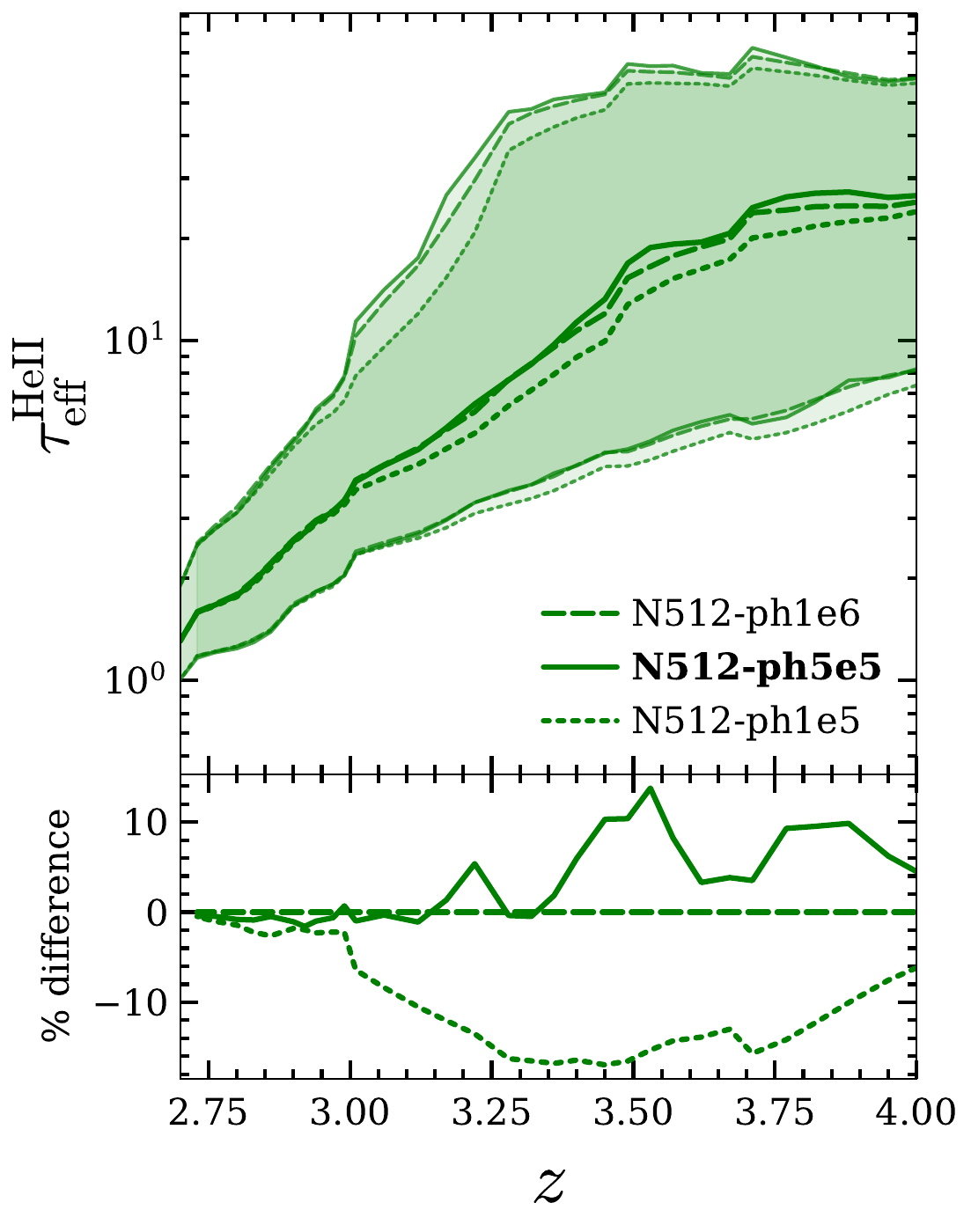}
\caption{{\it Top panel}: evolution of the He~{\sc ii} optical depth in simulations with  $N_{\gamma}$ = $10^6$ (dashed curve), $5 \times 10^5$ (solid) and $10^5$ (dotted). The shaded regions denote the 68$\%$ confidence intervals. {\it Bottom panel}: relative differences with respect to the highest resolution simulation \texttt{N512-ph1e6}. The fiducial simulation is marked in bold font.}
\label{fig:photon_pckt_convergence_tau}
\end{figure}

The convergence in global quantities like the reionization history is  however not indicative of convergence in local observed quantities. For this reason, in figure \ref{fig:photon_pckt_convergence_tau} I show the redshift evolution of the He~{\sc ii} effective optical depth for the three values of $N_{\gamma}$ employed. Also in this case \texttt{N512-ph5e5} and \texttt{N512-ph1e6} exhibit an excellent convergence (within 2\%) for \textit{z} $\lesssim$ 3.1. I note that this holds true not only for the median effective optical depth, but also for the entire distribution, as shown by the overlapping shaded regions (corresponding to the central 68\% of the data) in the figure. Unlike the reionization history case, the effective optical depth in \texttt{N512-ph1e5} is significantly different than in the runs with a larger $N_{\gamma}$, except for the tail end of reionization, when the entire volume is fully ionize and therefore the role of photon sampling is significantly reduced.

This analysis demonstrates that \texttt{N512-ph5e5} achieves a very good convergence in terms of photon sampling.

\subsection{Convergence with grid dimension}
\label{appendix:convergence_Ngrid}
\begin{figure} 
\centering
    \includegraphics[width=100mm]{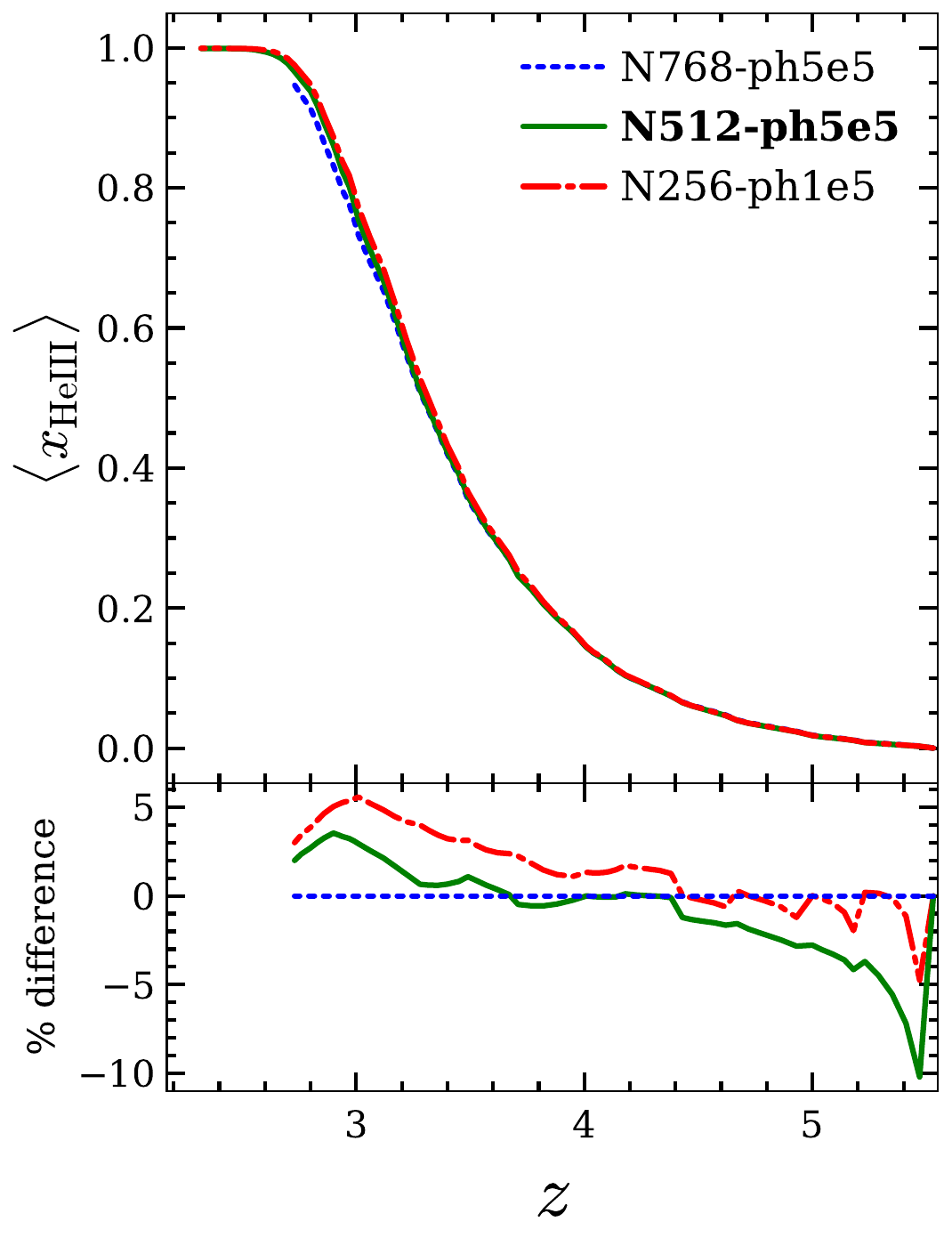}
    \caption{{\it Top panel}: evolution of the volume averaged He~{\sc iii} fraction in simulations with $N_{\rm{grid}}=768^{3}$ (blue dotted curve), $512^{3}$ (green solid) and $256^{3}$ (red dash-dotted). {\it Bottom panel}: corresponding 
    relative difference  with respect to \texttt{N768-ph5e5}. The fiducial simulation is marked in bold font.}
    \label{fig:Ngrid_convergence}
\end{figure}

In order to test the convergence of my fiducial simulation with respect to the physical resolution of the baryonic component, I have run a set of simulations systematically varying the number of grid points used to discretize the simulated volume, namely $N_{\rm{grid}}$= $256^{3}$, $512^{3}$, $768^{3}$. Figure \ref{fig:Ngrid_convergence} shows the volume-averaged He~{\sc iii} fraction in these simulations (top panel) and their differences relative to \texttt{N768-ph5e5} (bottom panel). Note that these runs employ a value of $N_{\gamma}$ that ensures convergence in the radiation sampling (see the previous section). 
The most remarkable feature is that $N_{\rm{grid}}$ guides the \textit{speed} of reionization, with lower resolution runs starting slower (i.e. with negative values in the bottom panel) but proceeding faster and eventually completing reionization earlier. This is a consequence of the fact that higher resolution simulations better resolve density contrast. The low-density channels enable faster photons escape in the early phases of reionization, but high-density regions are more resilient to reionization, slowing down this process. Nevertheless, the relative difference between my fiducial model and the higher-resolution one remains below $\approx$3\% at all redshift relevant for helium reionization, showing an excellent convergence. 

\begin{figure} 
\centering
    \includegraphics[width=100mm]{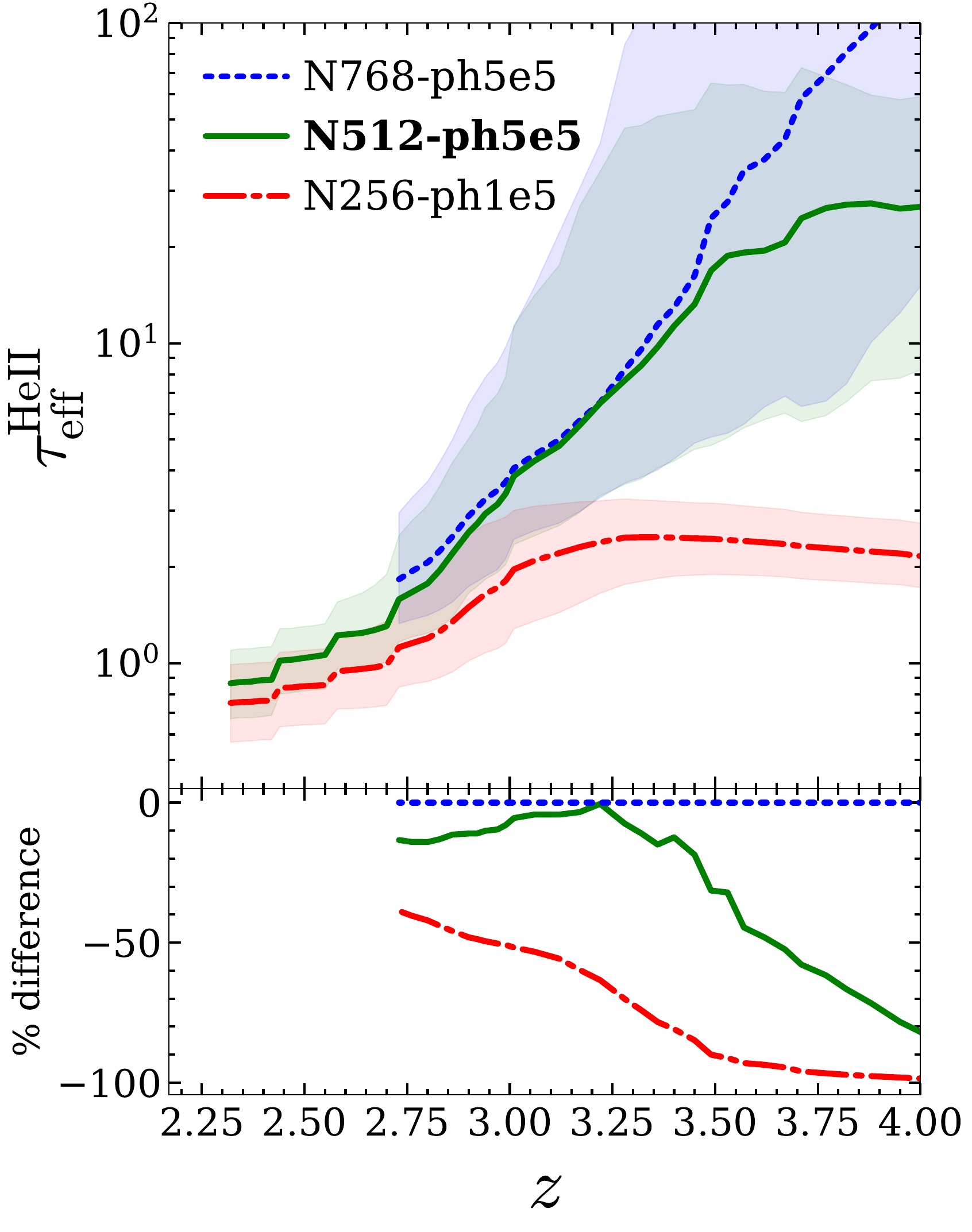}
    \caption{{\it Top panel}: evolution of the He~{\sc ii} optical depth in simulations with $N_{\rm{grid}}=768^{3}$ (blue dotted curve), $512^{3}$ (green solid) and $256^{3}$ (red dash-dotted). The shaded regions denote the 68$\%$ confidence intervals. {\it Bottom panel}: relative differences with respect to \texttt{N768-ph5e5}. The fiducial simulation is marked in bold font.}
    \label{fig:Ngrid_convergence_tau}
\end{figure}
In figure \ref{fig:Ngrid_convergence_tau} I show the evolution of the He~{\sc ii} effective optical depth in these different models (top panel, with solid lines indicating the median value and shaded regions marking the central 68\% of the data) and the relative difference of their medians (bottom panel). For this observable, my fiducial simulation is converged within 10\% at $z \lesssim 3.3$, while at earlier times it significantly under-predicts $\tau_\mathrm{eff}^\mathrm{HeII}$. Remarkably, however, the low-$\tau_\mathrm{eff}^\mathrm{HeII}$ part of the distribution is in very good agreement all the way to $z\lesssim 3.8$, since in the second half of reionization the increased gas resolution mostly affects the high-density (and therefore high-$\tau_\mathrm{eff}^\mathrm{HeII}$) regions, as described above. Since at $z\gtrsim3$ observations are only sensitive to the low-$\tau_\mathrm{eff}^\mathrm{HeII}$ part of the intrinsic distribution (see figure \ref{fig:taueff} and relative discussion), I deem this an acceptable convergence level. Finally, \texttt{N256-ph1e5} displays very poor convergence throughout the entire simulation evolution.

\subsection{Convergence with periodic boundary condition}
\label{appendix:convergence_PBC}

\begin{figure} 
\centering
    \includegraphics[width=100mm]{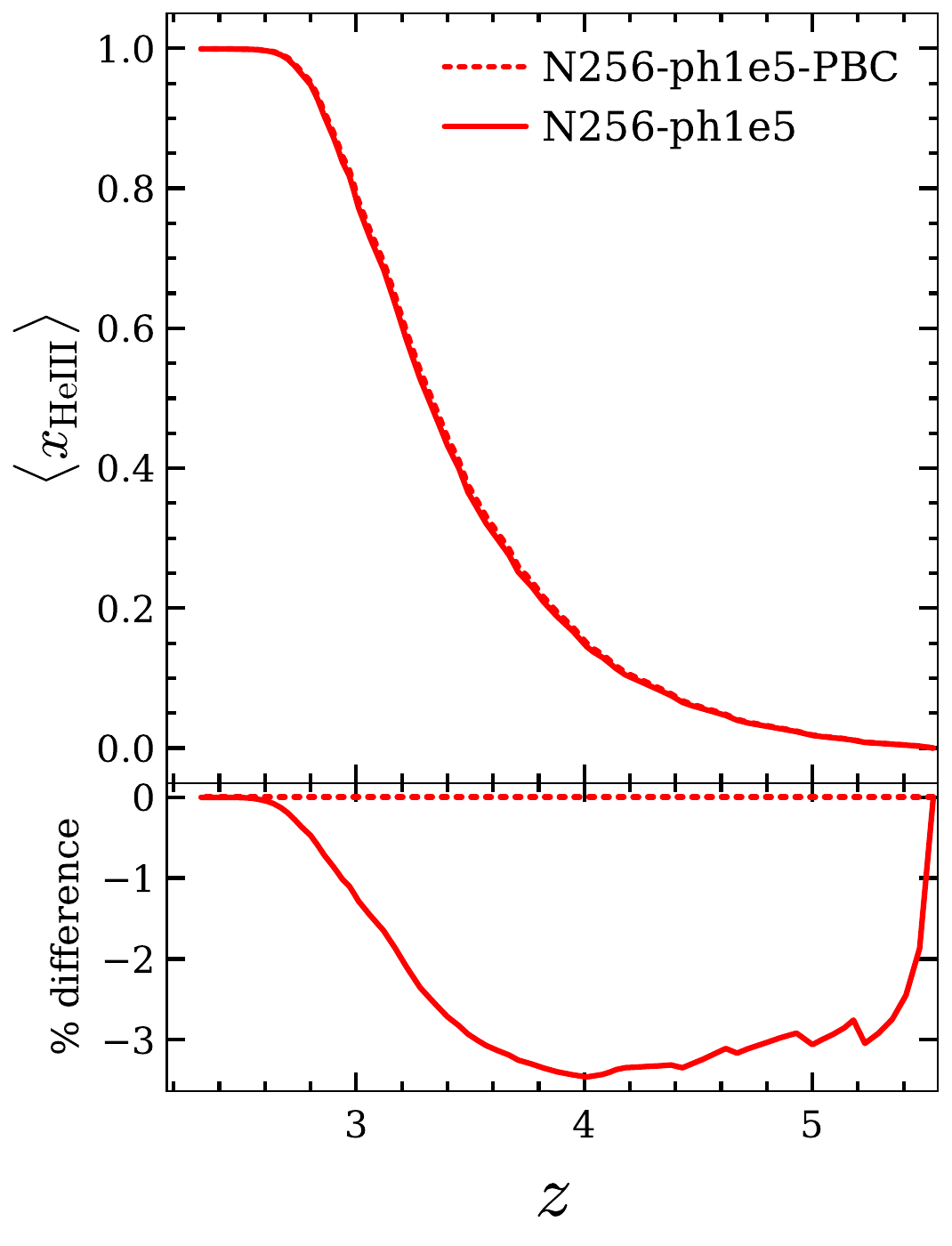}
    \caption{{\it Top panel}: evolution of the volume averaged He~{\sc iii} fraction with (dotted line) and without (solid) periodic boundary condition for simulations run with $N_{\rm{grid}}$= $256^{3}$. {\it Bottom panel}: relative difference with respect to \texttt{N256-ph1e5-PBC}. }
    \label{fig:PBC_convergence}
\end{figure}

\begin{figure} 
\centering
    \includegraphics[width=100mm]{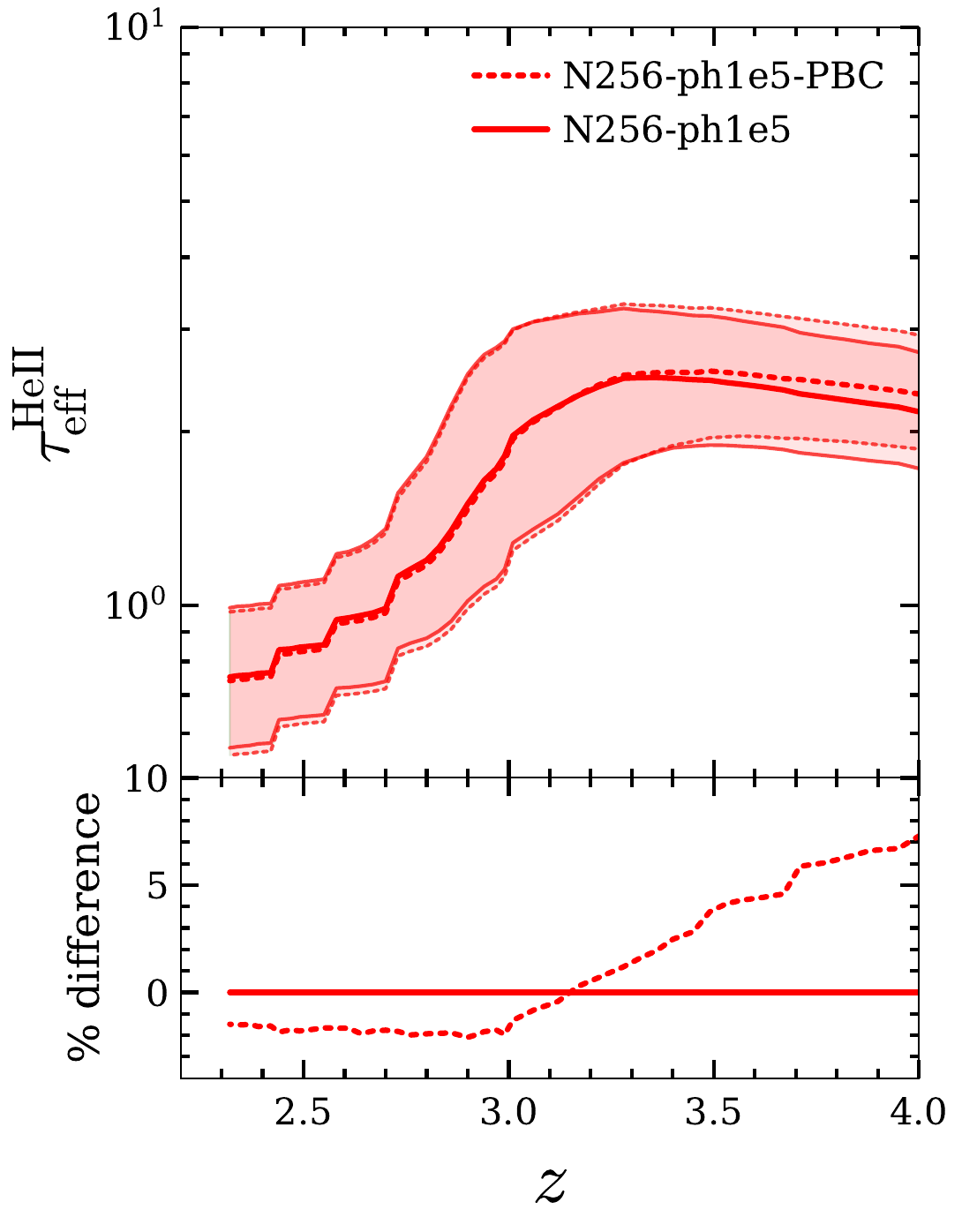}
    \caption{{\it Top panel}: evolution of the He~{\sc ii} optical depth with (dotted line) and without (solid) periodic boundary condition for simulations run with $N_{\rm{grid}}$= $256^{3}$. The shaded regions denote the 68 $\%$ confidence intervals. {\it Bottom panel}: relative differences with respect  to \texttt{N256-ph1e5-PBC}.}
    \label{fig:PBC_convergence_tau}
\end{figure}

Finally, I analyze the impact of periodic boundary conditions (PBC) in my simulations. Since I are purely interested in a comparative analysis, the actual convergence of their physical predictions is of little importance here. Therefore I have employed low-resolution runs that are computationally cheaper. These simulations have been run only until $z_\mathrm{final}=2.73$, as by then perfect convergence has been already established. 
The inclusion of PBC has a small impact on both the reionization history ($\lesssim$3\%) and $\tau_\mathrm{eff}^\mathrm{HeII}$ ($\lesssim$10\%) in the initial stages. Such difference is progressively reduced until perfect ($\lesssim$1\%) convergence is reached at $z\approx3$. 
In figure \ref{fig:PBC_convergence} I show the evolution of the volume averaged He~{\sc iii} fraction (top panel) and their relative difference (bottom panel) for simulations with (\texttt{N256-ph1e5-PBC}, dotted line) and without (\texttt{N256-ph1e5}, solid) PBC. In figure \ref{fig:PBC_convergence_tau} I report the He~{\sc ii} effective optical depth evolution and the relative difference for the same two simulations. Note that I have conducted the same test for \texttt{N512-ph5e5} (i.e. employing the same $N_{\gamma}$ as in my fiducial run) until $z_\mathrm{final} \sim 4$, finding similar results.

\begin{figure}
\centering
    \includegraphics[width=120mm]{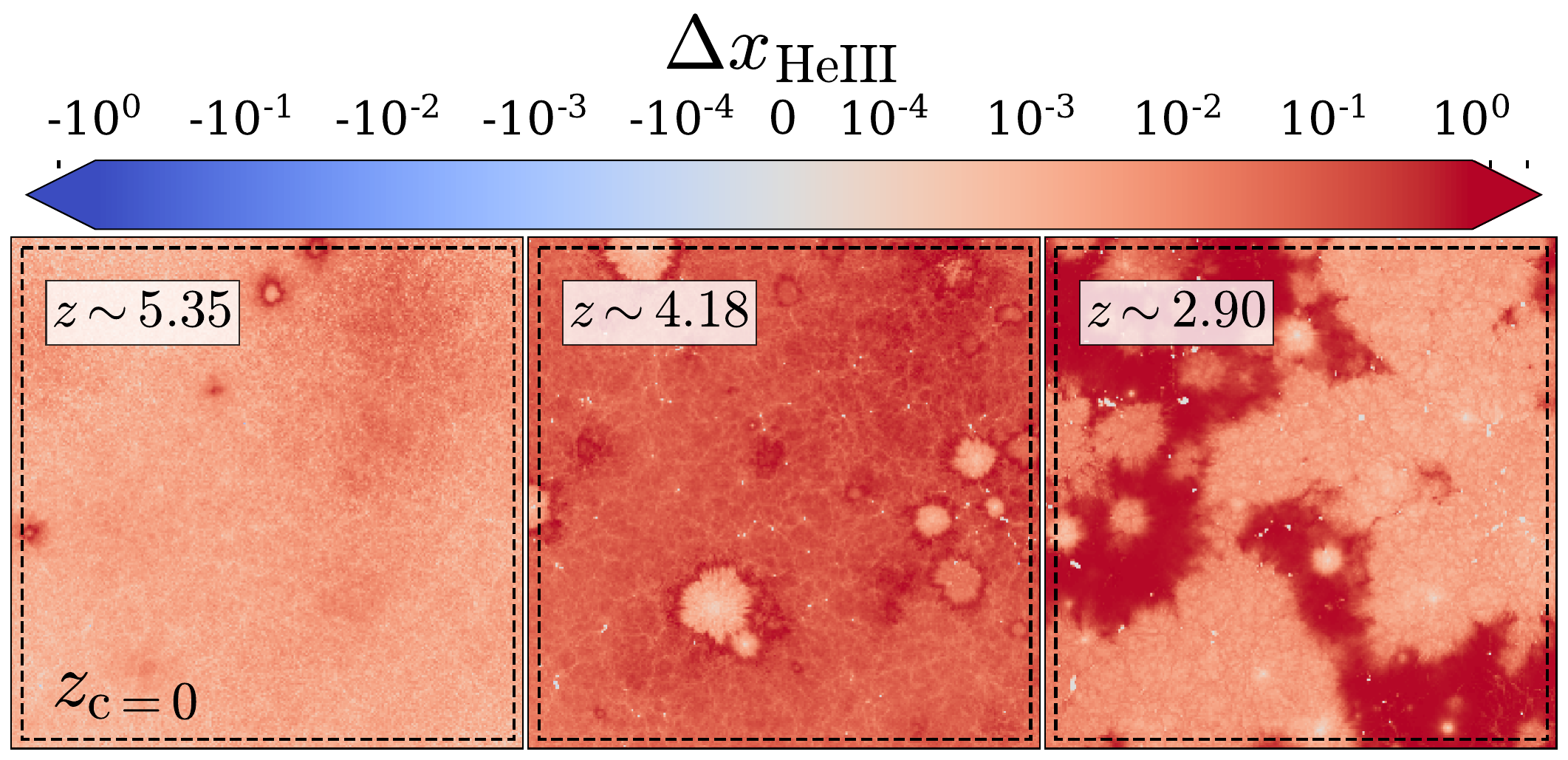}\\
    \includegraphics[width=120mm]{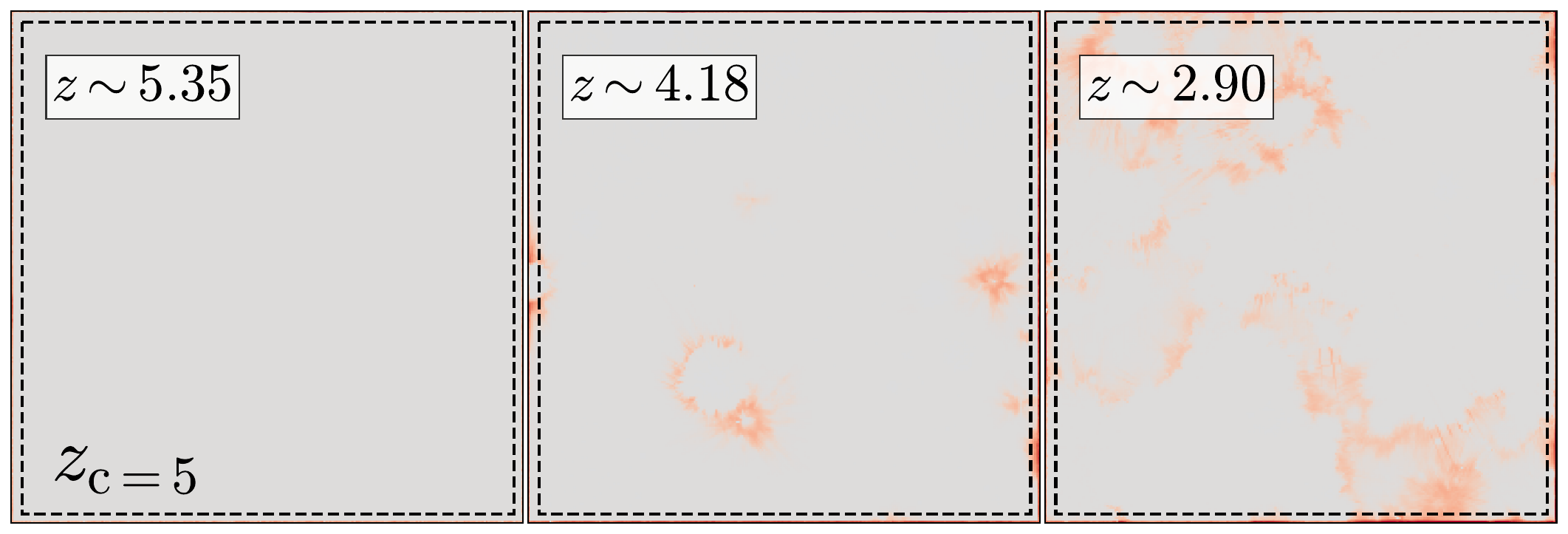}\\
    \includegraphics[width=120mm]{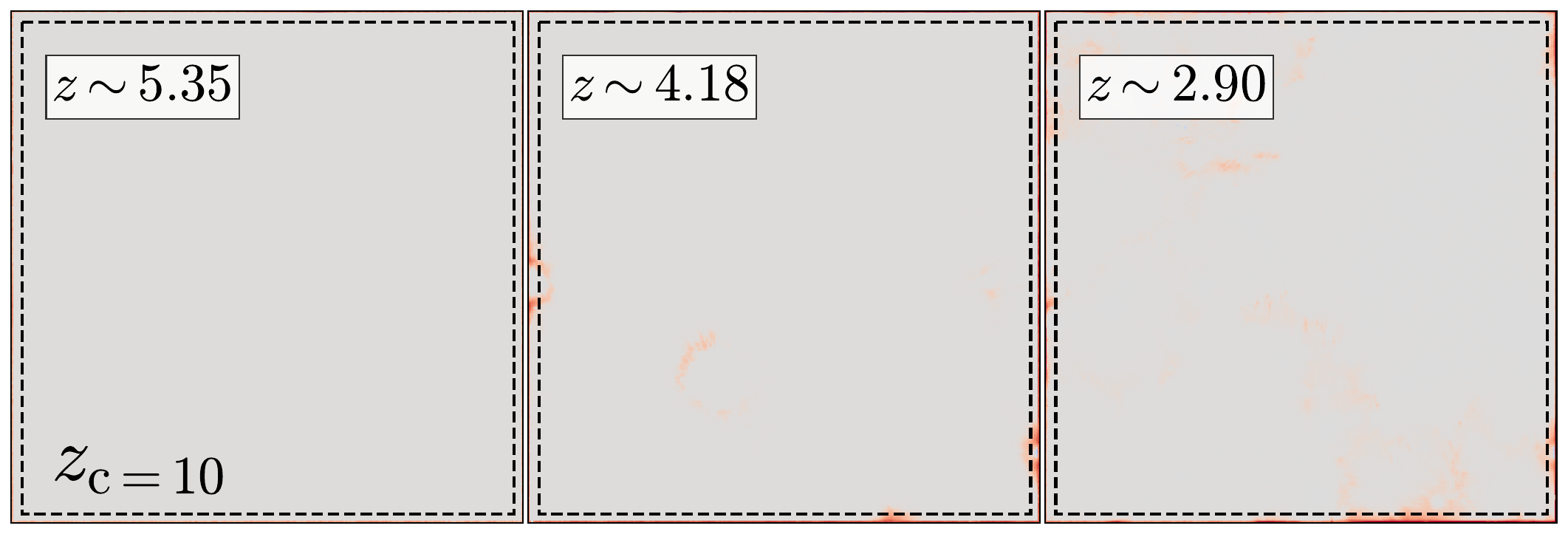}
    
    \caption{Maps showing the difference in He~{\sc iii} fraction between \texttt{N256-ph1e5-PBC} and \texttt{N256-ph1e5} at \textit{z} =5.35, 4.18, and 2.90 (from left to right). The rows refer to slices of thickness 800 $h^{-1}$~ckpc located at various distances along the $z$-direction (marked in grid unit) from the edge of the box. The outside region of the black dashed rectangle shows the parts of the edges of all slices which are finally removed from the analysis.}
    \label{fig:reion_history_pbc}
\end{figure}

As mentioned in section \ref{method}, I have removed from my analysis the 5 layers of cells closest to each edge of the simulation box. The reason is shown in figure \ref{fig:reion_history_pbc}, where I display the difference in He~{\sc iii} fraction between \texttt{N256-ph1e5-PBC} and \texttt{N256-ph1e5} in three slices (spanning the entire simulation box in two dimensions) placed at the edge of the simulation box (top row), 5 cells away (middle row) and 10 cells away (bottom row). Already in the middle panel, the effect of the PBC are negligible. Notice that the edge of each slice will also be removed as it is close to one of the edges of the simulation box, as indicated by black dashed rectangles in the slices shown.

\chapter{Dependence on BH-to-halo relation}
\label{appendix:BH_to_halo}

\begin{figure} 
\centering
    \includegraphics[width=100mm]{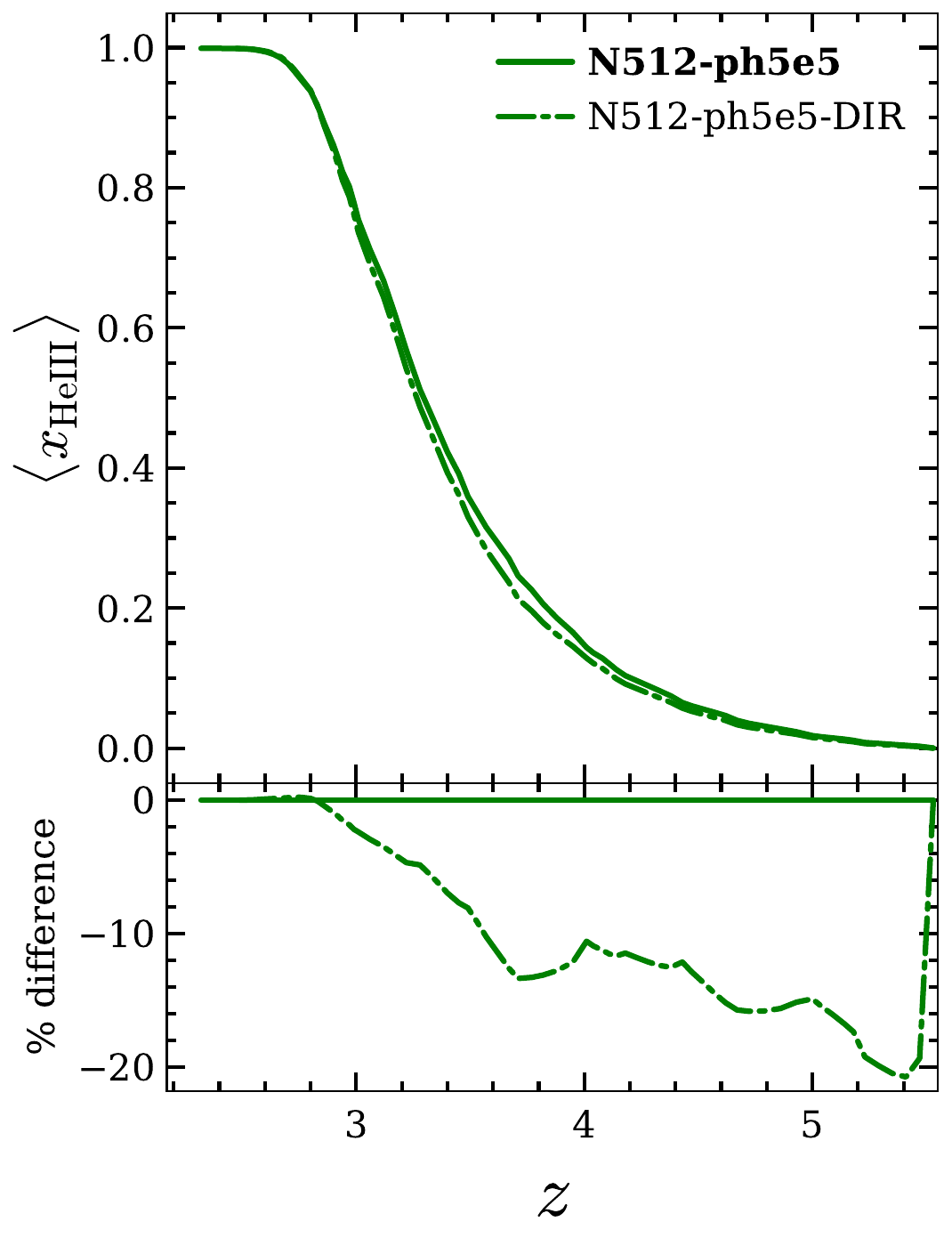}
    \caption{{\it Top panel:} Volume averaged He~{\sc iii} fraction for two different methods to assign quasars to the DM haloes: abundance matching with scatter (\texttt{N512-ph5e5}, solid curve) and  abundance matching without any scatter (\texttt{N512-ph5e5-DIR}, dashed curve). {\it Bottom panel}: percentage difference between the curves in the top panel. The fiducial simulation is marked in bold font.}
    \label{fig:two_approaches}
\end{figure}

\begin{figure}
\centering
    \includegraphics[width=100mm]{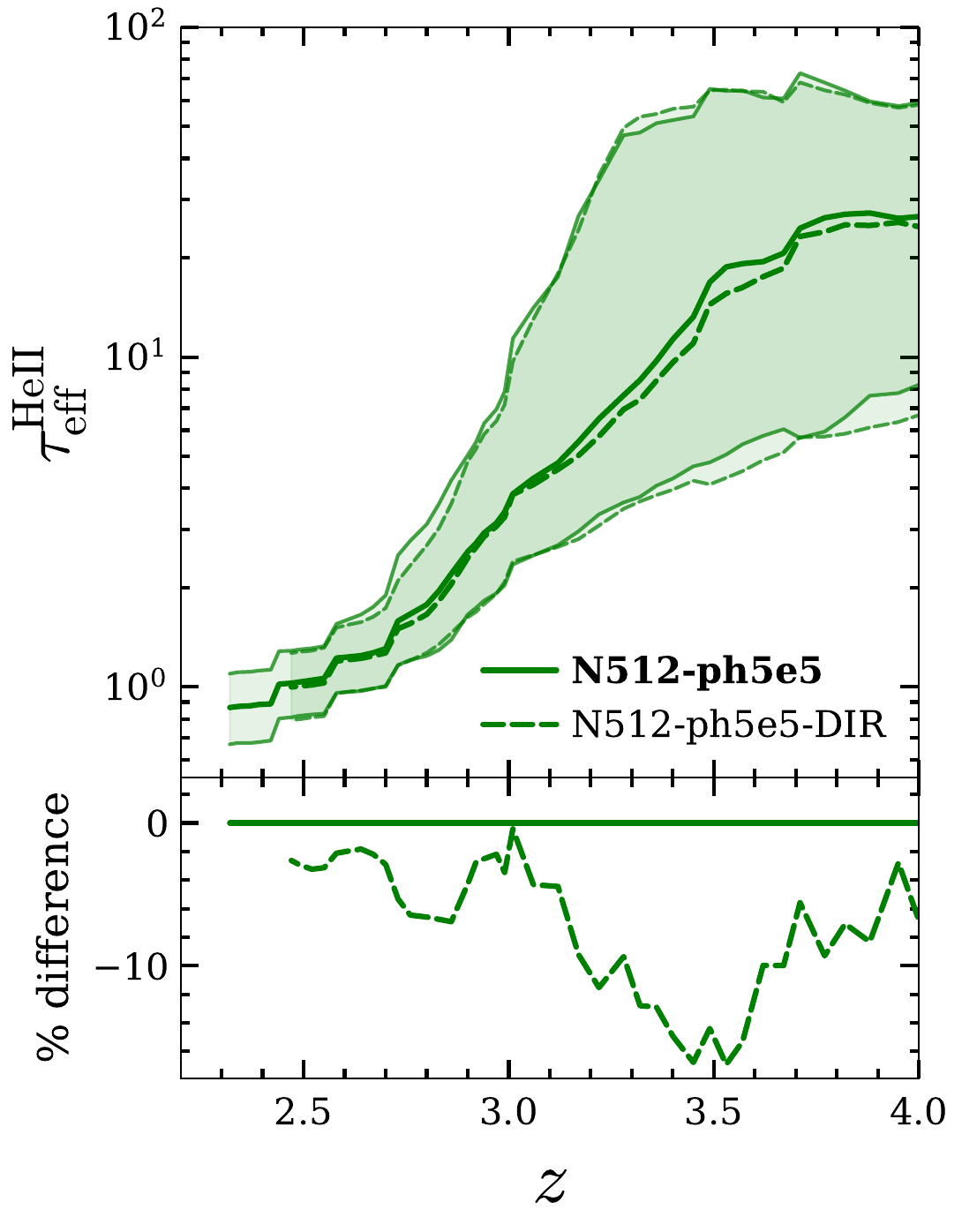}
    \caption{{\it Top Panel}: He~{\sc ii} effective optical depth for two different methods to assign quasars to the DM haloes: abundance matching with scatter (\texttt{N512-ph5e5}, solid curve) and  abundance matching without any scatter (\texttt{N512-ph5e5-DIR}, dashed curve). 68 $\%$ confidence intervals are denoted by the shaded regions. {\it Bottom panel}: percentage difference between the curves in the top panel. The fiducial simulation is marked in bold font.}
    \label{fig:two_approaches_tau}
\end{figure}

In this Appendix I investigate the impact on the results of my two BHs positioning schemes (dubbed `fiducial' and `direct' in Section \ref{tng+crash}). To do so, I run an additional simulation, \texttt{N512-ph5e5-DIR}, that differs from my fiducial run \texttt{N512-ph5e5} just in the fact that QSOs are placed in the simulation using my `direct' method. 
In figure \ref{fig:two_approaches} I compare the evolution of the volume-averaged He~{\sc iii} fraction obtained using these two approaches. 
AThe reionization process progresses slower in the `direct' method, as QSOs are placed in more massive haloes, and therefore --~on average~-- in higher density regions. Consequently, the higher recombination slows down the advancements of reionization fronts, and therefore of the IGM ionization. By $z\approx3$, helium reionization is completed and therefore these differences vanish. The situation is similar for the He~{\sc ii} effective optical depth shown in figure \ref{fig:two_approaches_tau}. The difference always remain within 20\% and decreases with time, until it settles on $\approx 5\%$ at $z\leq 3$. This is in line with the differences in reionization histories. Additionally, at $z\leq 3$ the residual difference is due to the fact that photons emitted by QSOs placed through the `direct' method are more strongly absorbed before reaching the IGM, because of the higher local gas densities, and hence residual neutral fractions.

  \backmatter
  \bibliographystyle{jkthesis}
  \cleardoublepage
  \addcontentsline{toc}{chapter}{Bibliography}
  \bibliography{literatur}
  \markboth{}{}

  \addcontentsline{toc}{chapter}{\protect Acknowledgements}

\chapter*{Acknowledgements}
I am deeply grateful to my supervisor, \textbf{Benedetta Ciardi}, for giving me the opportunity to work at MPA and for guiding me throughout the past four years. Her constant support, and her regular check-ins — always asking, ``\textit{Are you doing good? Is the work stressful?}" — truly helped me feel at ease and stay focused on my work. I would also like to thank my long-time collaborator, \textbf{Enrico Garaldi}, for helping me get started with the project from scratch and for always being there to explain things and remind me why the project is important. Without his support, it would have taken me much longer to make progress. 

A big thank you to my \textbf{high-redshift group members at MPA} — Anshuman, Benedetta (Benny), Aniket, Ivan, Katyayani, Pei, Seok-Jun, and Shubham — for making this journey meaningful. I’m also grateful to all the postdocs (especially Maria for helping with my zusammenfassung), PhD students, senior colleagues, and collaborators for the many scientific discussions that helped me grow.

Living in a new country came with its share of ups and downs. I am very thankful to the MPA secretaries, especially \textbf{Maria, Gabi, Sonja, Cornelia}, and later \textbf{Solvejg, Isabel, Marzia} and \textbf{Lena}, for always making sure things ran smoothly. I would also like to thank \textbf{Sonia} in HR, \textbf{Claudia} in the travel department, \textbf{Annette} at IMPRS, and \textbf{Anna-Serena} at the Dean’s office for helping me with all the administrative work during my PhD. I am also thankful for MPA’s generous travel funding, which allowed me to attend many important conferences, meetings, and workshops. These events were not just great opportunities to share my work, but also to connect with researchers from around the world.

My PhD years in Germany would not have been the same without the \textbf{friends who became my family}. To my \textbf{“best friend” Silvia (Silvi)} — I cannot thank you enough. You were always there for me, in good days and bad, with patience, encouragement, and warmth. You made the difficult moments lighter and the happy moments brighter. I don’t know how I would have survived these years without you. From daily chats to deep talks, from laughter to emotional support, you were truly my pillar here. \textbf{Anshuman}, thank you for being my rock — always ready with a smile, a helping hand, and company for everything from science discussions to spontaneous social plans. \textbf{Hitesh}, thank you for dragging me out of my comfort zone, making me explore the gym, badminton, hikes, table tennis, and perhaps climbing and bouldering in the future — you made life at MPA exciting and full of energy. \textbf{Abinaya}, your warmth and positivity made every dinner, every social gathering, and every little moment brighter. \textbf{Katyayani}, thank you for patiently listening to all my rants and for always being a reliable and understanding friend. \textbf{Nikita} — you were my constant companion for everything else. Thank you for giving me company in our casual plans and in all possible ways. Your presence always made me feel less alone. I am also grateful to my larger circle of friends — \textbf{Joanne, Géza, Alankar, Ritali, Gitanjali, Akash, Jelena, Simran, Bo, Aakash, Sreejita} and many others — for the joy, the Memes, the coffee breaks, the after-lunch crosswords, the board-game nights, the dancing evenings, the Yoga, and the countless moments of laughter. Special thanks to Hitesh and Abinaya, along with Nikita and Gitanjali, for organizing our much-needed drinking nights, and to Aakash for being both a gym motivator and an enthusiastic table tennis partner. Thank you also to \textbf{Satadru Da}, \textbf{Sanchita Di}, and \textbf{little `Titas'} for the mountain plans, the dinners, and the feeling of home away from home.

To \textbf{Eishica}, for our long-distance science talks and yearly meetups that reminded me that distance cannot weaken true friendship.
To \textbf{Soumik}, \textbf{Shankha}, and \textbf{Sarbari}, for the late-night calls and the never-quite-successful (once though) trip plans.
And to my \textbf{`Ninu' (Mimasha)}, for being there every single day during my PhD, listening to my rants, sharing life updates, and reminding me I was never alone — you have been my anchor throughout this journey. And finally, a heartfelt thank you to \textbf{Dipa}, \textbf{Sudo}, \textbf{Debu}, and the entire \textbf{Tamasha Theatre Group}, for giving me a space to continue my love for acting and for being my comfort zone outside of research.

The social life at MPA, MPE, and ESO added so much to my experience — from coffee chats and parties to barbecues and talks, the sense of community helped me grow as both a researcher and a person. I’m also thankful to the city of Munich — walking around, discovering cozy cafés, and enjoying the calm really helped me unwind.

And lastly, to \textbf{my parents, my brother, my sister-in-law} and \textbf{my extended family} — thank you for your endless support and encouragement throughout this journey. I couldn’t have done it without you.

\end{document}